\documentclass[pdftex,twocolumn,epjc3_preprint,runningheads]{svjour3}

\usepackage[T1]{fontenc}
\usepackage{lmodern}
\usepackage{calc}
\usepackage{graphicx}
\usepackage{booktabs}
\usepackage{textcomp}
\usepackage{xspace}
\usepackage{relsize}
\usepackage{amssymb}
\usepackage{amsmath}
\usepackage{listings}
\usepackage{microtype}
\usepackage{multirow}
\usepackage{tabularx}
\usepackage{array}
\usepackage{placeins}
\usepackage{cuted}
\usepackage{soul} % only for \st; delete if this causes you problems.
\usepackage{fixltx2e}
\usepackage{slashed}
\usepackage{bm}
\usepackage[numbers,sort&compress]{natbib}
\usepackage[labelfont=bf,font=small]{caption}
\usepackage[skip=-2pt]{subcaption}
\usepackage[colorlinks,citecolor=blue,urlcolor=blue,linkcolor=blue,breaklinks=breakall]{hyperref}
\usepackage{breakurl}
\usepackage[dvipsnames]{xcolor}
\usepackage[clockwise,figuresright]{rotating}
\usepackage{siunitx}
\usepackage{tikz}
\usepackage[normalem]{ulem}
\usepackage[utf8]{inputenc}

\usepackage{etoolbox}
\AfterEndEnvironment{strip}{\leavevmode}

\allowdisplaybreaks

\newcolumntype{L}{>{\raggedright\let\newline\\\arraybackslash\hspace{0pt}}X}
\newcolumntype{R}{>{\raggedleft\let\newline\\\arraybackslash\hspace{0pt}}X}
\newcolumntype{C}{>{\centering\let\newline\\\arraybackslash\hspace{0pt}}X}

\newcommand{\gambitinstitute}[1]{\expandafter\csname #1\endcsname \label{#1}}

\newcommand{\imperial}{Department of Physics, Imperial College London, Blackett Laboratory, Prince Consort Road, London SW7 2AZ, UK}

\newcommand{\oslo}{Department of Physics, University of Oslo, N-0316 Oslo, Norway}
\newcommand{\adelaide}{Department of Physics, University of Adelaide, Adelaide, SA 5005, Australia}
\newcommand{\glasgow}{School of Physics and Astronomy, University of Glasgow, Glasgow, G12 8QQ, UK}
\newcommand{\monash}{School of Physics and Astronomy, Monash University, Melbourne, VIC 3800, Australia}
\newcommand{\coepp}{Australian Research Council Centre of Excellence for Particle Physics at the Tera-scale}

\newcommand{\mcgill}{Department of Physics, McGill University, 3600 rue University, Montr\'eal, Qu\'ebec H3A 2T8, Canada}
\newcommand{\ucla}{Physics and Astronomy Department, University of California, Los Angeles, CA 90095, USA}

\newcommand{\kansas}{Department of Physics and Astronomy, Malott Hall, University of Kansas, Lawrence, Kansas 66045, USA}

\newcommand{\ubc}{Department of Physics, University of British Columbia, Vancouver BC, Canada}
\newcommand{\lpthe}{LPTHE-CNRS-UPMC, Bo\^ite 126, T13-14 4e \'etage, 4 place Jussieu 75252 Paris CEDEX 05, France}
\newcommand{\nanjing}{Department of Physics and Institute of Theoretical Physics, Nanjing Normal University, Nanjing, Jiangsu 210023, China}

\makeatletter

\newcommand{\preprintnumber}[1]{\gdef\@preprintnumber{\begin{flushright}{#1}\end{flushright}}}

\g@addto@macro\bfseries{\boldmath}
\makeatother

\newcommand{\subparagraph}{} %< workaround for svjour not defining subparagraph
\usepackage{titlesec}
\titleformat*{\paragraph}{\bfseries}
\journalname{Eur. Phys. J. C}
\smartqed
\let\underscore\_
\renewcommand{\_}{\discretionary{\underscore}{}{\underscore}}

\makeatletter
\let\orgdescriptionlabel\descriptionlabel
\renewcommand*{\descriptionlabel}[1]{%
  \let\orglabel\label
  \let\label\@gobble
  \phantomsection
  \protected@edef\@currentlabel{#1}%
  \let\label\orglabel
  \orgdescriptionlabel{#1}%
}
\makeatother

\newcommand\postnewlinemarker{\hbox{\ensuremath{\hookrightarrow}}}
\newcommand\cpp[1]{{\lstinline!#1!}}  % Apparently curly braces are only "experimental"

\newcommand\yaml[1]{{\lstset{style=yaml}\lstinline!#1!\lstset{style=cpp}}}

\newcommand\term[1]{{\lstset{style=terminal}\lstinline!#1!\lstset{style=cpp}}}
\newcommand\fortran[1]{{\lstset{style=fortran}\lstinline!#1!\lstset{style=cpp}}}
\newcommand\py[1]{{\lstset{style=python}\lstinline!#1!\lstset{style=cpp}}}
\newcommand\customtilde{{\raisebox{0.2ex}{\scalebox{0.6}{\boldmath$\sim$}}}}
\newcommand\mathematica[1]{{\lstset{style=Mathematica}\lstinline!#1!\lstset{style=cpp}}}

\lstnewenvironment{lstlistingyaml}{\lstset{style=yaml}}{\lstset{style=cpp}}
\lstnewenvironment{lstlistingterm}{\lstset{style=terminal}}{\lstset{style=cpp}}
\lstnewenvironment{lstlistingfortran}{\lstset{style=fortran}}{\lstset{style=cpp}}
\lstnewenvironment{lstcpp}{\lstset{style=cpp}}{\lstset{style=cpp}}
\lstnewenvironment{lstcppalt}{\lstset{style=cppalt}}{\lstset{style=cpp}}
\lstnewenvironment{lstcppnum}{\lstset{style=cppnum}}{\lstset{style=cpp}}
\lstnewenvironment{lstyaml}{\lstset{style=yaml}}{\lstset{style=cpp}}
\lstnewenvironment{lstterm}{\lstset{style=terminal}}{\lstset{style=cpp}}
\lstnewenvironment{lsttermalt}{\lstset{style=terminalalt}}{\lstset{style=cpp}}
\lstnewenvironment{lsttext}{\lstset{style=text}}{\lstset{style=cpp}}
\lstnewenvironment{lstfortran}{\lstset{style=fortran}}{\lstset{style=cpp}}
\lstnewenvironment{lstpy}{\lstset{style=python}}{\lstset{style=cpp}}
\lstnewenvironment{lstmathematica}{\lstset{style=mathematica}}{\lstset{style=cpp}}

\newcommand{\tmpname}{}
\newcommand{\tmplistingname}{}
\makeatletter
\newif\ifATOlabelname
\lst@Key{labelname}{Listing}{\def\ATOlabelname{#1}\global\ATOlabelnametrue}
\makeatother
\lstnewenvironment{lstcpplabel}[1][]{
  \lstset{style=cpp,#1} % #1 allows to add new options with [] same as for normal lstlistings environment
  \ifATOlabelname
    \renewcommand{\tmpname}{\lstlistingname}
    \renewcommand{\tmplistingname}{\lstlistlistingname}
    \renewcommand{\lstlistingname}{\ATOlabelname}% Listing -> labelname
    \renewcommand{\lstlistlistingname}{List of \lstlistingname s}% List of Listings -> List of labelname
  \fi
}{
  \renewcommand{\lstlistingname}{\tmpname}
  \renewcommand{\lstlistlistingname}{\tmplistingname}
  \lstset{style=cpp}
}
\definecolor{solarized@base03}{HTML}{002B36}
\definecolor{solarized@base02}{HTML}{073642}
\definecolor{solarized@base01}{HTML}{586e75}
\definecolor{solarized@base00}{HTML}{657b83}
\definecolor{solarized@base0}{HTML}{839496}
\definecolor{solarized@base1}{HTML}{93a1a1}
\definecolor{solarized@base2}{HTML}{EEE8D5}
\definecolor{solarized@base3}{HTML}{FDF6E3}
\definecolor{solarized@yellow}{HTML}{B58900}
\definecolor{solarized@orange}{HTML}{CB4B16}
\definecolor{solarized@red}{HTML}{DC322F}
\definecolor{solarized@magenta}{HTML}{D33682}
\definecolor{solarized@violet}{HTML}{6C71C4}
\definecolor{solarized@blue}{HTML}{268BD2}
\definecolor{solarized@cyan}{HTML}{2AA198}
\definecolor{solarized@green}{HTML}{859900}
\definecolor{darkred}{HTML}{550003}
\definecolor{darkgreen}{HTML}{00AA00}

\newcommand\YAMLstringstyle{\footnotesize\color{solarized@green}\mdseries}
\newcommand\YAMLkeystyle{\footnotesize\color{solarized@blue}\ttfamily}
\newcommand\YAMLvaluestyle{\footnotesize\color{blue}\mdseries}
\newcommand\ProcessThreeDashes{\llap{\color{cyan}\mdseries-{-}-}}
\newcommand\CPPcommentstyle{\color{solarized@violet}\footnotesize\ttfamily}
\newcommand\CPPdirectivestyle{\color{solarized@magenta}\footnotesize\ttfamily}
\newcommand\termplainstyle{\footnotesize\ttfamily}

\newcommand\processLongMacroDelimiter
{%
\CPPdirectivestyle%
\#define%
}

\lstdefinestyle{cpp}
{
  language=C++,
  basicstyle=\footnotesize\ttfamily,
  basewidth={0.53em,0.44em}, %Ben: experimenting a bit with the fixed-width width (first argument); feels a bit more readable to me with the slightly smaller width (was 0.6em by default)
  numbers=none,
  tabsize=2,
  breaklines=true,
  escapeinside={@}{@},
  showstringspaces=false,
  numberstyle=\tiny\color{solarized@base01},
  keywordstyle=\color{solarized@orange},
  stringstyle=\color{solarized@red}\ttfamily,
  identifierstyle=\color{solarized@blue},
  commentstyle=\CPPcommentstyle,
  directivestyle=\CPPdirectivestyle,
  emphstyle=\color{solarized@green},
  frame=single,
  rulecolor=\color{solarized@base2},
  rulesepcolor=\color{solarized@base2},
  literate={~} {\customtilde}1,
  moredelim=*[directive]\ \ \#,
  moredelim=*[directive]\ \ \ \ \#
}

\lstdefinestyle{cppalt}
{
  language=C++,
  basicstyle=\footnotesize\ttfamily,
  basewidth={0.53em,0.44em}, %Ben: experimenting a bit with the fixed-width width (first argument); feels a bit more readable to me with the slightly smaller width (was 0.6em by default)
  numbers=none,
  tabsize=2,
  breaklines=true,
  escapeinside={*@}{@*},
  showstringspaces=false,
  numberstyle=\tiny\color{solarized@base01},
  keywordstyle=\color{solarized@orange},
  stringstyle=\color{solarized@red}\ttfamily,
  identifierstyle=\color{solarized@blue},
  commentstyle=\CPPcommentstyle,
  directivestyle=\CPPdirectivestyle,
  emphstyle=\color{solarized@green},
  frame=single,
  rulecolor=\color{solarized@base2},
  rulesepcolor=\color{solarized@base2},
  literate={~}{\customtilde}1,
  moredelim=**[is][\processLongMacroDelimiter]{BeginLongMacro}{EndLongMacro} %special delimiter for long macros that go over several lines
}

\lstdefinestyle{cppnum}
{
  language=C++,
  basicstyle=\footnotesize\ttfamily,
  basewidth={0.53em,0.44em}, %Ben: experimenting a bit with the fixed-width width (first argument); feels a bit more readable to me with the slightly smaller width (was 0.6em by default)
  numbers=none,
  tabsize=2,
  breaklines=true,
  escapeinside={@}{@},
  numberstyle=\tiny\color{solarized@base01},
  showstringspaces=false,
  numberstyle=\tiny\color{solarized@base01},
  keywordstyle=\color{solarized@orange},
  stringstyle=\color{solarized@red}\ttfamily,
  identifierstyle=\color{solarized@blue},
  commentstyle=\CPPcommentstyle,
  directivestyle=\CPPdirectivestyle,
  emphstyle=\color{solarized@green},
  frame=single,
  rulecolor=\color{solarized@base2},
  rulesepcolor=\color{solarized@base2},
  literate={~} {\customtilde}1,
  moredelim=*[directive]\ \ \#,
  moredelim=*[directive]\ \ \ \ \#
}

\lstdefinestyle{python}
{
  language=Python,
  basicstyle=\footnotesize\ttfamily,
  basewidth={0.53em,0.44em},
  numbers=none,
  tabsize=2,
  breaklines=true,
  escapeinside={@}{@},
  showstringspaces=false,
  numberstyle=\tiny\color{solarized@base01},
  keywordstyle=\color{blue},
  stringstyle=\color{orange}\ttfamily,
  identifierstyle=\color{darkred},
  commentstyle=\color{purple},
  emphstyle=\color{green},
  frame=single,
  rulecolor=\color{solarized@base2},
  rulesepcolor=\color{solarized@base2},
  literate = {~}{\customtilde}1
             {\ as\ }{{\color{blue}\ as\ \color{black}}}3
}

\lstdefinestyle{fortran}
{
  language=Fortran,
  basicstyle=\footnotesize\ttfamily,
  basewidth={0.53em,0.44em},
  numbers=none,
  tabsize=2,
  breaklines=true,
  escapeinside={@}{@},
  showstringspaces=false,
  numberstyle=\tiny\color{solarized@base01},
  keywordstyle=\color{blue},
  stringstyle=\color{orange}\ttfamily,
  identifierstyle=\color{Periwinkle},
  commentstyle=\color{purple},
  emphstyle=\color{green},
  morekeywords={and, or, true, false},
  frame=single,
  rulecolor=\color{solarized@base2},
  rulesepcolor=\color{solarized@base2},
  literate={~}{\customtilde}1
}

\lstdefinestyle{terminal}
{
  language=bash,
  basicstyle=\termplainstyle,
  numbers=none,
  tabsize=2,
  breaklines=true,
  escapeinside={@}{@},
  frame=single,
  showstringspaces=false,
  numberstyle=\tiny\color{solarized@base01},
  keywordstyle=\color{solarized@orange},
  stringstyle=\color{solarized@red}\ttfamily,
  identifierstyle=\color{black},
  commentstyle=\color{solarized@violet},
  emphstyle=\color{solarized@green},
  frame=single,
  rulecolor=\color{solarized@base2},
  rulesepcolor=\color{solarized@base2},
  morekeywords={gambit, cmake, make, mkdir},
  deletekeywords={test},
  literate = {\ gambit}{{\ }{\color{black}}gambit}7
             {/gambit}{{/}{\color{black}}gambit}6
             {gambit/}{{\color{black}}gambit{/}}6
             {/include}{{/}{\color{black}}include}8
             {cmake/}{{\color{black}}cmake/}6
             {.cmake}{{.}{\color{black}}cmake}6
             {~}{\customtilde}1
}

\lstdefinestyle{terminalalt}
{
  language=bash,
  basicstyle=\footnotesize\ttfamily,
  numbers=none,
  tabsize=2,
  breaklines=true,
  escapeinside={*@}{@*},
  frame=single,
  showstringspaces=false,
  numberstyle=\tiny\color{solarized@base01},
  keywordstyle=\color{solarized@orange},
  stringstyle=\color{solarized@red}\ttfamily,
  identifierstyle=\color{black},
  commentstyle=\color{solarized@violet},
  emphstyle=\color{solarized@green},
  frame=single,
  rulecolor=\color{solarized@base2},
  rulesepcolor=\color{solarized@base2},
  morekeywords={gambit, cmake, make, mkdir},
  deletekeywords={test},
  literate = {\ gambit}{{\ }{\color{black}}gambit}7
             {/gambit}{{/}{\color{black}}gambit}6
             {gambit/}{{\color{black}}gambit{/}}6
             {/include}{{/}{\color{black}}include}8
             {cmake/}{{\color{black}}cmake/}6
             {.cmake}{{.}{\color{black}}cmake}6
             {~}{\customtilde}1
}

\lstdefinestyle{text}
{
  language={},
  basicstyle=\footnotesize\ttfamily,
  identifierstyle=\color{black},
  numbers=none,
  tabsize=2,
  breaklines=true,
  escapeinside={*@}{@*},
  showstringspaces=false,
  frame=single,
  rulecolor=\color{solarized@base2},
  rulesepcolor=\color{solarized@base2},
  literate={~}{\customtilde}1
}

\lstdefinestyle{yaml}
{
  language=bash,
  escapeinside={@}{@},
  keywords={true,false,null},
  otherkeywords={},
  keywordstyle=\color{solarized@base0}\bfseries,
  basicstyle=\footnotesize\color{black}\ttfamily,
  identifierstyle=\YAMLkeystyle,
  sensitive=false,
  commentstyle=\color{solarized@orange}\ttfamily,
  morecomment=[l]{\#},
  morecomment=[s]{/*}{*/},
  stringstyle=\YAMLstringstyle\ttfamily,
  moredelim=**[s][\YAMLkeystyle]{,}{:},   % switch to value style at : but back to key style at,
  moredelim=**[l][\YAMLvaluestyle]{:},    % switch to value style at :
  morestring=[b]',
  morestring=[b]",
  literate =    {---}{{\ProcessThreeDashes}}3
                {>}{{\textcolor{solarized@red}\textgreater}}1
                {|}{{\textcolor{solarized@red}\textbar}}1
                {\ -\ }{{\mdseries\color{black}\ -\ \negmedspace}}3
                {\}}{{{\color{black} \}}}}1
                {\{}{{{\color{black} \{}}}1
                {[}{{{\color{black} [}}}1
                {]}{{{\color{black} ]}}}1
                {~}{\customtilde}1,
  breakindent=0pt,
  breakatwhitespace,
  columns=fullflexible
}

\lstdefinestyle{mathematica}
{
  language={Mathematica},
  basicstyle=\footnotesize\ttfamily,
  basewidth={0.53em,0.44em},
  numbers=none,
  tabsize=2,
  breaklines=true,
  escapeinside={@}{@},
  numberstyle=\tiny\color{black},
  showstringspaces=false,
  numberstyle=\tiny\color{solarized@base01},
  keywordstyle=\color{solarized@orange},
  stringstyle=\color{solarized@red}\ttfamily,
  identifierstyle=\color{solarized@orange}\ttfamily,
  commentstyle=\color{solarized@gray}\ttfamily,
  directivestyle=\color{solarized@orange}\ttfamily,
  emphstyle=\color{solarized@green},
  frame=single,
  rulecolor=\color{solarized@base2},
  rulesepcolor=\color{solarized@base2},
  literate={~} {\customtilde}1,
  moredelim=*[directive]\ \ \#,
  moredelim=*[directive]\ \ \ \ \#,
  mathescape=true
}

\newcommand{\doublecross}[2]{\hyperref[#2]{\textbf{#1}}}
\newcommand{\doublecrosssf}[2]{\hyperref[#2]{\textbf{\textsf{#1}}}}

\newcommand{\startglossary}{\section{Glossary}\label{glossary}Here we explain some terms that have specific technical definitions in \GB.\begin{description}}
\newcommand{\finishglossary}{\end{description}}

\newcommand{\bcode}{\begin{lstlisting}}
\newcommand{\ecode}{\end{lstlisting}}

\newcommand{\eV}{\ensuremath{\text{e}\mspace{-0.8mu}\text{V}}\xspace}

\newcommand{\TeV}{\text{T\eV}\xspace}

\newcommand{\fb}{\text{fb}\xspace}

\newcommand{\invfb}{\ensuremath{\fb^{-1}}\xspace}

\newcommand{\MSBar}{\overline{MS}}

\newcommand{\eg}{e.g.\ }
\newcommand{\atlas}{ATLAS\xspace}
\newcommand{\cms}{CMS\xspace}
\newcommand{\gambit}{\textsf{GAMBIT}\xspace}

\newcommand{\darkbit}{\textsf{DarkBit}\xspace}
\newcommand{\colliderbit}{\textsf{ColliderBit}\xspace}

\newcommand{\specbit}{\textsf{SpecBit}\xspace}
\newcommand{\decaybit}{\textsf{DecayBit}\xspace}

\newcommand{\scannerbit}{\textsf{ScannerBit}\xspace}

\newcommand{\GB}{\gambit}

\newcommand{\buckfast}{\textsf{BuckFast}\xspace}
\newcommand{\delphes}{\textsf{Delphes}\xspace}

\newcommand{\pythia}{\textsf{Pythia}\xspace}
\newcommand{\pythiaeight}{\textsf{Pythia\,8}\xspace}

\newcommand{\prospino}{\textsf{Prospino}\xspace}

\newcommand{\madgraph}{\textsf{MadGraph}\xspace}
\newcommand{\MGaMCNLO}{\textsf{MadGraph5\_aMC@NLO}\xspace}
\newcommand{\fastjet}{\textsf{FastJet}\xspace}

\newcommand\flexiblesusy{\FlexibleSUSY}
\newcommand\FlexibleSUSY{\textsf{FlexibleSUSY}\xspace}

\newcommand\SOFTSUSY{\textsf{SOFTSUSY}\xspace}

\newcommand\HDECAY{\textsf{HDECAY}\xspace}

\newcommand\SDECAY{\textsf{SDECAY}\xspace}
\newcommand\SUSYHIT{\textsf{SUSY-HIT}\xspace}

\newcommand\SARAH{\textsf{SARAH}\xspace}

\newcommand\nulike{\textsf{nulike}\xspace}

\newcommand{\capgen}{\textsf{Capt'n General}\xspace}

\newcommand\twalk{\textsf{T-Walk}\xspace}
\newcommand\diver{\textsf{Diver}\xspace}
\newcommand\ddcalc{\textsf{DDCalc}\xspace}

\newcommand\beq{\begin{equation}}
\newcommand\eeq{\end{equation}}

\renewcommand{\url}[1]{\href{#1}{#1}}

\usepackage{makecell}

\newcommand{\pvalue}{$p$-value\xspace}
\newcommand{\pvalues}{$p$-values\xspace}

\newcolumntype{X}[1]{>{\centering\arraybackslash}p{#1}}

\begin{document}

\preprintnumber{CoEPP-MN-18-7}

\title{Combined collider constraints on neutralinos and charginos}

\author
{
The GAMBIT Collaboration:
Peter~Athron\thanksref{inst:a,inst:b} \and
Csaba~Bal\'azs\thanksref{inst:a,inst:b} \and
Andy~Buckley\thanksref{inst:d} \and
Jonathan~M.~Cornell\thanksref{inst:i} \and
Matthias~Danninger\thanksref{inst:x} \and
Ben~Farmer\thanksref{inst:q} \and
Andrew~Fowlie\thanksref{inst:a,inst:b,inst:z}\and
Tom\'as~E.~Gonzalo\thanksref{inst:c}\and
Julia~Harz\thanksref{inst:y}\and
Paul~Jackson\thanksref{inst:b,inst:k} \and
Rose~Kudzman-Blais\thanksref{inst:x} \and
Anders~Kvellestad\thanksref{inst:q,inst:c,e2} \and
Gregory~D.~Martinez\thanksref{inst:p} \and
Andreas~Petridis\thanksref{inst:b,inst:k}\and
Are~Raklev\thanksref{inst:c} \and
Christopher~Rogan\thanksref{inst:s} \and
Pat~Scott\thanksref{inst:q} \and
Abhishek~Sharma\thanksref{inst:b,inst:k}\and
Martin~White\thanksref{inst:b,inst:k,e4}\and
Yang~Zhang\thanksref{inst:a,inst:b}
}

\institute{%
  \monash\label{inst:a} \and
  \coepp\label{inst:b} \and
  \glasgow\label{inst:d} \and
  \mcgill\label{inst:i} \and
  \ubc\label{inst:x} \and
  \imperial\label{inst:q} \and
  \nanjing\label{inst:z} \and
  \oslo\label{inst:c} \and
  \lpthe\label{inst:y} \and
  \adelaide\label{inst:k} \and
  \ucla\label{inst:p} \and
  \kansas\label{inst:s}
}

\newcommand{\af}[1]{\textbf{AF: #1}}
\newcommand{\ben}[1]{\textbf{BF: #1}}
\newcommand{\yz}[1]{\textbf{Yang: #1}}

\thankstext{e2}{anders.kvellestad@fys.uio.no}
\thankstext{e4}{martin.white@adelaide.edu.au}

\titlerunning{Electroweakinos with GAMBIT}
\authorrunning{The GAMBIT Collaboration}

\date{Received: date / Accepted: date}

\maketitle

\begin{abstract}
Searches for supersymmetric electroweakinos have entered a crucial phase, as the integrated luminosity of the Large Hadron Collider is now high enough to compensate for their weak production cross-sections. Working in a framework where the neutralinos and charginos are the only light sparticles in the Minimal Supersymmetric Standard Model, we use \gambit to perform a detailed likelihood analysis of the electroweakino sector.  We focus on the impacts of recent ATLAS and CMS searches with 36 fb$^{-1}$ of 13\,TeV proton-proton collision data. We also include constraints from LEP and invisible decays of the $Z$ and Higgs bosons.  Under the background-only hypothesis, we show that current LHC searches do not robustly exclude any range of neutralino or chargino masses. However, a pattern of excesses in several LHC analyses points towards a possible signal, with neutralino masses of $(m_{\tilde{\chi}_1^0}, m_{\tilde{\chi}_2^0}, m_{\tilde{\chi}_3^0}, m_{\tilde{\chi}_4^0})$ =  (8--155, 103--260, 130--473, 219--502)\,GeV and chargino masses of $(m_{\tilde{\chi}_1^{\pm}}, m_{\tilde{\chi}_2^{\pm}})$ = (104--259, 224--507)\,GeV at the 95\% confidence level. The lightest neutralino is mostly bino, with a possible modest Higgsino or wino component. We find that this excess has a combined local significance of $3.3\sigma$, subject to a number of cautions. If one includes LHC searches for charginos and neutralinos conducted with 8\,TeV proton-proton collision data, the local significance is lowered to 2.9$\sigma$. We briefly consider the implications for dark matter, finding that the correct relic density can be obtained through the Higgs-funnel and $Z$-funnel mechanisms, even assuming that all other sparticles are decoupled. All samples, \gambit input files and best-fit models from this study are available on \textsf{Zenodo}.

\end{abstract}

\vspace*{-4ex}
\tableofcontents

%%%%%%%%%%%%%%%%%%%%%%%%%%%%%%%%%%%%%%%%%%%%%%%%%%%%%%%
%%%%%%%%%%%%%%%%%%%%%%%%%%%%%%%%%%%%%%%%%%%%%%%%%%%%%%%
\section{Introduction}
\label{intro}
%%%%%%%%%%%%%%%%%%%%%%%%%%%%%%%%%%%%%%%%%%%%%%%%%%%%%%%

Supersymmetry (SUSY) provides well-justified extensions of the
Standard Model (SM) of particle physics that can stabilise the
electroweak scale against quantum
corrections \cite{Dimopoulos:1981au, Witten:1981nf, Dine:1981za,
Dimopoulos:1981zb, Sakai:1981gr, Kaul:1981hi}, radiatively break
electroweak
symmetry \cite{Ibanez:1982fr,Inoue:1982pi,Ellis:1982wr,Ibanez:1982qk}
and provide a dark matter (DM) candidate
with the right abundance \cite{Ellis:1983ew,
Jungman:1995df}. Supersymmetric models are, however, increasingly
challenged by null observations at a number of experiments,
including searches for supersymmetric particles (sparticles) in proton--proton collisions at the Large
Hadron Collider (LHC), and direct and indirect searches for DM.

In the minimal
supersymmetric standard model (MSSM), the
superpartners of the electroweak gauge and Higgs bosons
mix to form \emph{electroweakinos}.  These consist of four
Majorana fermions (neutralinos $\tilde{\chi}_i^0$, with $i=1,2,3,4$ in
order of increasing mass), and two Dirac fermions (charginos $\tilde{\chi}_i^\pm$, with $i=1,2$).
The two mass matrices that mix these
states contain only four parameters: the soft-breaking bino mass, $M_1$, the
soft-breaking wino mass, $M_2$, the Higgsino superpotential mass parameter, $\mu$, and the ratio
of the two Higgs vacuum expectation values, $\tan\beta$.

Although the masses of the neutralinos and charginos are unknown,
there are theoretical reasons to expect them to be light. The $\mu$
parameter, which governs Higgsino masses, enters tadpole cancellations
required for electroweak symmetry breaking.  Were $\mu$ significantly
greater than the weak scale, other parameters would need to be
fine-tuned in order to satisfy these relations.  Indeed, according to
some measures of fine tuning presented in the literature, it is
possible to have low fine tuning when the sfermions and gluino are
heavy, provided that the Higgsinos (and therefore $\mu$) remain
light \cite{Chan:1997bi,Feng:1999mn,Feng:1999zg,Feng:2012jfa,Feng:2011aa,Delgado:2014vha,Akula:2011jx,Liu:2013ula,Baer:2012up,Baer:2012mv,Baer:2012cf,Baer:2013vpa,Baer:2013gva,Baer:2013xua,Baer:2015rja,Baer:2016bwh,Kim:2013uxa,Fowlie:2014faa,Athron:2017fxj,CahillRowley:2012rv,Casas:2014eca,Athron:2015tsa}.
SUSY models with electroweakino states significantly lighter than the
other SUSY states have been presented as natural
SUSY \cite{Brust:2011tb, Papucci:2011wy, Cao:2012rz, Han:2013kga,
Han:2013usa, Ding:2015vla, Buckley:2016kvr, Kim:2016rsd} and in models
where naturalness has been abandoned as a guiding
principle \cite{Wells:2004di, ArkaniHamed:2004fb, Giudice:2004tc,
ArkaniHamed:2004yi, Hall:2011jd, Ibe:2011aa, Ibe:2012hu,
Arvanitaki:2012ps, ArkaniHamed:2012gw}.  In the latter, other
motivations such as DM, where the lightest neutralino may
play the role of DM even if the rest of the SUSY spectrum is heavy,
are used as the guiding principles.\footnote{Note that in many models
described as ``natural SUSY'' the stop is still relatively light,
though significantly heavier than the electroweakinos.  Similarly, in
some of the explicitly un-natural models, the gluino is often much
lighter than the sfermions, but again remains significantly heavier
than the electroweakinos.}

In this paper we take an agnostic approach to the questions of fine tuning and whether
or not the neutralino plays the role of DM. Instead,
we attempt to present a precise picture of current experimental knowledge of
the electroweakino sector (which we
call the \emph{EWMSSM}) from direct collider searches for sparticles.

Constraints on electroweakinos have commonly been calculated under
restrictive assumptions about their masses or compositions,
or only over restricted slices of parameter space.  For example, lower limits on the mass of the lightest neutralino
from LEP \cite{ALEPH:sleptons_gauginos,OPAL:gauginos} are based on assumptions about the unification of gaugino masses at high scales. The purpose of this work is to determine whether
the current suite of direct searches allows some range of the
electroweakino masses (and/or couplings) to be robustly excluded -- or alternatively, preferred.

Previous studies have investigated the combined impacts of various
DM and collider constraints on
the electroweakino sector, in the limit that other sparticles are
decoupled~\cite{Huang:2017kdh, Profumo:2017ntc, Duan:2017ucw,Fuks:2017rio, Badziak:2017the, Bramante:2015una, Cheung:2012qy, Choudhury:2016lku, Chakraborti:2015mra, Chakraborti:2014gea, Cao:2015efs,arXiv:1510.05378,Calibbi:2014lga,Han:2016qtc,Han:2013kza}.
Here, we carry out a more detailed, model-independent study,
performing a global fit of the EWMSSM using only collider
constraints from LEP, ATLAS and CMS arising either from direct
searches for electroweakinos, or SM particle decays into them.  Having a complete picture of the constraints on this sector from LEP and the LHC, independent of any assumptions about DM or Higgs physics, is of great interest.  It may be the case, for example, that $R$-parity violation renders the $\tilde{\chi}_1^0$ metastable, or that the true Higgs sector is far more complex than that of the MSSM.

Electroweakino constraints from the LHC were first considered in detail in Ref.~\cite{Martin:2014qra}, which our study extends in a number of ways. First, we consider LEP searches in detail, plus constraints arising from measurement of the $Z$ and $h$ invisible widths. Second, we perform a convergent global statistical fit of the parameter space, with Monte Carlo (MC) event generation for LHC processes at each sampled parameter point, rather than simply performing a rectangular grid scan of the parameter space (and we generate at least twice as many MC events per parameter point as the previous study). Our statistical treatment is also superior, as we recreate the ATLAS and CMS limit-setting procedures for each analysis rather than comparing the predicted number of signal events to the ATLAS and CMS 95\% CL exclusions on the numbers of signal events. This allow us to combine continuous likelihood terms from each analysis, and thus explore possible tensions between analyses in a rigorous fashion. Most significantly, Ref.~\cite{Martin:2014qra} is based on searches for electroweakinos using the 8\,TeV proton--proton collision dataset of the LHC, which have been all but superseded by new ATLAS and CMS searches that use 36 fb$^{-1}$ of 13\,TeV proton--proton collision data. This dramatically extends the possible discovery and exclusion reach of the LHC searches.

We begin in Sec.~\ref{sec:phys} by introducing the model and parameters
over which we scan, followed by our sampling methodology, adopted priors and statistical framework.
In Sec.~\ref{sec:lnL}, we then give a brief summary of the observables and
likelihoods that we employ.  We present our main results in Sec.\ \ref{sec:results}
and briefly consider the implications for DM in Sec.\ \ref{sec:darkmatter} before presenting final conclusions in Sec.~\ref{sec:conc}. Appendix \ref{app} provides additional details for the interested reader on the impact of 8\,TeV data on our results, and Appendix \ref{appb} provides best-fit signal predictions for all signal regions of all analyses that we consider.
All \GB input files, generated likelihood
samples and best-fit benchmarks for this paper are publicly available online
through \textsf{Zenodo} \cite{the_gambit_collaboration_2018_1410335}.

%%%%%%%%%%%%%%%%%%%%%%%%%%%%%%%%%%%%%%%%%%%%%%%%%%%%%%%
%%%%%%%%%%%%%%%%%%%%%%%%%%%%%%%%%%%%%%%%%%%%%%%%%%%%%%%
\section{Model and fitting framework}
\label{sec:phys}
%%%%%%%%%%%%%%%%%%%%%%%%%%%%%%%%%%%%%%%%%%%%%%%%%%%%%%%

%%%%%%%%%%%%%%%%%%%%%%%
\subsection{Model definition}
%%%%%%%%%%%%%%%%%%%%%%%
In this study we investigate the electroweakino sector of the MSSM.
This sector is composed of Higgsinos ($\tilde{H}_u^0, \tilde{H}_u^+, \tilde{H}_d^-, \tilde{H}_d^0
$) and electroweak gauginos: the bino ($\tilde{B}$) and winos ($\tilde{W}^0,\tilde{W}^+,\tilde{W}^- $).  The neutral states mix together to form neutralinos, while the
charged states mix to form charginos. The Lagrangian density therefore includes
\begin{equation}
\mathcal{L}_{\textrm{EWino}} = -\frac12 (\psi^0)^T M_N \psi^0 - \frac12 (\psi^\pm)^T M_C \psi^\pm + c.c.
%\mathcal{L}_{\textrm{EWino}} = -\frac12 (\psi^0)^T M_{\chi^0} \psi^0 - \frac12 (\psi^\pm)^T M_C \psi^\pm + c.c.
\end{equation}
where
\begin{equation}
\psi^0 = (\tilde{B},\tilde{W}^0,\tilde{H_d^0},\tilde{H_u^0} ),~ \psi^\pm = (\tilde{W}^+,\tilde{H}_u^+,\tilde{W}^-,\tilde{H}_d^- ),
\end{equation}
and the neutralino mass matrix is
\begin{equation}
%M_{\tilde\chi^0} =
M_N =
\begin{pmatrix}
M_1 & 0 & -\frac12 g' v c_\beta & \frac12 g' v s_\beta \\
0 & M_2 & \frac12 g  v c_\beta & -\frac12 g v s_\beta\\
-\frac12 g' v c_\beta & \frac12 g v c_\beta & 0 & -\mu \\
\frac12 g' v s_\beta & -\frac12 g v s_\beta& -\mu & 0
\end{pmatrix}.
\label{Eq:mneut}
\end{equation}
Here $s_\beta = \sin\beta$ and $c_\beta = \cos\beta$, and the $SU(2)$ and $U(1)_Y$ gauge couplings, $g$ and $g'$, and the electroweak VEV, $v$ are fixed from data while the ratio $\tan\beta = v_u / v_d$ is a free parameter.

Similarly, the chargino mass matrix may be written as
\begin{equation}
M_C =
\begin{pmatrix}
0 & X^T\\
X & 0
\end{pmatrix},
\;\;
\textrm{where} %% are obtained by performing a bi-unitary transformation to diagionalise,
\;\;
X =
\begin{pmatrix}
M_2 & \frac{g v s_\beta}{\sqrt{2}} \\
\frac{g v c_\beta}{\sqrt{2}} & \mu
\end{pmatrix}.
\label{Eq:mcha}
\end{equation}

Therefore the electroweakinos can be described using just the four
electroweakino parameters mentioned in the introduction: $M_1$, $M_2$,
$\mu$ and $\tan\beta$.

 An electroweakino effective field theory (EFT) can be constructed by
including additional light states, namely the SM fermions, gauge
bosons and a SM-like Higgs boson. As with $g$ and $g^\prime$, the
$SU(3)$ gauge coupling and SM Yukawa couplings can be fixed from data.
The Higgs potential parameters can be fixed by imposing the
minimisation condition and requiring that the Higgs mass is fixed to
its measured value $m_h = 125.09\,\text{GeV}$~\cite{PDG17}.

Note that in the MSSM, the quartic couplings in the Higgs potential
are fixed by SM gauge couplings, allowing the Higgs mass to be calculated
given a value of $\tan\beta$.  To find $m_h \simeq 125\,\text{GeV}$
over a range of input $\tan\beta$, one would then have to vary
additional MSSM parameters. We choose to instead fix the Higgs mass,
in the spirit of interpreting the results in an
electroweakino EFT rather than any specific MSSM ultraviolet completion.  This avoids introducing additional
degrees of freedom that are not part of the electroweakino sector.

In principle it is possible to perform all calculations
in such an electroweakino EFT.  In practise, it is simpler to use an MSSM model
where the rest of the states are heavy and decoupled, and make use of
existing MSSM tools for computing e.g.\ electroweakino decays.  We
implement this model within the \GB MSSM model hierarchy,
in which the user may define child models of more general
scenarios. The \GB SUSY models include a chain of scenarios in which
the MSSM soft SUSY-breaking Lagrangian parameters are defined at some
scale $Q$, which one typically sets to be near the weak scale.  The
most general model has 63 free parameters: the gaugino masses $M_1$,
$M_2$, and $M_3$, the trilinear coupling matrices
$\mathbf{A}_u, \mathbf{A}_d$ and $\mathbf{A}_e$ (9 parameters each),
the squared soft sfermion mass matrices $\mathbf{m}^\mathbf{2}_Q$,
$\mathbf{m}^\mathbf{2}_u$, $\mathbf{m}^\mathbf{2}_d$,
$\mathbf{m}^\mathbf{2}_L$ and $\mathbf{m}^\mathbf{2}_e$ (6 parameters
each), and three additional parameters describing the Higgs sector.

In this work we define the dimensionful parameters at the SUSY
scale $Q = M_{\textrm{SUSY}} = 3$\,TeV. We set
all trilinear couplings to zero.  We take all diagonal entries of the squared soft
sfermion mass matrices to be $M^2_{\textrm{SUSY}}$, and all off-diagonal entries to be zero.   We adopt a value of 5\,TeV for
both the pseudo-scalar Higgs mass $m_A$ and the gluino mass parameter
$M_3$. We choose these values in order to effectively decouple
all sparticles except for the electroweakinos.  Their precise
values are not significant, and simply serve to push the
model into the decoupling regime.  In this way, we fix all MSSM parameters except the four free parameters of the EWMSSM given in
Table~\ref{tab:param}.

In this model we also assume that $R$-parity is either conserved or
broken sufficiently weakly that the lightest supersymmetric particle
(LSP) is metastable on detector timescales; we thus discard all parameter
combinations where the LSP is not a neutralino.

\begin{table}
\begin{center}
\begin{tabular}{l@{\ }c c c}
\hline
Parameter & Minimum & Maximum & Priors    \\
\hline
$M_1(Q)$               & $-$2\,TeV& 2\,TeV & hybrid, flat  \\

$M_2(Q)$               & 0\,TeV& 2\,TeV & hybrid, flat  \\
$\mu(Q)$                & $-2\,\mathrm{TeV}$ & $2\,\mathrm{TeV}$  & hybrid, flat \\
$\tan\beta(m_Z)$           & 1         & 70      & flat          \\
\hline
$Q$                     & \multicolumn{2}{c}{3\,TeV} & fixed         \\
\hline
$\alpha_s^{\MSBar}(m_Z)$ & \multicolumn{2}{c}{0.1181} & fixed         \\
Top quark pole mass & \multicolumn{2}{c}{$171.06\,\mathrm{GeV}$} & fixed \\
\end{tabular}
\caption{\label{tab:param} Parameters, ranges and priors adopted in the scans of this paper.  The ``hybrid'' prior is flat where $|x| < 10\,\mathrm{GeV}$, and logarithmic elsewhere.
All other soft SUSY-breaking parameters are decoupled; see the text for details.}
\end{center}
\end{table}

%%%%%%%%%%%%%%%%%%
\subsection{Global fitting framework}
%%%%%%%%%%%%%%%%%%

The fits that we present in this paper are done with \GB \cite{gambit,ColliderBit,DarkBit,FlavBit,SDPBit,ScannerBit} \textsf{1.2.0}.  The LHC and LEP constraints that we apply come from \colliderbit \cite{ColliderBit} and the invisible width constraints are from \decaybit \cite{SDPBit}.  Both rely on spectrum calculations carried out with \specbit \cite{SDPBit}. All sampling is driven by \scannerbit \cite{ScannerBit}.  We later explore DM implications (Sec.\ \ref{sec:darkmatter}) with \darkbit \cite{DarkBit}.

Compared to \GB \textsf{1.1}, version \textsf{1.2} offers a number of new features.  Those of most relevance for this study are updates to \decaybit to include the invisible $Z$ width and theory errors on the invisible Higgs width (Sec.\ \ref{sec:invisible widths}), and to \colliderbit to include \begin{itemize}
\item many new 13\,TeV analyses
\item a LEP search for degenerate chargino--neutralino pairs (Sec.\ \ref{sec:LHCanalyses}),
\item the ability to account for background correlations in different signal regions via simplified likelihoods (Sec.\ \ref{sec:simplified likelihoods}),
\item a dynamic convergence test of LHC Monte Carlo simulations designed to achieve a specific fractional signal uncertainty,
\item explicit output of individual LHC likelihood components, and
\item the ability to simultaneously include likelihood components from multiple uncorrelated signal regions in a single analysis.
\end{itemize}

Other updates include \begin{itemize}
\item the ability to call backends written in \textsf{Python},
\item an interface to the \textsf{polychord} sampler \cite{Handley:2015},
\item improved parallelism and shutdown handling in the \textsf{hdf5} printers and the \twalk sampler,
\item a standalone \textsf{hdf5} combination utility,
\item a new \textsf{cout} printer that sends outputs directly to the system standard output,
\item support for DM semi-annihilation processes and related models in \darkbit and \specbit \cite{SSDM2},
\item a wider range of Higgs portal models \cite{SSDM2,HiggsPortal},
\item a number of new MSSM parameterisations (using $\mu$ and $m_A$ instead of $m_{H_u}^2$ and $m_{H_d}^2$), and
\item support for a number of new and updated external packages, including \flexiblesusy \textsf{2.0} \cite{Athron:2017fvs}, \nulike \textsf{1.0.6} \cite{IC22Methods, IC79_SUSY}, \ddcalc \textsf{2.0.0} \cite{HiggsPortal}, \capgen \textsf{1.0.0} \cite{HiggsPortal} and \textsf{fjcore} \textsf{3.2.0} \cite{Cacciari:2011ma}.
\end{itemize}

%%%%%%%%%%%%%%%%%%%%%%%
\subsection{Parameters and priors}
%%%%%%%%%%%%%%%%%%%%%%%

Table~\ref{tab:param} summarises the ranges over which we scan
the EWMSSM parameters, along with the priors that we assume.\footnote{As ours
is a frequentist analysis, the priors merely define a metric upon the
parameter space that we scan. They do not reflect any prior
beliefs about the EWMSSM, and are chosen only to thoroughly and
efficiently map the likelihood surface.}  Except for $\tan\beta$,
which we sample using a flat prior, our main scan employs a ``hybrid'' prior on
each of the parameters $x$, which is flat where $|x| < 10\,\mathrm{GeV}$, and logarithmic elsewhere.

To ensure that we include all possible mass hierarchies, we allow the three
dimensionful parameters $M_1$, $M_2$ and $\mu$ to vary up to a
magnitude of $2$\,TeV.  This is well beyond the LHC reach for electroweak
states.  Without loss of generality, we restrict $M_2$ to positive
values, as is commonly done in the literature (see
e.g.\ \cite{MasterCodeMSSM10,Bagnaschi:2017tru}), while allowing both
positive and negative signs for both $\mu$ and $M_1$.  Although we do not expect our results to be very sensitive to $\tan\beta$,
we consider a large range of possible values for this parameter (1--70), as
previous work \cite{CMSSM,MSSM} has shown a preference for large $\tan\beta$.

For the purposes of mapping the profile likelihood, we sample the parameter
space of the EWMSSM using the differential evolution
sampler \diver \textsf{1.0.4}~\cite{ScannerBit}, employing the self-adaptive
jDE version of the algorithm~\cite{Brest06}. We set the population size \fortran{NP} to
18\,700, and the convergence threshold \fortran{convthresh} to
$10^{-3}$.  In order to sample the final high-likelihood region more efficiently, we performed two additional targeted scans, one for $|\mu| < 500$\,GeV and another for $M_2 < 500$\,GeV, using flat priors for the dimensionful parameters and the same \diver settings as the full-range scan.

A critical factor in the scanning strategy is the number of MC events generated per point to determine the LHC likelihood. This is particularly important for electroweak supersymmetry searches, since the acceptance of the analyses is often very small, due to very stringent kinematic selections that are designed to reject SM backgrounds that otherwise swamp the tiny SUSY signal. This problem is made worse by the necessity for some analyses of pre-selecting the signal region to use for a given parameter point, according to which of the available signal regions is expected to have the best sensitivity to the model.  As the MC statistics are increased, the signal region with the best expected sensitivity to a given parameter point may change abruptly. When the level of agreement between data and background expectations differs notably between signal regions, a switch in which signal region is pre-selected can cause a large change in the likelihood assigned to the parameter point.

To combat this we perform our initial scan of the full parameter space with 100\,000 generated events per parameter point, and the targeted scans with 500\,000 events. We then carry out a
sequential post-processing of the scan results to increase the MC statistics for points within the parameter regions preferred by our fit. Through this post-processing we ensure a minimum of 4 million generated events for all points in the $2\sigma$ region, 16 million events for all points inside the $1\sigma$ region, and 64 million events for the 500 points with highest likelihood. In total, we process $2.4\times10^5$ of the original scan samples with at least 4 million MC events. All the results that we present in this paper are based on this set of post-processed samples, unless otherwise stated.

\subsection{Electroweakino spectrum and decays}

In the course of our scans, model parameter values
are sampled by \scannerbit and passed to
an MSSM \FlexibleSUSY \cite{Athron:2014yba,Athron:2017fvs} spectrum
generator\footnote{\FlexibleSUSY uses \SARAH \cite{Staub:2008uz,Staub:2010jh} and numerical routines from \SOFTSUSY \cite{Allanach:2001kg,Allanach:2013kza} to create the spectrum generator. }, which determines $\overline{\textrm{DR}}$ couplings and
computes the predicted electroweakino masses and mixings. It computes neutralino
and chargino masses at the full one-loop level, performing a
fixed-order calculation at the SUSY scale $Q = 3$\,TeV.  The separation
of scales implies somewhat large fractional corrections to the masses:
\mbox{$\sim g^2/(4\pi)^2 \ln (M^2_{\textrm{SUSY}}/m^2_Z)$} for
gaugino-like states and \mbox{$\sim y_t^2/(4\pi)^2 \ln (M^2_{\textrm{SUSY}}/m^2_t)$} for
Higgsinos.

A more precise calculation of the masses could be achieved by using
effective field theory techniques of matching and running to resum
logs, or by including two-loop dominant $\mathcal{O}(\alpha_t \alpha_s)$ and $\mathcal{O}(\alpha \alpha_s)$
corrections to the neutralino and chargino
masses \cite{Schofbeck:2006gs,Schofbeck:2007ib}. However, such
improvements would not have a significant impact on our conclusions
about the implications of experimental searches for the electroweakino
sector.

We extract the electroweak gauge couplings at one-loop level using a fixed-order calculation at
scale $m_Z$, and thus these also receive electroweak
corrections with logarithms between the SUSY scale and $m_Z$.\footnote{The Yukawa
interactions of electroweakinos always involve very heavy sfermions,
so such processes do not play a significant role in our calculations.}

We calculate neutralino and chargino decay branching fractions
with \SUSYHIT \textsf{1.5} \cite{Djouadi:2006bz}, which
incorporates \HDECAY \cite{Djouadi:1997yw}
and \SDECAY \cite{Muhlleitner:2003vg}. The resulting total widths and branching ratios are passed to the \pythiaeight event generator~\cite{Sjostrand:2006za,Sjostrand:2014zea} which performs the decays. Since \pythiaeight is in most instances limited to phase space decays, the kinematics of three-body decays of electroweakinos through off-shell gauge bosons,  $\tilde\chi_1^\pm\to \tilde\chi_1^0 W^*$  and $\tilde\chi_2^0\to \tilde\chi_1^0 Z^*$, is not perfectly described. This is a limitation inherent to our fast simulation of LHC events. As will be clear below, the problematic region of the parameter space is not preferred by our scans.

%%%%%%%%%%%%%%%%%%%%%%%%%%%%%%%%%%%%%%%%%%%%%%%%%%%%%%%
\section{Observables and likelihoods}
\label{sec:lnL}
%%%%%%%%%%%%%%%%%%%%%%%%%%%%%%%%%%%%%%%%%%%%%%%%%%%%%%%

Having chosen to investigate only the constraints provided by collider
data on the electroweakino sector, our study includes a variety of
direct searches for charginos and neutralinos from the OPAL and L3 experiments at the LEP collider, and the ATLAS and CMS experiments at the LHC, plus constraints on the invisible widths of the $Z$ and
Higgs bosons.

%%%%%%%%%%%%%%%%%%%%%%%%%
\subsection{Higgs and $Z$ boson invisible width}
\label{sec:invisible widths}
%%%%%%%%%%%%%%%%%%%%%%%%%%

We calculate the $Z$ boson decay width to neutrinos
$\Gamma(Z\to\nu\nu)$ at two loops in terms of SM nuisance parameters,
using a parametric formula from Ref.~\cite{Dubovyk:2018rlg}. To
calculate the invisible width, we add this width to the tree-level
decay width to the LSP, $\Gamma(Z\to \tilde\chi^0_1 \tilde\chi^0_1)$.
Indirect LEP measurements~\cite{PDB16} require that the invisible
width,
\begin{equation}
\Gamma(Z\to\text{inv.}) = 499.0 \pm 1.5 \,\text{MeV}.
\end{equation}
We use a Gaussian likelihood for this measurement, including in
quadrature a $10\%$ theoretical error in
$\Gamma(Z\to \tilde\chi^0_1 \tilde\chi^0_1)$ and an error
of $0.048 \,\text{MeV}$ accounting
for missing higher-order corrections in $\Gamma(Z\to\nu\nu)$~\cite{Dubovyk:2018rlg}.
This indirect measurement is stronger than constraints from monophoton
searches at
LEP near the $Z$ pole \cite{Adeva:1991rp,Buskulic:1993ke,Akers:1994vh,Acciarri:1998vf},
which we did not include.\footnote{In Sec.~\ref{sec:LEP} we do include a monophoton search from LEP, but for the production of invisibly decaying charginos at higher centre-of-mass energies.}

Higgs measurements at ATLAS, CMS and the Tevatron constrain the
invisible branching fraction of the Higgs,
$\text{BF}(h\to\text{inv.})$. Assuming SM-like couplings for the
Higgs, Ref.~\cite{Belanger:2013xza} found that a combination of such
measurements requires
\begin{equation}
\text{BF}(h\to \text{inv.}) \le 0.19,
\end{equation}
at $95\%$ confidence.  This combined limit remains stronger than more recent single-experiment limits (e.g.\ \cite{CMS-PAS-HIG-17-023}).  More recent combinations (e.g.\ \cite{Khachatryan:2016vau}) do not assume SM-like couplings, allowing all Higgs couplings to vary freely in their fits.  We employ the likelihood for the Higgs invisible branching fraction described in Ref.\ \cite{SDPBit}, based on the chi-squared as a function of
invisible branching fraction extracted from Ref.~\cite{Belanger:2013xza}.  Here we apply this likelihood to Higgs decays to the LSP,
$\text{BF}(h\to \tilde\chi^0_1\tilde\chi^0_1)$, bearing in mind that heavier neutralinos are unstable and therefore not invisible.

We calculate the decay widths to (all) charginos and neutralinos at
tree level \cite{Djouadi:2005gj}, and then add them to the decay width in the SM \cite{YellowBook13}
to estimate the total width of the Higgs in our simplified electroweakino scenario.  Because we consider
such a simplified scenario, we do not include one-loop corrections to
the decay widths to charginos or neutralinos. We therefore include a conservative $50\%$ log-normal
theory uncertainty on our prediction of the invisible branching fraction,
based on findings from one-loop calculations in
Ref.~\cite{Heinemeyer:2015pfa}.

%%%%%%%%%%%%%%%%%%%%%%%%%%%%%
\subsection{LEP searches for electroweakino production}
\label{sec:LEP}
%%%%%%%%%%%%%%%%%%%%%%%%%%%%%

Electroweakino production provides an excellent example of a case where
limits from the LEP experiment remain competitive with LHC searches,
particularly for light, degenerate spectra. The \colliderbit module
of \GB includes individual cross-section limits on the pair production
of neutralinos and charginos from the L3 and OPAL experiments,
expressed as a function of the sparticle masses. For each point in
the EWMSSM parameter space, we calculate the LEP pair-production
cross-sections for the processes given in Table~\ref{tab:SUSY_LEP},
and calculate the product of the cross-section and branching fraction for
each process (using the \decaybit interface
to \SUSYHIT \textsf{1.5}). These are then compared to digitised, and
interpolated, LEP cross-section limits from the analyses listed in
Table~\ref{tab:SUSY_LEP} to form a Gaussian likelihood term, as
described in~\cite{ColliderBit,CMSSM}. The likelihoods from each
channel and experiment are multiplied, on the assumption that they are
independent measurements.

\begin{table}[tb]
  \centering
  \begin{tabular}{lll}
   \toprule
    Production & Signature & Experiment  \\
    \midrule
    $\tilde\chi^0_i\tilde\chi^0_1$ & $\tilde\chi^0_i\to q\bar q\tilde\chi^0_1$  & OPAL~\cite{OPAL:gauginos}\\
    ($i=2,3,4$) &  $\tilde\chi^0_i\to \ell\bar \ell \tilde\chi^0_1$ & L3~\cite{L3:gauginos}\\
    $\tilde\chi^+_i\tilde\chi^-_i$ & $\tilde\chi^+_i\tilde\chi^-_i\to q\bar q'q\bar q'\tilde\chi^0_1 \tilde\chi^0_1$ & OPAL~\cite{OPAL:gauginos}\\
    ($i=1,2$)  & $\tilde\chi^+_i\tilde\chi^-_i\to q\bar q' \ell\nu\tilde\chi^0_1\tilde\chi^0_1$ & OPAL~\cite{OPAL:gauginos}\\
    & $\tilde\chi^+_i\tilde\chi^-_i\to \ell\nu \ell\nu\tilde\chi^0_1\tilde\chi^0_1$ & OPAL~\cite{OPAL:gauginos}, L3~\cite{L3:gauginos}\\
    & ISR $\gamma$ + missing energy & OPAL \cite{Abbiendi:2002vz} \\
    \bottomrule
  \end{tabular}
  \caption{Results from LEP on sparticle pair production used in the scans.}
  \label{tab:SUSY_LEP}
\end{table}

The selection of searches originally included in the \colliderbit
module are only sensitive down to electroweakino mass differences of 3\,GeV. We have therefore also included the OPAL search for a degenerate
chargino--neutralino pair \cite{Abbiendi:2002vz} in \colliderbit. This is
sensitive to mass differences from 320\,MeV to 5\,GeV, and is
important for constraining wino and Higgsino LSP scenarios from 45\,GeV
up to the kinematic limit of 95\,GeV.

The implementation follows that of the other electroweakino searches
from LEP: the pair-production cross-section of the (lightest) chargino is
calculated and compared to the digitised OPAL limit in the
plane of chargino mass versus chargino--neutralino mass difference to
find the likelihood contribution. This particular search does not rely on the decay of the chargino, because it is based on missing energy plus the emission of a photon as initial state radiation (ISR).

%%%%%%%%%%%%%%%%%%%%%%%%%%%%%
\subsection{LHC searches for electroweakino production}
%%%%%%%%%%%%%%%%%%%%%%%%%%%%%

\subsubsection{Analyses}
\label{sec:LHCanalyses}

There is a long list of searches for supersymmetry from the ATLAS and CMS
experiments of the LHC, conducted using proton--proton collision data
taken at $\sqrt{s}=7$, $8$ and $13$\,TeV. Searches for strongly-coupled
supersymmetric particles are conducted in final states with jets (including
$b$-jets), missing transverse energy $E_T^{\text{miss}}$ and/or some number of leptons, and are
specifically optimised on simplified models of gluino and squark
production. This includes dedicated searches for third generation
squark production. Models involving only chargino and neutralino
production are not expected to pass the stringent multiplicity and
kinematic selections required by these analyses.

\begin{figure*}
  \centering
  \includegraphics[width=0.25\textwidth]{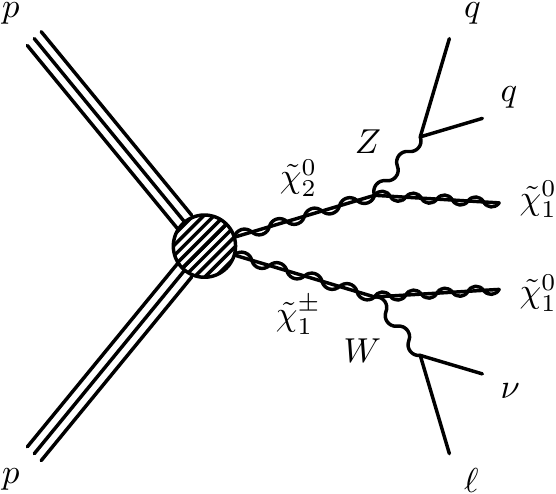}
  \hspace{1cm}
  \includegraphics[width=0.25\textwidth]{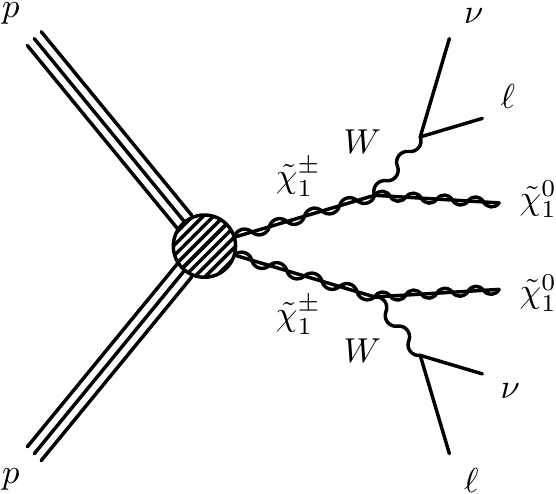}
  \hspace{1cm}
  \includegraphics[width=0.25\textwidth]{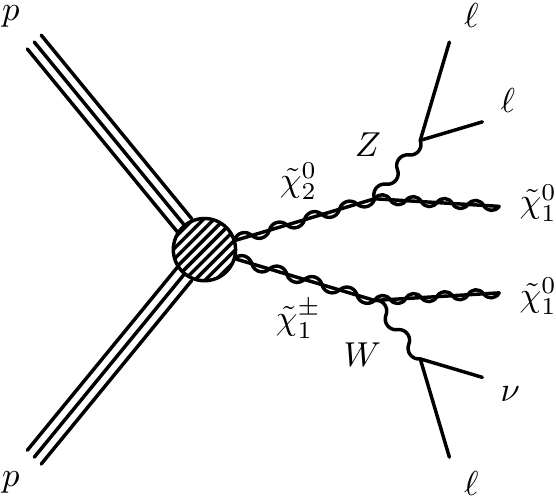}
  \caption{The simplified model used for the optimisation of ATLAS and CMS searches targeting on-shell $W$ and $Z$ production.
  Neutralinos and charginos are pair-produced, resulting in final states with leptons, jets and missing energy.
  }
  \label{fig:simplifiedModelWZ}
\end{figure*}

Searches for weakly-produced sparticles are generally challenging due
to the small production cross-sections, and the dominant constraints
come from final states rich in leptons, but relatively poor in jets. Searches are typically optimised on simplified models, with the most relevant model for our work shown in Figure~\ref{fig:simplifiedModelWZ}. This model assumes that $\tilde{\chi}_1^+ \tilde{\chi}_1^-$ and $\tilde{\chi}_1^\pm \tilde{\chi}_2^0$ production are the only available SUSY production processes at the LHC, and that the decay of the electroweakinos involves on-shell $W$ and $Z$ production. It is further assumed that the $\tilde{\chi}_2^0$ and $\tilde{\chi}_1^\pm$ are degenerate in mass and are wino-dominated, and that the $\tilde{\chi}_1^0$ is bino-dominated. This sets the production cross-sections for these processes, whilst ensuring that there are only two parameters remaining in the simplified model: $m_{\tilde{\chi}_1^\pm}$ (or, equivalently, $m_{\tilde{\chi}_2^0}$) and  $m_{\tilde{\chi}_1^0}$. Each analysis that uses the simplified model is then optimised by generating a grid of simulated signal events in the $(m_{\tilde{\chi}_1^\pm},m_{\tilde{\chi}_1^0})$ mass plane, and defining a number of signal regions that exploit differences between the expected kinematics of the signal and the expected kinematics of the dominant SM background processes for each region of the mass plane. Null observations are interpreted in terms of a 95\% confidence-level exclusion contour in the plane.

Many of the signal regions that we use in this paper are optimised for finding the simplified model of Figure~\ref{fig:simplifiedModelWZ}.  However, we also make use of signal regions optimised on (and interpreted using) an extension containing additional intermediate sleptons.  Despite not being obviously relevant to a model with decoupled sleptons, it is still possible for these regions to have \emph{some} sensitivity to the EWMSSM. We also use analyses that have been optimised on a number of other models, e.g.\ general gauge mediation, in case they have sensitivity to our model of interest; we explain below why one might expect this to be the case.

In the main section of this paper, we include only LHC analyses based on the full 36\,fb$^{-1}$ of data from Run II at $\sqrt{s}=13$\,TeV.  These are discussed below, and are far more sensitive than the earlier 7 or 8\,TeV results.  For the sake of completeness, in Appendix \ref{app} we also consider the relatively small impact of also including 8\,TeV data.

\noindent {\bf The ATLAS search for chargino and neutralino production in two- and three-lepton final states~\cite{Aaboud:2018jiw}:} This has search regions optimised in three channels. The two-lepton and zero-jets channel targets $\tilde{\chi}^+_1 \tilde{\chi}^-_1$ production and $\tilde{\ell}\tilde{\ell}$ production in signal regions with no jets, optimised using the dilepton invariant mass $m_{\ell\ell}$ and the ``stransverse mass'' $m_{T2}$ (see Table 1 of~\cite{Aaboud:2018jiw}, and note that we use the inclusive signal region definitions). The two-lepton and jets channel targets  $\tilde{\chi}^+_1 \tilde{\chi}^0_2$ production with decays via gauge bosons into two same-flavour, opposite-sign leptons (assumed to come from an on-shell $Z$ boson), and at least two jets (assumed to come from an on-shell $W$ boson). The signal regions in this case are split into dedicated categories for high, intermediate and low $\tilde{\chi}^+_1/\tilde{\chi}^0_2$--$\tilde{\chi}^0_1$ mass differences, and use a variety of variables including the dilepton invariant mass $m_{\ell\ell}$, the dijet invariant mass $m_{jj}$, the missing transverse energy, a list of angular distances, the $W$ and $Z$ boson transverse momenta  and $m_{T2}$ (see Table 2 of Ref.~\cite{Aaboud:2018jiw}). Finally, the three-lepton channel targets $\tilde{\chi}^+_1 \tilde{\chi}^0_2$ production with decays via intermediate $\tilde{\ell}$ or gauge bosons into final states with three leptons. The signal regions use the invariant mass of the same-flavour, opposite-sign lepton pair in the events, the missing transverse energy, the $p_T$ of the third lepton, the number of jets, the transverse mass, the $p_T$ of the three-lepton system, and the $p_T$ of the leading jet (see Table 4 of~\cite{Aaboud:2018jiw}). It is important to note in particular that the jet multiplicity in this analysis splits the $3\ell$ regions targeting on-shell $W$ and $Z$ production into a region with no jets, and a region with at least one jet. No significant excess was reported in any signal region, although there are modest excesses in some regions. The most significant of these has a local significance of 1.8$\sigma$, occurring in a region, {\sf SR3\_WZ\_1Jc}, that requires three leptons and at least one jet, along with a same-flavour, opposite-sign lepton pair with an invariant mass consistent with a $Z$ boson, $E_T^\text{miss}>200$\,GeV, and other kinematic cuts on the lepton and jet systems. Taken as whole, this analysis should be very sensitive to parts of the EWMSSM parameter space, with the most sensitivity occurring in the regions targeting $W$ and $Z$ production.

\noindent {\bf  The ATLAS search for chargino and neutralino production using recursive jigsaw reconstruction in final states with two or three leptons~\cite{Aaboud:2018sua}:} This analysis has four signal regions dedicated to high, intermediate, and low mass splittings, along with an ISR-initiated search region, in both the two- and three-lepton final states. The two-lepton and three-lepton regions select leptonic $Z$-boson decays, with hadronic $W$-boson decays being chosen for the former (via a cut on the dijet mass) and leptonic $W$-boson decays for the latter (via a transverse mass selection). Only minimal event selection is applied on object momenta and multiplicity criteria, with variables arising from the application of the recursive jigsaw reconstruction technique used instead~\cite{Jackson:2017gcy}. This provides so-called hemisphere variables that test the scale and balance of events using a specific \emph{decay tree} formulation designed to test whether a given event looks more signal- or background-like. The signal regions are constructed such that the low-mass and ISR regions in the two-lepton and three-lepton searches are non-overlapping. The ISR regions use a specific formulation of the recursive jigsaw method outlined in Ref.~\cite{Jackson:2016mfb}, which requires at least one hadronic jet associated with a strong ISR system, making the ISR regions orthogonal to the low-mass regions. The results were compatible with the SM background expectation in all signal regions targeting large and intermediate $\tilde{\chi}^\pm_1/\tilde{\chi}^0_2-\tilde{\chi}^0_1$ mass splittings (leading to the best exclusion limits to date in that mass range), but revealed excesses in four signal regions targeting low mass splittings, with local significances of 2.0, 3.0, 1.4 and 2.1$\sigma$.

\noindent {\bf The ATLAS search for pair production of Higgsinos in the $hh$ final state~\cite{Aaboud:2018htj}: } This consists of two separate analyses with 24.3 and 36.1 fb$^{-1}$, focused on light and heavy Higgsinos, respectively. The signature is in both cases four jets that are kinematically consistent with two SM Higgs boson candidates, with three or four $b$-jet tags present. This is sensitive to the pair production of two Higgsinos -- charged or neutral -- where any charged Higgsino decays to the neutral with very soft SM decay products,  and the resulting pair of neutral Higgsinos each decay to a Higgs boson and a light neutral sparticle. The search is motivated by gauge-mediated supersymmetry-breaking scenarios, where the light sparticle is a gravitino.  We include this search here because the light sparticle may just as well be a lighter neutralino. Each analysis has a large number of signal regions. For the low-mass search, ATLAS set exclusion limits on the basis of the two-dimensional distribution of events in a histogram with bins of missing energy $E_T^\text{miss}$ and effective mass $m_\text{eff}$.  We use all 42 bins from the original analysis as signal regions. Similarly, the high-mass search uses seven orthogonal signal regions, optimised for exclusion sensitivity. In addition to these exclusion-optimised signal regions, two discovery regions were defined for each analysis. Because of overlaps between the low-mass and high-mass signal regions, we have chosen to use only the low-mass signal regions in this study, so as to maximise the exclusion power in the most interesting (i.e.\ low-mass) region.

\noindent {\bf The ATLAS search for supersymmetry in final states with four or more leptons~\cite{Aaboud:2018zeb}:} This examined final states with four or more leptons, including up to two hadronically decaying taus. The search was optimised on simplified models of General Gauge-Mediated (GGM) SUSY breaking with $R$-parity conservation, and on simplified models with $R$-parity violation. However, the model dependence of the search was reduced by making the requirements on the effective mass and transverse missing momentum in the selected events fairly loose; these were applied along with a requirement of the presence or absence of a $Z$-boson candidate. This search should be sensitive to certain EWMSSM models through the production of multi-gauge-boson final states, which are capable of producing events with four leptons. Note that we here only include the search regions with at least four {\em light} leptons. The ATLAS results showed no significant excess in any of the signal regions, except for a modest one (2.3$\sigma$ local) in {\sf SR0D}, which required two $Z$ boson candidates and $E_T^\text{miss}>100$\,GeV.

\noindent {\bf The CMS search for chargino and neutralino production in the $Wh$ final state~\cite{CMS:2017fth}:} This was optimised on a simplified model that assumed $\tilde{\chi}_1^\pm \tilde{\chi}_2^0$ production, followed by the decays $\tilde{\chi}^\pm_1\rightarrow W^\pm \tilde{\chi}^0_1$ and $\tilde{\chi}^0_2\rightarrow h \tilde{\chi}^0_1$. Events were selected to have $E_T^\text{miss}>125$\,GeV, two $b$-jets with an invariant mass close to the Higgs boson mass, a transverse mass of the lepton-$E_T^\text{miss}$ system greater than 150\,GeV, and a ``contranverse mass'' $M_{CT}>170$\,GeV~\cite{Tovey:2008ui,Polesello:2009rn}. No significant excess was reported in the two signal regions, which were defined using different bins of $E_T^\text{miss}$. This analysis should be sensitive to the EWMSSM, which is more than capable of producing $Wh$ final states.

\noindent {\bf The CMS search for degenerate charginos and neutralinos in final states with two low-momentum opposite-sign leptons~\cite{Sirunyan:2018iwl}:} This search targets $\tilde{\chi}_1^\pm \tilde{\chi}_2^0$ production with a mass-degenerate $\tilde{\chi}_1^\pm$ and $\tilde{\chi}_2^0$ that are assumed to decay to the $\tilde{\chi}_1^0$ via virtual $W$ and $Z$ bosons (note that there are also search regions defined for stop squark pair production, which we ignore). The results were optimised on and interpreted in two variants of the $\tilde{\chi}_1^\pm \tilde{\chi}_2^0$ simplified model, in which the $\tilde{\chi}_1^\pm$ and $\tilde{\chi}_2^0$ are either wino-dominated or Higgsino-dominated. A second Higgsino model considers $\tilde{\chi}_1^0 \tilde{\chi}_2^0$ production, where the mass of the chargino is set to $m_{\tilde{\chi}_1^\pm}=(m_{\tilde{\chi}_2^0}+m_{\tilde{\chi}_1^0})/2$. The selected events have two opposite-sign leptons and at least one jet. A pre-selection includes requirements that the transverse mass of both lepton-$E_T^\text{miss}$ combinations is less than 70\,GeV, that the $E_T^\text{miss}$ is greater than 125\,GeV, that the dilepton invariant mass must be less than 50\,GeV, and that the lepton transverse momenta must be less than 30\,GeV. Thus, this analysis would be sensitive to off-shell gauge boson production in the EWMSSM in cases of compressed mass spectra, but would rapidly lose sensitivity to on-shell production. Signal regions are defined in bins of $E_T^\text{miss}$ and $m_{\ell\ell}$, and we use the simplified composite likelihood treatment to combine the bins as described in Sec.~\ref{sec:simplified likelihoods}.

\noindent {\bf The CMS search in states with jets and two opposite-sign same-flavour leptons~\cite{Sirunyan:2017qaj}:} This analysis uses the invariant mass of the lepton pair, searching for a kinematic edge or a resonant-like excess compatible with the $Z$-boson mass. We deal with the latter search only, since the former is designed to target strong sparticle production. The electroweakino search was optimised on the wino-dominated $\tilde{\chi}_1^\pm \tilde{\chi}_2^0$ production model shown in Figure~\ref{fig:simplifiedModelWZ}, and a second model based on gauge-mediated SUSY breaking. In the electroweakino search, selected events are required to have a dilepton invariant mass close to the $Z$-boson mass, at least two jets, and a missing transverse energy in excess of 100\,GeV. Multiple signal regions are defined with bins of the dijet mass, $M_{T2}$ and $E_T^\text{miss}$. Regions with two $b$-jets are also defined, in order to target $hZ$ final states. We use the simplified composite likelihood treatment to combine the bins as described in Sec.~\ref{sec:simplified likelihoods}. This search should be very sensitive to models in the EWMSSM.

\noindent {\bf The CMS search for chargino and neutralino production in final states with two or three leptons~\cite{CMS-PAS-SUS-16-039}:} This search targeted various scenarios of direct $\tilde{\chi}_1^\pm \tilde{\chi}_2^0$ production, with a wino-dominated $\tilde{\chi}_1^\pm$ and $\tilde{\chi}_2^0$. One set of simplified models included light sleptons, whilst the other was essentially that shown in Figure~\ref{fig:simplifiedModelWZ}, but with an extra model in which the $\tilde{\chi}_2^0$ produces an $h$ boson rather than a $Z$ boson. CMS searched events with two same-sign light leptons, in which they binned the events in the transverse mass, the transverse momentum of the dilepton system, and the $E_T^\text{miss}$, for a total of 30 bins. They also performed a three-lepton search using bins of the transverse mass, $E_T^\text{miss}$, and the dilepton invariant mass, with 44 bins defined for the case where two of the leptons form an opposite-sign, same-flavour pair, and six additional regions defined for the opposite case. Further regions were defined for the case where there was at least one hadronically-decaying tau. To facilitate reinterpretation of the results, they defined aggregated signal regions (i.e.\ signal regions with a wider selection on the kinematic properties than the single bins), most of which require a missing transverse energy of at least 200\,GeV. We provide a thorough discussion of the difference between using the aggregated signal regions and the full set of bins below.

\phantom{A hidden paragraph.}

Additionally, in test scans, we investigated the impact of the CMS monojet analysis, which may be sensitive to $\tilde{\chi}_1^0\tilde{\chi}_1^0$ production~\cite{Sirunyan:2017jix}. We found that this had no sensitivity in any region of the parameter space. This matches the naive expectation based on the literature, so we exclude this analysis from our final results.

% For reference, here are the ColliderBit names for the above analyses:\tabularnewline
% - ATLAS_13TeV_MultiLEP
% - ATLAS_13TeV_RJ3L
% - ATLAS_13TeV_3b
% - ATLAS_13TeV_4LEP
% - CMS_13TeV_1LEPbb
% - CMS_13TeV_2LEPsoft
% - CMS_13TeV_2OSLEP
% - CMS_13TeV_MultiLEP
% - Not included: CMS_13TeV_MONOJET_36invfb

\begin{table*}[h!]
  \centering
  \begin{tabular}{ll}
   \toprule
    Likelihood label & Source \\
    \midrule
    \textsf{ATLAS\_4b} & ATLAS Higgsino search~\cite{Aaboud:2018htj} \\
    \textsf{ATLAS\_4lep} & ATLAS $4\ell$ search~\cite{Aaboud:2018zeb} \\
    \textsf{ATLAS\_MultiLep\_2lep\_0jet} & ATLAS multilepton EW search~\cite{Aaboud:2018jiw} \\
    \textsf{ATLAS\_MultiLep\_2lep\_jet} & ATLAS multilepton EW search~\cite{Aaboud:2018jiw} \\
    \textsf{ATLAS\_MultiLep\_3lep} & ATLAS multilepton EW search~\cite{Aaboud:2018jiw} \\
    \textsf{ATLAS\_RJ\_2lep\_2jet} & ATLAS recursive jigsaw EW search~\cite{Aaboud:2018sua} \\
    \textsf{ATLAS\_RJ\_3lep} & ATLAS recursive jigsaw EW search~\cite{Aaboud:2018sua} \\
    \textsf{CMS\_1lep\_2b} & CMS $Wh$ search~\cite{CMS:2017fth} \\
    \textsf{CMS\_2lep\_soft} & CMS 2 soft opposite-charge lepton search~\cite{Sirunyan:2018iwl} \\
    \textsf{CMS\_2OSlep} & CMS 2 opposite-charge lepton search~\cite{Sirunyan:2017qaj} \\
    \textsf{CMS\_MultiLep\_2SSlep} & CMS multilepton EW search~\cite{CMS-PAS-SUS-16-039} \\
    \textsf{CMS\_MultiLep\_3lep} & CMS multilepton EW search~\cite{CMS-PAS-SUS-16-039} \\
    \bottomrule
  \end{tabular}
  \caption{Labels for the independent likelihood terms included in our LHC likelihood, along with the analyses from which they are derived.}
  \label{tab:lhclike}
\end{table*}

A typical LHC search includes quantifying the impact of a long list of systematic uncertainties, including those related to the jet energy scale and resolution, lepton identification and reconstruction, trigger efficiency, $b$-tagging, MC modelling (such as the choice of renormalisation and factorisation scales, plus uncertainties related to the choice of parton distribution function), pileup modelling, and particle production cross-sections. These are often correlated across signal regions, and this must be taken into account in determining the likelihood of a SUSY model given the observed data and expected SM background contribution. In addition, for searches with non-orthogonal signal region selections, there will be a correlated number of events in overlapping regions.

For most of the analyses that we use, no detailed information is provided by the experiments regarding the correlation of event numbers and uncertainties between the different signal regions (the exceptions will be discussed below). Best practise in this case is to take the signal region expected to give the highest exclusion power for a given point in the SUSY parameter space, and use that region to calculate a likelihood contribution using the observed LHC data. In previous \GB studies \cite{CMSSM,MSSM,ColliderBit}, our approach has been to select a single such ``best expected'' signal region across those contained in a given paper, for each point in the SUSY parameter space. However, the division of experimental results into different papers does not always make this a sensible procedure, given that several papers summarise the results of multiple analyses that are thematically similar, but actually orthogonal from the point of view of selecting events. Therefore, in this study, we instead divide the signal regions by final state, and assume that the ``best expected'' region in each final state can be used to obtain a likelihood contribution independently of other final states (and, of course, a final state in the ATLAS data yields an independent likelihood term from the same final state in the CMS data). This gives a series of independent likelihood terms whose origin is summarised in Table~\ref{tab:lhclike}.

A possible flaw in this approach is the inclusion of two recent ATLAS searches
for two- and three-lepton final states (\cite{Aaboud:2018jiw} and~\cite{Aaboud:2018sua})
as independent contributions in our scan likelihood function. In this case, however, ATLAS have published plots showing that the overlap in the selected events for the two analyses is small, and we have performed our own checks that our final conclusions do not change substantially when the ATLAS recursive jigsaw electroweak (EW) analysis is supplemented by the earlier analysis that uses conventional variables.

We have added all the above searches to the \colliderbit module in \GB.  \colliderbit implements
LHC constraints by performing a Monte Carlo (MC) simulation of sparticle
production at the 13\,TeV LHC for each point in the parameter space
(using the \pythiaeight MC generator~\cite{Sjostrand:2006za,Sjostrand:2014zea}), before passing the events
through a custom fast detector parameterisation of the ATLAS and CMS detectors, and an
implementation of the relevant analysis cuts. This gives the expected
yield of signal events in each analysis which, for most analyses, is
used to define a Poisson likelihood term marginalised over
statistical and systematic uncertainties, based on the signal region
with the best expected exclusion power. Further details can be found
in~\cite{ColliderBit,CMSSM}. The likelihoods for different analyses
are treated as independent, and are multiplied together. In the above
analyses, we have implemented new efficiencies for leptons and
$b$-jets in certain analyses, in order to better reproduce the
published cutflows.

A potential weakness in our approach is that we use leading order (LO) cross-sections plus leading logarithmic (LL) corrections from \pythia, due to the prohibitive computational cost of next-to-leading order (NLO) and next-to-leading logarithmic (NLL) calculations. We return to the expected effect of this approximation in our results discussion.

\subsubsection{Validation}
\label{sec:LHCvalidation}

Example cut-flows are shown in
Tables~\ref{tab:cutflow_3b}--\ref{tab:cutflow_2_OSSF_L}, for
the ATLAS search for two Higgs bosons and $E_T^\text{miss}$~\cite{Aaboud:2018htj},
the CMS two low-momentum
opposite-sign leptons and $E_T^\text{miss}$ search~\cite{Sirunyan:2018iwl},
and the CMS two opposite-sign same-flavour leptons and $E_T^\text{miss}$ search~\cite{Sirunyan:2017qaj}. The agreement is in general good, rising to a maximum discrepancy of $\sim$40\%
in the worst case.

\begin{table}
\begin{center}
\begin{tabularx}{\linewidth}{Lrrr}
\toprule
Cut & ATLAS & \gambit & Ratio\\
\midrule
All events								& 14028	& 14028	& 1.00 \\
Trigger, 4 jets         	& 1455	& 1906	& 1.31 \\
\ \ ($p_T > 40$\,GeV, 2 $b$-tags) & & & \\
$\ge 4$ $b$-tags						& 163.0	& 161.0	& 0.99 \\
$\ge 2$ Higgses						& 126.4	& 140.8	& 1.11 \\
Lepton veto							& 126.1	& 140.3	& 1.11 \\
$X_{Wt} > 1.8$							& 108.4	& 132.8	& 1.23 \\
$X_{hh}^{SR} < 1.6$						& 53.4	& 52.47	& 0.98 \\
\midrule
SR1: $m_\text{meff}>440$\,GeV				& 37.0	& 43.58	& 1.18 \\
SR2: $m_\text{meff}>440$\,GeV +     & 14.2	& 16.27	& 1.15 \\
\ \ $E_T^\text{miss}>150$\,GeV & &       & \\
\bottomrule
\end{tabularx}
\caption{Example comparison of \gambit and \atlas \cite{Aaboud:2018htj} cutflows for two signal regions targeting low-mass Higgsinos in a search for new physics in events with two Higgs bosons decaying
  into $\bar b b$. Shown are the numbers of events expected in 24.3\,\invfb of 13\,\TeV \atlas data for Higgsino pair production with a signal cross-section of 0.577 pb, $m_{\tilde{H}}=250$\,GeV and a massless gravitino, assuming 100\% branching fraction for $\tilde H\to h\tilde G$.
  }
\label{tab:cutflow_3b}
\end{center}
\end{table}

\begin{table}
\begin{center}
\begin{tabularx}{\linewidth}{Lrrr}
\toprule
Cut & CMS & \gambit & Ratio\\
\midrule
All events                              					& 172000	& 172000	& 1.00 \\
2 reconstructed muons with	& 1250	& 1212	& 0.97 \\
\ \ $5 < p_{T} < 30$\,GeV & & & \\
muons oppositely charged                       			& 1200	& 1099	& 0.91 \\
$p_{T}(\mu\mu) > 3$\,GeV     					& 1176	& 1067	& 0.97 \\
$M(\mu\mu) \in [4,50]$\,GeV                      			& 1095 	& 1062	& 1.02 \\
$M(\mu\mu) \in [9,10.5]$\,GeV veto     			& 988.5 	& 1011	& 0.99 \\
$125 < p^\text{miss}_{T} < 200$\,GeV			& 46.8 	& 46.4	& 0.98 \\
Trigger efficiency							& 30.7  	& 30.2	& 1.07 \\
ISR jet									& 27.9  	& 29.9	& 1.17 \\
$H_{T} > 100$\,GeV							& 23.6  	& 27.7 	& 1.40 \\
$0.6 < p^\text{miss}_{T}/H_{T} < 1.4$			& 17.2 	& 24.0	& 1.42 \\
$b$-tag veto								& 14.0	& 19.8	& 1.25 \\
$M(\tau\tau)$ veto							& 12.3	& 15.4	& 1.25\\
$M_{T}(\mu_{x},p^\text{miss}_{T}) < 70$\,GeV		& 9.3		& 10.3	& 1.11\\
\bottomrule
\end{tabularx}
\caption{Comparison of the \gambit and \cms \cite{Sirunyan:2018iwl} cutflows for a $WZ$ signal model ($m_{\tilde{\chi}_1^{\pm}}=150$\,GeV, $m_{\tilde{\chi}_1^{0}}=130$\,GeV) in
  a search for new physics in events with two low-momentum opposite-sign leptons and missing transverse momentum. Shown are the numbers of events expected in 33.2\,\invfb of 13\,\TeV \cms data for a signal cross-section of 5.18 pb~\cite{SUS-16-048_cutflow}. Both the \cms cutflow and \gambit cutflow are generated for production of $\tilde{\chi}_1^{\pm}\tilde{\chi}_2^{0}$ in a simplified model with decays via off-shell $W/Z$.
  }
\label{tab:cutflow_2_soft_L}
\end{center}
\end{table}

\begin{table}%[h!]
\begin{center}
\begin{tabularx}{\linewidth}{Lrrr}
\toprule
Cut & \cms & \gambit & Ratio\\
\midrule
All events                              & 109.35& 1084.18 & 9.91 \\
2 SFOS leptons                          & 24.21 & 30.00   & 1.24 \\
Extra lepton vetoes                      & 18.37 & 25.07   & 1.36 \\
$m_{\ell\ell} \in [86,96]$\,GeV          & 14.13 & 15.97   & 1.13 \\
2-3 Jets                                & 11.98 & 9.83    & 0.82 \\
$\Delta\Phi(E_T^\text{miss},j_{1,2}) > 0.4$  & 10.95 & 9.07    & 0.83 \\
B-tag veto                              & 9.92  & 8.86    & 0.89 \\
$M_{\rm T2}(\ell\ell)>80$\,GeV           & 8.04  & 7.27    & 0.90 \\
$M_{\ell\ell}<150$\,GeV                  & 5.62  & 5.26    & 0.94 \\
\midrule
SR1: $E_T^\text{miss}>100$\,GeV          & 5.41 & 5.05 & 0.93 \\
SR2: $E_T^\text{miss}>150$\,GeV          & 4.96 & 4.76 & 0.96 \\
SR3: $E_T^\text{miss}>250$\,GeV          & 3.59 & 3.49 & 0.97 \\
SR4: $E_T^\text{miss}>350$\,GeV          & 1.94 & 1.95 & 0.96 \\
\bottomrule
\end{tabularx}
\caption{Comparison of the \gambit and published \cms cutflows \cite{Sirunyan:2017qaj} in four signal regions of a search for new physics in events with two
  opposite-charge same-flavor leptons and missing transverse momentum, for a $WZ$ signal model ($m_{\tilde{\chi}_1^{\pm}}=550$\,GeV, $m_{\tilde{\chi}_1^{0}}=200$\,GeV).  Shown are the numbers of
  events expected in 35.9\,\invfb of 13\,\TeV \cms data, and the ratio of the
  \gambit and \cms numbers. Note that
  the \cms cutflow is generated for a $\tilde{\chi}_1^{\pm}\tilde{\chi}_2^{0}$ simplified model decaying via $W/Z$ where the $Z$ boson decays leptonically, while the \gambit cutflow is generated without specifying $Z$ boson decay mode. This explains the discrepancy at the ``All events'' cut.}
    \label{tab:cutflow_2_OSSF_L}
\end{center}
\end{table}

\begin{figure*}
  \centering
  \includegraphics[width=0.48\textwidth]{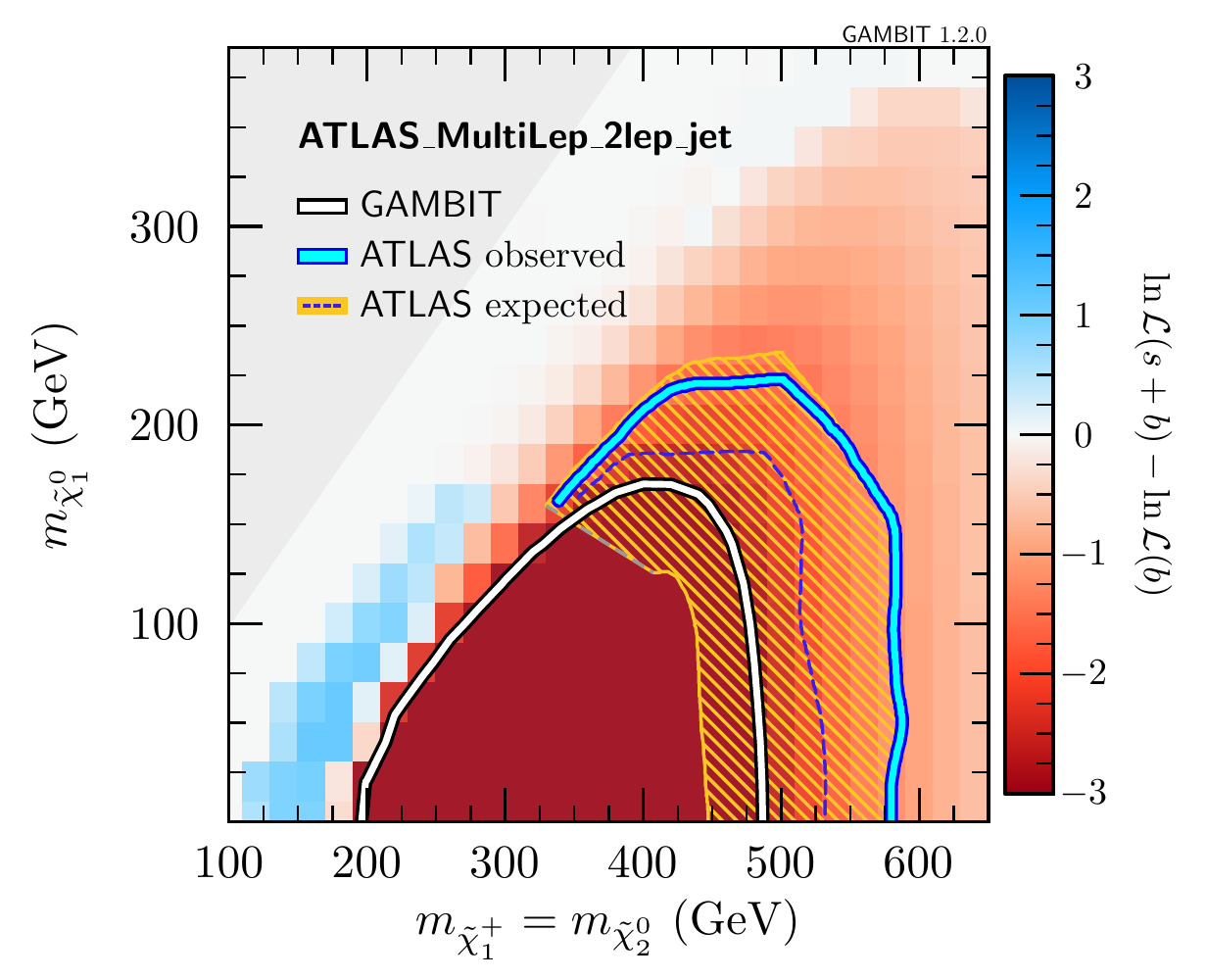}
  \includegraphics[width=0.48\textwidth]{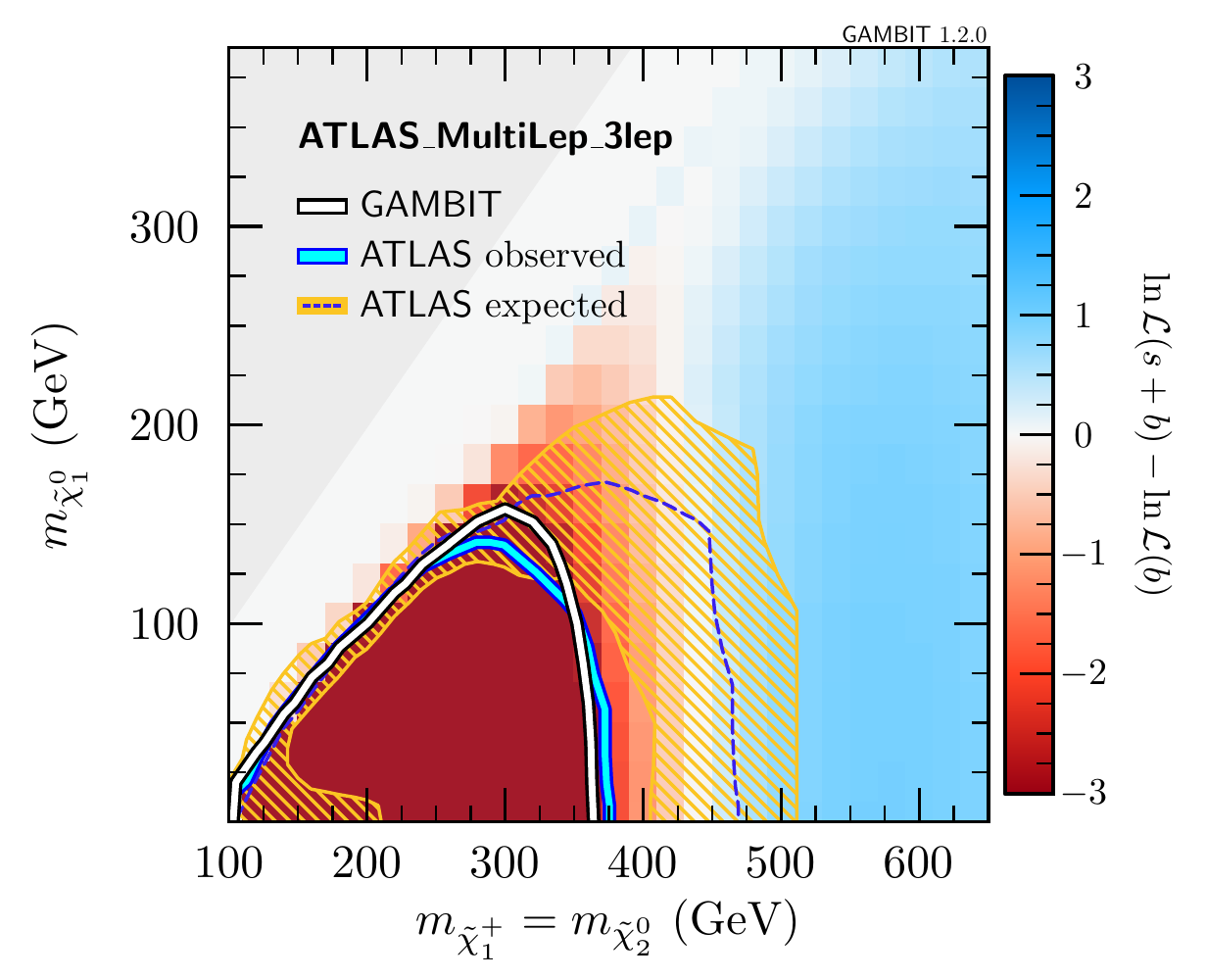}\\
  \includegraphics[width=0.48\textwidth]{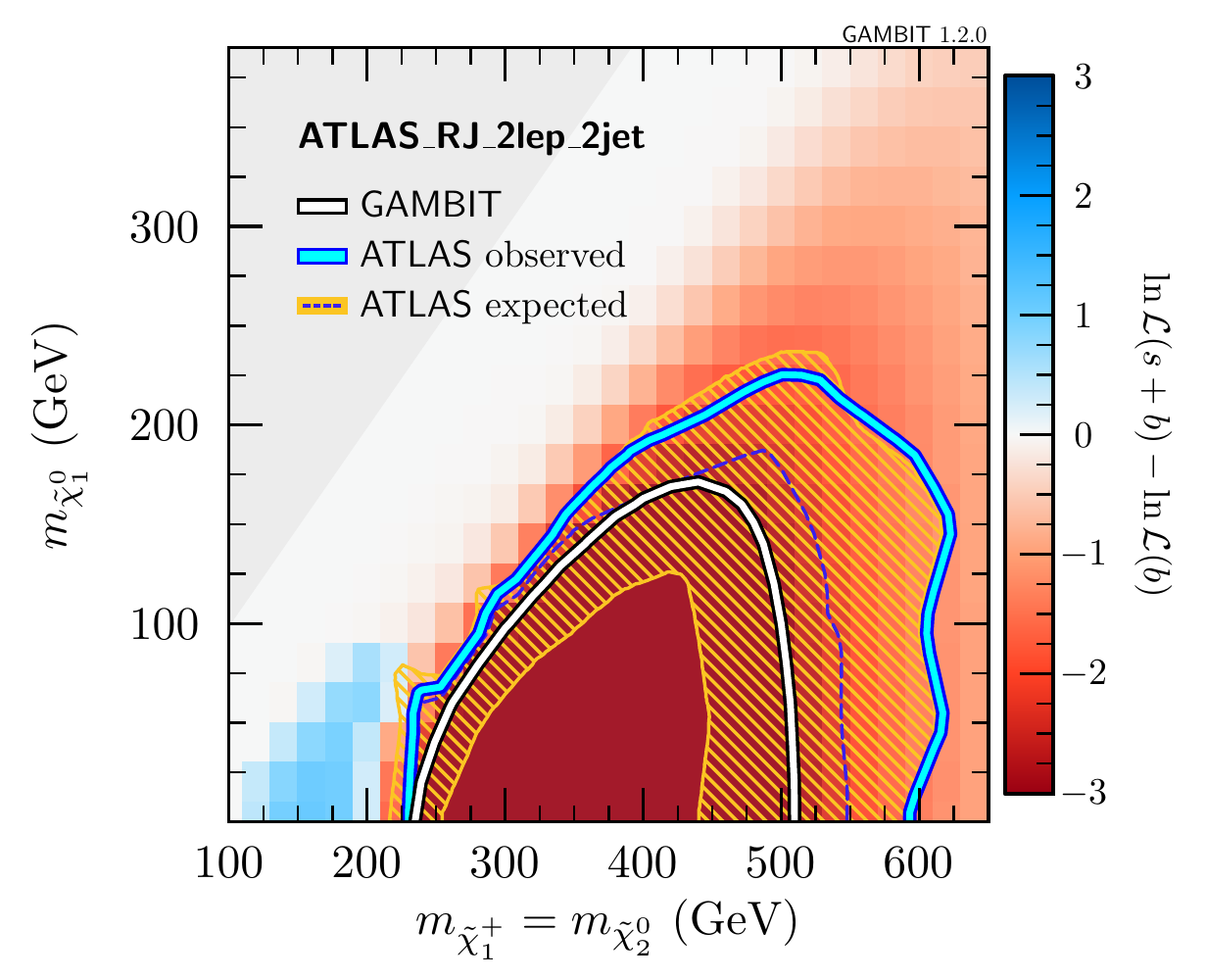}
  \includegraphics[width=0.48\textwidth]{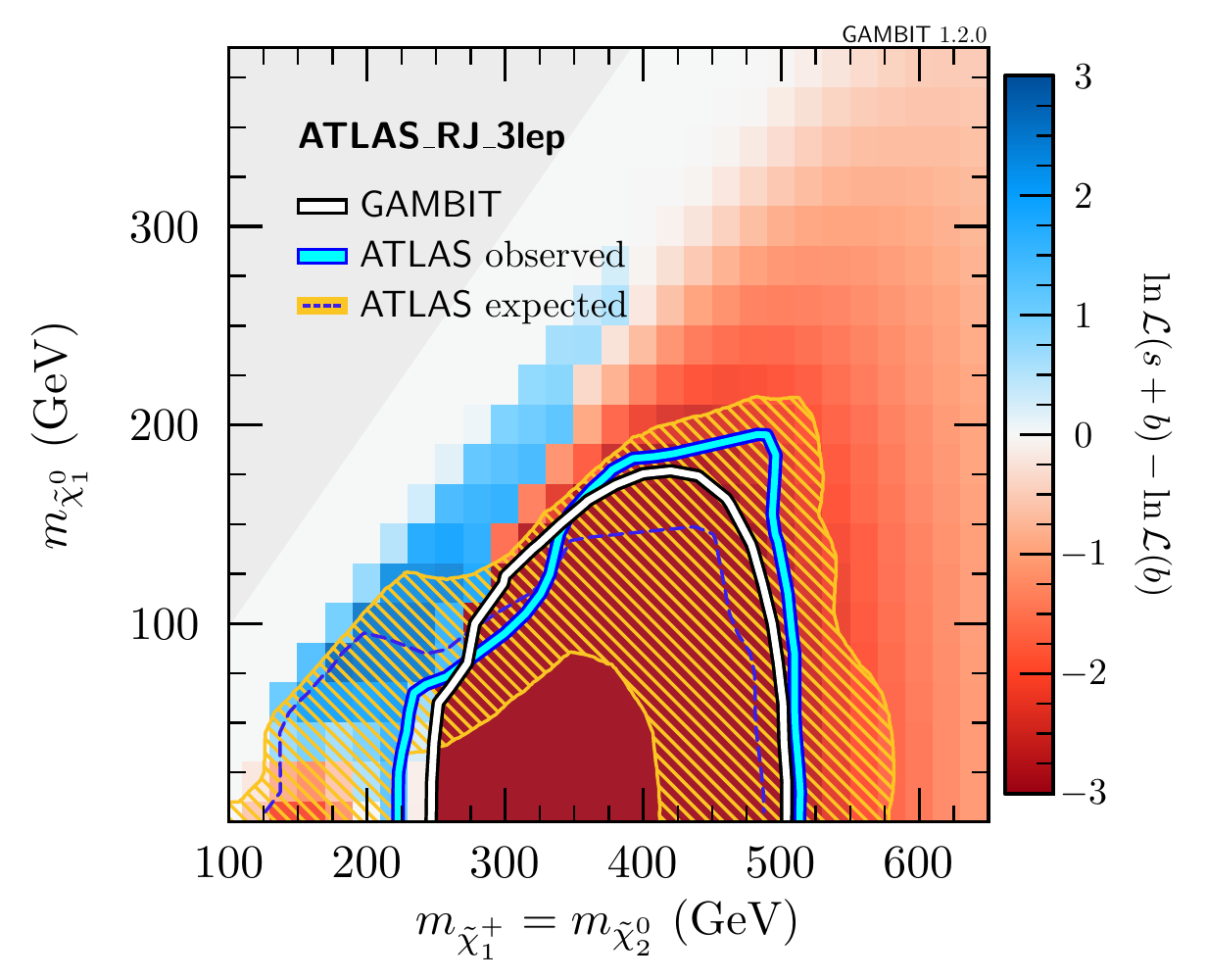}\\
  \caption{\gambit reproductions of 95\% CL ATLAS exclusion limits for a simplified model of wino production. The results for the ``conventional'' multilepton analysis~\cite{Aaboud:2018jiw} and the recursive jigsaw analysis~\cite{Aaboud:2018sua} are shown in the top and bottom rows, respectively. In both cases results are given separately for 2-lepton (left) and 3-lepton (right) signal regions. The ATLAS observed (light blue) and expected (dashed, dark blue) limits, along with the $\pm 1\sigma$ uncertainty band (hatched, yellow) on the expected limit, are obtained from the published auxiliary materials~\cite{Aaboud:2018jiw_aux,Aaboud:2018sua_aux}. The signal predictions from \gambit have been scaled to the NLO+NLL cross-sections for wino pair production~\cite{SUSYCrossSections13TeVn2x1wino}. The underlying heatmap depicts the full log-likelihood function obtained from the \gambit simulations.}
  \label{fig:limitComparison}
\end{figure*}

To provide further validation,
Figure~\ref{fig:limitComparison} displays a \gambit version of the
exclusion limit in the $(m_{\tilde{\chi}_1^\pm},m_{\tilde{\chi}_1^0})$
mass plane arising from the conventional ATLAS multilepton analysis~\cite{Aaboud:2018jiw},
and the ATLAS RJ analysis~\cite{Aaboud:2018sua}, for a simplified model in which production
of the wino-dominated $\tilde{\chi}_1^\pm$ and $\tilde{\chi}_1^0$ is
followed by decays to $W$ and $Z$ gauge bosons and neutralinos.
For these reproductions we have scaled the signal predictions from \gambit using the NLO+NLL cross-sections for wino pair production taken from~\cite{SUSYCrossSections13TeVn2x1wino}.
We see that the overall agreement is good, particularly at low masses. Some differences exist for heavy $\tilde{\chi}_2^0$ (and $\tilde{\chi}_1^\pm$) in the two-lepton searches, however, this is not so surprising given the low number of signal events in this area, which makes the exclusion limit very sensitive to small details of the analysis. Despite this, the agreement indicates
that our implementations of these particular analyses are capable of supplying a similar exclusion to that reported by ATLAS when used on the same simplified model.

We note that it is difficult to reproduce the reported
exclusion from the equivalent CMS multilepton analysis in the simplified model defined in that analysis~\cite{CMS-PAS-SUS-16-039}, as that limit is obtained
using a combination of many bins for which covariance information is
not supplied. For this analysis, we use the aggregated signal regions
defined in the original version of the analysis in
Ref.~\cite{CMS-PAS-SUS-16-039}. These are recommended for
reinterpretation purposes by the CMS collaboration, on the grounds
that the aggregated region with the best-expected exclusion should
be more constraining than the single bin with the best expected
exclusion in the multibin analysis, i.e. the extra power of the
multibin analysis comes from the combination of bins, not the
individual bins.

Another reason for making this choice is that taking the single bin with the best expected sensitivity is not very robust against statistical fluctuations, both in the original data and MC fluctuations in the signal evaluation.  This is because in the full combination of bins, a bad fit to the data in one bin can be compensated for by a sufficiently good fit to the data in other bins.

We have compared the result obtained with the
aggregated regions to a naive sum of the log-likelihoods for all bins
used in the CMS exclusion limit derivation for this analysis, and find that we get very
large differences for simplified model points that are well within the
CMS exclusion contour. Whilst these differences may be mitigated by
the use of the relevant covariance information, it is impossible to
quantify the size of this effect without access to that
information. This is therefore a case where best practise does not
allow us to fully estimate the likelihood of the CMS search, and we
will revisit this point in the final presentation of our results.

A common theme in the included electroweak searches is the requirement of one or more ISR jets to isolate the signal. Given that our simulation with \pythia, unlike the signal description in the original ATLAS and CMS analysis, does not include extra hard jets in the matrix element description, the efficiency of the signal in our simulation should be smaller. This is to some degree borne out in the cut-flow
shown in Table~\ref{tab:cutflow_2_OSSF_L}, but not in Table~\ref{tab:cutflow_2_soft_L}.

\begin{figure*}[h!]
  \centering
  \includegraphics[height=0.85\columnwidth]{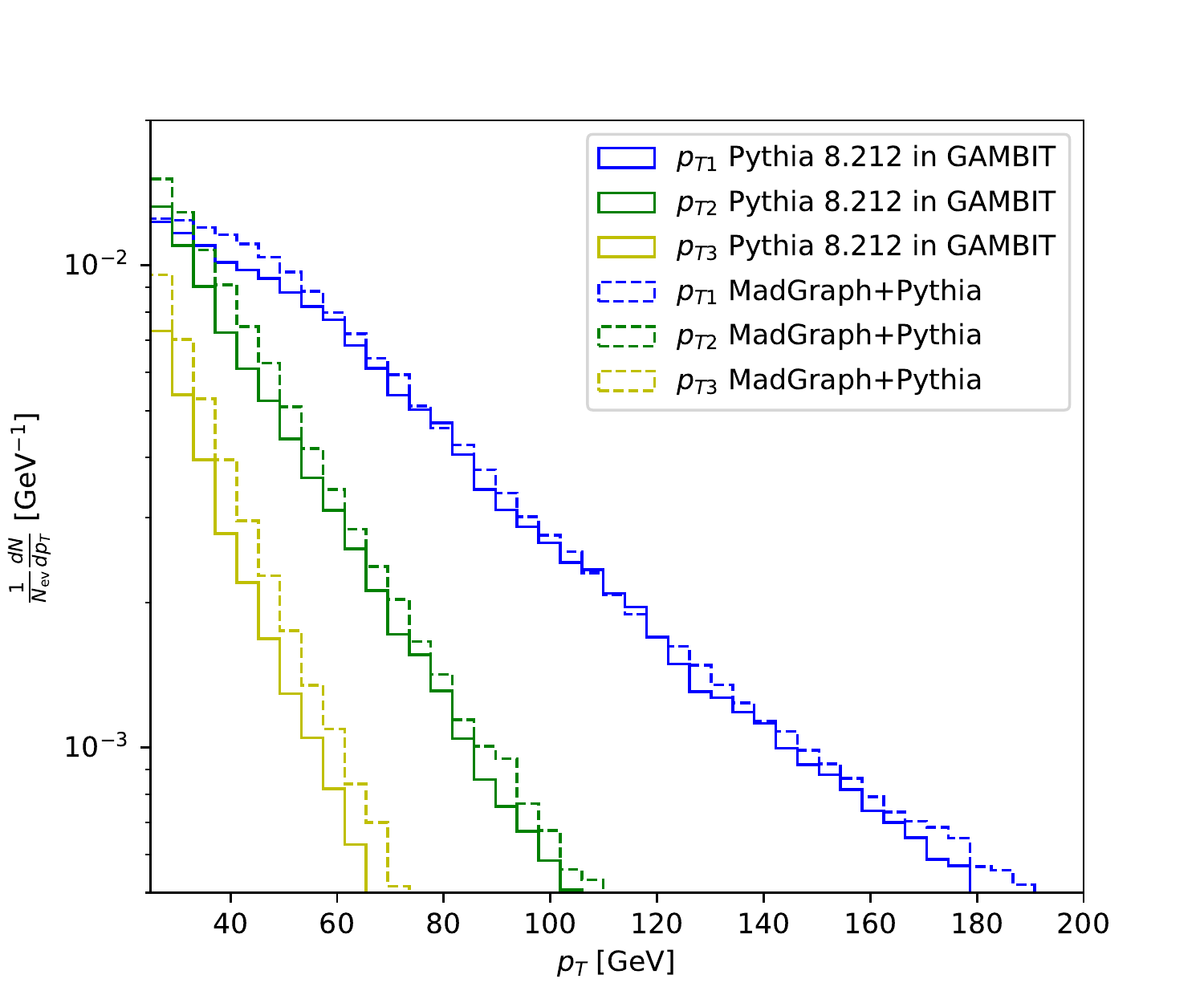}
  \includegraphics[height=0.85\columnwidth]{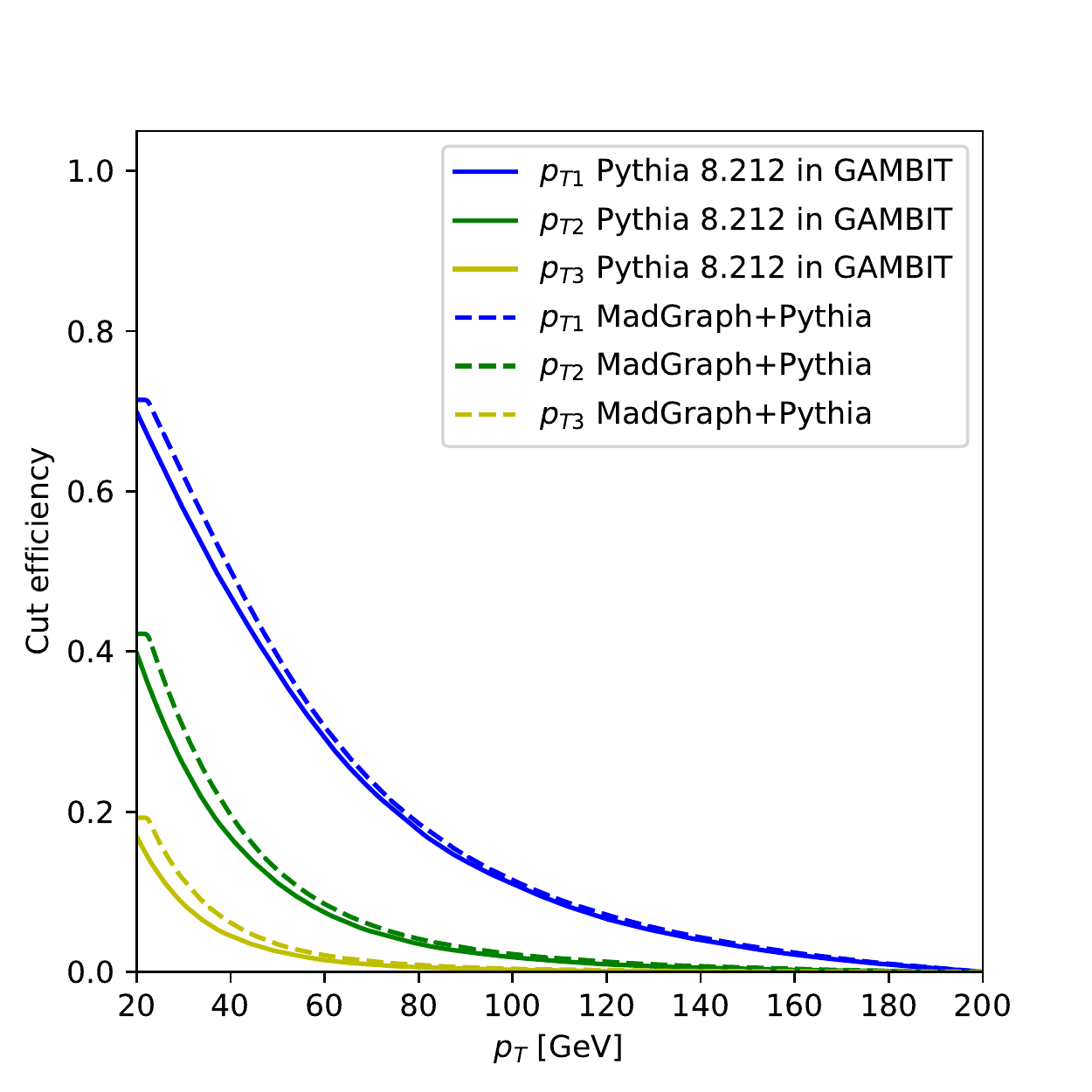}
  \caption{$p_T$ distributions (left) for the three hardest jets in a benchmark model with production of chargino--neutralino pairs with $m_{\tilde{\chi}_2^0,\tilde{\chi}_1^{\pm}}=200$\,GeV, as well as the corresponding cut-efficiencies (right).}
  \label{fig:ISR_comp}
\end{figure*}

In Fig.\ \ref{fig:ISR_comp} (left) we show the $p_T$ distributions for the three hardest jets in signal events simulated in \gambit with \pythia \textsf{8.212}, compared to a simulation using \MGaMCNLO~\cite{Alwall:2011uj,Alwall:2014hca} and \pythia with a matching procedure including up to two extra hard jets in the matrix element, which copies the signal simulation used by the experiments. The chosen benchmark point features production of wino-dominated $\tilde{\chi}_2^0\tilde{\chi}_1^{\pm}$ pairs with  $m_{\tilde{\chi}_2^0,\tilde{\chi}_1^{\pm}}=200$\,GeV, which decay into a bino $\tilde{\chi}_1^0$ with $m_{\tilde{\chi}_1^{0}}=100$\,GeV and vector bosons with 100\% branching fraction. The  vector bosons in turn decay leptonically. The latter choice maximises any difference between the simulations as there are no extra jets from hadronic decays of the vector bosons. In both cases, jets are reconstructed using the anti-$k_T$ algorithm with $R=0.4$~\cite{Cacciari:2008gp}, as implemented in \fastjet~\cite{Cacciari:2011ma}. For the \gambit sample, we reconstruct jets and apply jet energy smearing and lepton isolation criteria using the \buckfast \cite{ColliderBit} detector output. For the \madgraph sample, we use the \delphes~\cite{deFavereau:2013fsa} simulation package.

The relatively small differences between the jet spectra in Fig.\ \ref{fig:ISR_comp}, in particular for the hardest jet, show that our simulation of signal events provides reasonable fidelity. Together with the existence of extra jets in hadronic vector boson decays this also explains the small  or absent decrease in efficiency observed for the jet cut in Tables~\ref{tab:cutflow_2_soft_L} and~\ref{tab:cutflow_2_OSSF_L}. While this result may seem somewhat surprising, it has been noted before that the ISR shower together with the implemented matrix element corrections in \pythia do quite well up to $p_T^\text{jet}\sim \mu_F/2$, where $\mu_F=\sqrt{p_T^2+{\hat m}^2}$ is the factorization scale used (given in terms of the $p_T$ of the produced sparticles and their average mass $\hat m$ \cite{Plehn:2005cq}).

In the final results this lower efficiency should result in a small systematic shift of the highest likelihoods towards lower masses with higher cross-sections in order to compensate. In Sec.\ \ref{sec:benchmark} we include this effect in the kinematical distributions for the best-fit point by running the same simulation as above with up to two extra hard jets in the matrix element.

\subsubsection{Simplified likelihoods}
\label{sec:simplified likelihoods}

Without correlation information, the conservative approach to likelihood
construction from multiple signal regions is to choose the \emph{single} signal region with the highest
expected signal significance for each model point.  This is the approach that we took in
earlier \gambit papers~\cite{MSSM,CMSSM,ColliderBit}, and is discussed above in Sec.~\ref{sec:LHCanalyses}.

As a result, what we refer to as the likelihood from a given LHC analysis, $\mathcal{L}_i$, is in fact a ratio between the signal-plus-background and the background-only likelihoods,
\begin{equation}
 \label{like_ratio}
 \mathcal{L}_i = \frac{\mathcal{L}_\text{marg}(n_i|s_i+b_i)}{\mathcal{L}_\text{marg}(n_i|b_i)},
\end{equation}
where $n_i$, $s_i$ and $b_i$ respectively refer to the number of events measured, predicted due to signal, and predicted due to background, in this expected best signal region.  We divide the signal-plus-background likelihood by the background likelihood in order to avoid the large likelihood normalization changes from point to point in parameter space that would otherwise occur when switching between signal regions. The total LHC likelihood from \colliderbit is then the direct product of these individual analysis likelihoods.  Here the numerator and denominator of Eq.\ \ref{like_ratio} are Poisson likelihoods, marginalised over a log-normally distributed nuisance parameter $\xi$, which accounts for fractional background and (where relevant) signal uncertainties characterised by $\sigma_\xi$
\begin{equation}
  \label{marg_poisson_likelihood}
  \begin{split}
  \mathcal{L}_\text{marg}(n|p) = & \int_0^\infty \frac{\left[\xi p\right]^{n}e^{-\xi p }}{n!} \\
                                 & \times \frac{1}{\sqrt{2\pi}\sigma_\xi} \frac{1}{\xi}\exp\left[-\frac{1}{2}\left(\frac{\ln\xi}{\sigma_\xi}\right)^2\right] \mathrm{d}\xi\,.
  \end{split}
\end{equation}
Further details on this one-dimensional marginalised likelihood can be found in Refs.\ \cite{Conrad03,Scott09c,ColliderBit}.

A new feature now available in the \colliderbit code is the ability to construct a
``simplified'' composite likelihood~\cite{Collaboration:2242860}, when the relevant
information about background correlations in different signal regions is available.
The simplified likelihood formalism steers a course between the pessimistic
approach of taking only one signal region, and the unavailable full experimental likelihood. The latter typically
makes use of interpolations between template yield histograms representing the
effects of each elementary systematic uncertainty, and hence requires
substantially more information to be published than just expected yields and
uncertainties. Simplified likelihoods replace this detailed likelihood with a
standard convolved Poisson-Gaussian form, in which the systematic uncertainties
on expected background yields are treated as Gaussian distributions, with
correlations encoded via a covariance matrix $\Sigma$:
\begin{equation}
  \label{eq:simplike}
  \begin{split}
    \mathcal{L}(\bm{s}, \bm{\gamma}) %\equiv& \, P(\bm{n}|\bm{s}, \bm{b}, \bm{\gamma}, \bm{\Sigma})\\
%    =& p(\bm{n}|\bm{s}, \bm{b}, \bm{\gamma}) \times p(\bm{\gamma}|\bm{\Sigma})\\
    =& \prod_i^{N_\text{bin}} \left[ \frac{(s_i + b_i + \gamma_i)^{n_i} e^{-(s_i + b_i + \gamma_i)}}{n_i!} \right]\\
    & \hphantom{\int} \times \frac{1}{\sqrt{\det2\pi\Sigma}} e^{-\frac{1}{2} \bm{\gamma}^T \bm{\Sigma^{-1}} \bm{\gamma}} \, .
  \end{split}
\end{equation}
Here, $n_i$, $s_i$, and $b_i$ are respectively the observed yield and the
nominal expected signal and background yields in signal region $i$, and
$\gamma_i$ is the background deviation from nominal due to systematic
uncertainties\footnote{We follow current CMS experiment procedure by treating the
$\gamma_i$ nuisance parameters directly as linear corrections to the background
expectations $b_i$.}.

In \colliderbit analyses where the simplified-likelihood correlation/covariance
matrices are published -- currently limited to some publications by the CMS experiment --
the full set of $N_\text{bin}$ signal regions is used to construct the
composite likelihood.  This is currently evaluated by marginalising the likelihood over the
background uncertainties $\gamma_i$, distributed as the
$N_\text{bin}$-dimensional Gaussian $G(\bm{0}, \bm{\Sigma})$:
\begin{equation}
  \label{eq:simplike}
  \begin{split}
    \mathcal{L}(\bm{s}) \equiv& \int \mathrm{d}\bm{\gamma} \mathcal{L}(\bm{s}, \bm{\gamma})\\
    =& \int \mathrm{d}\bm{\gamma} \, p(\bm{n}|\bm{s}, \bm{b}, \bm{\gamma}) \times G(\bm{\gamma}|\bm{0}, \bm{\Sigma}).
  \end{split}
\end{equation}
In practice this marginalisation is performed by sampling $\bm{\gamma}$ vectors
from the Gaussian, calculating the Poisson $p(\bm{n}|\bm{s},\bm{b},\bm{\gamma})$
for each, and averaging over the set of samples. For computational speed,
\colliderbit performs this sampling in parallel using \textsf{OpenMP}, and skips
it entirely if the signal prediction from the event generator run is exactly
zero in all signal regions.  Numerical convergence of the sampling is ensured by
iterative doubling of the number of samples $N_\text{samp}$, starting from $10^5$,
until the marginalised likelihood estimator is stable within 5\%, or the
absolute variation in the likelihood estimate drops below 0.05. In this study we
use the simplified likelihood approach for the likelihood contributions from the
CMS two-lepton searches in Refs.~\cite{Sirunyan:2018iwl} and \cite{Sirunyan:2017qaj}.

%%%%%%%%%%%%%%%%%%
\subsection{\pvalue calculations}
\label{sec:pvalue}
%%%%%%%%%%%%%%%%%%

To quantify the significance of deviations from the SM across multiple
LHC and LEP searches for sparticles, as well as to quantify the
absolute goodness-of-fit of our EWMSSM best-fit point, we compute
\pvalues via likelihood-ratio tests. These computations are performed by dedicated Monte-Carlo simulations outside of the main GAMBIT software framework. The `local significance' test
and the `goodness-of-fit' test each use a different form of likelihood
ratio, so we describe them separately below.

\subsubsection{Local significance}
\label{sec:local_pvals}

Computing the significance of any excesses in the data is done by
attempting to exclude the background-only hypothesis across all analyses
simultaneously. We construct this test by assigning a single ``signal strength''
parameter $\mu$
across all analyses,\footnote{This is of course completely unrelated to the $\mu$ parameter in the MSSM superpotential.}
where the nominal ($\mu=1$) signal is
obtained via the predictions of the best-fit point found in our scan. We then
attempt to exclude the $\mu=0$ null hypothesis.

For example, consider the simplified likelihood of Eq.~\ref{eq:simplike}. The signal predictions for each analysis bin $s_i$ become $\mu s_i$, and this scaling is applied consistently across all components of the joint likelihood. By setting $\mu=1$ we obtain a `nominal' signal hypothesis for a given parameter point, whilst $\mu=0$ retrieves the joint background-only hypothesis.

The test statistic we construct is then
\begin{equation}
    q_\mathrm{LS} = -2 \log \frac{\mathcal{L}_\mathrm{joint}(\mu=1,\hat{\eta})}
                     {\mathcal{L}_\mathrm{joint}(\mu=0,\hat{\hat{\eta}})},
\end{equation}
where $\mathcal{L}_\mathrm{joint}$ is the joint likelihood for all analyses (with $\mu=0$
setting the signal to zero in the denominator case), and $\hat{\eta}$
and $\hat{\hat{\eta}}$ are the best-fit (i.e. profiled) values of nuisance
parameters under each hypothesis (for example the $\gamma_i$ in Eq.~\ref{eq:simplike}). When
the null hypothesis $\mu=0$ is true, this test statistic is (asymptotically)
distributed as a Gaussian \cite[Sec.\ 3.8]{Cowan:2010js}.
However, because some analyses involve
few events and may jeopardise the asymptotic assumptions, we determine the test
statistic distribution by Monte Carlo simulation.

For the LHC analyses, $\eta$ represents nuisance parameters that characterise
uncertainties in the background estimates. In our scan we marginalised
over these (see Sec. \ref{sec:simplified likelihoods}) due to better numerical
stability, however, for our p-value calculations we have chosen to profile them so that our Monte Carlo output could
be validated by comparison with the predictions of asymptotic theory, and to maintain a frequentist intepretation of the resulting p-values.

It is of great importance to note that this test performs only a {\em local}
significance test at chosen parameter points.
In principle a ``trial'' correction should be computed, as choosing to test the best-fit EWMSSM point after analysing the data
constitutes a form of ``cherry-picking''. This problem is
also known as the ``look-elsewhere effect'', or, in statistics, the ``problem of multiple
comparisons''.

Unfortunately, it is incredibly computationally demanding to correct
for this in parameter spaces larger than one or two dimensions, and
is beyond our means at present.\footnote{A fully rigorous trial correction would require us
to MC the entire global fit under many pseudodata realisations, as we
need to know the distribution of the best-fit local \pvalues, see e.g. \cite{Fowlie:2017fya}. A compromise approach
would be to reweight a sufficiently dense set of parameter samples under many pseudodata realisations and
find the distribution of best-fit \pvalues in just that chain, e.g.\ as discussed in
\cite{Fowlie:2013aar,Fittinocoverage}. However, our chains are not large enough
that they would reliably contain points close to the best fits under pseudodata, because our scans concentrate around the observed best fit but are sparse in other parts of the parameter space where good fits to the pseudodata might lie. Approximate procedures can be applied in lower dimensions, e.g.\ \cite{Gross:2010qma,1748-0221-11-12-P12010,2018arXiv180303858A}, however, they are are mainly aimed at reducing the number of pseudodata realisations that are required to perform the trial correction, which is not the issue here. Our problem is instead that obtaining sufficiently good sampling of the possible signal predictions in the EWMSSM is hard.}
We nevertheless can get some idea of a `global'
significance by computing the goodness-of-fit of the background-only
(SM) hypothesis in a test against a fully general signal
hypothesis. We discuss this further in section \ref{sec:gof_test}. The
results of applying this test to each analysis individually, and to
their combination, are listed in Table \ref{tab:pvalues}.

\subsubsection{Goodness-of-fit}
\label{sec:gof_test}
Aside from the joint significance of excesses, we are interested in quantifying
the absolute goodness-of-fit of points in the EWMSSM.  Profile likelihood
contours do not have the power to exclude the best-fit point in a global fit, as
they are computed based on likelihood ratios {\em relative} to the best fit.  Their
stated coverage is also often somewhat incorrect, as they are computed based on
asymptotic theory relying on Wilks' theorem, whose regularity assumptions are often violated in complicated parameter spaces such
as the EWMSSM \cite{SBcoverage,Akrami11coverage,Strege12}.

To formulate this test, we take the predictions of the best-fit point of our scan
and embed them in a larger ``proxy'' hypothesis space, where the possible signals
are allowed to vary in a more general way. For example in the likelihood of
Eq.~\ref{eq:simplike} we simply take the signal predictions $s_i$ in each bin
as independent free parameters. We can thus test the goodness-of-fit
of any EWMSSM point by seeing whether a sufficiently better-fitting point can be
found in the more general hypothesis space. The method is similar to a common chi-squared
test used to measure goodness-of-fit in histograms \cite{BAKER1984437}, as each of
our signal regions may be thought of as one bin in a histogram. Such a test has
much less statistical power to detect signals than a more targeted test like the one we use
to compute local significances.  However, its false positive rate is better controlled, because it is less susceptible to the
look-elsewhere effect.\footnote{Some smaller level of look-elsewhere effect will remain
due to the pre-selection of which signal regions to use for the test. This effect
would be avoided completely if correlation information was available for all analyses
and we were able to remove the step of pre-selecting signal regions based on their expected sensitivity.}

For a more explicit example, let us consider the LHC analyses for which we have no correlation information.
In these analyses we pre-select the signal region with the best expected sensitivity to the
signal predictions of the parameter point of interest (see Sec. \ref{sec:LHCanalyses}). The simplified pdfs
for these analyses then reduce to a single Poisson distribution times a Gaussian constraint on a nuisance
parameter:
\begin{align}
    \mathrm{Pr}(n,\hat{\gamma}) = \mathrm{Poisson} (n | s+b+\gamma) \cdot \mathrm{Normal}(\hat{\gamma}|\gamma),
\end{align}
where $n$ is the number of events observed in that signal region, and $\hat{\gamma}$ is the maximum likelihood
estimator for $\gamma$ obtained from control measurements. For the observed data $\hat{\gamma}$ is zero by definition, however, it is a random variable from the point of view
of pseudodata generation, as we keep $b$ fixed. As in the case of Eq.~\ref{eq:simplike}, the signal expectation
$s$ is allowed to vary freely (over both positive and negative values), for each pre-selected
signal region in every analysis.

When correlation information is available (the Eq.~\ref{eq:simplike} case), a free signal parameter is assigned to every signal
region in the analysis. In the case of Gaussian likelihoods (the Higgs and $Z$ invisible width likelihoods), the expected value is allowed
to vary as a free parameter.

We then construct the test statistic
\begin{equation}
    q_\mathrm{GOF} = -2 \log \frac{\mathcal{L}_\mathrm{joint}(\mathbf{s}(\theta),\hat{\eta})}
                     {\mathcal{L}_\mathrm{joint}(\hat{\hat{\mathbf{s}}},\hat{\hat{\eta}})},
\end{equation}
where $\mathbf{s}(\theta)$ are the predictions of EWMSSM point $\theta$ (or SM)
and form the null hypothesis, whilst $\hat{\hat{\mathbf{s}}}$ are the global best-fit
values of the parameters $\mathbf{s}$ in the free-signal parameter space. $\hat{\eta}$ and
$\hat{\hat{\eta}}$ likewise represent vectors of nuisance parameters fit to the null hypothesis
and free-signal, respectively.

When data is generated under $\mathbf{s}(\theta)$ this test statistic is
asymptotically distributed as a $\chi^2$ variable, whose degrees of freedom are equal
to the dimension of $\mathbf{s}$. This parameter space has good regularity properties so
Wilks' theorem applies well, meaning the theoretical
distribution should be quite reliable. However, discretisation and boundary effects can
still enter for signal regions with low expected count numbers, so we also compute
these distributions via Monte Carlo simulation.

We use this test to assess the goodness-of-fit of both the SM and our best-fit EWMSSM point to the data observed in each analysis individually, as
well as jointly. The results are given in Table \ref{tab:pvalues}.

\begin{figure*}[h!]
  \centering
  \includegraphics[height=0.8\columnwidth]{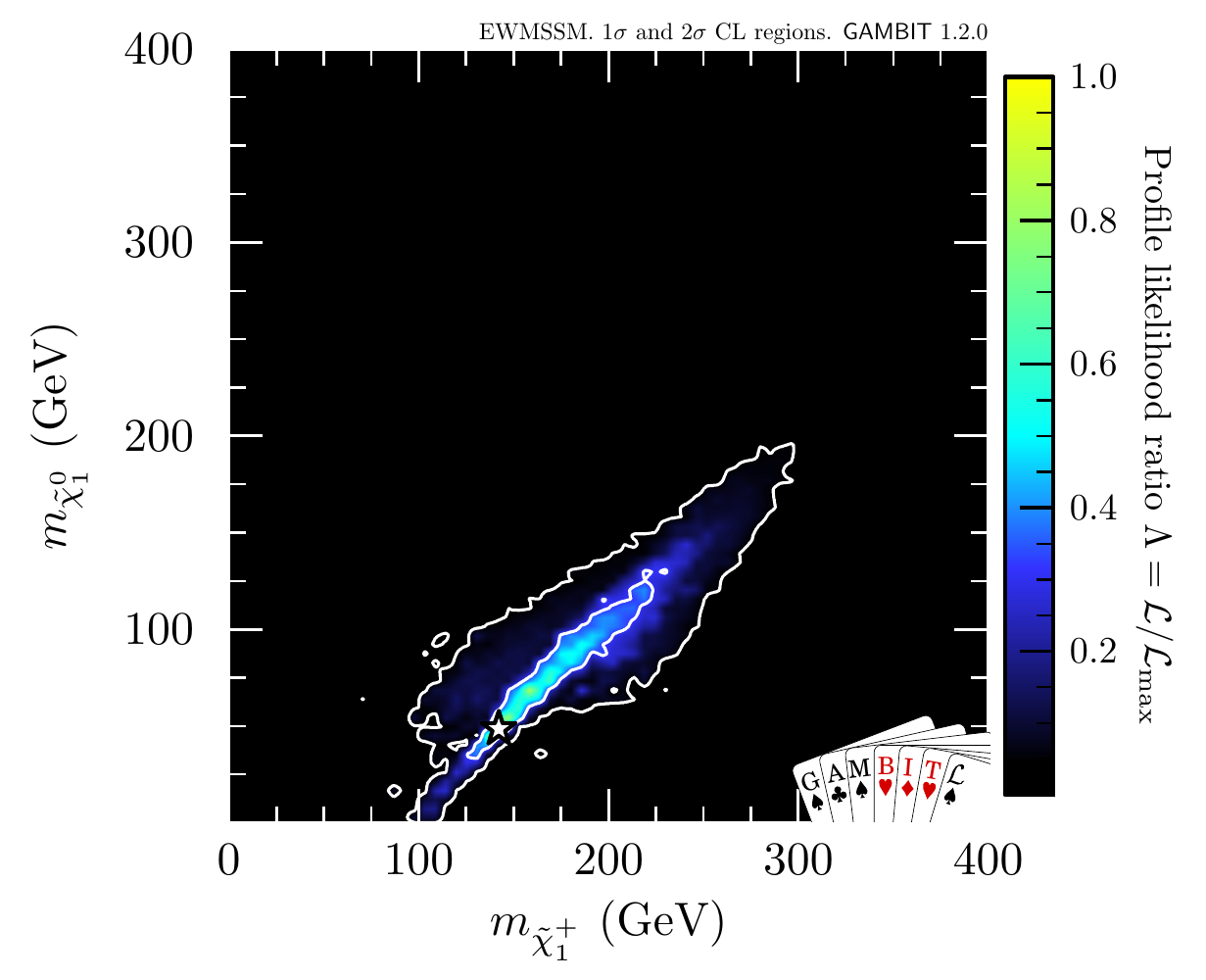}
  \includegraphics[height=0.8\columnwidth]{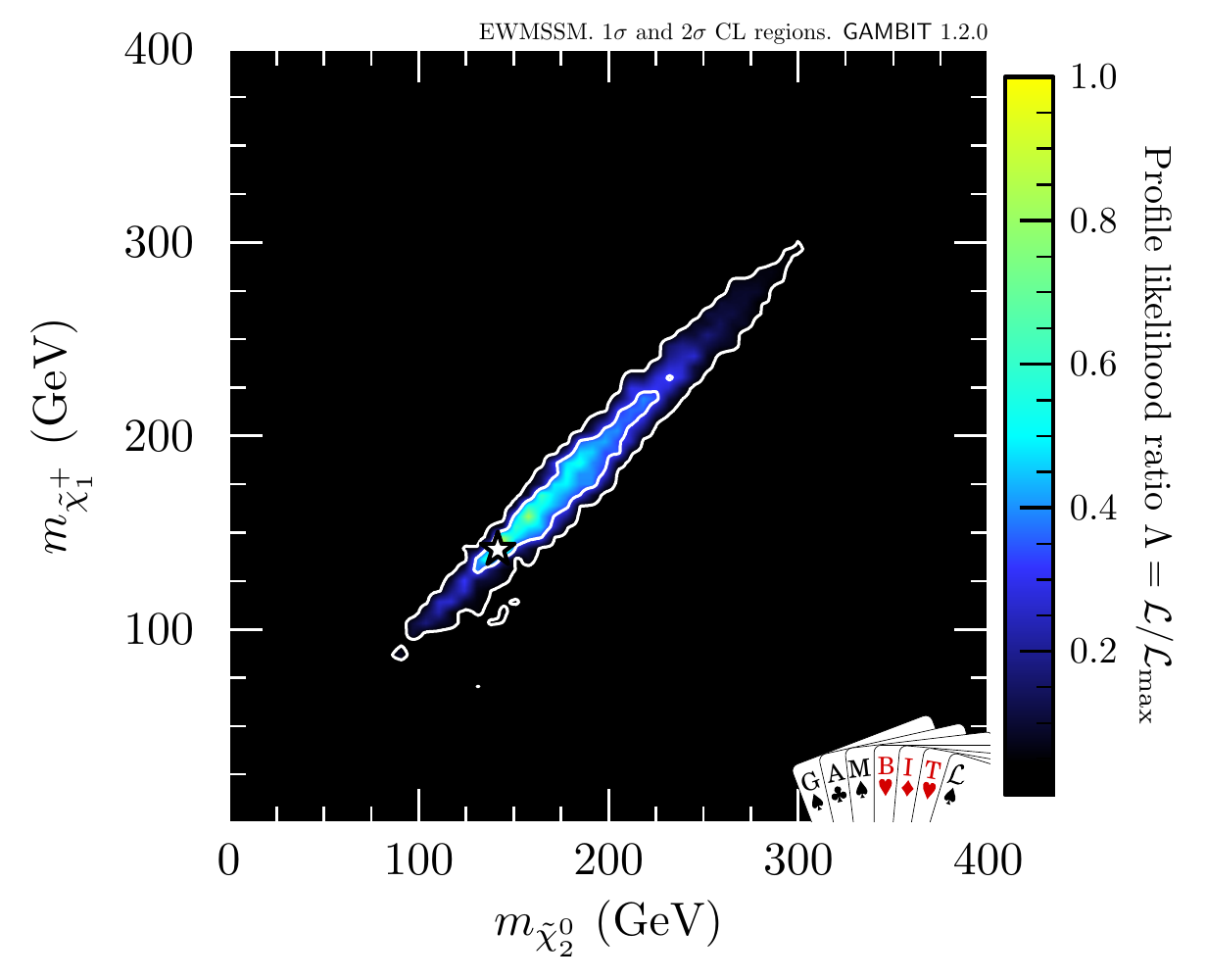}\\
  \includegraphics[height=0.8\columnwidth]{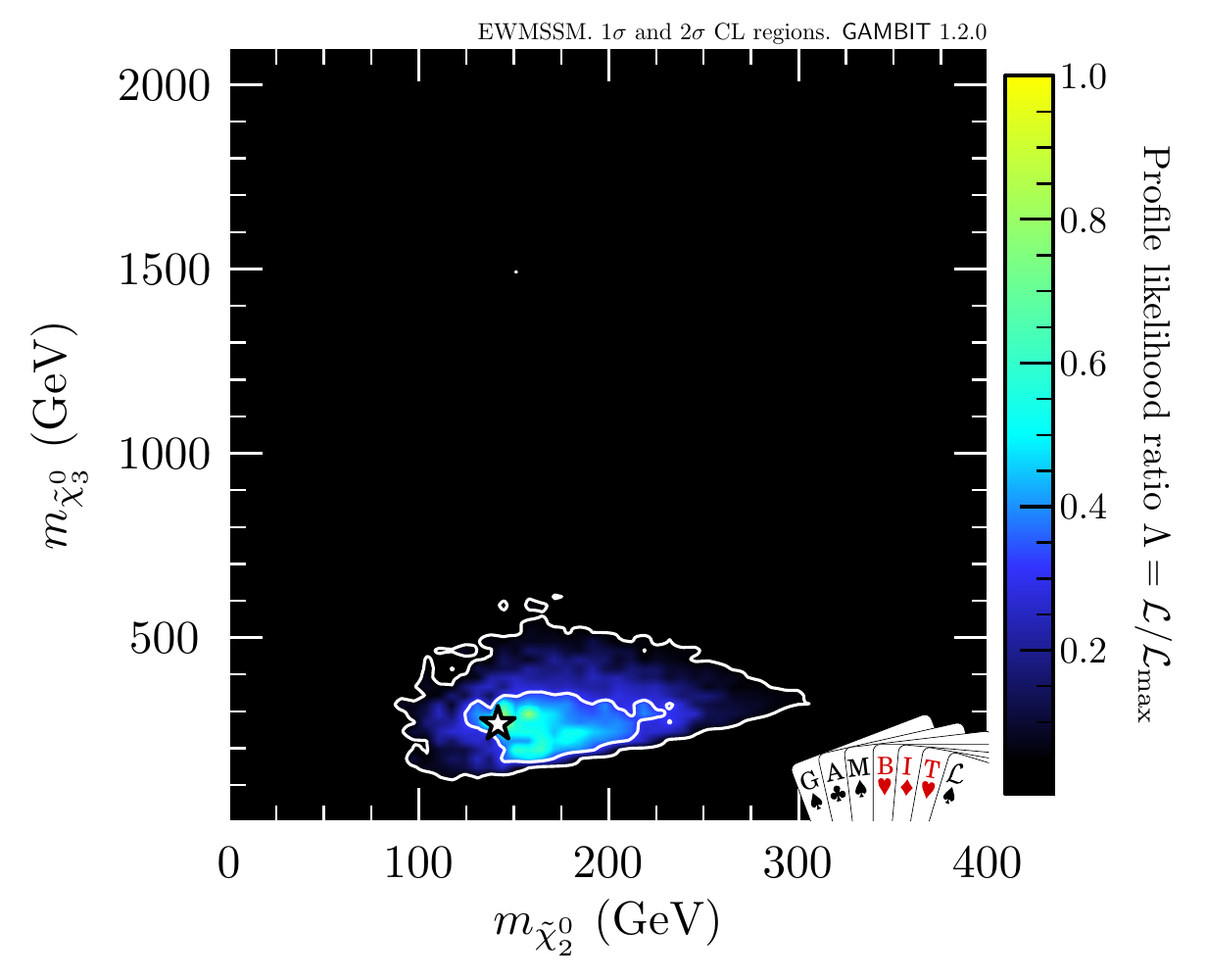}
  \includegraphics[height=0.8\columnwidth]{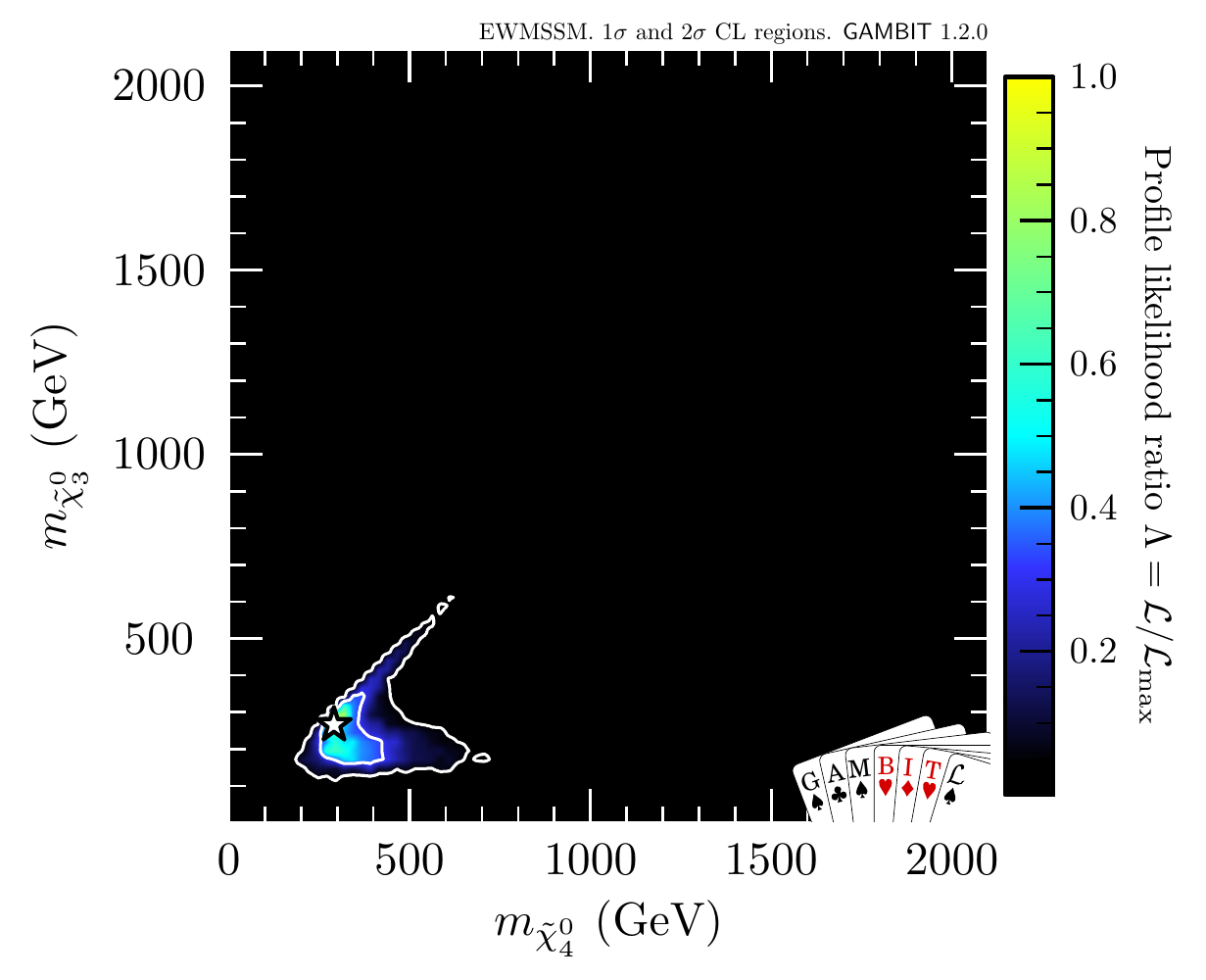}
	\caption{Profile likelihood in the $(m_{\tilde{\chi}_1^{\pm}},m_{\tilde{\chi}_1^0})$ plane (upper left), the  $(m_{\tilde{\chi}_2^0},m_{\tilde{\chi}_1^{\pm}})$ plane (upper right), the $(m_{\tilde{\chi}_2^0},m_{\tilde{\chi}_3^0})$ plane (lower left) and the $(m_{\tilde{\chi}_4^0},m_{\tilde{\chi}_3^{0}})$ plane (lower right). The contour lines show the $1\sigma$ and $2\sigma$ confidence regions. The best-fit point is marked by the white star.
  }
  \label{fig:pole_masses_2D}
\end{figure*}
%%%%%%%%%%%%%%%%%%%%%%%%%%%%%%%%%%%%%%%%%%%%%%%%%%%%%%%
%%%%%%%%%%%%%%%%%%%%%%%%%%%%%%%%%%%%%%%%%%%%%%%%%%%%%%%
\section{Results}
\label{sec:results}
\subsection{Profile likelihood maps}
\begin{figure*}[h!]
  \centering
  \includegraphics[height=0.8\columnwidth]{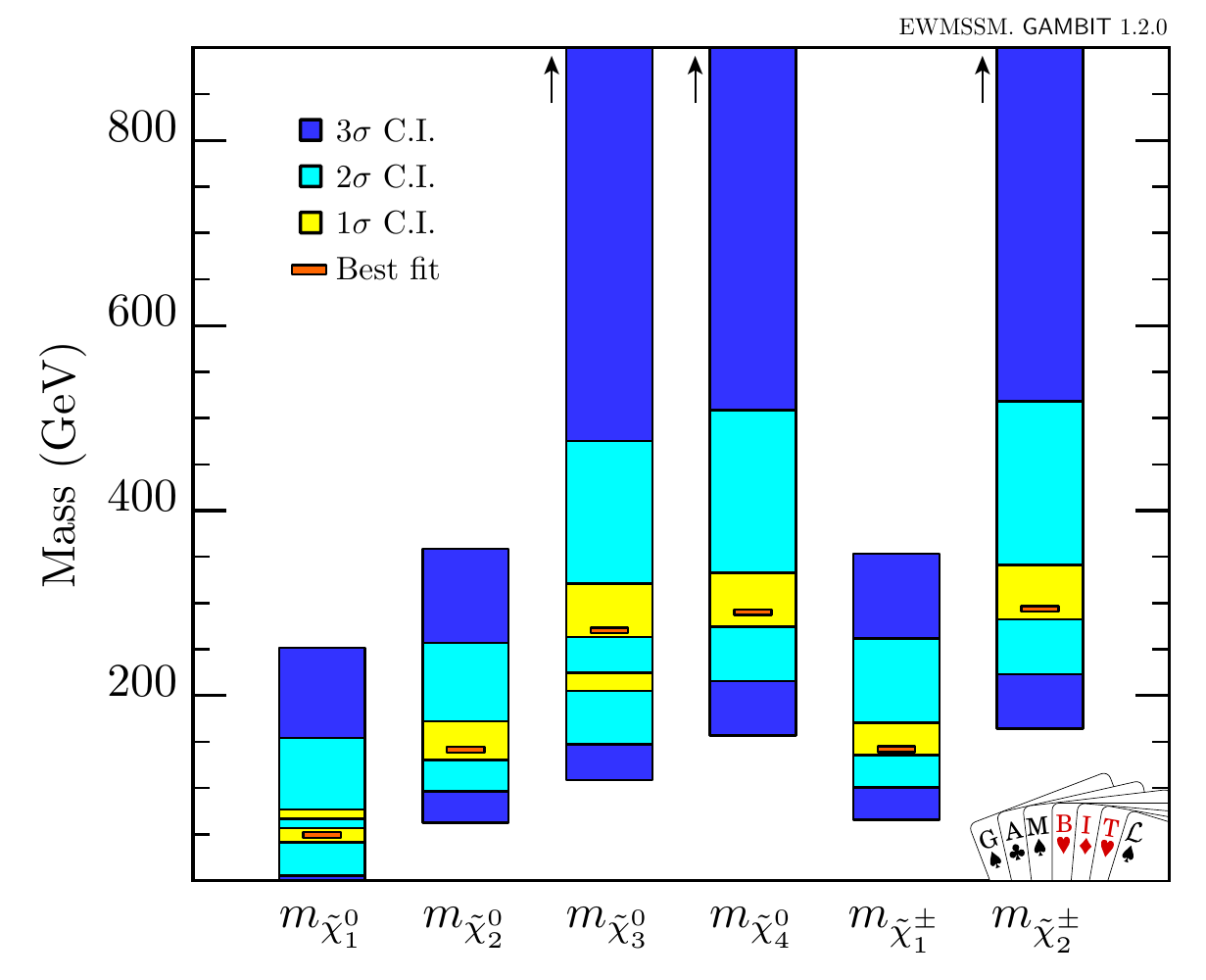}
  \caption{Summary of the one-dimensional $1\sigma$, $2\sigma$ and $3\sigma$ confidence intervals for the neutralino and chargino masses. The orange lines mark the best-fit values. For $m_{\tilde{\chi}_3^0}$, $m_{\tilde{\chi}_4^0}$ and $m_{\tilde{\chi}_2^\pm}$, the $3\sigma$ confidence intervals extend up to the $2$~TeV upper limit on the mass parameters in our scan.}
  \label{fig:mass_summary_plot}
\end{figure*}

Figure~\ref{fig:pole_masses_2D} shows our results for the profile
likelihood in various electroweakino mass planes. There is a clear preference for a mass scale in the $(m_{\tilde{\chi}_1^{\pm}},m_{\tilde{\chi}_1^0})$ plane (top-left), centered on $m_{\tilde{\chi}_1^{\pm}}\approx$~150\,GeV, and  $m_{\tilde{\chi}_1^{0}}\approx$~50\,GeV. We also find that $m_{\tilde{\chi}_1^{\pm}} \lesssim 300$\,GeV and $m_{\tilde{\chi}_1^{0}}\lesssim 200 $\,GeV at the $2 \sigma$ level. This preference is driven by the small number of coincident excesses in a variety of ATLAS and CMS searches, which we discuss in detail below. One can also see that the best-fitting solutions lie far from the line $m_{\tilde{\chi}_1^{\pm}}=m_{\tilde{\chi}_1^{0}}$, indicating a preference for a predominantly bino LSP.

The top right panel of Figure~\ref{fig:pole_masses_2D} shows results
in the $(m_{\tilde{\chi}_1^{\pm}},m_{\tilde{\chi}_2^0})$ plane and
indicates that the best-fitting solutions exhibit an approximate
degeneracy between the $\tilde{\chi}_1^{\pm}$ and
$\tilde{\chi}_2^{0}$ masses, such as would be expected if they were
dominantly composed of Higgsinos, winos or a mixture of the two.  As such we also find
$m_{\tilde{\chi}_2^{0}} \lesssim 300$\,GeV within the $2 \sigma$
contours.

In the bottom left panel of Figure~\ref{fig:pole_masses_2D} the results are displayed in the $(m_{\tilde{\chi}_2^0}, m_{\tilde{\chi}_3^0})$ plane, which clearly shows that $m_{\tilde{\chi}_2^0} \lesssim 300$\,GeV, as in the top right panel, and $m_{\tilde{\chi}_3^0} \lesssim 700$\,GeV within the $2\sigma$ region. There is a slight preference for $m_{\tilde{\chi}_2^0} \ll m_{\tilde{\chi}_3^0}$, as represented by the best-fit point, corresponding to the scenario where winos are lighter than Higgsinos. The opposite scenario, where Higgsinos are lighter than winos and $m_{\tilde{\chi}_2^0} \sim m_{\tilde{\chi}_3^0}$, is also present within $1\sigma$ of the best fit, albeit for somewhat higher $\tilde{\chi}_2^0$ masses.

The bottom right panel shows results in the mass
planes of the heaviest neutralinos, $\tilde{\chi}_3^{0}$ and $\tilde{\chi}_4^0$. Within the $2 \sigma$ contours
the masses of these states are bounded by $m_{\tilde{\chi}_3^{0}} \lesssim 700$\,GeV and $m_{\tilde{\chi}_4^{0}} \lesssim 700$\,GeV. For even heavier $\chi^0_3$ or $\chi^0_4$, the profile likelihood function flattens out beyond the $2\sigma$ contour and becomes indifferent to the specific mass. One therefore obtains a better fit to the LHC data when the entire neutralino and chargino spectrum is light, but the heavier electroweakinos are not constrained at the 3$\sigma$ level. We do not show results for $m_{\tilde{\chi}_2^\pm}$, as our results indicate that it is nearly degenerate in mass with $m_{\tilde{\chi}_4^0}$ for the full $2\sigma$ region.

Our findings for the electroweakino masses are neatly summarised in
Figure~\ref{fig:mass_summary_plot}, where we show the $1\sigma$,
$2\sigma$ and $3\sigma$ bands for each electroweakino mass.\footnote{We emphasise that these are now the 1D $n\sigma$ regions, and thus are not directly comparable to the contours of the 2D plots.}
This shows that we find $3\sigma$ upper limits on the masses of the two lightest neutralinos and the lightest
chargino.  At this confidence level, the heavier neutralino and chargino masses saturate
the upper limits set by the allowed range for the input parameters.

Let us first assume that this pattern of excesses arises from statistical fluctuations,
and that there is no production of electroweakinos (or any other sparticle) at the LHC.
Under this assumption, it is interesting to determine what limits the present data from LHC direct SUSY searches put on charginos and neutralinos in the EWMSSM.
A simple way to do this is to consider a capped version of our LHC likelihood,
\begin{equation}
\label{lcap}
\mathcal{L}_\text{cap} = \min[\mathcal{L}_\text{LHC}(\bm{s}+\bm{b}), \mathcal{L}_\text{LHC}(\bm{b})],
\end{equation}
where $\mathcal{L}_\text{LHC}$ is the combined likelihood from all the simulated 13\,TeV SUSY searches.
This construction ensures that no EWMSSM parameter point can achieve a likelihood higher than the background-only expectation.  This makes it only possible to exclude EWMSSM models.  Note that the `capping' in Eq.\ \ref{lcap} is done on the final composite likelihood \textit{for all analyses}, not on the likelihood contribution from each analysis individually.\footnote{Something
similar was done in our previous fits of supersymmetry \cite{CMSSM,MSSM}.}
The profile likelihood ratio of the capped likelihood to its best possible value, $\mathcal{L}_\text{cap}/\mathcal{L}_\text{cap,max} \equiv \mathcal{L}_\text{cap}/\mathcal{L}_\text{LHC}(\bm{b})$
is thus a measure of how much \textit{worse} a given EWMSSM parameter point does in fitting the data than the SM does. A likelihood ratio of 1 means that the EWMSSM does at least as well as the SM, whereas a ratio of less than 1 means that the SM fits the data better than the EWMSSM point.
To obtain a ratio of 1 for a given point in the EWMSSM parameter space, it must either be the case that no analysis is sensitive to the given parameter point (\eg $\bm{s} = \bm{0}$), or that a bad fit to some of the analyses is completely offset by a sufficiently good fit
to other analyses.

\begin{figure}[t]
  \centering
  \includegraphics[height=0.8\columnwidth]{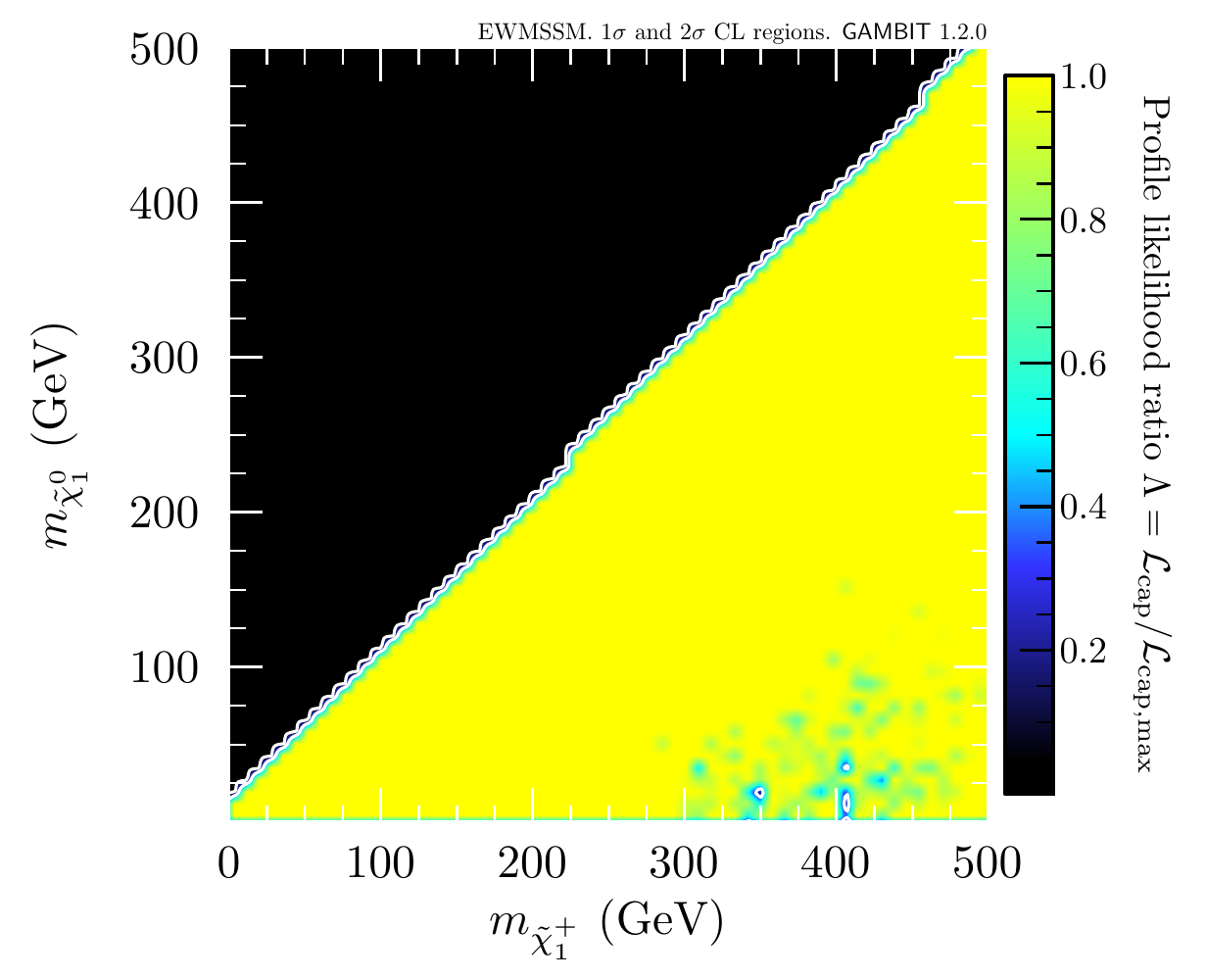}
  \caption{Capped profile likelihood in the $(m_{\tilde{\chi}_1^{\pm}},m_{\tilde{\chi}_1^0})$ plane.  The capped likelihood function (Eq.\ \ref{lcap}) is based solely on the joint likelihood for the 13\,TeV LHC direct SUSY searches.  The contour lines show the $1\sigma$ and $2\sigma$ confidence regions.}
  \label{fig:capped_log-like}
\end{figure}

In Figure \ref{fig:capped_log-like} we plot this profile likelihood ratio in
the $(m_{\tilde{\chi}_1^{\pm}},m_{\tilde{\chi}_1^0})$ plane.
The result shows little variation across the entire mass plane, indicating
that the combined results from the 13\,TeV LHC direct searches in fact do not produce any significant general
constraint on the masses of neutralinos or charginos. Naively, this conclusion would seem to be
in conflict with published ATLAS and CMS results.  However, the ATLAS and CMS analyses are all optimised and interpreted in terms of simplified models. The full electroweakino sector of the
MSSM has a far richer phenomenology than the simplified models.  When the likelihoods from this multi-dimensional space are profiled
onto the neutralino-chargino plane, there is only a very weak constraint remaining,
on some isolated islands in the mass plane.

Such a lack of exclusion has been noted
before~\cite{vanBeekveld:2016hug}, and can be understood
physically. For example, non-wino dominated $\tilde{\chi}^{\pm}_1$ and
$\tilde{\chi}^0_2$ pairs have a lower production cross-section compared to a scenario with pure winos. Also, the prevalence of other production and decay modes changes the typical final states, so that for a given EWMSSM parameter point the signal regions with the best expected sensitivity may differ from the signal regions with best sensitivity to a simplified model with similar masses for the light electroweakinos. We emphasise that, in a frequentist approach, this lack of exclusion must be interpreted literally. In a Bayesian framework, one could instead marginalise over the dimensions not appearing on the axes of each plane, to determine the posterior mass in excluded scenarios; we leave such an analysis for future work.

We note that in order to obtain a large enough dataset to
produce Figure~\ref{fig:capped_log-like}, we include
all parameter samples with at least $500\,000$ MC events in the LHC likelihood calculation.
This should be contrasted with the other results in this paper, where only samples with
at least $4$~million MC events are used. Because the profile likelihood picks out
the least constrained parameter sample for every point in
the $(m_{\tilde{\chi}_1^{\pm}},m_{\tilde{\chi}_1^0})$ plane, this larger MC
uncertainty implies that the result in Figure \ref{fig:capped_log-like}
should be viewed as a somewhat conservative estimate of the constraining
power of the combined data.

Let us now remove the assumption that there are no sparticles within
reach of the LHC, and return to a consideration of the complete, uncapped
profile likelihood. In this case, the observed results are not
surprising in light of the ATLAS recursive jigsaw (RJ) search described in Sec.~\ref{sec:LHCanalyses}, which
saw excesses in four signal regions targeting chargino plus neutralino
production, with decays to $W$ and $Z$ bosons and lightest
neutralinos.

Note that an excess in a search for electroweakinos that is optimised
for on-shell $W$ and $Z$ production effectively sets one
chargino-neutralino mass difference to be somewhere near the $W$ mass,
whilst also setting a neutralino-neutralino mass difference to be at
least equal to the $Z$ mass, after which the overall mass scale is
forced to the value with a cross-section that is able to reproduce the
size of the excess. In the simplified model approach, these mass
differences would be defined between the $\tilde{\chi}_1^{\pm}$ and
$\tilde{\chi}_1^0$, and between the $\tilde{\chi}_2^0$ and
$\tilde{\chi}_1^0$, but we see departures from this behaviour due to
the fact that other electroweakino production and decay processes are
able to produce on-shell $W$ and $Z$ bosons. Nonetheless, in
Fig.~\ref{fig:pole_masses_2D} we still see a mild preference for a
mass difference of around 100\,GeV between $\tilde{\chi}_1^{\pm}$ and
$\tilde{\chi}_1^0$.

\begin{figure*}[th!]
  \centering
  \includegraphics[width=0.32\textwidth]{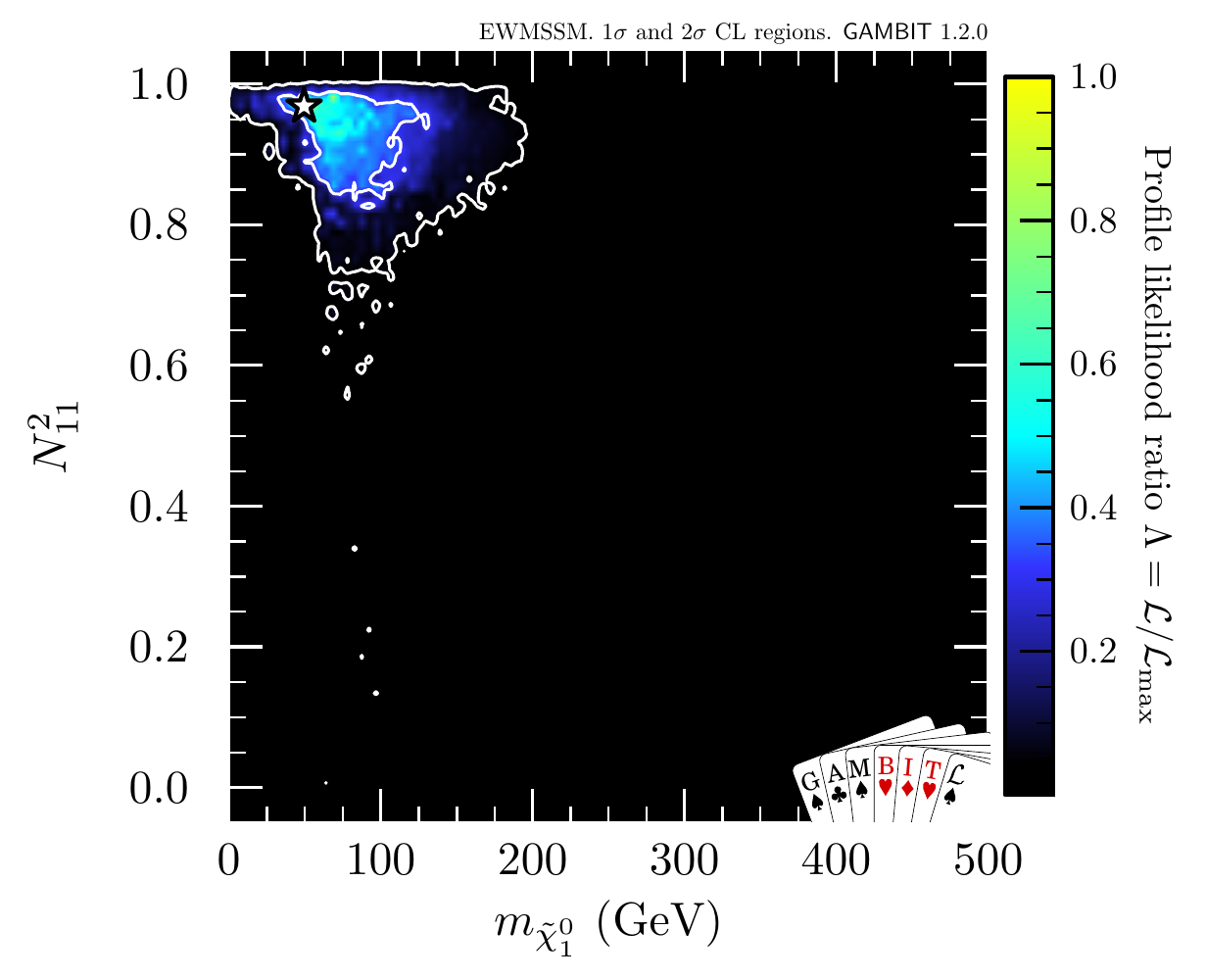}
  \includegraphics[width=0.32\textwidth]{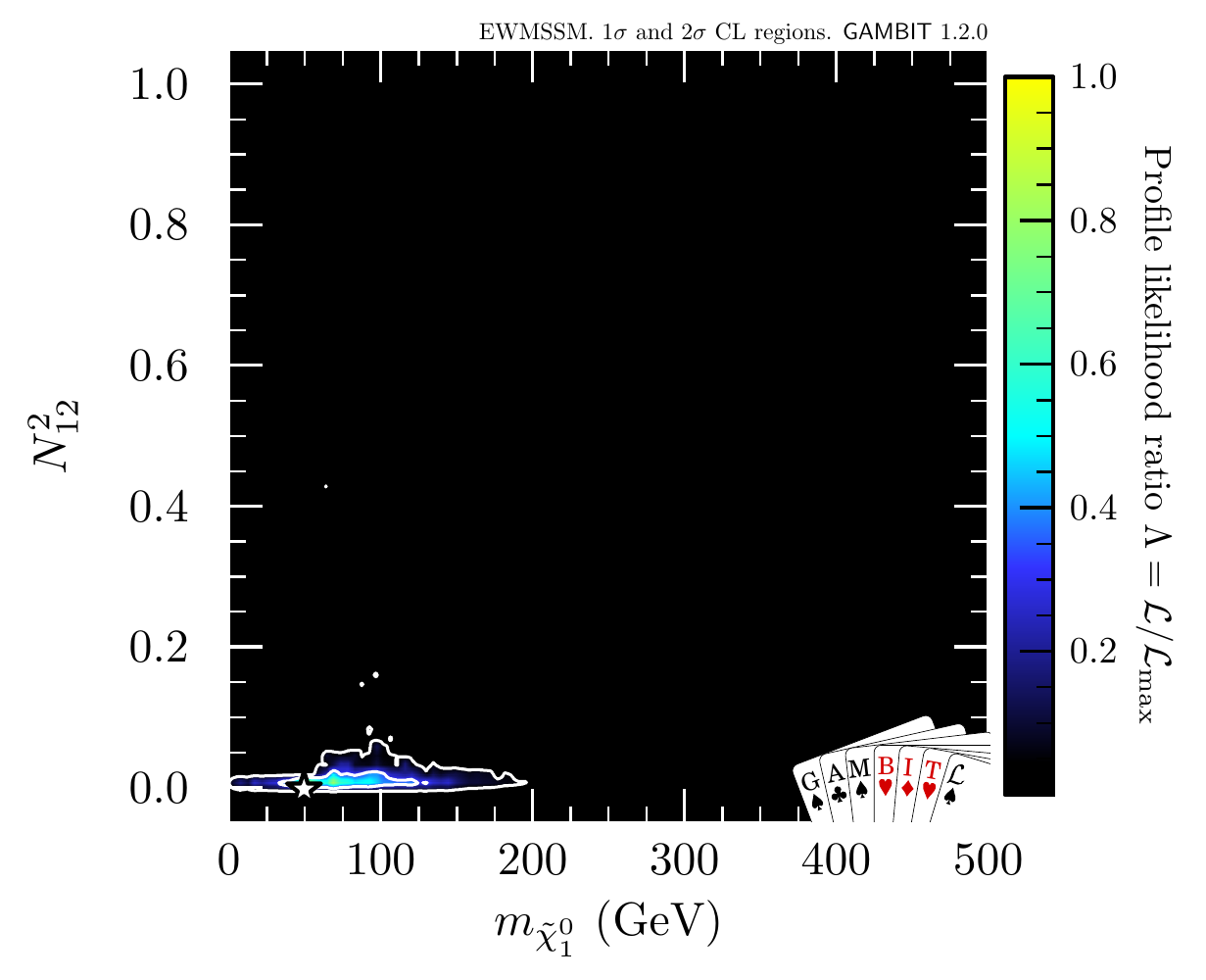}
  \includegraphics[width=0.32\textwidth]{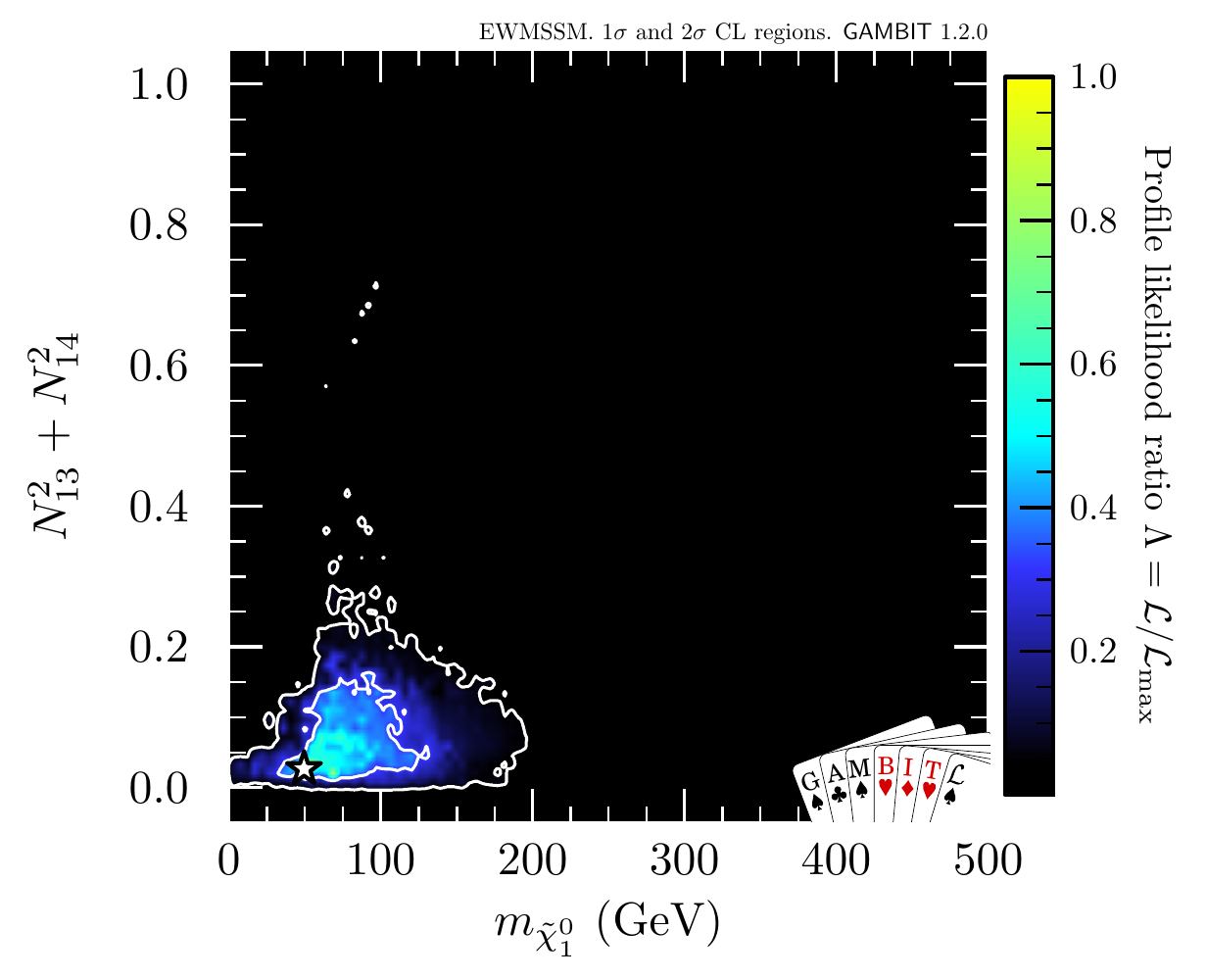}\\
  \includegraphics[width=0.32\textwidth]{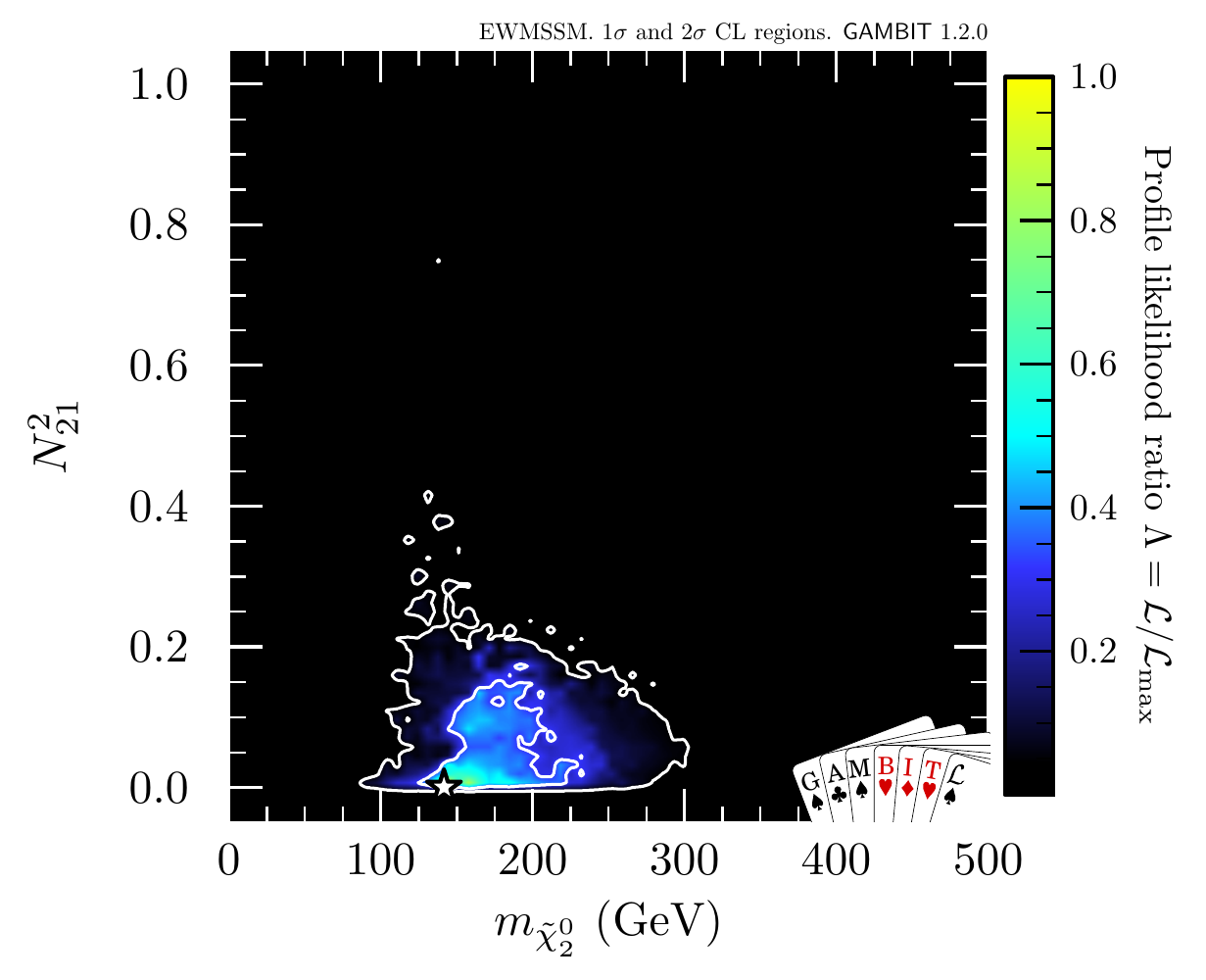}
  \includegraphics[width=0.32\textwidth]{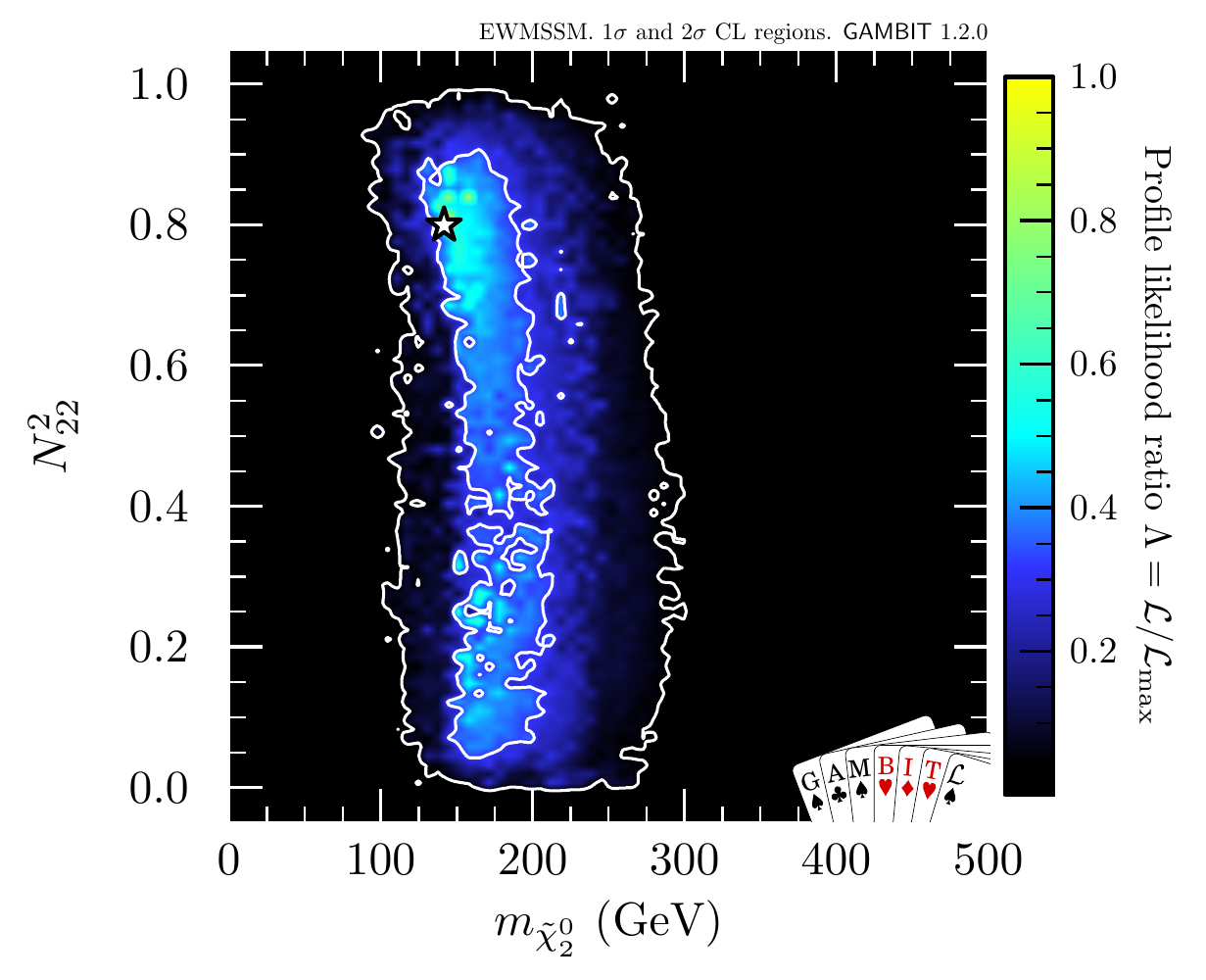}
  \includegraphics[width=0.32\textwidth]{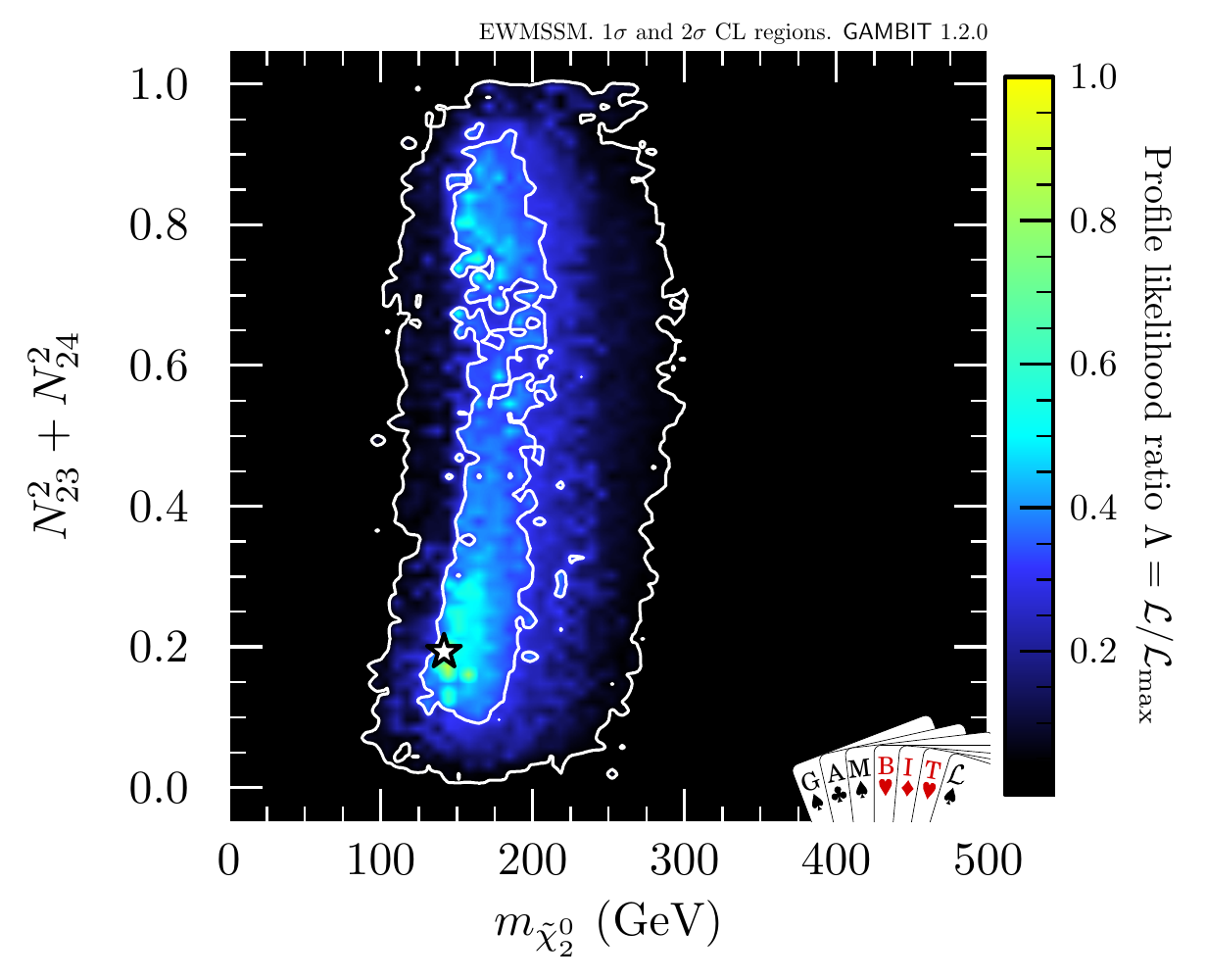}\\
  \includegraphics[width=0.32\textwidth]{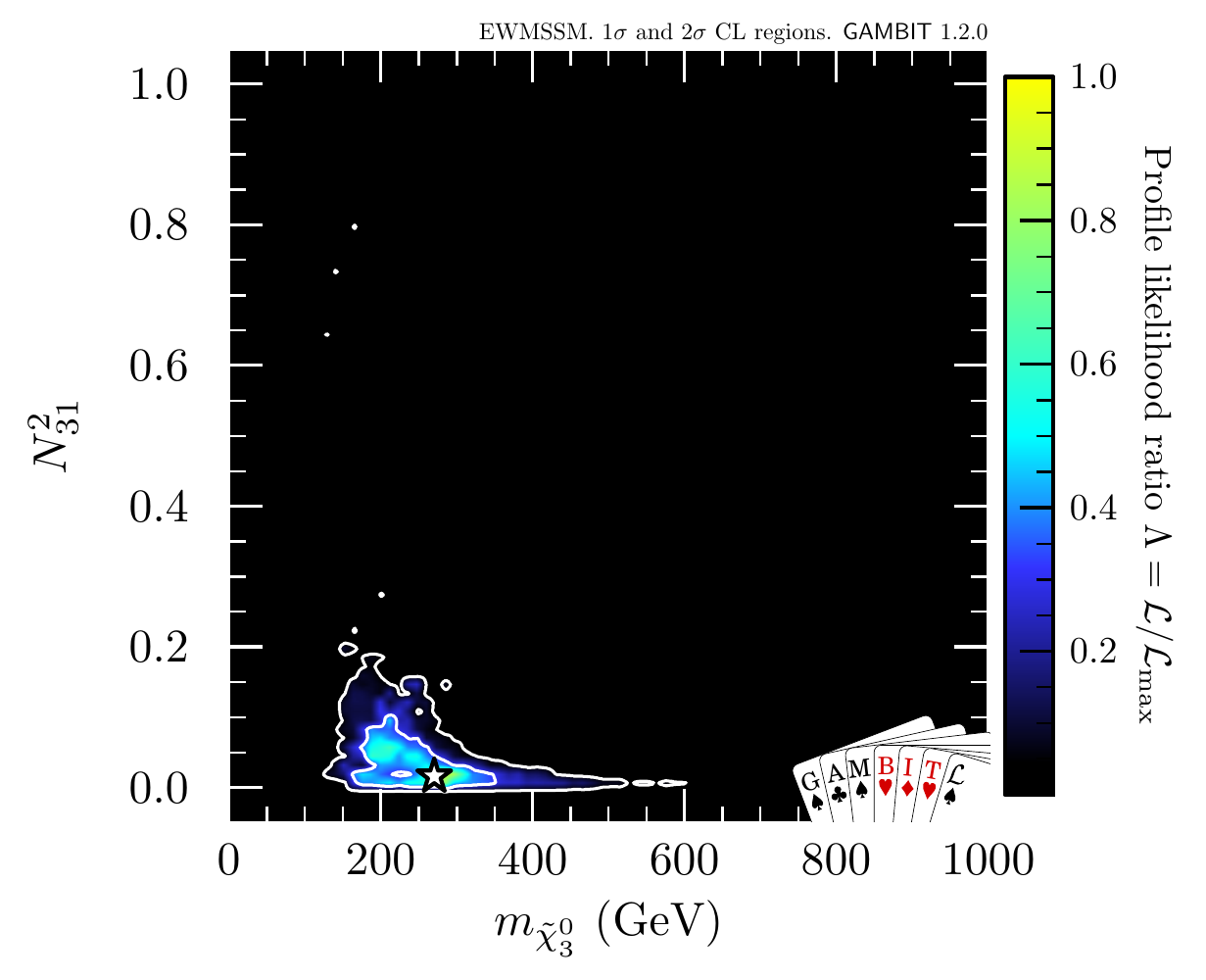}
  \includegraphics[width=0.32\textwidth]{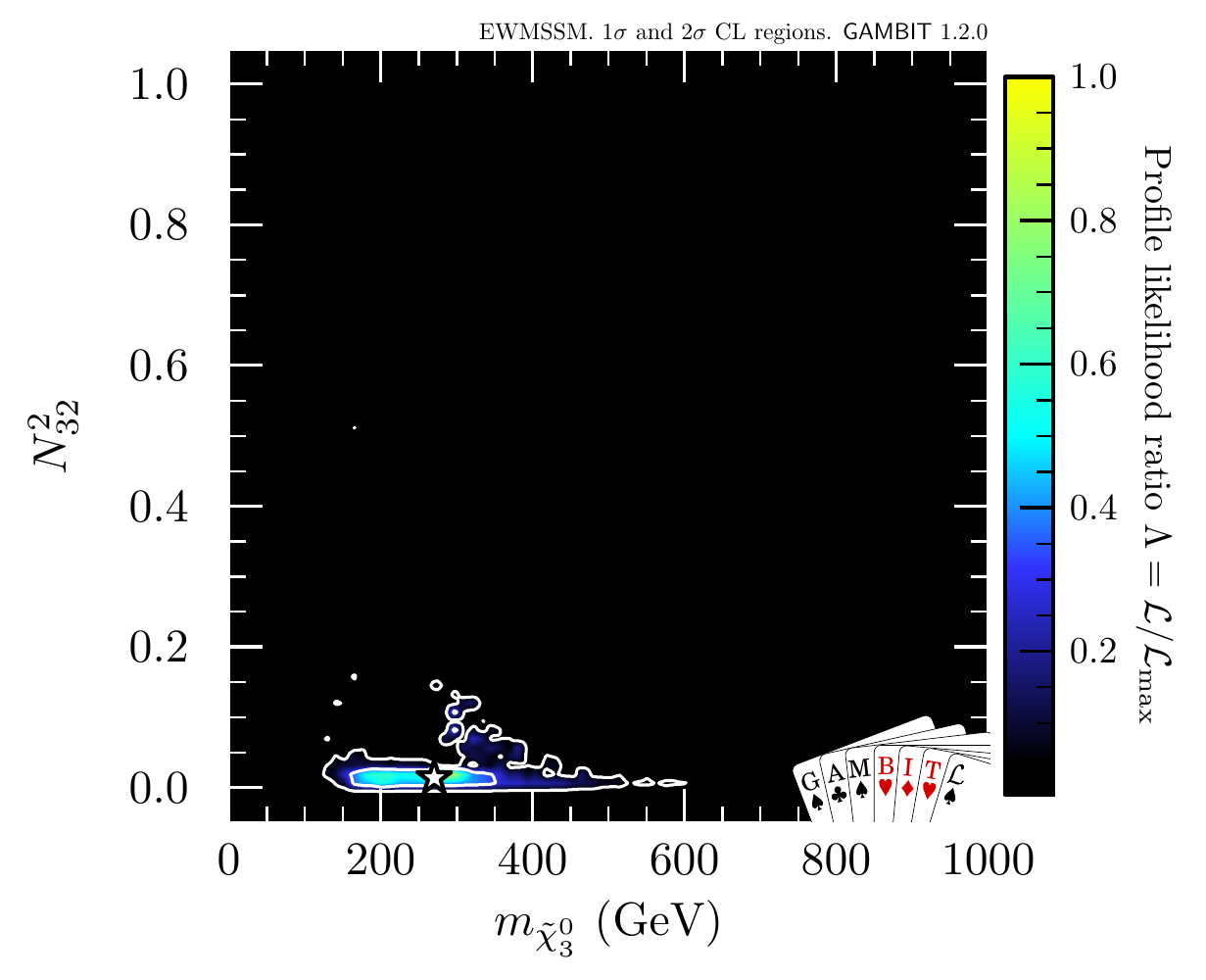}
  \includegraphics[width=0.32\textwidth]{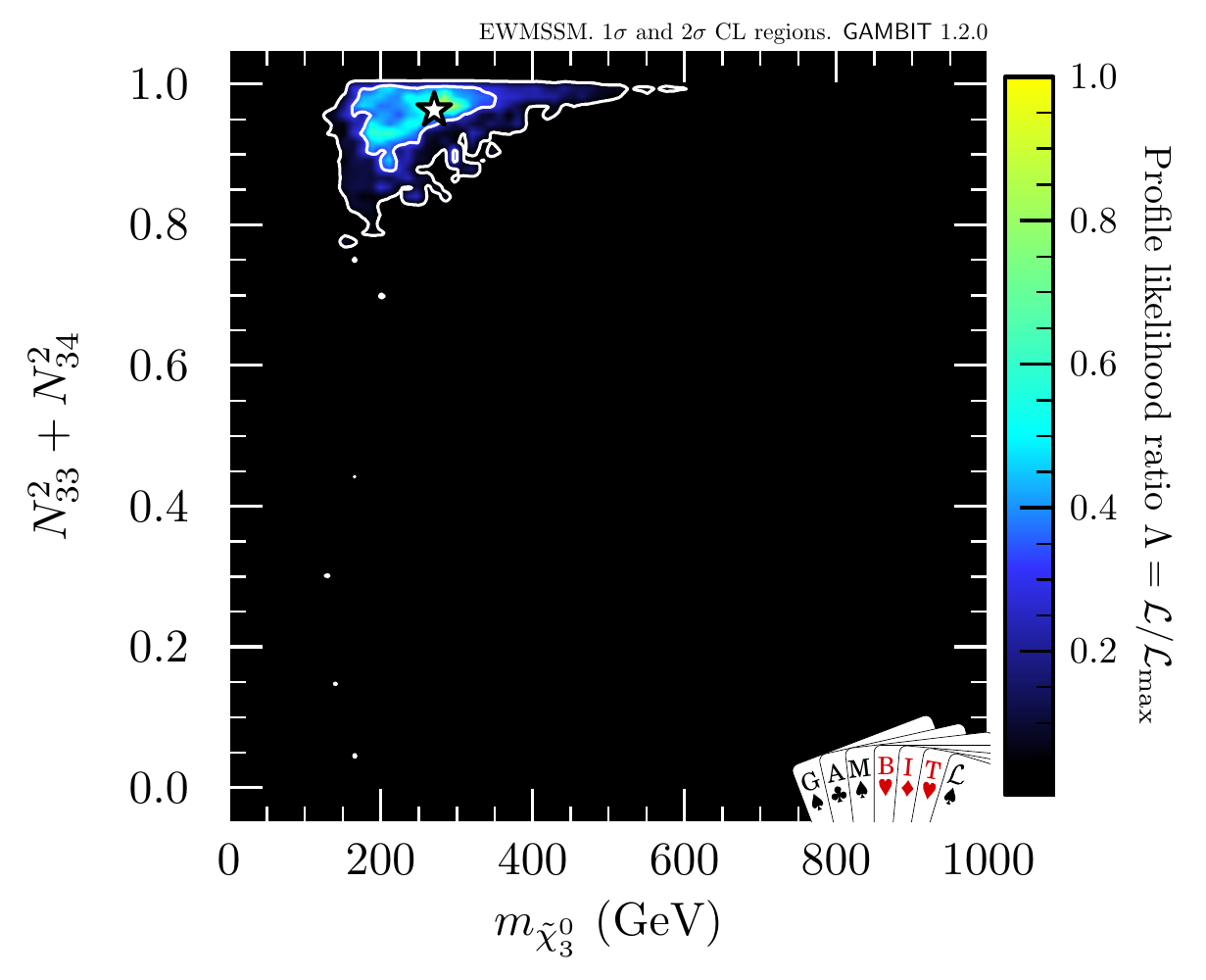}\\
  \includegraphics[width=0.32\textwidth]{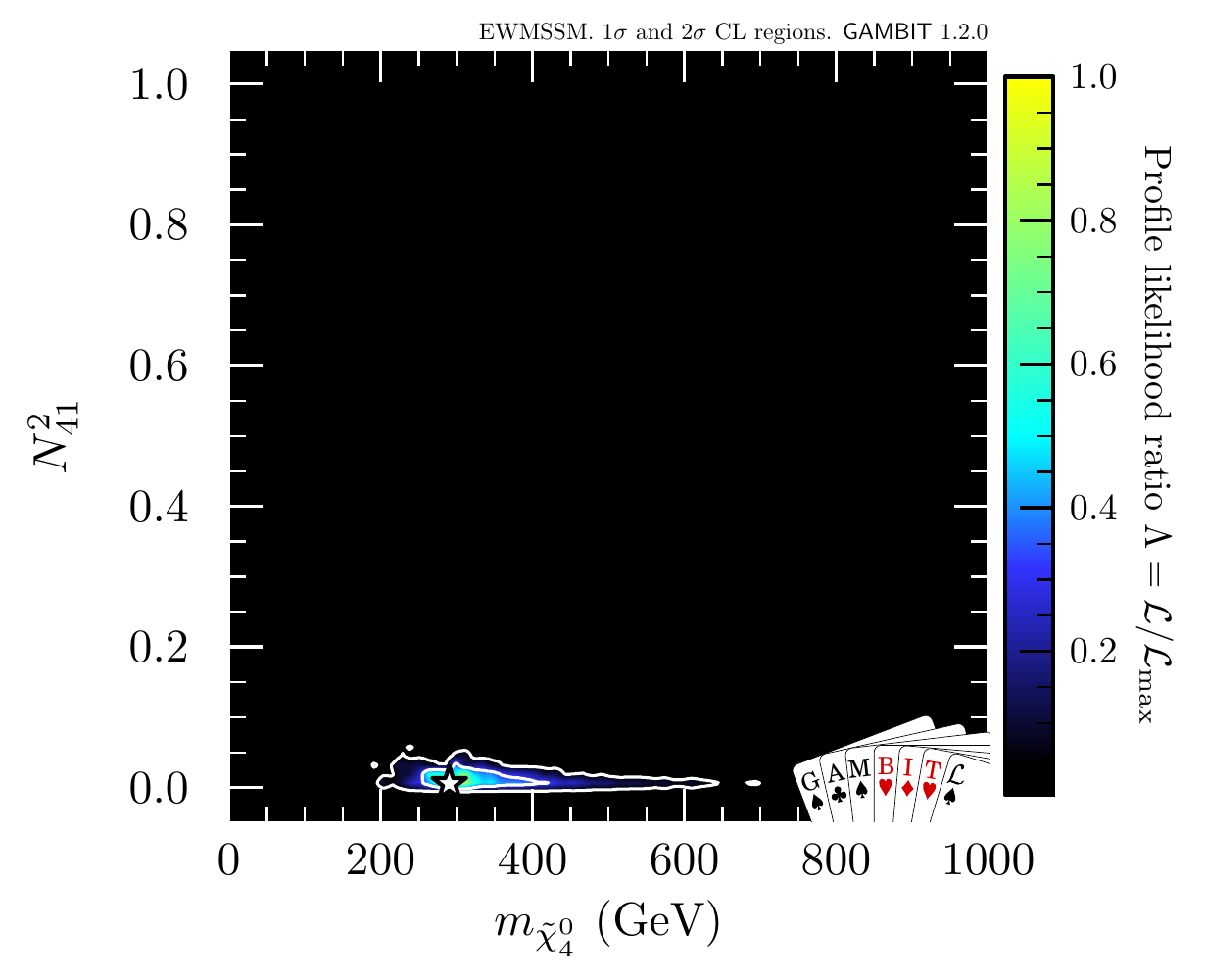}
  \includegraphics[width=0.32\textwidth]{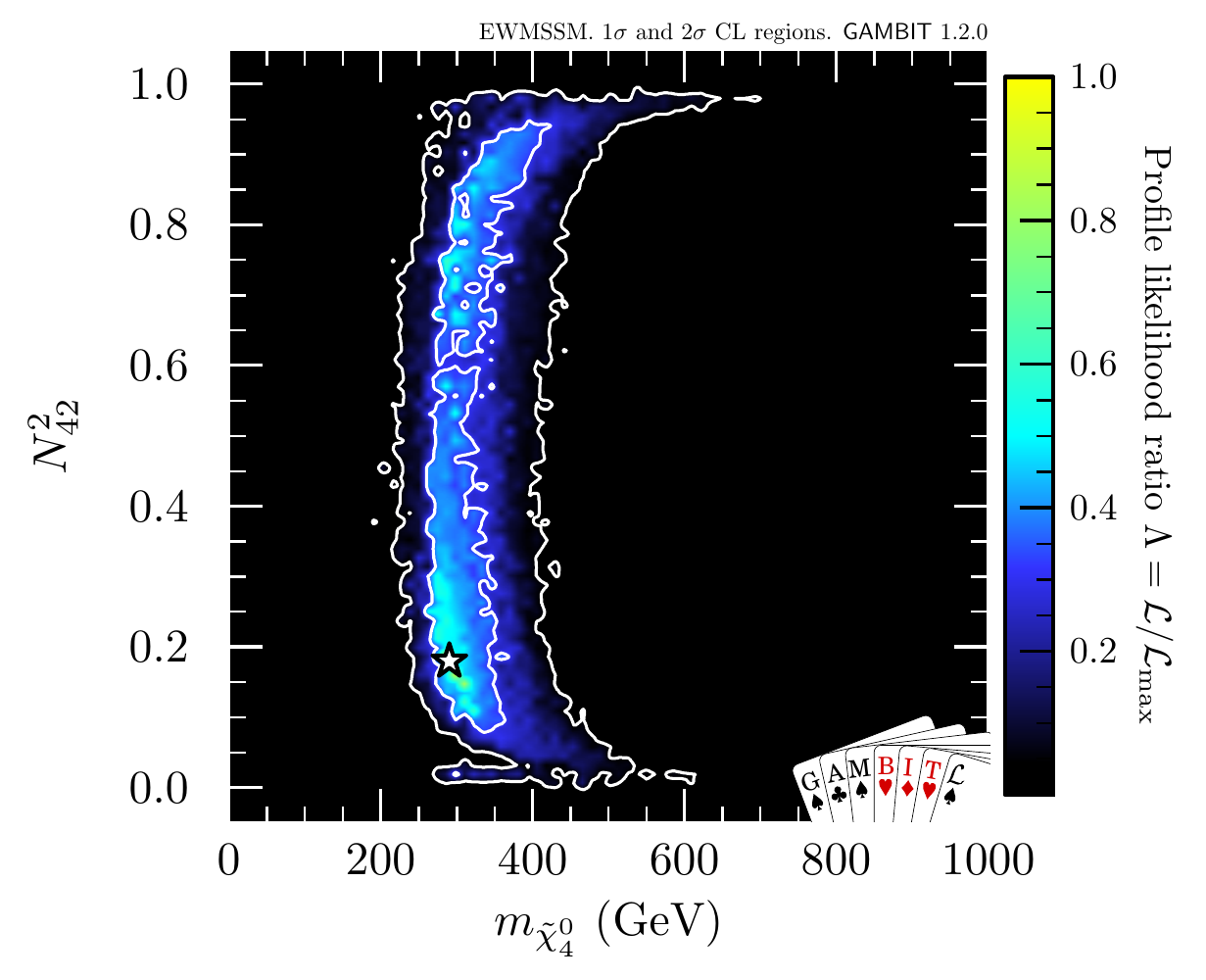}
  \includegraphics[width=0.32\textwidth]{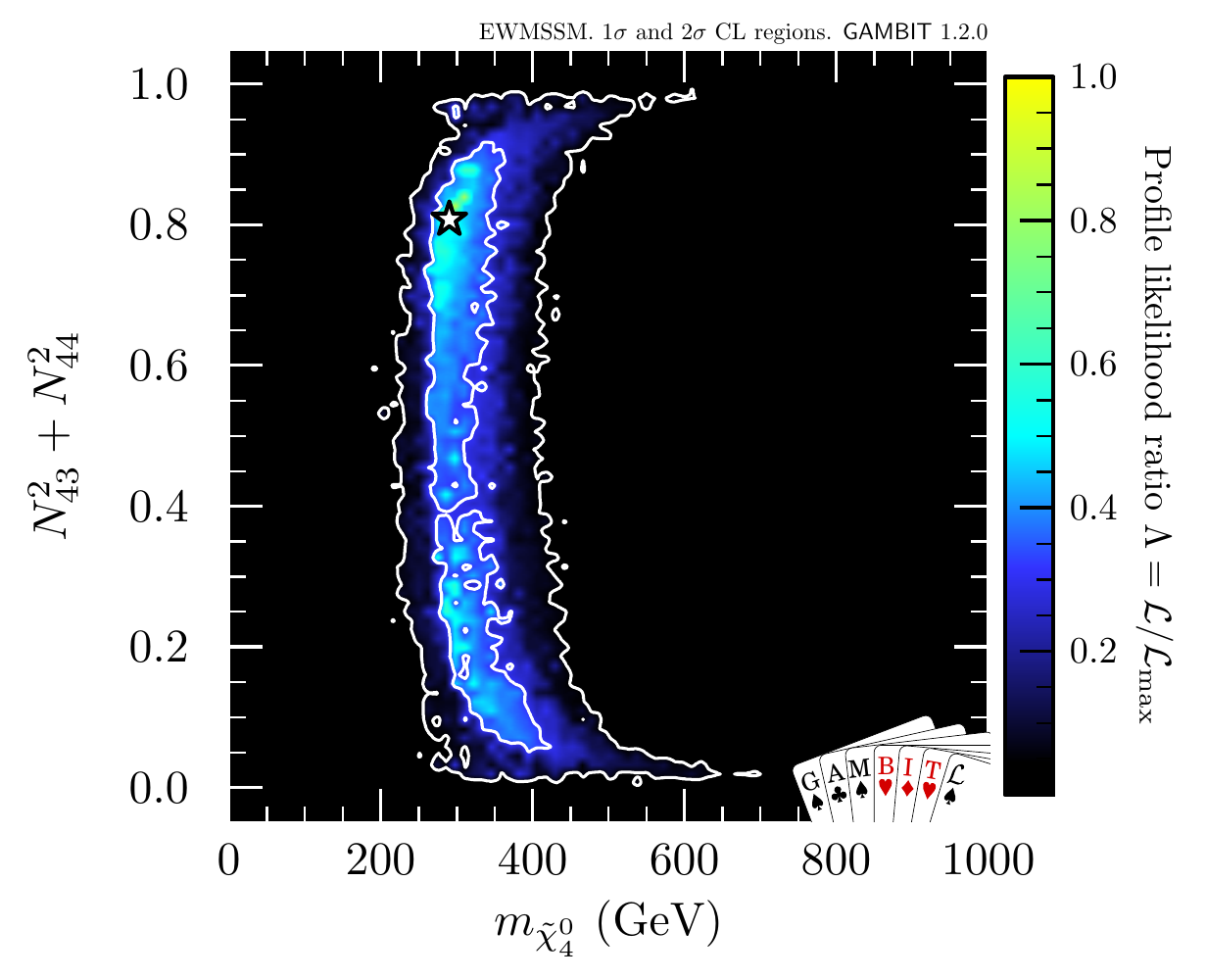}
  \caption{Profile likelihood of the bino (left), wino (middle) and Higgsino (right) content of the four neutralinos (starting from the lightest in the top row), plotted against the mass of the respective neutralino.  Contour lines show the $1\sigma$ and $2\sigma$ confidence regions.  The best-fit point is marked by the white star.}
  \label{fig:n1content}
\end{figure*}

It is also
true that the gaugino contents are heavily constrained by the
observation of the $W$ and $Z$ decay modes, which can provide more
information about the electroweakino sector than would have been
possible given an excess in another channel. In
Figure~\ref{fig:n1content}, we show plots of the fraction of bino,
wino and Higgsino in each neutralino, plotted against the mass of
that neutralino, and with the profile likelihood shown as a colour
contour.  The first row confirms the previous hint that the best-fitting points have a predominantly bino LSP, with a small admixture
of Higgsino and/or wino. The maximum allowed Higgsino contribution exceeds the maximum allowed wino contribution.

Figure~\ref{fig:n1content} also shows that the data has little preference
between wino, Higgsino or mixed scenarios for the $\tilde{\chi}_2^0$ and
$\tilde{\chi}_4^0$, though due to the mass relations between Higgsinos there is a preference for $\tilde{\chi}_3^0$ to be
Higgsino at the $2 \sigma$ level. As expected, when the heavier neutralinos
$\tilde{\chi}_{3,4}^0$ are pushed up in mass they tend to be pure gauge
eigenstates.

There is no preference in the data for the content of the charginos,
which may be wino-like, Higgsino-like or a mixture of the
two.  This is to be expected, given that the data likewise allow any wino-Higgsino admixture for the $\tilde{\chi}^0_2$, and prefer solutions where $\tilde{\chi}^0_2$ and $\tilde{\chi}^\pm_1$ are essentially degenerate in mass.
The only exception is that we again see a tendency for pure
states to arise at high masses, as can be deduced
from the corresponding pure neutralino states.  We therefore omit
plots of the chargino composition.

\subsection{Discussion of excesses}
An important question is how the pattern of these
excesses can be consistent with other published searches, which are a mix of null results, and modest excesses that were not previously thought to be significant. In this section, we investigate whether the different LHC results are consistent with each other for our best-fit models, or whether there are tensions between different analyses.

First we show in Figure~\ref{fig:mass_plane_contributions} the
contribution of each analysis to the total combined likelihood, inside
the interesting $1\sigma$, $2\sigma$ and $3\sigma$ preferred regions, which are
bounded by orange contour lines. These plots show the relative
contribution to the best likelihood in each bin of the 2D profile likelihood map when all analyses are
included. The log-likelihood on the $z$-axis favours a signal
if it is greater than zero. Thus, blue regions indicate analyses
that contribute positively to the combined likelihood, white regions indicate that the analyses have no sensitivity, and red regions indicate tension with the model with the highest likelihood in each bin. We can thus divide the analyses into the following categories:

\begin{figure*}[t]
  \centering
  \includegraphics[width=0.32\textwidth]{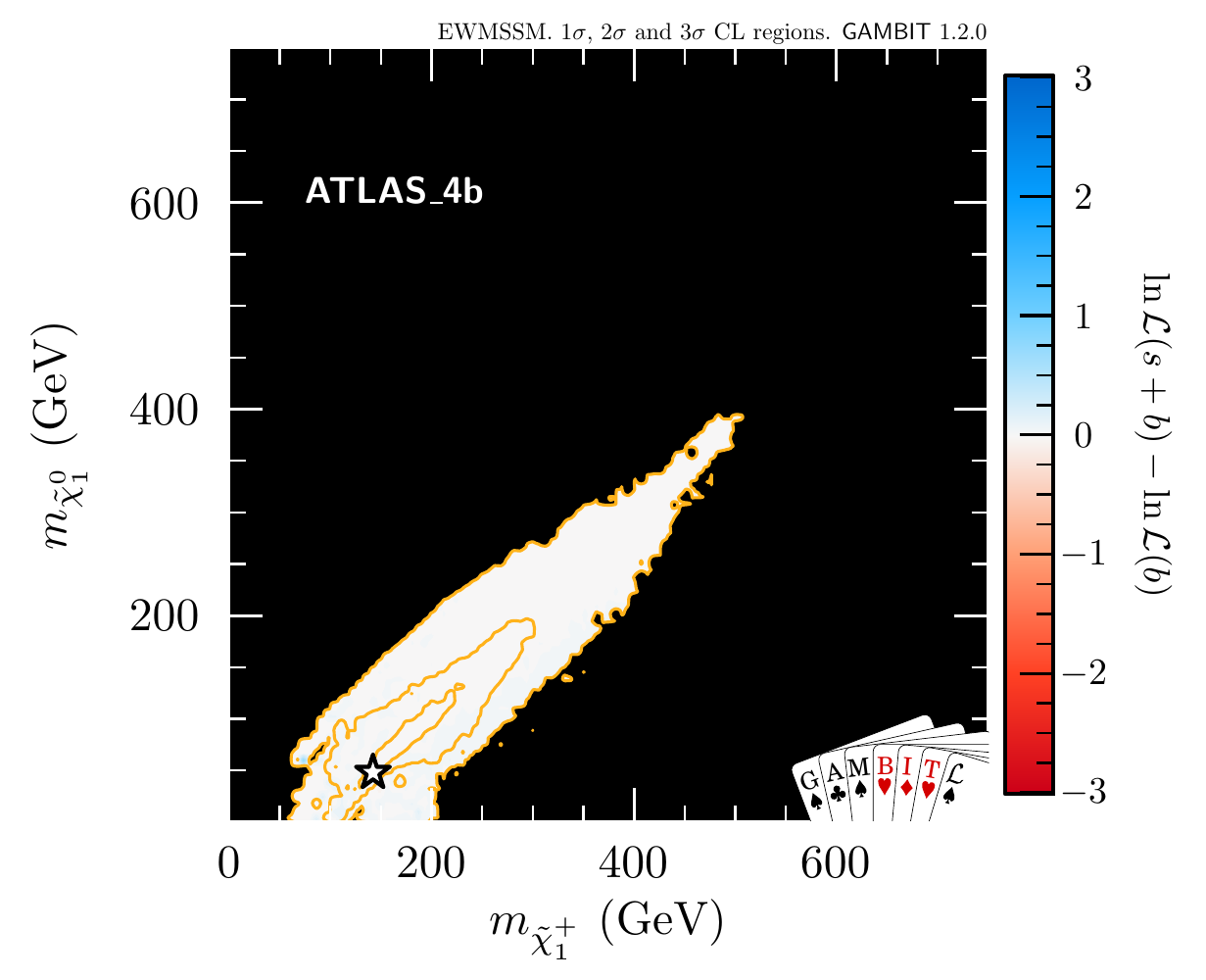}
  \includegraphics[width=0.32\textwidth]{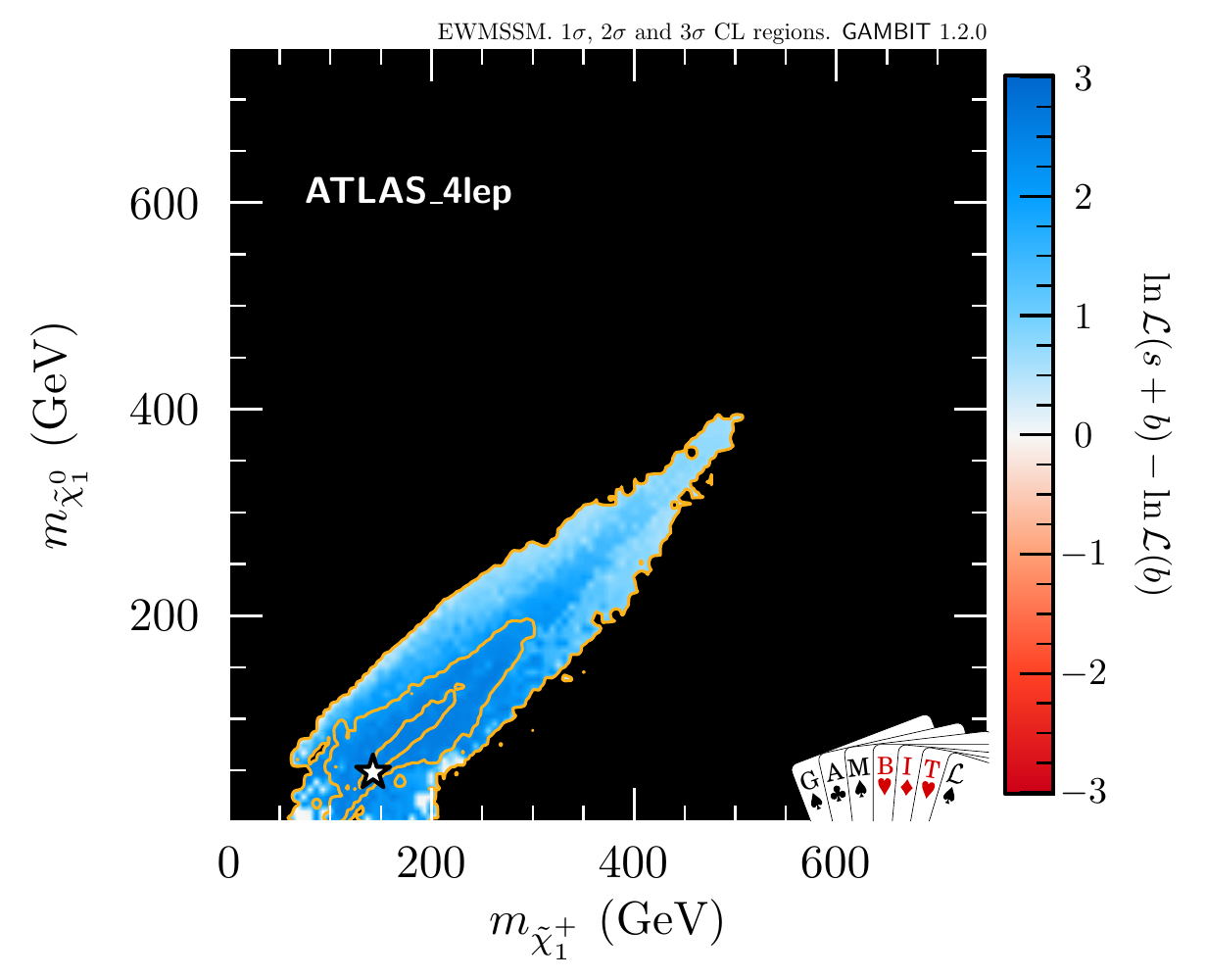}
  \includegraphics[width=0.32\textwidth]{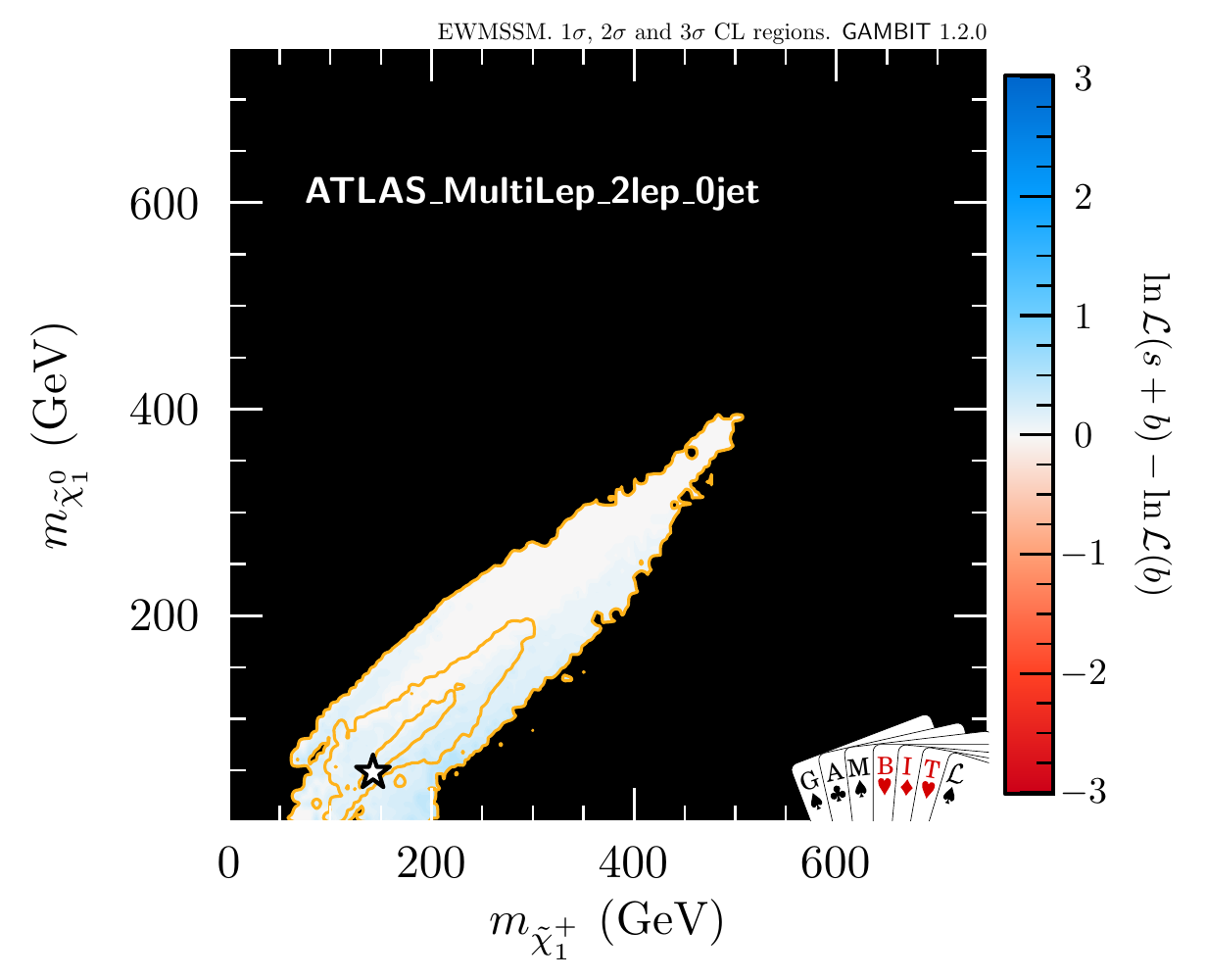}\
  \includegraphics[width=0.32\textwidth]{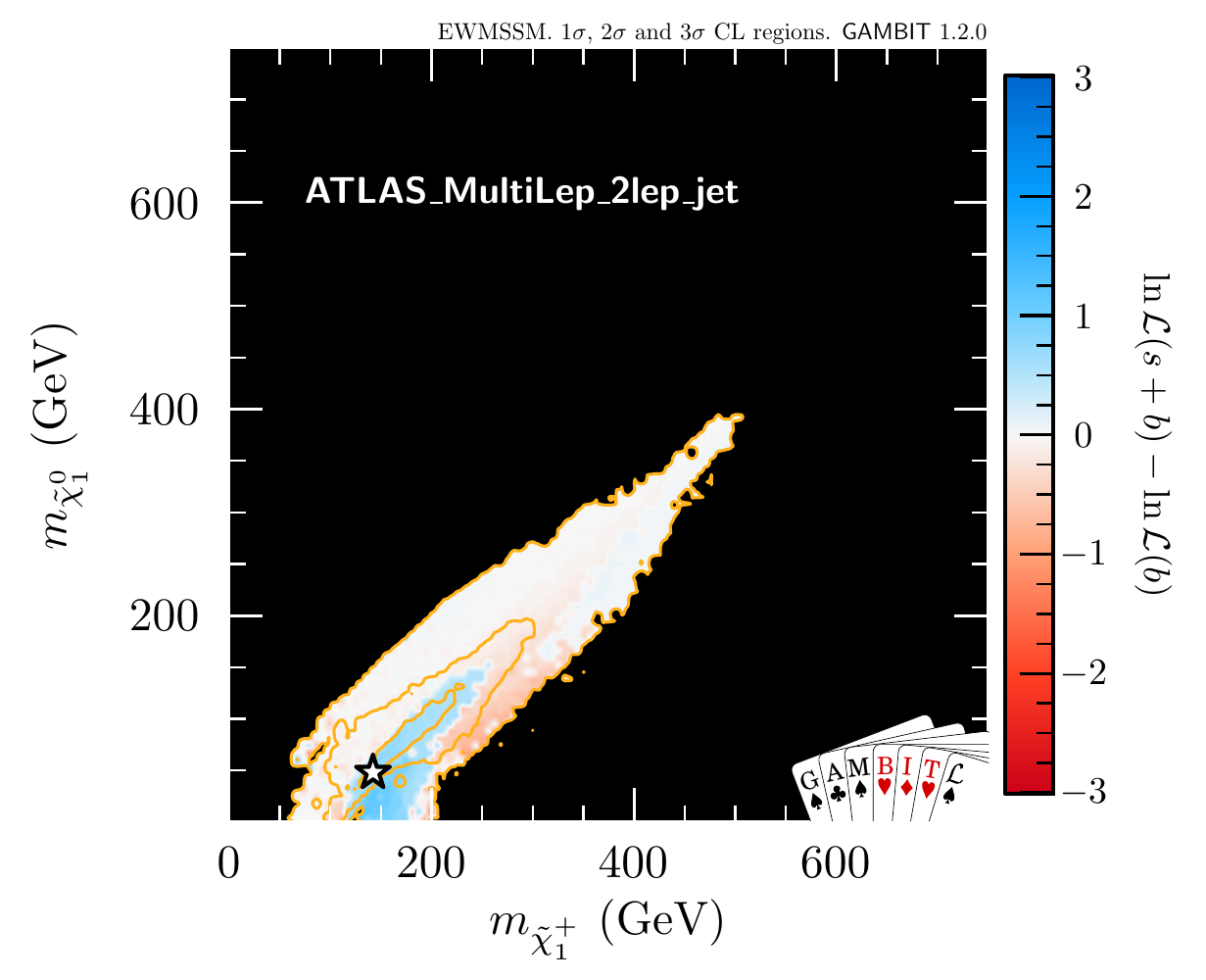}
  \includegraphics[width=0.32\textwidth]{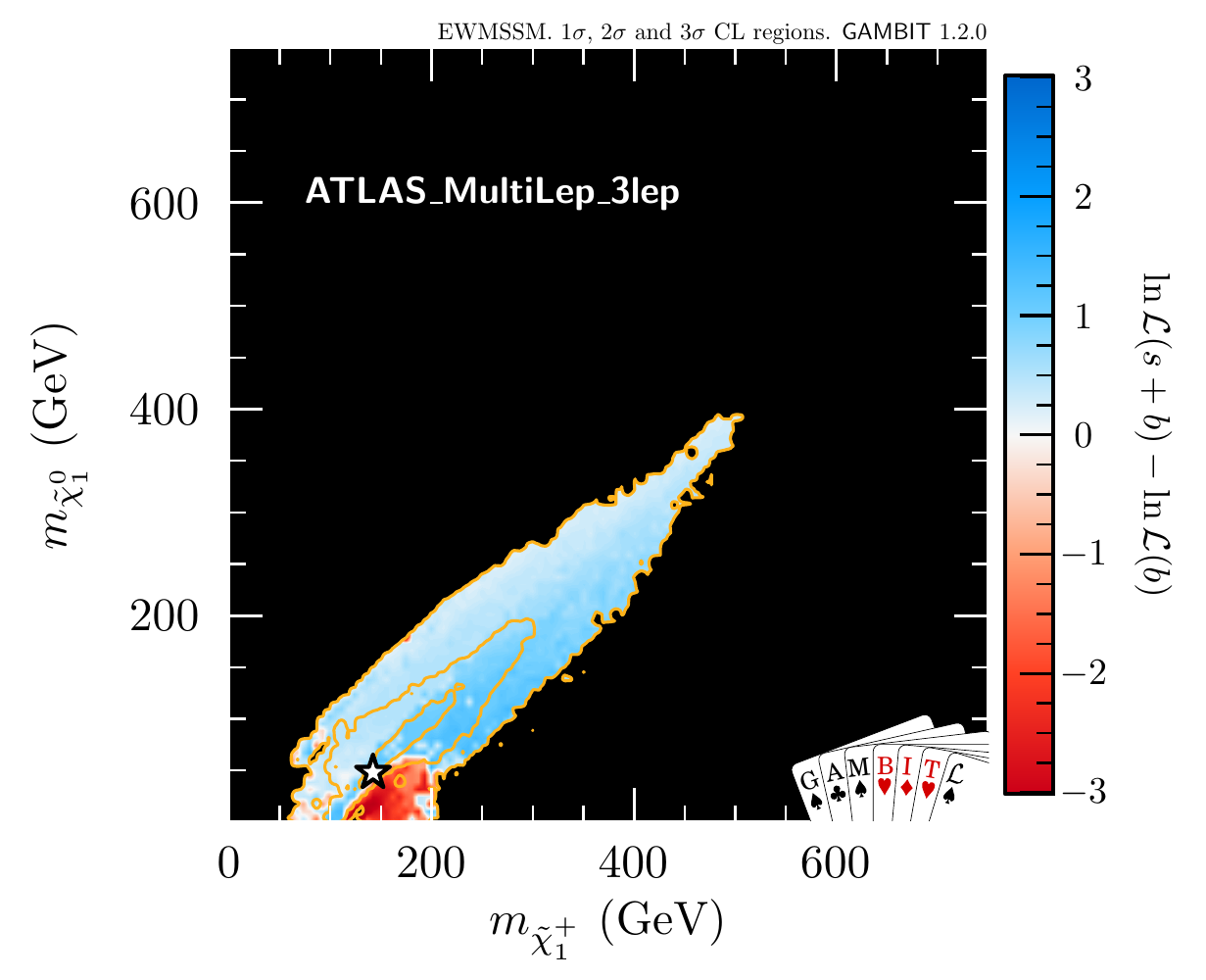}
  \includegraphics[width=0.32\textwidth]{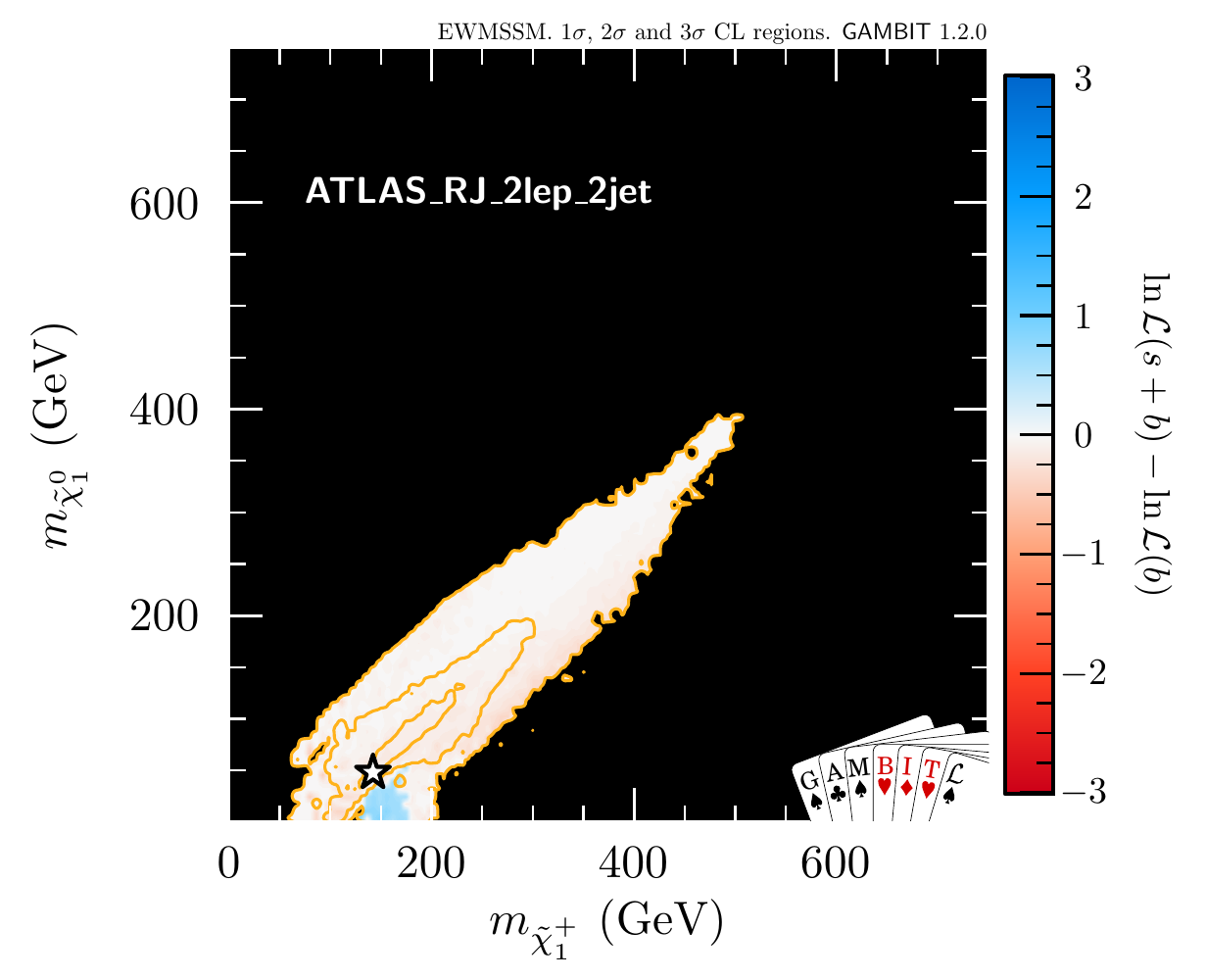}\
  \includegraphics[width=0.32\textwidth]{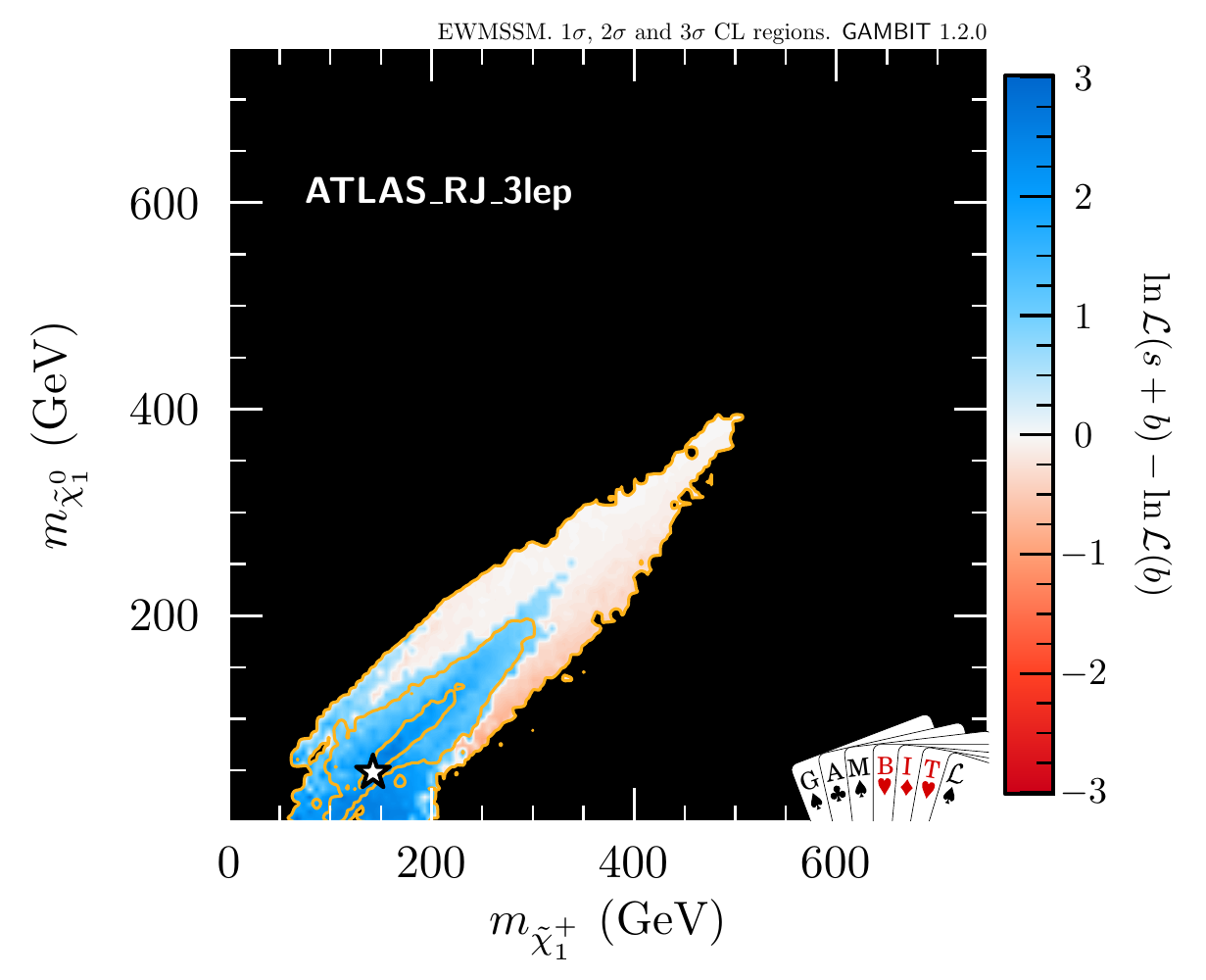}
  \includegraphics[width=0.32\textwidth]{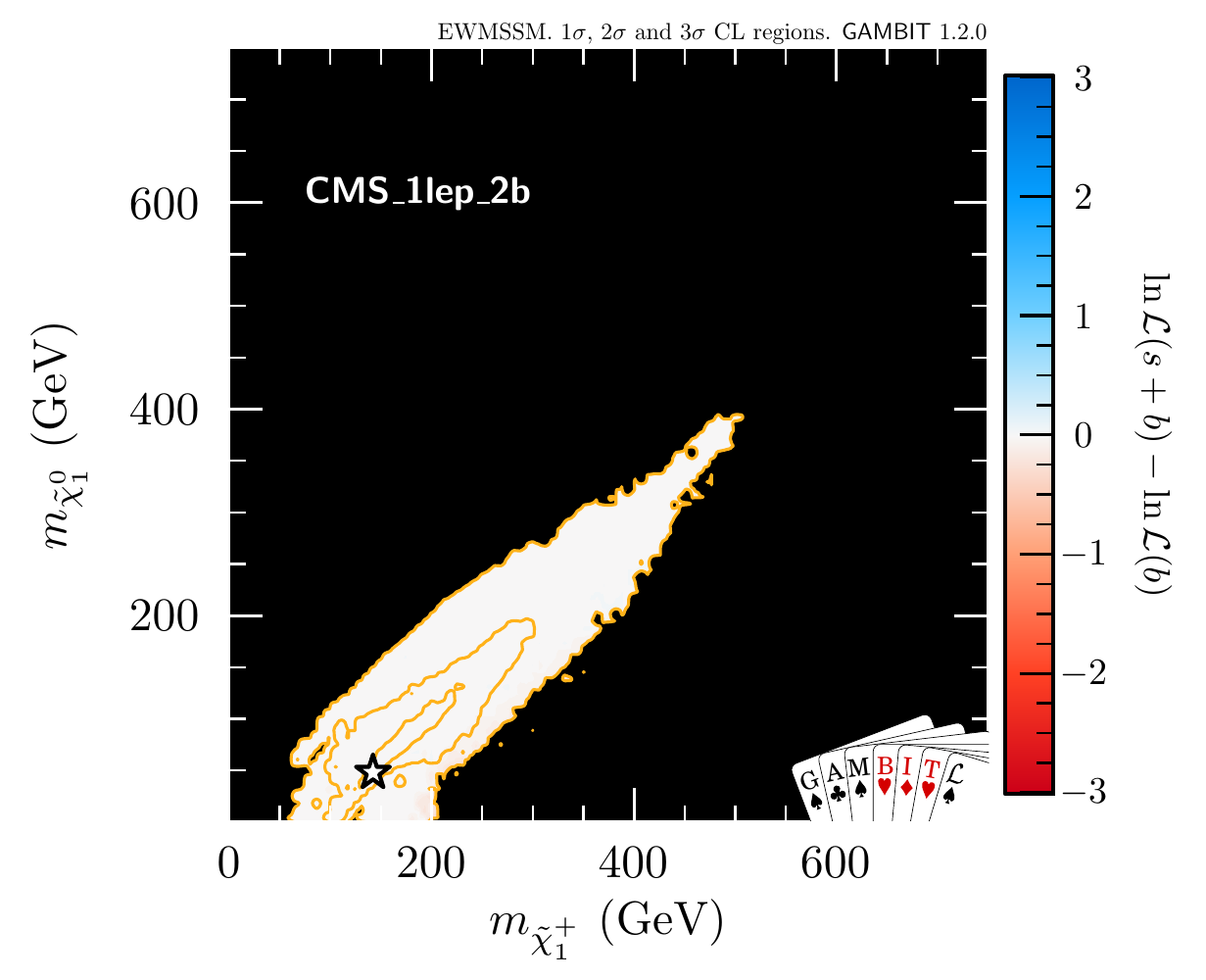}
  \includegraphics[width=0.32\textwidth]{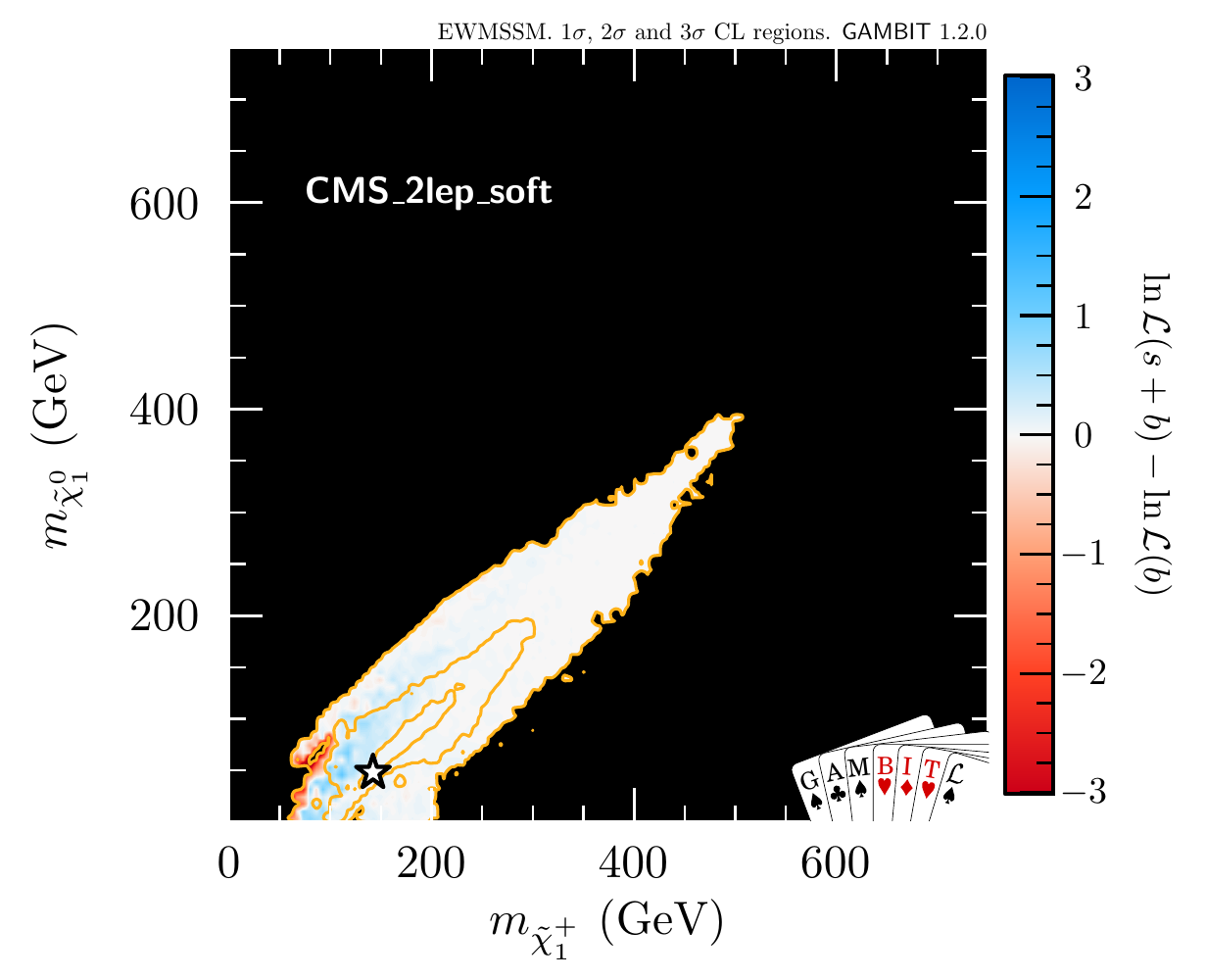}\
  \includegraphics[width=0.32\textwidth]{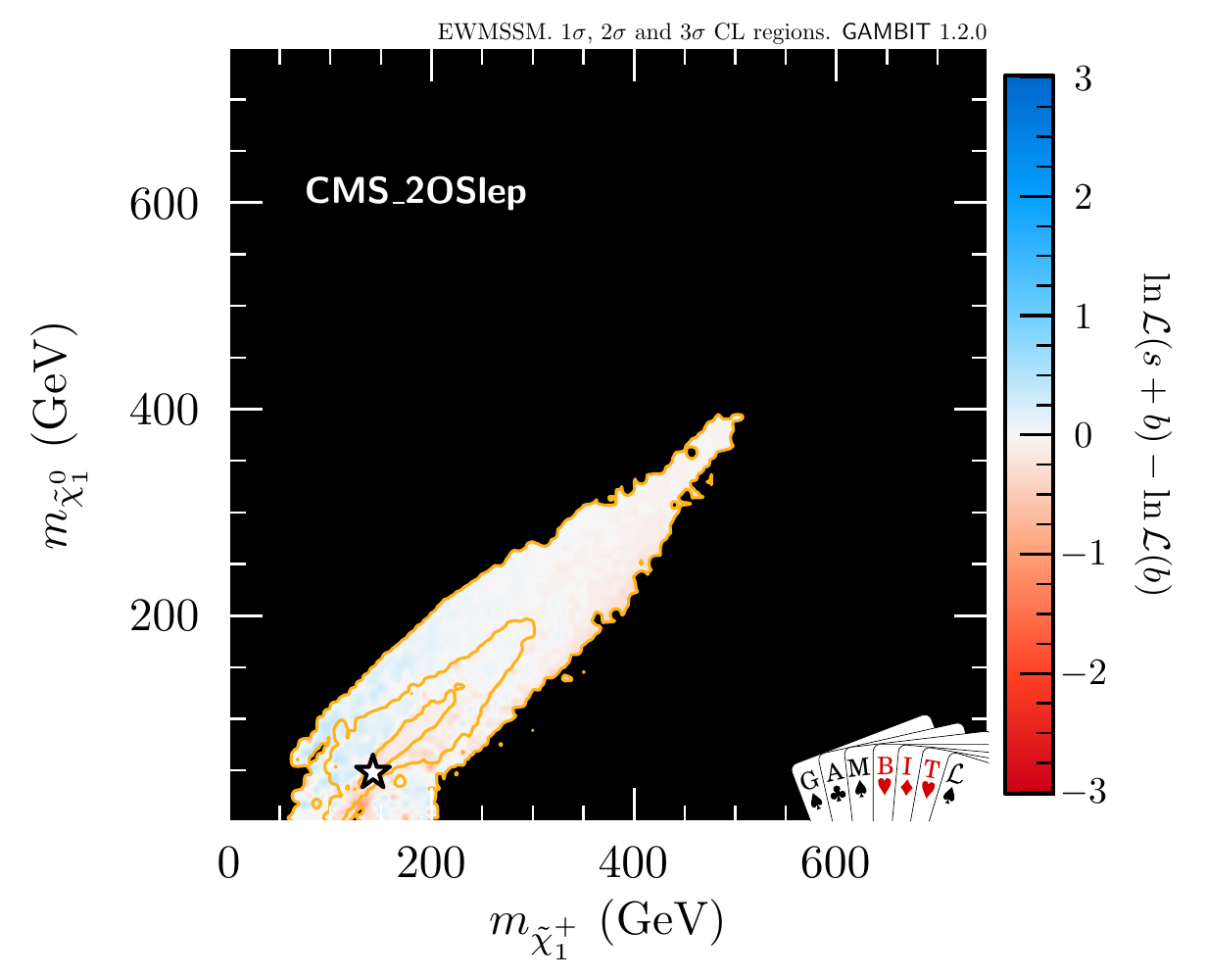}
  \includegraphics[width=0.32\textwidth]{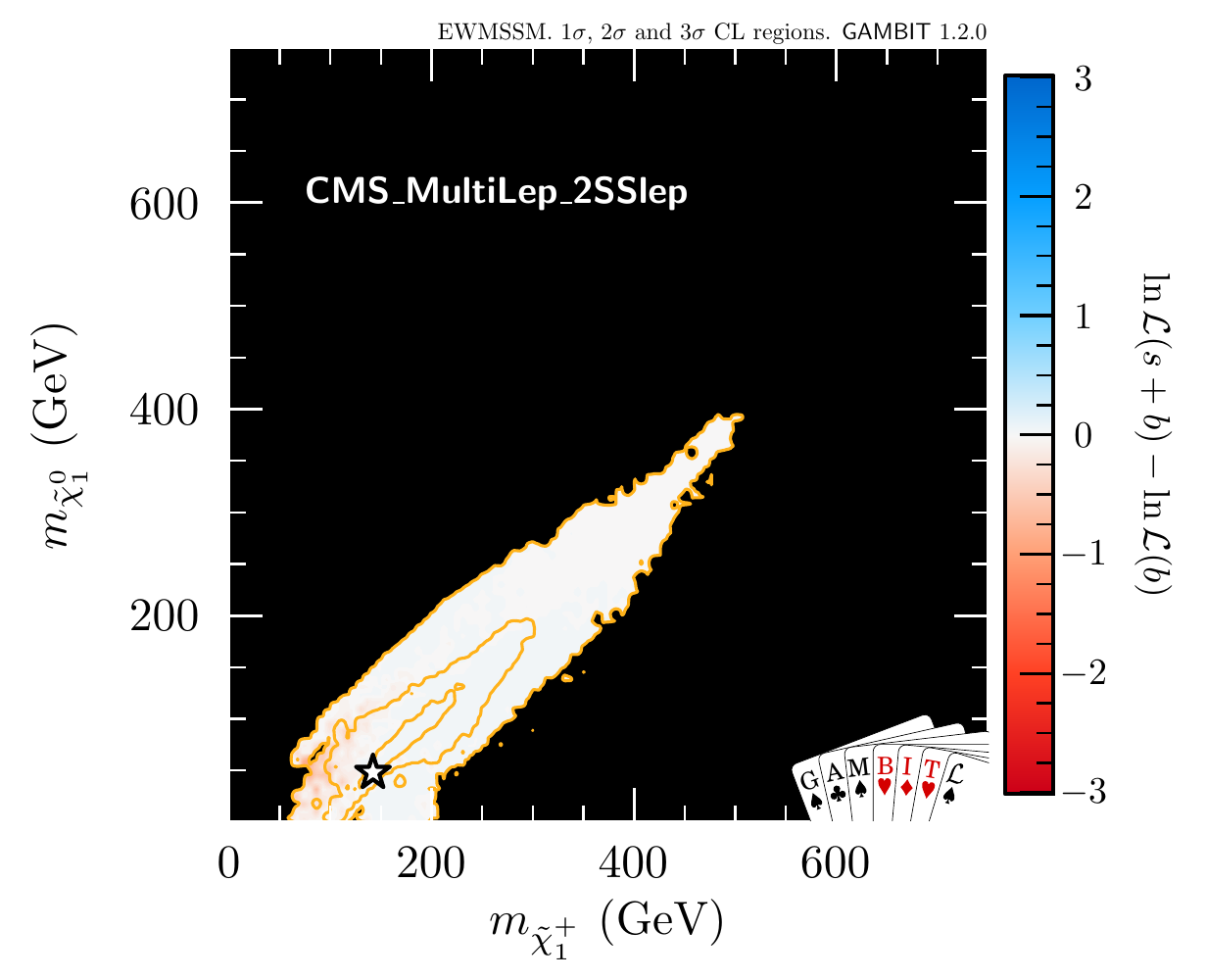}
  \includegraphics[width=0.32\textwidth]{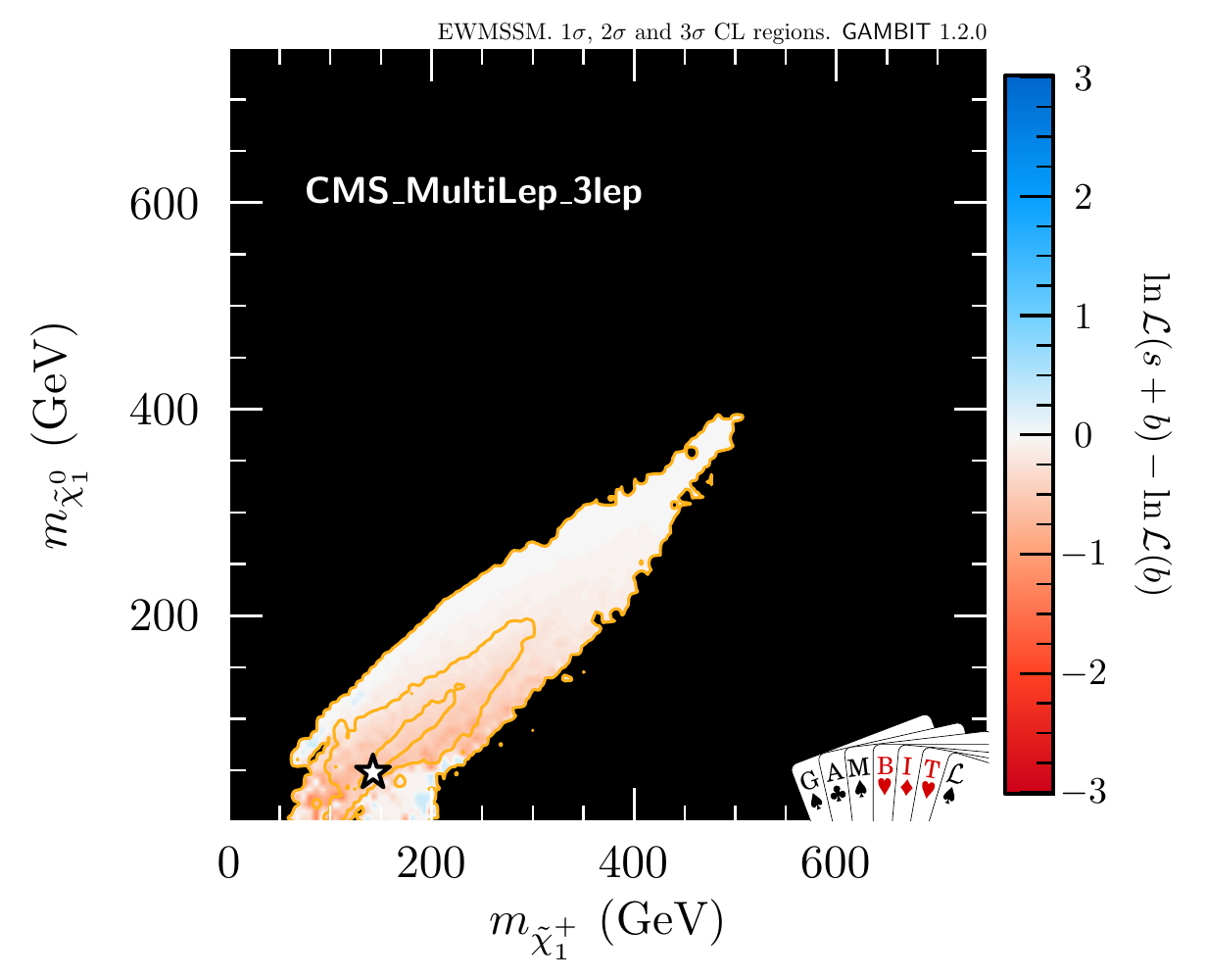}
  \caption{The $1\sigma$, $2\sigma$ and $3\sigma$ regions (orange lines) preferred by our
  combination of searches in the
  $(m_{\tilde{\chi}_1^0}, m_{\tilde{\chi}_1^{\pm}})$ plane.
  For each of the twelve panels, the colors (where present) show
  the contribution to the total log-likelihood from a different search (white
  text). Blue
  indicates that the signal improves the fit to that search
  and red that it worsens it.}
  \label{fig:mass_plane_contributions}
\end{figure*}

\begin{itemize}
\item {\bf Favours background only (red): }A mild tension results from the CMS
multilepton analysis (in the three-soft-lepton signal region, \textsf{CMS\_MultiLep\_3lep}),
which persists across most of our best-fit region. The fact that this search has exclusionary power for our best-fit region
makes sense on the grounds that, of all the signal regions in that
paper, it is the most sensitive to on-shell $W$ and $Z$ production. As
noted earlier, the likelihood of this analysis is hard to estimate
given the lack of published covariance information for the multibin
analysis, and we are forced to use aggregated signal regions that
might not have similar exclusion power. Our results suggest that
there is a mild tension between this analysis and the other analyses
in our combined likelihood, but it is impossible to quantify the
effect precisely, and we will therefore leave this as an open
question. Note that we also see a stronger tension in a small region beyond our $2\sigma$ contour in the three lepton signal region of the conventional ATLAS multilepton analysis (\textsf{ATLAS\_MultiLep\_3lep}), which is important in shaping our final 2$\sigma$ contour.
\item{\bf No sensitivity (white): }The \textsf{ATLAS\_4b} analysis has no sensitivity to our best-fit models, which is to be expected given that it is optimised for scenarios with two on-shell Higgs bosons present in the final state. Although our best-fit models will include some Higgs production, they must feature copious production of $W$ and $Z$ bosons in order to fit the observed excesses in the searches targeted at on-shell $W$ and $Z$ production. The CMS one lepton plus two $b$-jet analysis, targeting $Wh$ final states, and the two same-sign lepton regions of the CMS multilepton analysis also show no sensitivity to our highest-likelihood models. This makes sense given that the ATLAS excesses require an on-shell $Z$ boson to be produced most of the time in our models. An alternative option is that there are in fact $hh$ and $Wh$ final states produced relatively often in our models, but the kinematics of the final state particles differ from those on which the CMS analyses were optimised. This makes our benchmark points, provided in Sec.~\ref{sec:benchmark} below, particularly interesting for the optimisation of future searches. We note with interest that the two-lepton zero-jet region of the conventional ATLAS multilepton analysis appears white in these plots at the best-fit region, indicating no tension with the analyses that show positive log-likelihood contributions. This indicates that there is no tension between the analyses containing excesses and these signal regions.  We expand on this point below.
\item {\bf Favours signal (blue): }The strongest positive contributions to our log-likelihood come from the conventional ATLAS multilepton analyses (in
the four-or-more-lepton, three-lepton and two-lepton plus jets final states, i.e., \textsf{ATLAS\_4lep}, \textsf{ATLAS\_MultiLep\_3lep} and \textsf{ATLAS\_MultiLep\_2lep\_jet}), and the ATLAS recursive jigsaw analysis (\textsf{ATLAS\_RJ\_3lep} and \textsf{ATLAS\_RJ\_2lep\_2jet}).
A weaker positive contribution near the best-fit region is evident in the CMS two soft lepton analysis (\textsf{CMS\_2lep\_soft}) and the CMS two opposite-sign lepton analysis (\textsf{CMS\_2OSlep}).
\end{itemize}

The fact that the conventional ATLAS multilepton analysis shows
evidence of an excess is naively in conflict with the published
exclusion limits.  However, we have already shown (left panel of
Figure~\ref{fig:limitComparison}) that our \colliderbit treatment of this
analysis can reproduce the exclusion in the same simplified model (which assumes $\tilde{\chi}_2^0\tilde{\chi}_1^\pm$ production and subsequent decay to $W$ and $Z$ bosons).
This analysis prefers a signal in our
results instead of an exclusion because
our electroweakino model differs from the ATLAS simplified model.

\begin{figure*}[t]
  \centering
  \includegraphics[width=0.32\textwidth]{figures/MSSMEW_155_151_obs2D_605.pdf}
  \includegraphics[width=0.32\textwidth]{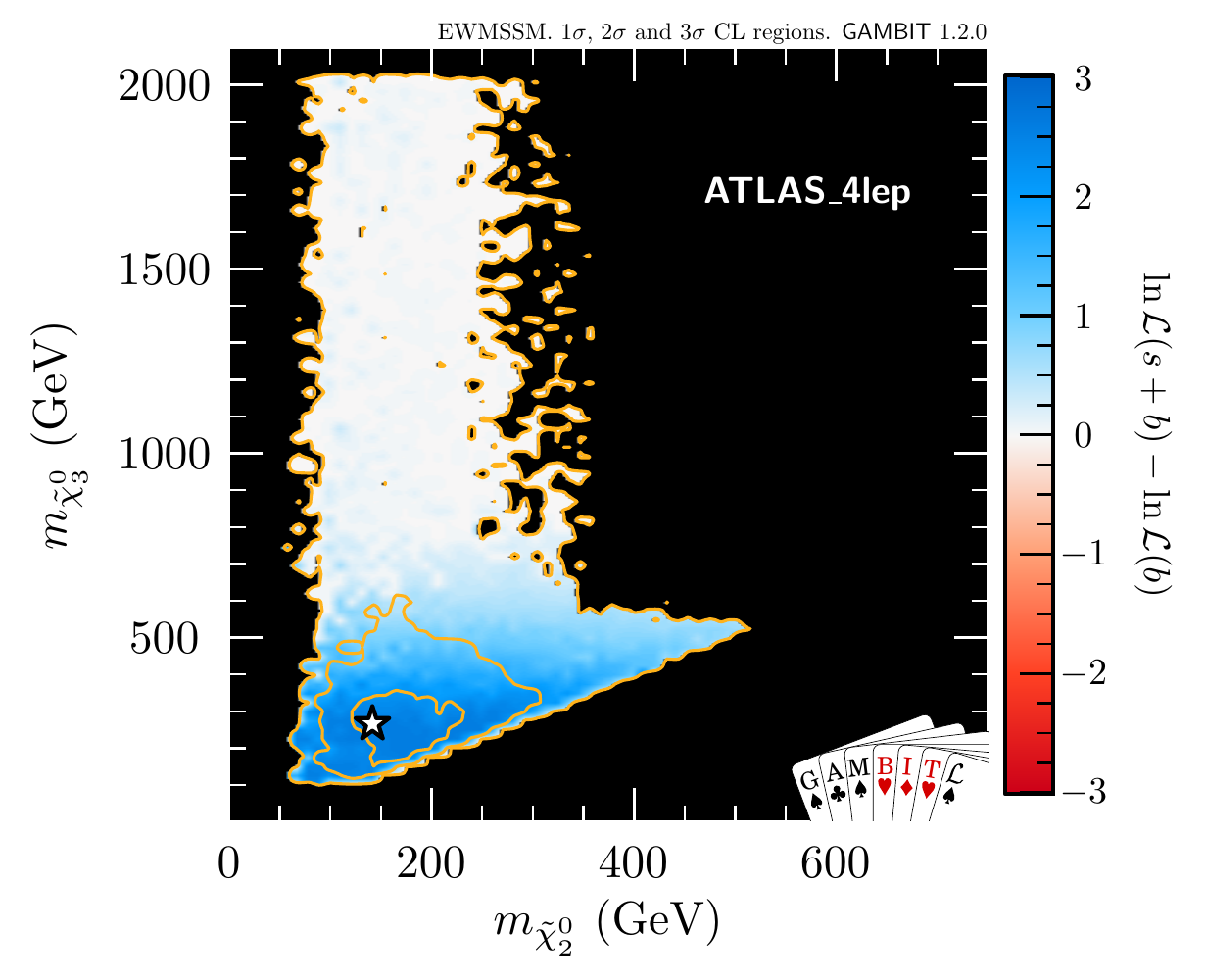}
  \includegraphics[width=0.32\textwidth]{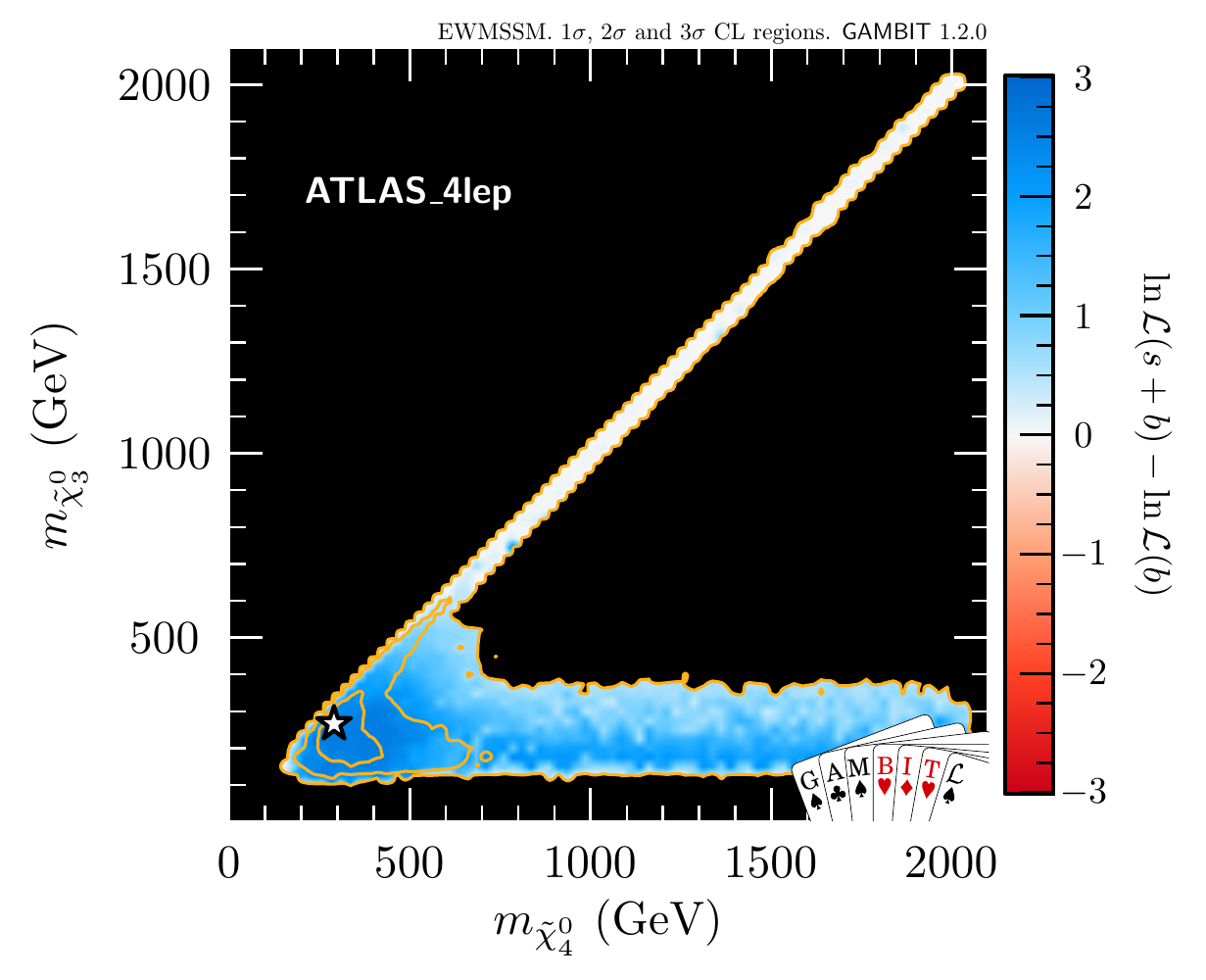}
  \includegraphics[width=0.32\textwidth]{figures/MSSMEW_155_151_obs2D_607.pdf}
  \includegraphics[width=0.32\textwidth]{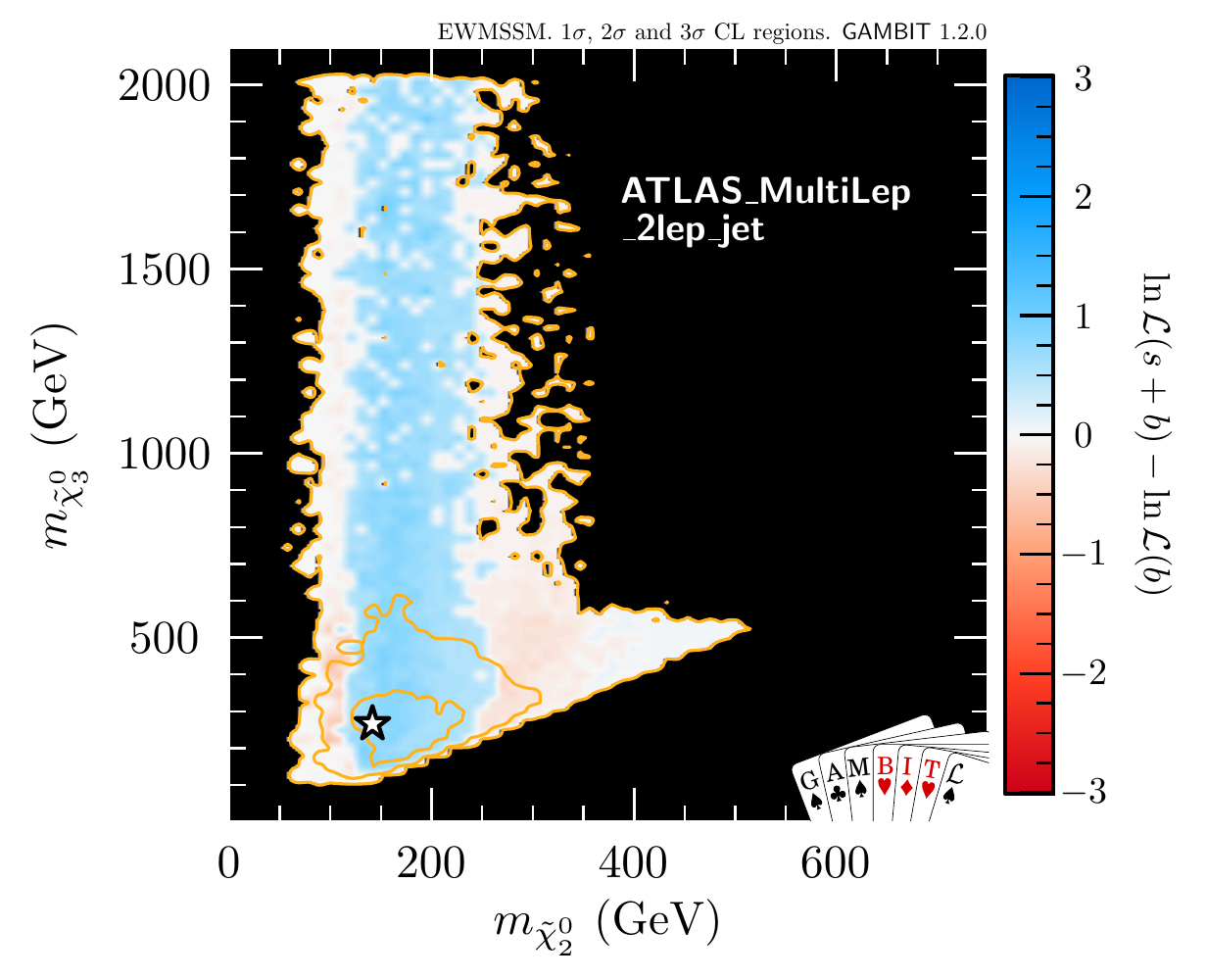}
  \includegraphics[width=0.32\textwidth]{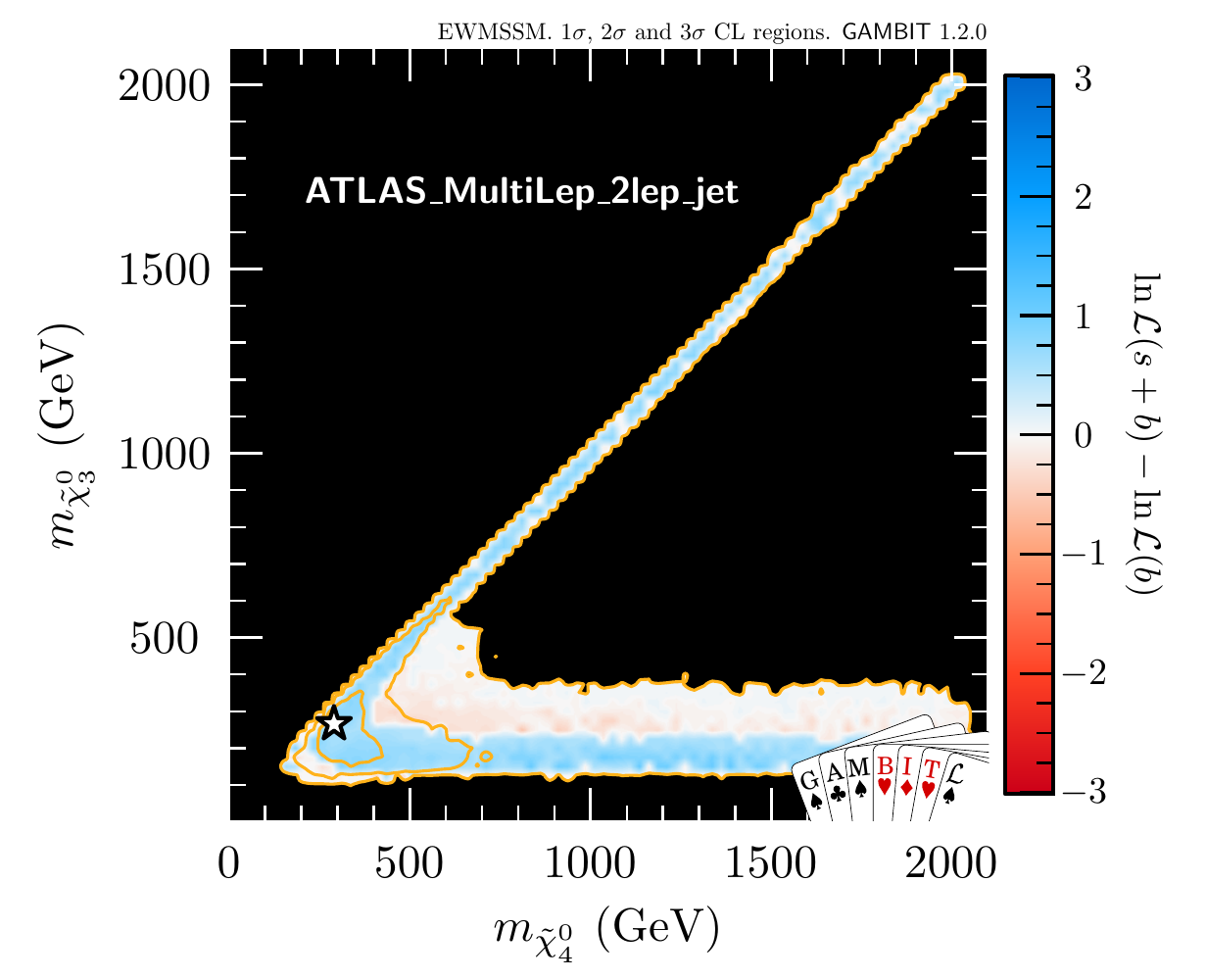}
  \includegraphics[width=0.32\textwidth]{figures/MSSMEW_155_151_obs2D_608.pdf}
  \includegraphics[width=0.32\textwidth]{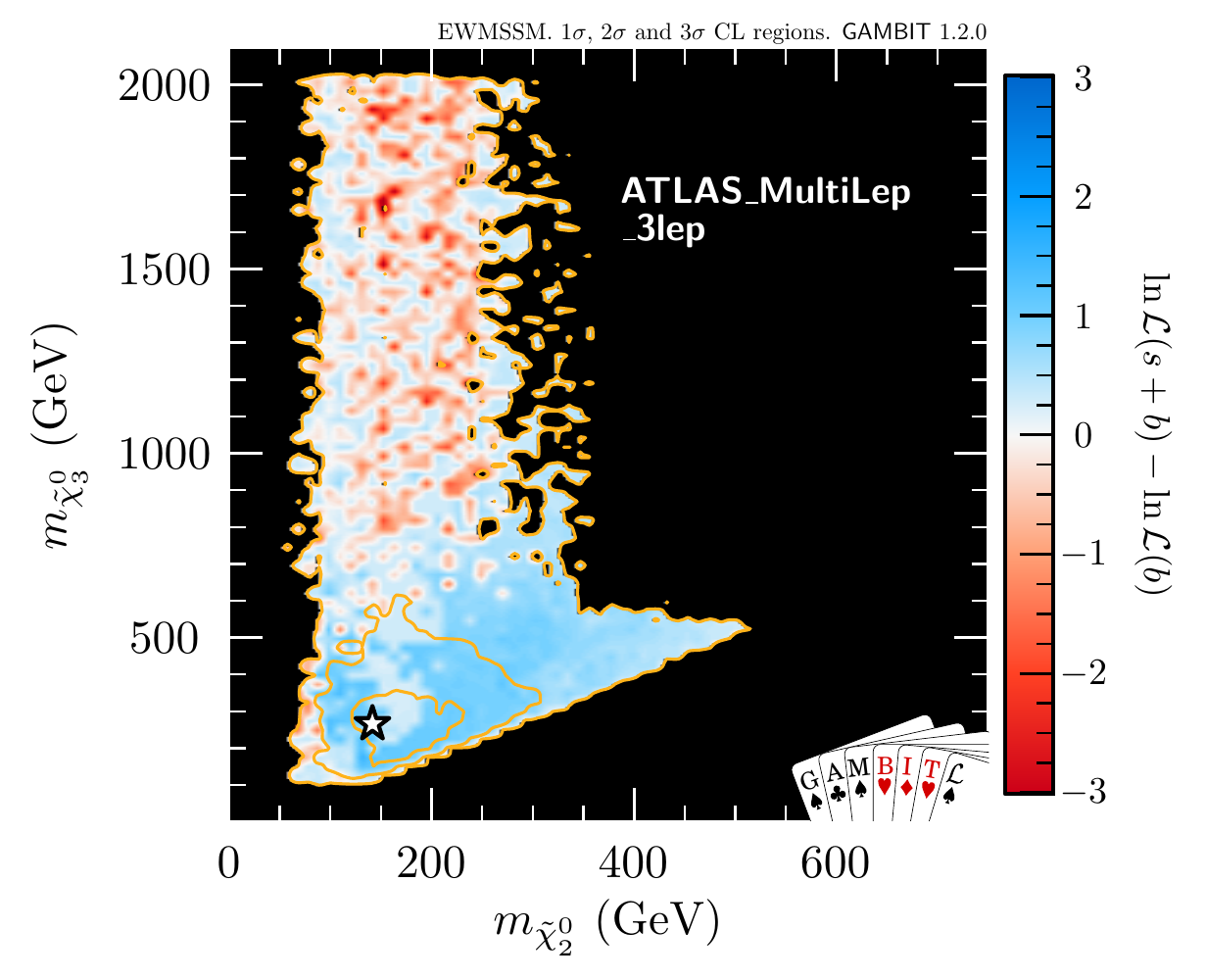}
  \includegraphics[width=0.32\textwidth]{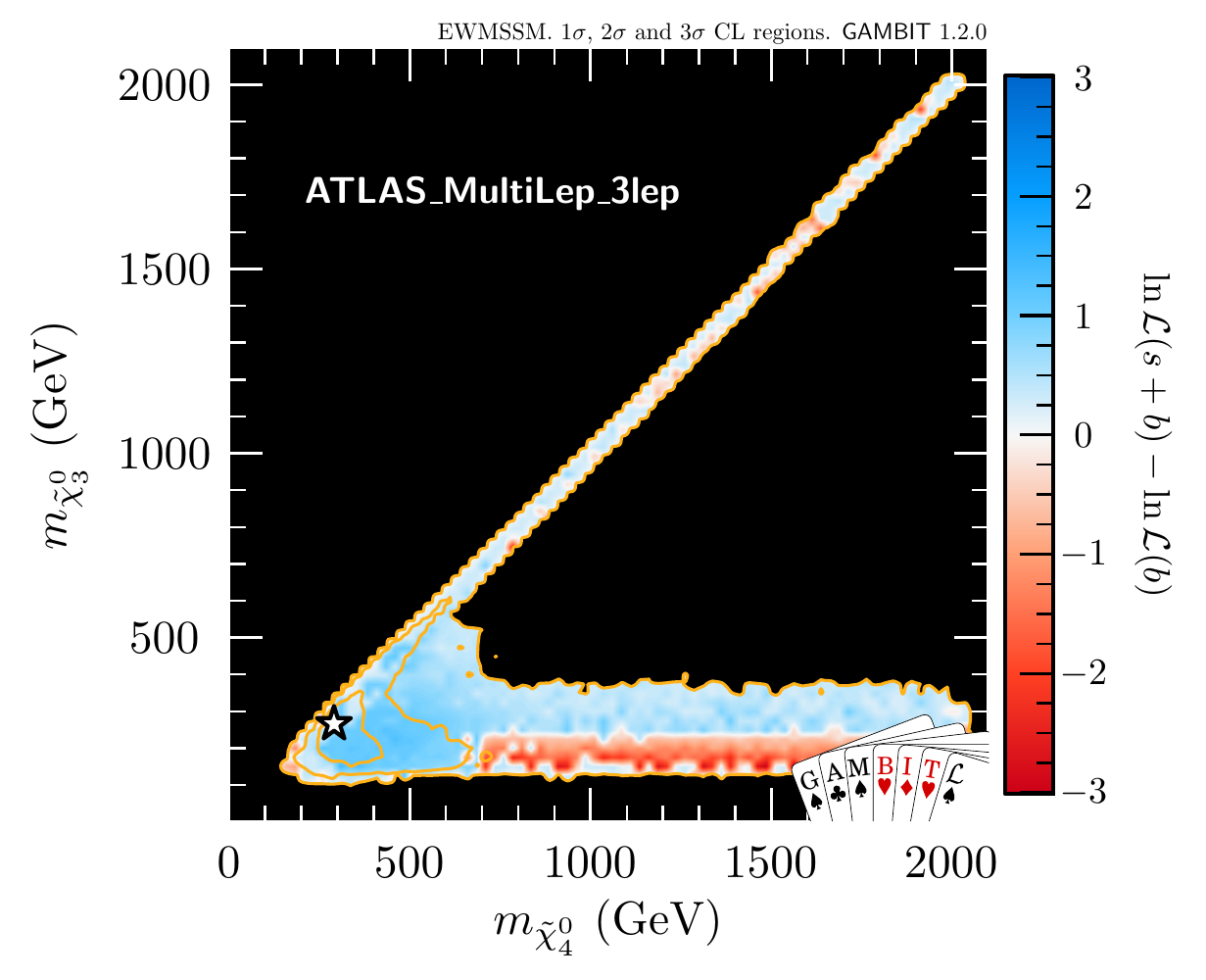}
  \includegraphics[width=0.32\textwidth]{figures/MSSMEW_155_151_obs2D_610.pdf}
  \includegraphics[width=0.32\textwidth]{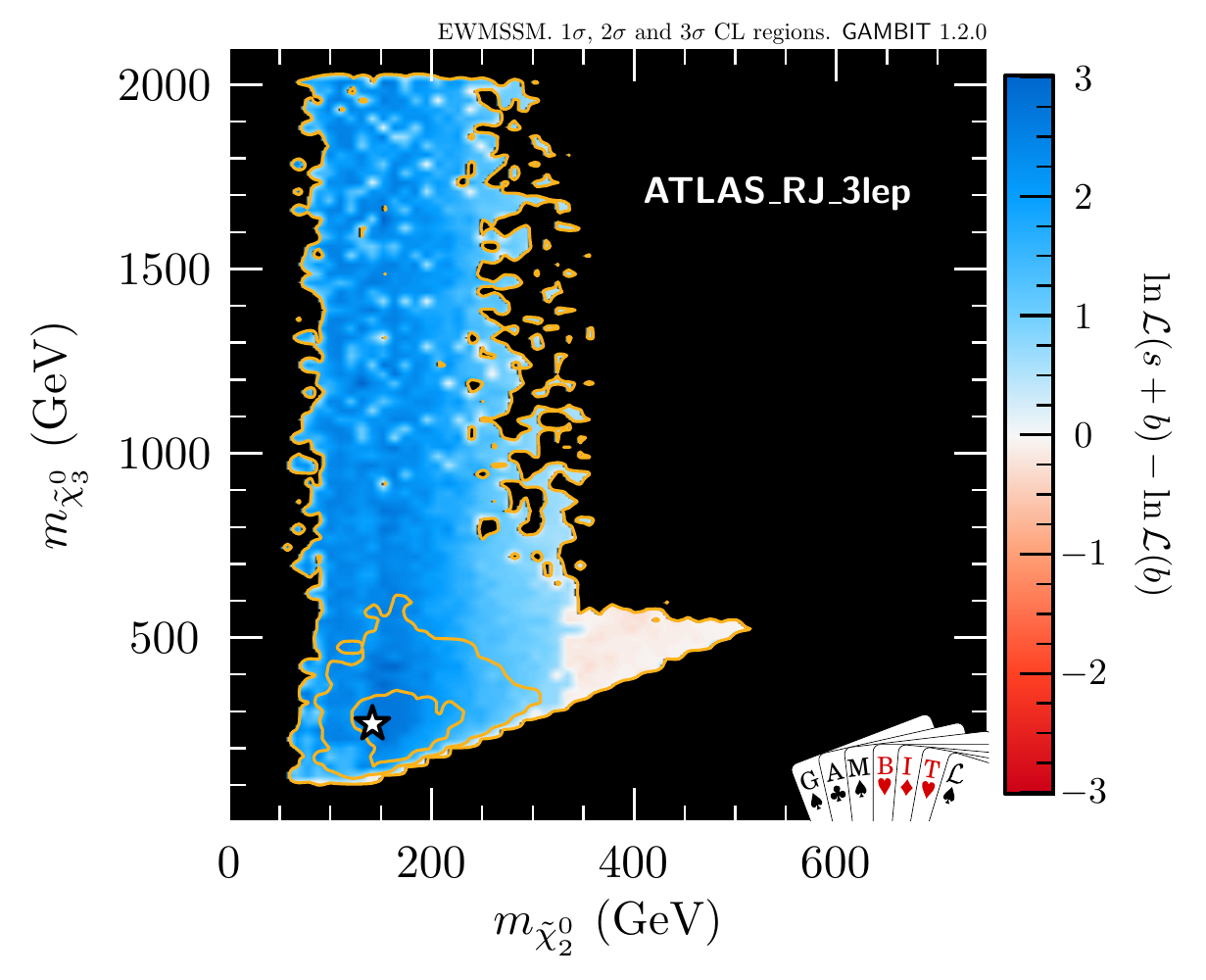}
  \includegraphics[width=0.32\textwidth]{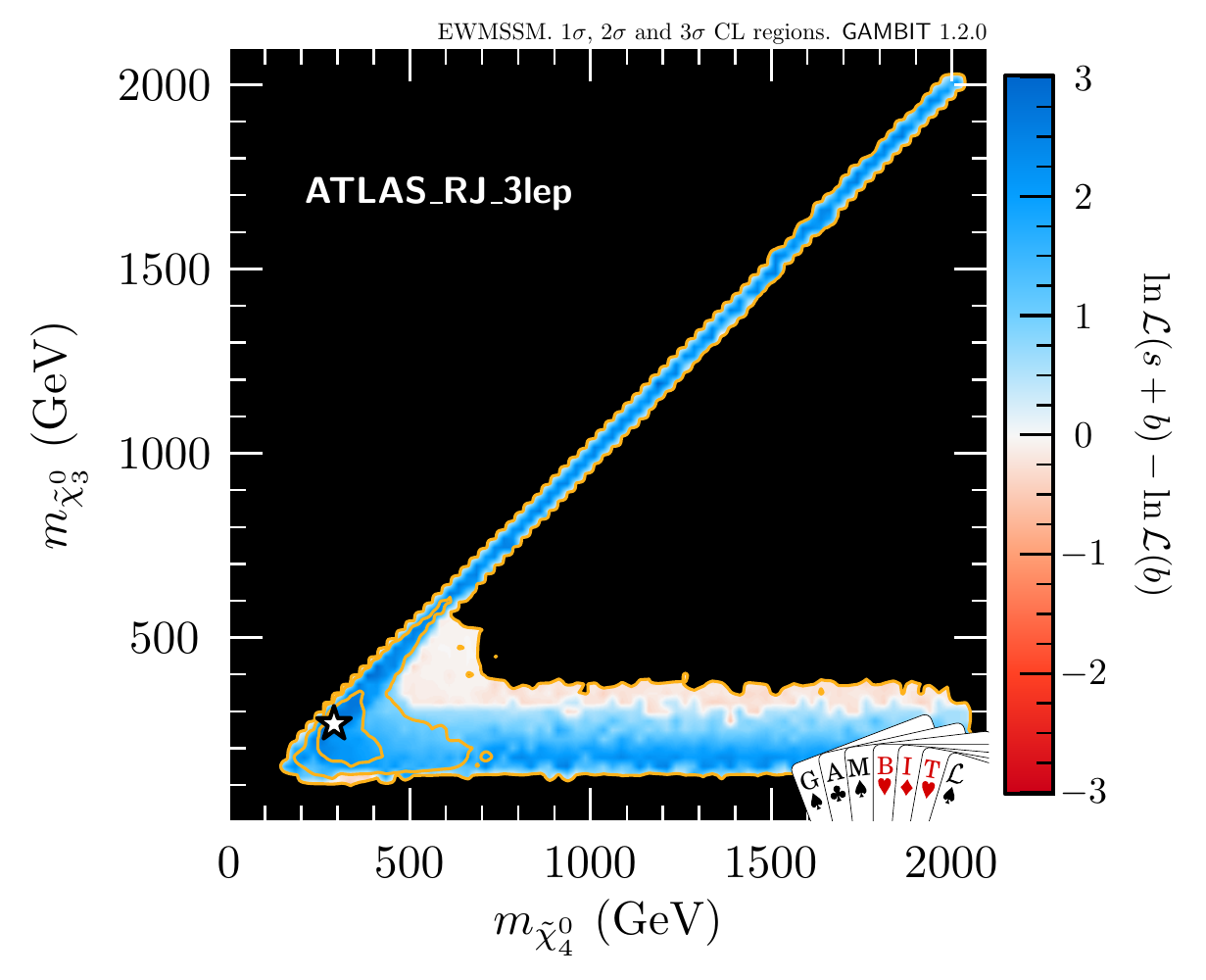}
  \caption{
  The  $1\sigma$, $2\sigma$ and $3\sigma$ regions (orange lines) preferred by our
  combination of searches in the
  $\tilde{\chi}^\pm_1$ versus $\tilde{\chi}^0_1$ (left),
  $\tilde{\chi}^0_2$ versus $\tilde{\chi}^0_3$ (middle) and
  $\tilde{\chi}^0_4$ versus $\tilde{\chi}^0_3$ (right)
  mass planes.
  The colors (where present) show the contribution to the total log-likelihood from
  the
  \textsf{ATLAS\_4lep} (top),
  \textsf{ATLAS\_MultiLEP\_2lep\_2jet} (second row),
  \textsf{ATLAS\_MultiLEP\_3lep}  (third row),
  and \textsf{ATLAS\_RJ\_3lep} (bottom)
  searches. Blue
  indicates that the signal improves the fit to that search
  and red that it worsens the fit.}
  \label{fig:mass_planes_selected_analyses}
\end{figure*}

To further understand the interplay between the analyses driving our fit result,
we show, in Figure~\ref{fig:mass_planes_selected_analyses},
log-likelihood contributions for selected analyses in the mass planes
$(m_{\tilde{\chi}_2^{0}},m_{\tilde{\chi}_3^0})$ and
$(m_{\tilde{\chi}_3^{0}},m_{\tilde{\chi}_4^0})$, and for easy
comparison, show again the
$(m_{\tilde{\chi}_1^{\pm}},m_{\tilde{\chi}_1^0})$ plane alongside. As in
the previous plot, the log-likelihood shown on the $z$-axis favours a
signal if it is greater than zero and has some exclusionary power if
it is less than zero.
These plots show that the contribution from an
analysis may depend very strongly on whether there are additional
light electroweakinos beyond the set of states included in the
simplified model, as shown by the dependence on the mass of $\tilde{\chi}^0_3$
in the middle panels and the dependence on the $\tilde{\chi}^0_4$ mass in the lower
part of the right-hand panel for \textsf{ATLAS\_MultiLEP\_3lep} (third row of third column).

Before discussing Figure~\ref{fig:mass_planes_selected_analyses} in more detail, we
note that some care must be taken in interpreting these plots.
First, as described earlier, there can be large errors from Monte Carlo statistics.  To reduce statistical fluctuations, we post-processed our results to have more Monte Carlo events in higher-likelihood regions.
In particular, we ensured that at least 4 million events were generated for all points within the $2 \sigma$ and $3 \sigma$ regions, at least 16 million events for points in the $1 \sigma$ region and 64 million events for the 500 highest-likelihood points.
Second, our parameter scans have sampled the interesting $2\sigma$ region -- where the
total likelihood function is peaked -- more densely than the very broad $3\sigma$ region.
The combination of the comparatively low MC statistics and less dense sampling
implies that there are significantly larger uncertainties on the
profile likelihood in the $3 \sigma$ region, compared to the $2 \sigma$ and $1\sigma$ regions.
When we profile over the dimensions not shown in the plane, we are selecting
points with the highest total likelihood, which biases the results
towards larger values for the log-likelihood contribution shown on the
$z$-axis of this plot, i.e.\ the MC and sampling uncertainties tend to lead to an
overestimate of the log-likelihood contribution when profiling.
For instance, one should interpret the small patches of blue (i.e.\
positive log-likelihood) that appear for higher values of
$m_{\tilde{\chi}_3^0}$ in the three lepton signal region (third row of second column)
with care, due to the larger statistical fluctuations in the
$3 \sigma$ region and the bias towards positive values from profiling.
Negative log-likelihood contributions between the $2 \sigma$ and $3 \sigma$
contours should therefore be considered somewhat more robust than positive ones.

We focus first on the plots for the ATLAS search in the four or more lepton final
state (\textsf{ATLAS\_4lep}), shown in the top row
of Figure~\ref{fig:mass_planes_selected_analyses}. The positive log-likelihood
contribution that our best-fit region gets from this search originates from the modest
excess seen in the \textsf{SR0D} signal region (described in Sec.~\ref{sec:LHCanalyses}) with two reconstructed $Z$-boson
candidates and missing energy. This can clearly not come from
$\tilde{\chi}_1^+\tilde{\chi}_1^-$ or $\tilde{\chi}_2^0\tilde{\chi}_1^\pm$ production
in a simplified model, but relies on the production of heavier neutralinos with decays
 $\tilde{\chi}_i^0\to Z\tilde{\chi}_{1,2}^0$. We see that the search
 prefers a $\tilde{\chi}_3^0$ lighter than around 500--600\,GeV, but does not
 significantly constrain the $\tilde{\chi}_4^0$ when this is a bino or a wino
 (the horizontal band at high $m_{\tilde{\chi}_4^0}$ in the right-hand plot).

Moving on to the plots for \textsf{ATLAS\_MultiLEP\_2lep\_jet}, \textsf{ATLAS\_MultiLEP\_3lep} and \textsf{ATLAS\_RJ\_3lep} in rows two, three and four, respectively, we notice that a preference for $m_{\tilde{\chi}_3^0} < 600$\,GeV is also seen in \textsf{ATLAS\_MultiLEP\_3lep}, in addition to a clear preference for $m_{\tilde{\chi}_4^0} < 700$\,GeV when $m_{\tilde{\chi}_3^0} < 250$\,GeV. This suggests that additional light neutralinos play a very important role in evading the limits placed by \textsf{ATLAS\_MultiLEP\_3lep} on the ATLAS simplified model (which we reproduced in the top right panel of Figure \ref{fig:limitComparison}). In contrast, both \textsf{ATLAS\_MultiLEP\_2lep\_jet} and \textsf{ATLAS\_RJ\_3lep} can provide positive log-likelihood contributions in the limit of decoupling $\tilde{\chi}_3^0$ and $\tilde{\chi}_4^0$ (both Higgsino), as long as the mass-splitting between the wino pair $\tilde{\chi}_2^0$/$\tilde{\chi}_1^\pm$ and the bino $\tilde{\chi}_1^0$ is larger than, but close to, $m_Z$. This was already evident in the simplified model likelihood maps for these analyses in Figure \ref{fig:limitComparison} (top left and bottom row panels). In Figure~\ref{fig:mass_planes_selected_analyses} this manifests as the positive log-likelihood contribution along the high-mass diagonals in the right-hand plots for \textsf{ATLAS\_MultiLEP\_2lep\_jet} and \textsf{ATLAS\_RJ\_3lep}.

The sharp changes in likelihood contribution visible in several of the plots are due to sudden changes in what scenarios are being picked out by the profiling, and consequently, which signal regions get to determine the analysis likelihoods.
One example of this can be seen in the horizontal band of high $\tilde{\chi}_4^0$ mass in the right-hand plots for \textsf{ATLAS\_MultiLEP\_2lep\_jet} and \textsf{ATLAS\_MultiLEP\_3lep}. In the region with $m_{\tilde{\chi}_3^0} > 250$\,GeV, the scenarios that are picked out by the profiling have a large $\tilde{\chi}_3^0$--$\tilde{\chi}_2^0$ mass splitting and a $\sim$$30$\,GeV $\tilde{\chi}_2^0$--$\tilde{\chi}_1^0$ splitting, suggesting that the $\tilde{\chi}_2^0$ and $\tilde{\chi}_1^0$ are predominantly Higgsino in these scenarios. For $m_{\tilde{\chi}_3^0} < 250$\,GeV, however, the profiling selects scenarios where the $\tilde{\chi}_2^0$ and $\tilde{\chi}_3^0$ are a Higgsino pair of similar mass, with a large mass gap down to a $\tilde{\chi}_1^0$ below $100$\,GeV. The scenario in this region looks a lot like the simplified model from ATLAS, except that there is production of three Higgsinos ($\tilde{\chi}_3^0$, $\tilde{\chi}_2^0$, $\tilde{\chi}_1^\pm$) instead of two winos ($\tilde{\chi}_2^0$, $\tilde{\chi}_1^\pm$). An important reason for the similarity with the simplified model is that the Higgsino nature of the produced sparticles ensures large branching ratios for decays to on-shell $Z$ bosons, even when the $\tilde{\chi}_{3,2}^0$--$\tilde{\chi}_1^0$ mass splittings are larger than $m_h$. This matches the assumption $BR(\tilde{\chi}_2^0 \rightarrow \tilde{\chi}_1^0 Z) = 100\%$ that is commonly employed for simplified models.\footnote{In contrast, for a pure wino $\tilde{\chi}_2^0$ decaying to a bino $\tilde{\chi}_1^0$, the $h \tilde{\chi}_1^0$ decay channel dominates for $m_{\tilde{\chi}_{2}^0}-m_{\tilde{\chi}_1^0} > m_h$.} Therefore, the tension between the likelihood contributions from \textsf{ATLAS\_MultiLEP\_3lep} and \textsf{ATLAS\_RJ\_3lep} that can be seen in this low-$m_{\tilde{\chi}_3^0}$, high-$m_{\tilde{\chi}_4^0}$ region is a manifestation of the tension one naively would expect based on the corresponding simplified model results.

The overall result is that models in the vicinity of our best fit have appreciable amounts of
$\tilde{\chi}_3^{0}$, $\tilde{\chi}_4^{0}$ and $\tilde{\chi}_2^{\pm}$
production in addition to $\tilde{\chi}_2^{0}$ and
$\tilde{\chi}_1^{\pm}$ production, whilst the ATLAS simplified model
only includes $\tilde{\chi}_2^{0}$ and $\tilde{\chi}_1^{\pm}$
production. Our models can thus produce richer final states than the ATLAS
simplified model, typically generating more gauge bosons,
which in turn produce leptons that allow the events to pass a three-lepton selection whilst also producing additional jets.

Examining the event record for the MC simulation of our best-fit point shows a variety of extra processes
that will lead either to a higher multiplicity or a change in the
missing $E_T$ distribution.  These include:
\begin{itemize}
\item $\tilde{\chi}_2^{0}\tilde{\chi}_3^{0}$ production, with e.g. \\
$\tilde{\chi}_2^{0}\rightarrow Z + \tilde{\chi}_1^{0}$, $\tilde{\chi}_3^{0}\rightarrow W^- + \tilde{\chi}_1^{+} \rightarrow W^- + W^+ + \tilde{\chi}_1^{0}$
\item $\tilde{\chi}_2^{\pm}\tilde{\chi}_2^{\mp}$ production, with e.g. \\
$\tilde{\chi}_2^{\pm}\rightarrow W^\pm + \tilde{\chi}_2^{0} \rightarrow W^\pm + Z + \tilde{\chi}_1^{0}$
\item $\tilde{\chi}_2^{\pm}\tilde{\chi}_3^{0}$ production, with e.g. \\
$\tilde{\chi}_2^{\pm}\rightarrow W^\pm + \tilde{\chi}_1^{0}$, $\tilde{\chi}_3^{0}\rightarrow Z + \tilde{\chi}_2^{0} \rightarrow Z + Z + \tilde{\chi}_1^{0}$
\item $\tilde{\chi}_2^{\pm}\tilde{\chi}_3^{0}$ production, with e.g. \\
$\tilde{\chi}_2^{\pm}\rightarrow W^\pm + \tilde{\chi}_2^{0}\rightarrow W^\pm + Z+ \tilde{\chi}_1^{0}$,\\
$\tilde{\chi}_3^{0}\rightarrow W^- + \tilde{\chi}_1^{+} \rightarrow W^- + W^+ + \tilde{\chi}_1^{0}$
\item $\tilde{\chi}_2^{\pm}\tilde{\chi}_4^{0}$ production, with e.g. \\
$\tilde{\chi}_2^{\pm}\rightarrow W^\pm + \tilde{\chi}_2^{0} \rightarrow  W^\pm +Z+ \tilde{\chi}_1^{0}$, $\tilde{\chi}_4^{0}\rightarrow Z + \tilde{\chi}_1^{0}$
\item $\tilde{\chi}_2^{\pm}\tilde{\chi}_2^{0}$ production, with e.g. \\
$\tilde{\chi}_2^{\pm}\rightarrow h + \tilde{\chi}_1^{\pm} \rightarrow h+W^\pm  + \tilde{\chi}_1^{0}$, $\tilde{\chi}_2^{0}\rightarrow Z + \tilde{\chi}_1^{0}$
\item $\tilde{\chi}_1^{\pm}\tilde{\chi}_3^{0}$ production, with e.g. \\
$\tilde{\chi}_1^{\pm}\rightarrow W^\pm + \tilde{\chi}_1^{0}$, $\tilde{\chi}_3^{0}\rightarrow W^- + \tilde{\chi}_1^{+}\rightarrow W^+ + W^- + \tilde{\chi}_1^{0}$
\item $\tilde{\chi}_2^{\pm}\tilde{\chi}_4^{0}$ production, with e.g. \\
$\tilde{\chi}_2^{\pm}\rightarrow Z + \tilde{\chi}_1^{\pm} \rightarrow Z+ W^\pm  + \tilde{\chi}_1^{0}$,\\
$\tilde{\chi}_4^{0}\rightarrow h + \tilde{\chi}_2^{0}\rightarrow  h + Z+  \tilde{\chi}_1^{0}$
\end{itemize}
Note that it is quite common to have four gauge bosons produced for our best-fit model. For the best fit point the $\tilde{\chi}_1^{\pm}$ and $\tilde{\chi}_2^{0}$ have a $\sim20\%$ Higgsino component, resulting in a smaller $\tilde{\chi}_1^{\pm}\tilde{\chi}_2^{0}$ production cross-section compared to a scenario with pure wino $\tilde{\chi}_1^{\pm}$ and $\tilde{\chi}_2^{0}$.
In the conventional 3-lepton ATLAS analysis (\textsf{ATLAS\_MultiLep\_3lep}), this has the effect of reducing the predicted signal in the {\sf SR3\_WZ\_0Ja} and {\sf SR3\_WZ\_0Jb} signal regions. At the same time, the additional production processes made relevant by the relatively light $\tilde{\chi}_3^{0}$, $\tilde{\chi}_4^{0}$ and $\tilde{\chi}_2^{\pm}$ increase the signal prediction for the {\sf SR3\_WZ\_1Jc} signal region, which in contrast to {\sf SR3\_WZ\_0Ja} and {\sf SR3\_WZ\_0Jb} requires $\ge 1$ jet, with a leading jet $p_T$ of at least 70\,GeV. The ATLAS results for these signal regions are:
\begin{itemize}
\item {\sf SR3\_WZ\_0Ja}: expected background $21.7 \pm 2.9$, observed 21.
\item {\sf SR3\_WZ\_0Jb}: expected background $2.7 \pm 0.5$, observed 1.
\item {\sf SR3\_WZ\_1Jc}: expected background $1.3 \pm 0.3$, observed 4.
\end{itemize}
Thus a reduction in the {\sf SR3\_WZ\_0Ja} and {\sf SR3\_WZ\_0Jb}
predicted signal yields, plus a simultaneous increase in the predicted
{\sf SR3\_WZ\_1Jc} yield, clearly helps a model fit the data
better. This change in what is the most sensitive signal region is
responsible for the switch from negative to positive log-likelihood
contribution from \textsf{ATLAS\_MultiLep\_3lep} in the
$m_{\tilde{\chi}_3^0}<250$\,GeV region when $m_{\tilde{\chi}_4^0}$
lowered below $\sim700$\,GeV.
In this case,
the \emph{same} light electroweakinos are also able to provide a good fit
to the results from \textsf{ATLAS\_4lep}, \textsf{ATLAS\_MultiLep\_2lep\_jet}
and \textsf{ATLAS\_RJ\_3lep}, allowing all analyses to contribute
positively to the combined log-likelihood.

As this work was nearing completion, a new CMS search for chargino pair production was made public, in which evidence for chargino production and decay to either $W$ bosons or intermediate sleptons is searched for in events with two opposite sign leptons, for different jet and $b$-jet multiplicities and lepton flavour configurations~\cite{Sirunyan:2018lul}. A large number of bins in $p_T^\text{miss}$ and $M_{T2}$ are used to determine exclusion limits on a variety of simplified model scenarios. When interpreted in the context of a model with decoupled sleptons, the observed exclusion limit on the cross-section $\sigma(pp\rightarrow \tilde{\chi}_1^+\tilde{\chi}_1^-)$ is weaker than the median expected limit at the $2\sigma$ level. While it is tempting to speculate about the connection between this and the pattern of excesses presented in this paper, a detailed treatment of this analysis is beyond the scope of the present work.

\subsection{Benchmark points}
\label{sec:benchmark}

\begin{table*}[t]
\begin{center}
\begin{tabular}{c c c c c c c}
\hline
Parameter & \#1 Best fit & \#2 Heavy winos & \#3 Highest mass & \#4 DM \\
\hline
$M_1(Q)$ & $-$50.6\,GeV & $-$79.2\,GeV & 133.4\,GeV & $-$45.6\,GeV \\
$M_2(Q)$ & 149.3\,GeV & 263.0\,GeV & 243.5\,GeV & 143.7\,GeV \\
$\mu(Q)$ & 252.7\,GeV & $-$187.3\,GeV & $-$293.2\,GeV & 260.8\,GeV\\
$\tan\beta(m_Z)$ & 28.7 & 40.4 & 41.5 & 16.4 \\
\hline
$m_{\tilde{\chi}_1^{0}}$ & $-$49.4\,GeV & $-$73.9\,GeV & 129.4\,GeV & $-$45.1\,GeV\\
$m_{\tilde{\chi}_2^{0}}$ & 141.6\,GeV & 165.7\,GeV & 230.6\,GeV & 136.5\,GeV\\
$m_{\tilde{\chi}_3^{0}}$ & $-$270.3\,GeV & $-$208.5\,GeV & $-$308.8\,GeV & $-$277.8\,GeV\\
$m_{\tilde{\chi}_4^{0}}$ & 290.2\,GeV & 292.6\,GeV & 344.6\,GeV & 297.2\,GeV\\
$m_{\tilde{\chi}_1^{\pm}}$ & 142.1\,GeV & 168.7\,GeV & 230.2\,GeV & 136.8\,GeV\\
$m_{\tilde{\chi}_2^{\pm}}$ & 293.9\,GeV & 294.2\,GeV & 345.8\,GeV & 300.5\,GeV\\
\hline
Collider log-likelihood & 10.8 & 10.3 & 9.7 & 10.4 \\
\end{tabular}
\caption{\label{tab:bestfit} Parameter values and sparticle masses for a variety of benchmark points. Point \#1 is our best-fit model, for which the Higgsinos are heavier than the winos. Point \#2 is a solution with the winos heavier than the Higgsinos with similar likelihood. Point \#3 is the point within the $1\sigma$ region with the highest LSP mass. Point \#4 has the best combined DM and collider likelihood.}
\end{center}
\centering
\end{table*}

\begin{figure*}
  \centering
  \includegraphics[width=0.45\textwidth]{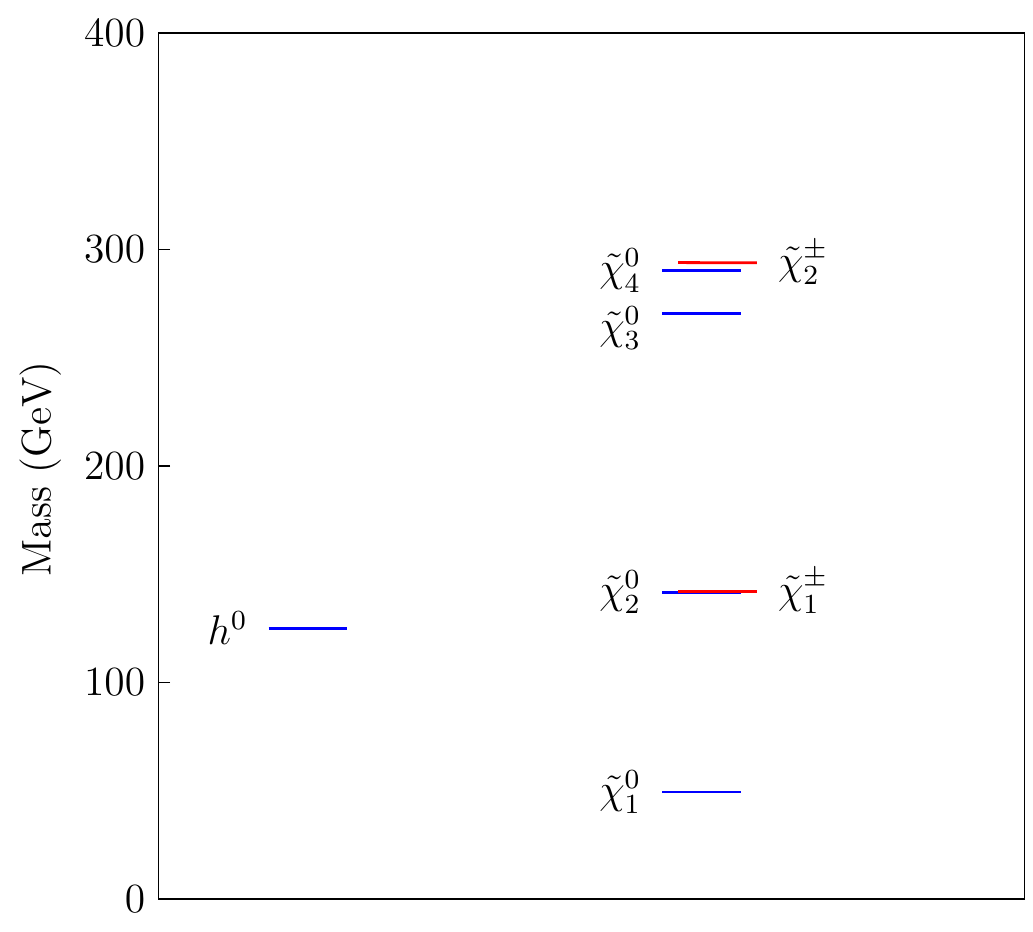}\hspace{0.05\textwidth}
  \includegraphics[width=0.45\textwidth]{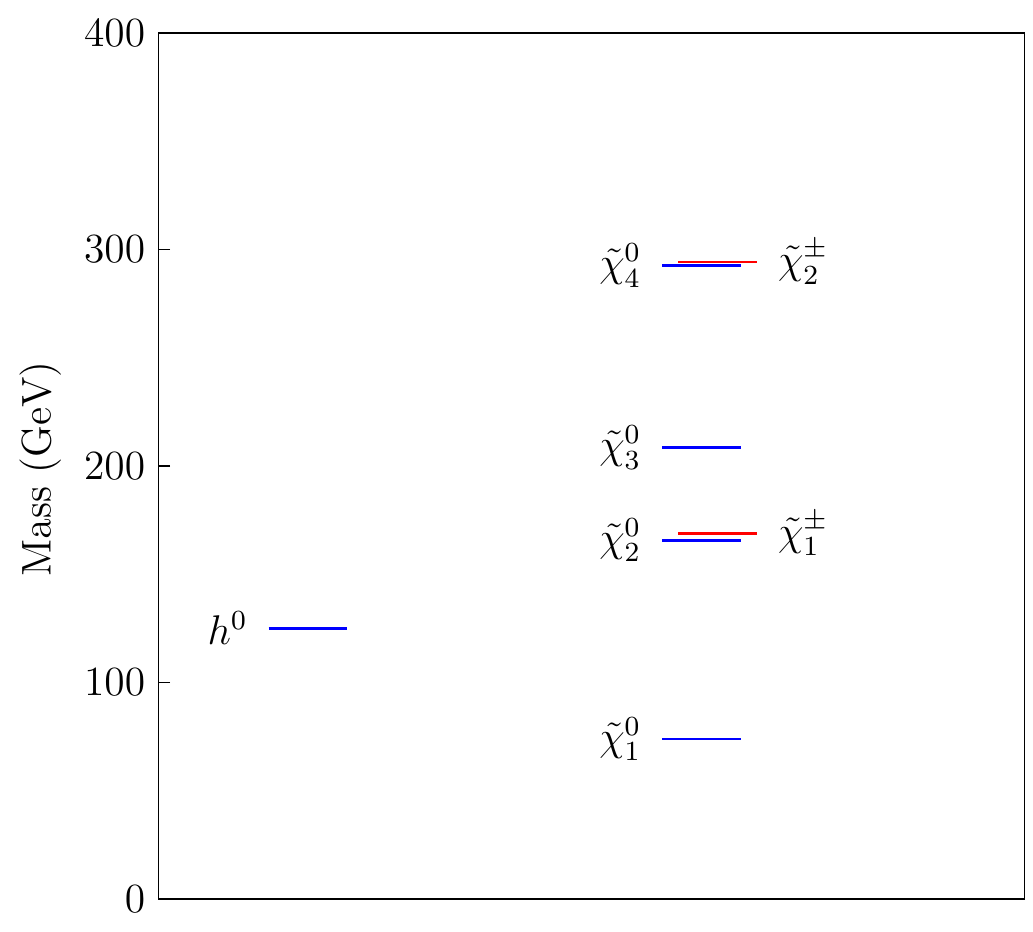}
\caption{\label{fig:specplot}The electroweakino mass spectra~\cite{Buckley:2013jua} for our best-fit point (benchmark point \#1; left) and another point with a similar likelihood but with heavier winos (benchmark point \#2; right).  The benchmarks are defined in Table.~\ref{tab:bestfit}.}
\end{figure*}

In Table~\ref{tab:bestfit}, we show the parameter values and
electroweakino masses for a number of benchmark points. The first of these corresponds to our best-fit point.  As can be seen in Figs.\ \ref{fig:pole_masses_2D} and \ref{fig:n1content}, there are many points
that give likelihoods that are very close to the best-fit value. In
particular, these include models where winos are lighter than Higgsinos (as occurs at our best-fit
point), models where they have similar masses, and models where the winos are heavier. We show a
second benchmark in
Table~\ref{tab:bestfit} where the latter is true, with a slightly smaller likelihood than the best fit.
In both benchmarks, all electroweakinos have masses less than about $300$\,GeV.
It is worth noting that this second benchmark point is also the highest likelihood point with negative $\mu$, often a preferred scenario for improving $g-2$ for the muon, in spite of this feature not being included in the analysis.  We show mass spectra
for both these possibilities in Figure~\ref{fig:specplot}.

Table~\ref{tab:bestfit} also shows a third benchmark point: that with the highest LSP mass within the $1\sigma$ region. This point features a bino LSP, as our best-fit point does, and all electroweakino masses are below $350$\,GeV.

Lastly, benchmark scenario 4 in Table~\ref{tab:bestfit} is the point with the best combined DM and collider likelihood. The value of the likelihood shown in Table~\ref{tab:bestfit} corresponds to the collider likelihood, clearly within $1\sigma$ of the best-fit point. The combined DM likelihood for this point constitutes essentially a perfect fit, showing no tension with direct nor indirect detection, and a relic density well below the observed value.  With a lightest neutralino mass of $m_{\tilde{\chi}^0_1} = 45.1$\,GeV, the LSP at this point falls right on the $Z$ funnel, explaining how it is able to avoid saturating the DM relic density despite being predominantly bino.

In Table \ref{tab:pvalues}, we give the local \pvalue estimates for
each analysis separately, confirming the picture presented in
Figure~\ref{fig:mass_plane_contributions}.  We also present generalised
goodness-of-fit
estimates for both the background-only hypothesis and our best-fit signal.
The total significance is
dominated by contributions from the \textsf{ATLAS\_4lep}
and \textsf{ATLAS\_RJ\_3lep}
analyses, whilst other analyses do not
disfavour this point.

\begin{table*}[t]
\centering
\begin{tabular}{lrrrrr|rrrr}
\toprule
 & & \multicolumn{4}{c}{Best expected SRs} & \multicolumn{4}{|c}{All SRs; neglect correlations} \\ 
 \midrule
Analysis & \multicolumn{2}{c}{\makecell{Local \\ signif. ($\sigma$)}} & \makecell{SM \\ fit ($\sigma$)} & \makecell{EWMSSM \\ fit ($\sigma$)} & \#SRs & \makecell{Local \\ signif. ($\sigma$)} & \makecell{SM \\ fit ($\sigma$)} & \makecell{EWMSSM \\ fit ($\sigma$)} & \#SRs \\
\midrule
Higgs invisible width &                                  & 0 & 0 & 0 &  1 & 0 & 0 & 0 &  1 \\
$Z$ invisible width\phantom{$\mathrm{fb}^{-1}$} &        & 0 & 1.3 & 1.3 &  1 & 0 & 1.3 & 1.3 &  1 \\
\textsf{ATLAS\_4b} &                                     & 0.7 & 0 & 0 &  1 & 1.5 & 0 & 0 &  $2^{*}$ \\
\textsf{ATLAS\_4lep} &                                   & 2.3 & 1.9 & 0 &  1 & 2.5 & 1.0 & 0 &  4 \\
\multicolumn{2}{l}{\textsf{ATLAS\_MultiLep\_2lep\_0jet}} & 0.9 & 0.3 & 0.1 &  1 & 1.3 & 0 & 0 &  6 \\
\multicolumn{2}{l}{\textsf{ATLAS\_MultiLep\_2lep\_jet}}  & 0 & 0 & 0.5 &  1 & 0.8 & 0.5 & 0.2 &  3 \\
\textsf{ATLAS\_MultiLep\_3lep} &                         & 1.8 & 1.5 & 0.7 &  1 & 1.2 & 0.4 & 0.3 &  11 \\
\textsf{ATLAS\_RJ\_2lep\_2jet} &                         & 0 & 0.3 & 0.5 &  1 & 1.5 & 1.8 & 1.5 &  4 \\
\textsf{ATLAS\_RJ\_3lep} &                               & 2.7 & 2.5 & 1.1 &  1 & 3.4 & 2.5 & 0.7 &  4 \\
\textsf{CMS\_1lep\_2b} &                                 & 0.8 & 0.3 & 0.3 &  1 & 0 & 0 & 0 &  2 \\
\textsf{CMS\_2lep\_soft} &                               & 0.1 & 0.2 & 0.2 &  12 & 0.1 & 0.2 & 0.2 &  12 \\
\textsf{CMS\_2OSlep} &                                   & 0.1 & 0.5 & 0.5 &  7 & 0 & 0.4 & 0.5 &  7 \\
\textsf{CMS\_MultiLep\_2SSlep} &                         & 0.2 & 0 & 0 &  1 & 0.2 & 0 & 0 &  2 \\
\textsf{CMS\_MultiLep\_3lep} &                           & 0 & 0 & 0.4 &  1 & 0 & 0 & 0 &  6 \\
\midrule
Combined &                                                                 & 3.3 & 1.4 & 0.2 &  31 & 4.1 & 1.2 & 0 &  65 \\
\bottomrule
\end{tabular}

\caption{
Combined significance of excesses in all analyses considered in this
paper, along with significances considering each analysis
individually. Local significances are computed with respect to the
nominal signal predictions of our EWMSSM best-fit point via the
method described in Sec. \ref{sec:local_pvals}. The SM and EWMSSM (using the best-fit
point found by our scan) goodness-of-fit are
computed via the method described in Sec. \ref{sec:gof_test}.
Significances displayed as `0' correspond to \pvalues
greater than $0.5$.
The ``Best expected SRs'' columns show significances computed using only the expected
most sensitive signal region in each analysis, except when correlation information is
available, as is done during our scan (this procedure is discussed in Sec. \ref{sec:LHCanalyses}).
The ``All SRs; neglect correlations'' columns show significances computed
using the same tests as described above, but assuming that no correlations exist between
signal regions within the analyses for which correlation information is unavailable.
We compute these to check that our best-fit signal
predictions are not in obvious tension with the observations in the
signal regions that were not pre-selected for analysis at our best-fit
point.
\\
$~^{*}${\footnotesize In our scan we have used the full set of 46 signal regions from the \textsf{ATLAS\_4b} analysis \cite{Aaboud:2018htj}, however for the `neglect correlations' goodness-of-fit tests this would increase the degrees of freedom of the test so much that the test would be extremely insensitive to any excesses in the other analyses. We therefore only use the two `discovery' signal regions in this test, which ATLAS has defined for very similar reasons.}
}
\label{tab:pvalues}
\end{table*}

We estimate the combined local \pvalue to be
3.3$\sigma$, but urge caution in its interpretation as it neglects
necessary look-elsewhere corrections.  For this reason, we have
also performed goodness-of-fit tests constructed with
less \textit{a priori} information about our best-fit signal (see Sec. \ref{sec:gof_test}).
This test has much less statistical power for discovery, due to its greater number of degrees of freedom compared to our local \pvalue test;
there is only about a $20\%$ probability to observe a $2\sigma$ or greater excess
under our best-fit model in this test, as opposed to over $95\%$ probability to observe a
$2 \sigma$ or greater excess in the local significance test.\footnote{We have investigated
this by Monte Carlo simulation.} However, our goodness-of-fit test has
a false positive rate much closer to the stated significance, due to a reduced
look-elsewhere effect. Under this test, we estimate the significance with which the background-only hypothesis is excluded to be $1.4\sigma$.

Performing the same test for the best-fit signal hypothesis, we see that our best-fit EWMSSM model is indeed a good fit to the data, showing only a $0.2\sigma$ significance in Table\ \ref{tab:pvalues}, with no significant tension evident between analyses. The worst
fit to our best-fit signal in an individual LHC analysis occurs in the
\textsf{ATLAS\_RJ\_ 3lep} analysis, at $1.1\sigma$; this
is because our EWMSSM best-fit point slightly under-predicts the excess
observed in this analysis.

A limitation of our significance estimates is that we can only perform rigorous tests using the selected signal
regions at our best-fit point (as described in Sec. \ref{sec:LHCanalyses}).  This is due to a lack of information about correlations
between signal regions for many analyses. One may therefore be concerned that our conclusions could
be significantly altered by the observations in signal regions that were deemed less sensitive to our best-fit point by Monte Carlo simulation, and therefore not included in our likelihood calculation for this point. To address this concern,
we have also computed the results that would be obtained using all signal regions in all analyses,
assuming them to be independent where no correlation information is available. These are
shown in the righthand side of Table.\ \ref{tab:pvalues}.
Neglecting unknown correlations is of course not fully correct, however it is sufficient to detect large tensions between signal regions that we might
have missed in our main analysis. The results do not indicate any large qualitative difference to
the main results; the local combined significance is mildly increased (from $3.3\sigma$ to $4.1\sigma$), whilst the EWMSSM and
SM goodness-of-fits are mildly improved ($0.2\sigma$ to $0\sigma$, and $1.4\sigma$ to $1.2\sigma$ respectively). Note, however,
that the goodness-of-fit tests have decreased statistical power due to the increased number of degrees of freedom that result
from combining more signal regions; this is the main reason for the improved goodness-of-fit in both cases.

With regards to other features in Table \ref{tab:pvalues},
one may notice that the $Z$ invisible
width measurement has zero local significance, but nevertheless shows
a $1.3\sigma$ tension with the background-only hypothesis. This is
because our EWMSSM best fit predicts zero new physics
contribution to the $Z$ invisible width, meaning it has a trivial
likelihood ratio and zero significance with respect to the SM.
However, the LEP measurement is slightly above the SM prediction, which means that the completely free hypothesis in our
goodness-of-fit test can improve upon the SM by a
small amount.

Our other LEP likelihoods
are not included in this combination because we implement them in our
scan via
interpolated limits, rather than fully simulating the analyses as we do
for the LHC likelihoods, and these limits are not sufficient to reconstruct pdfs that
can be used to produce pseudo-data. However, our best-fit point predicts
zero contribution from these likelihoods due to the fact that all the electroweakinos except the $\tilde\chi_1^0$ are out of the kinematic reach of LEP, so we do not expect them to
have much effect on the \pvalues presented here.

\begin{figure*}[t]
  \centering
  \includegraphics[width=0.9\columnwidth]{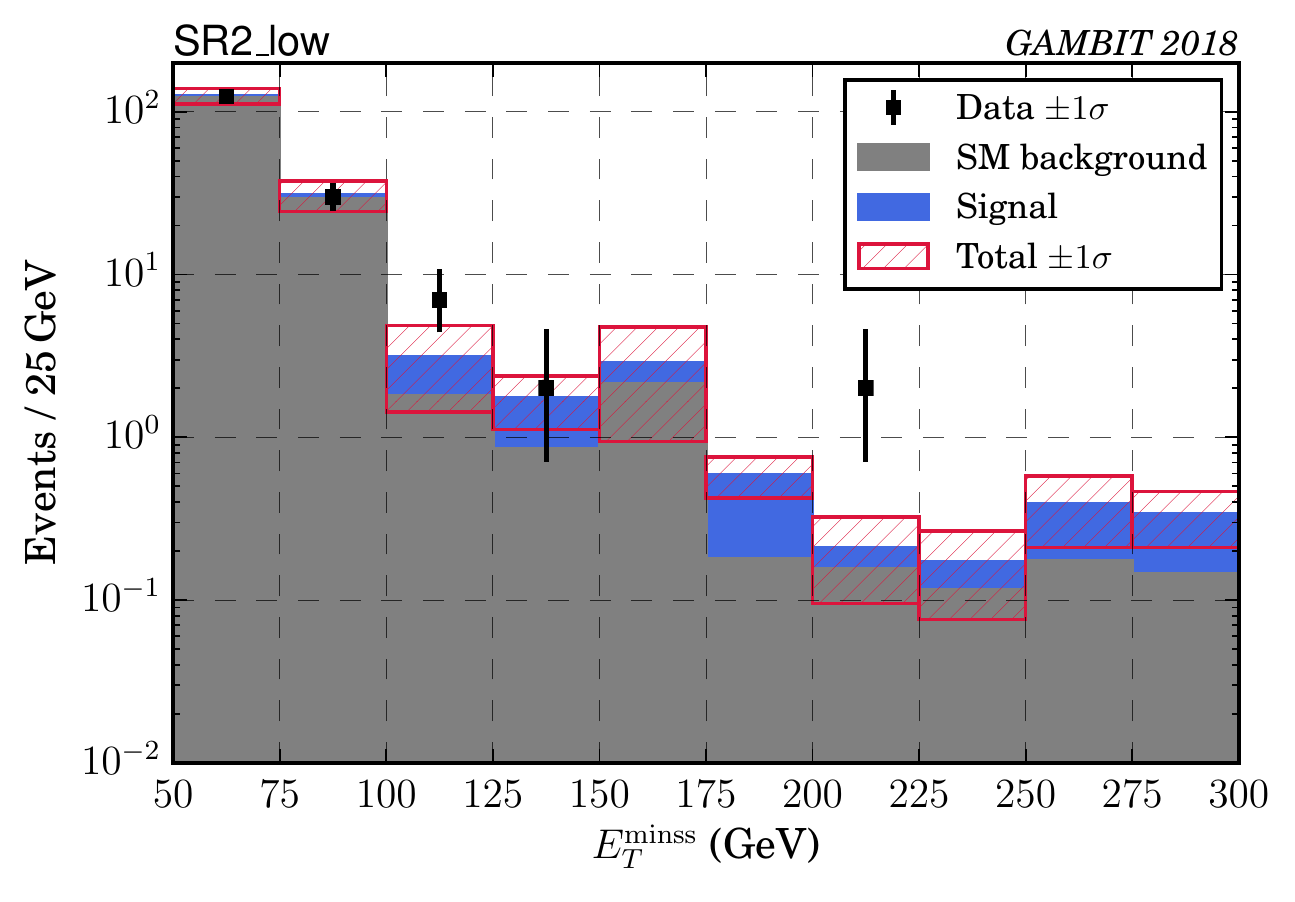}\hspace{0.1\columnwidth}
  \includegraphics[width=0.9\columnwidth]{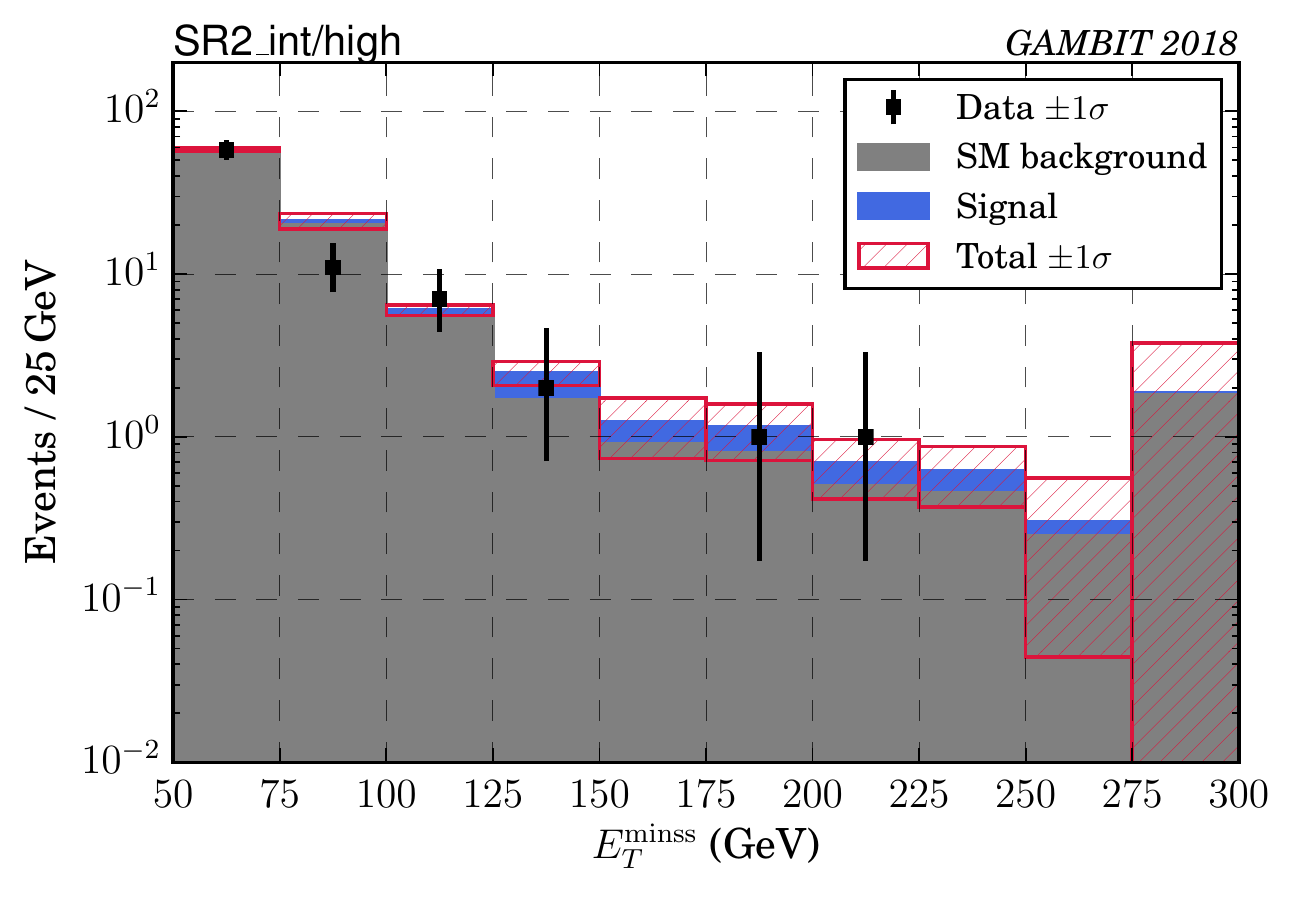}\\
  \caption{Distribution of missing transverse energy in the 2 lepton plus jets signal regions of the traditional ATLAS multilepton analysis, after applying all selection requirements. The grey bars show the total SM background (taken from Ref.~\cite{Aaboud:2018jiw}) and the stacked blue bars show the signal for our best-fit point. The hatched red bands show the $1\sigma$ uncertainty on the total number of expected events, found by summing in quadrature the background uncertainty and the signal statistical uncertainty for our best-fit point. The black points show the ATLAS data.}
  \label{fig:atlasComparison1}
\end{figure*}

\begin{figure*}[t]
  \centering
  \includegraphics[width=0.9\columnwidth]{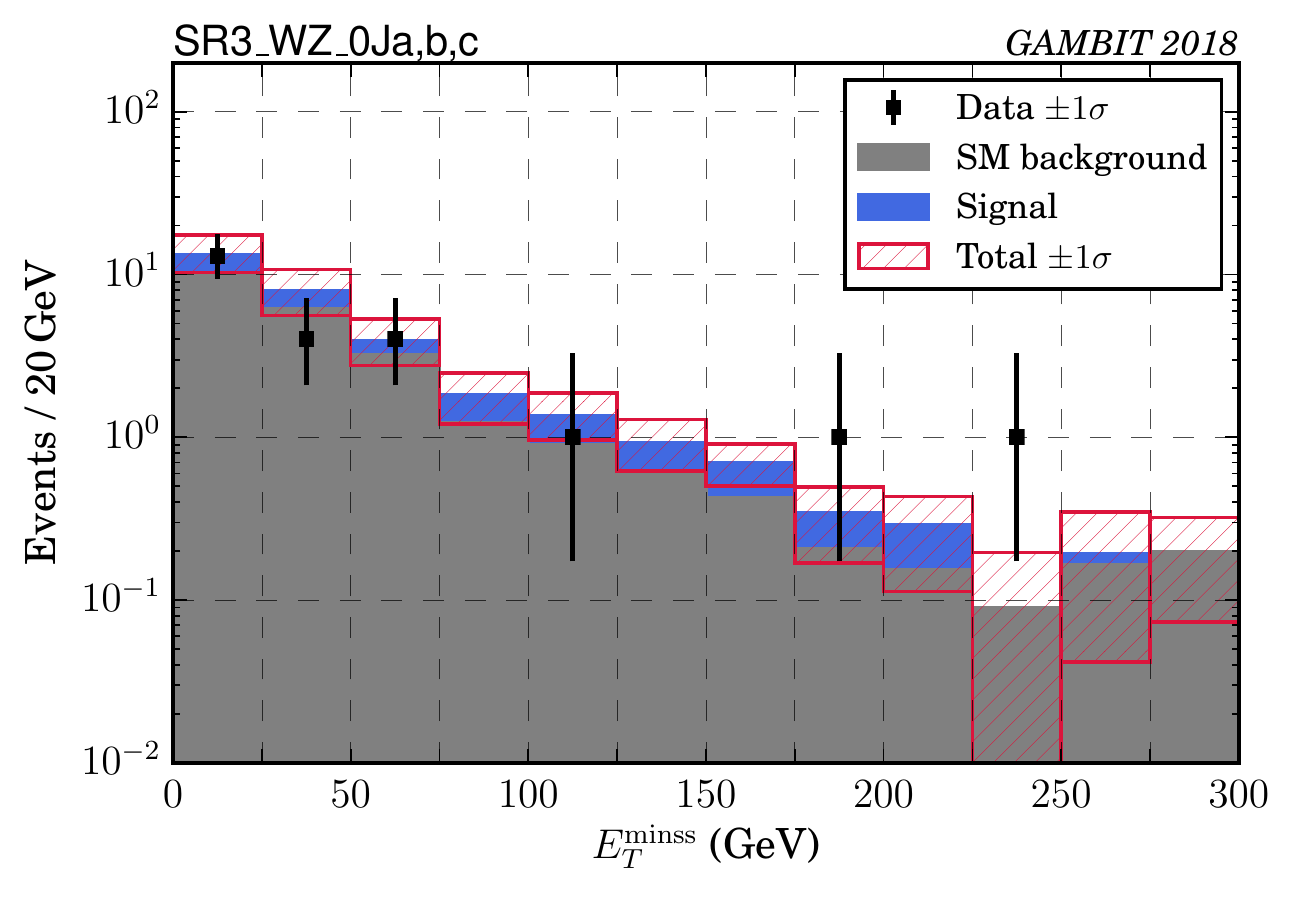}\hspace{0.1\columnwidth}
  \includegraphics[width=0.9\columnwidth]{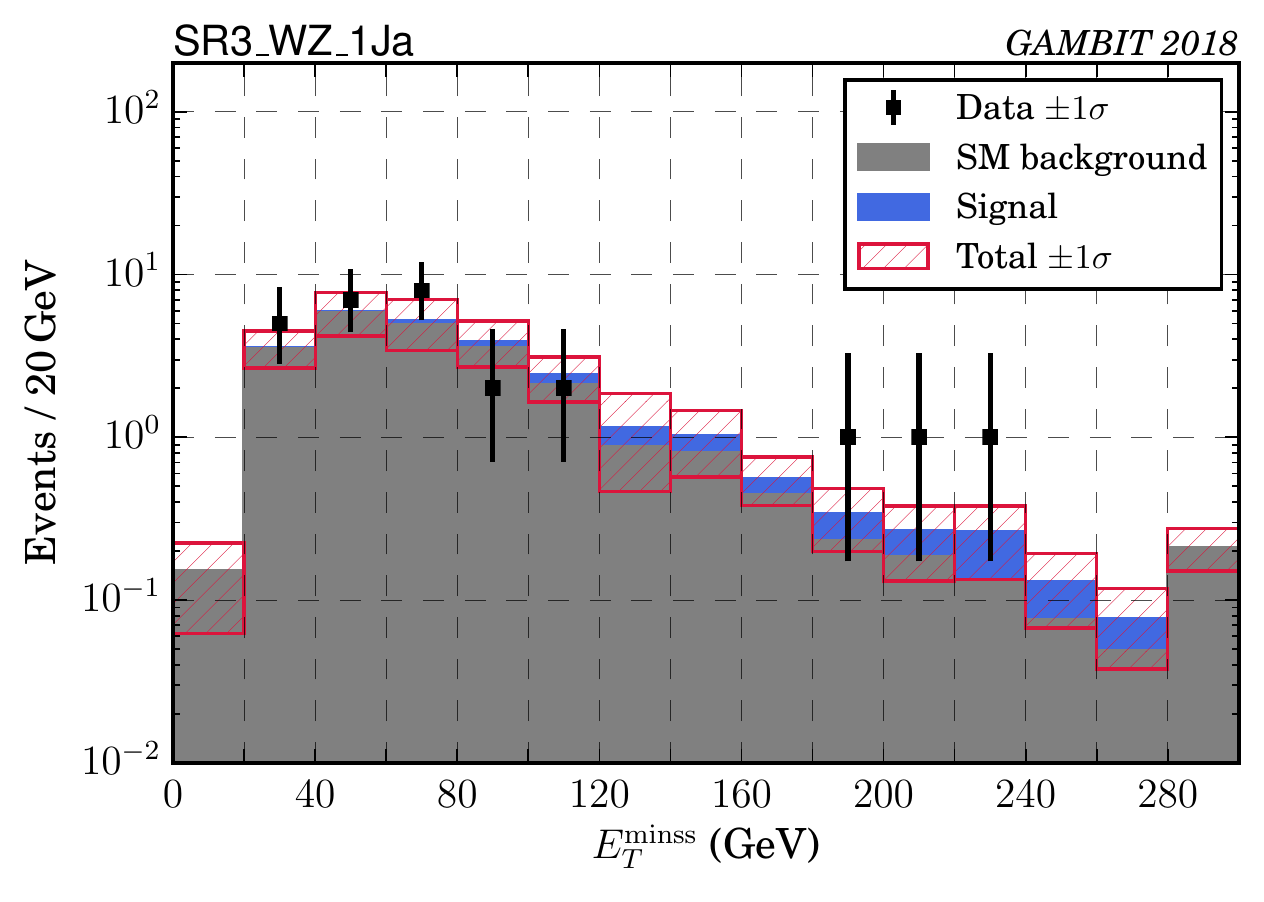}\\
  \includegraphics[width=0.9\columnwidth]{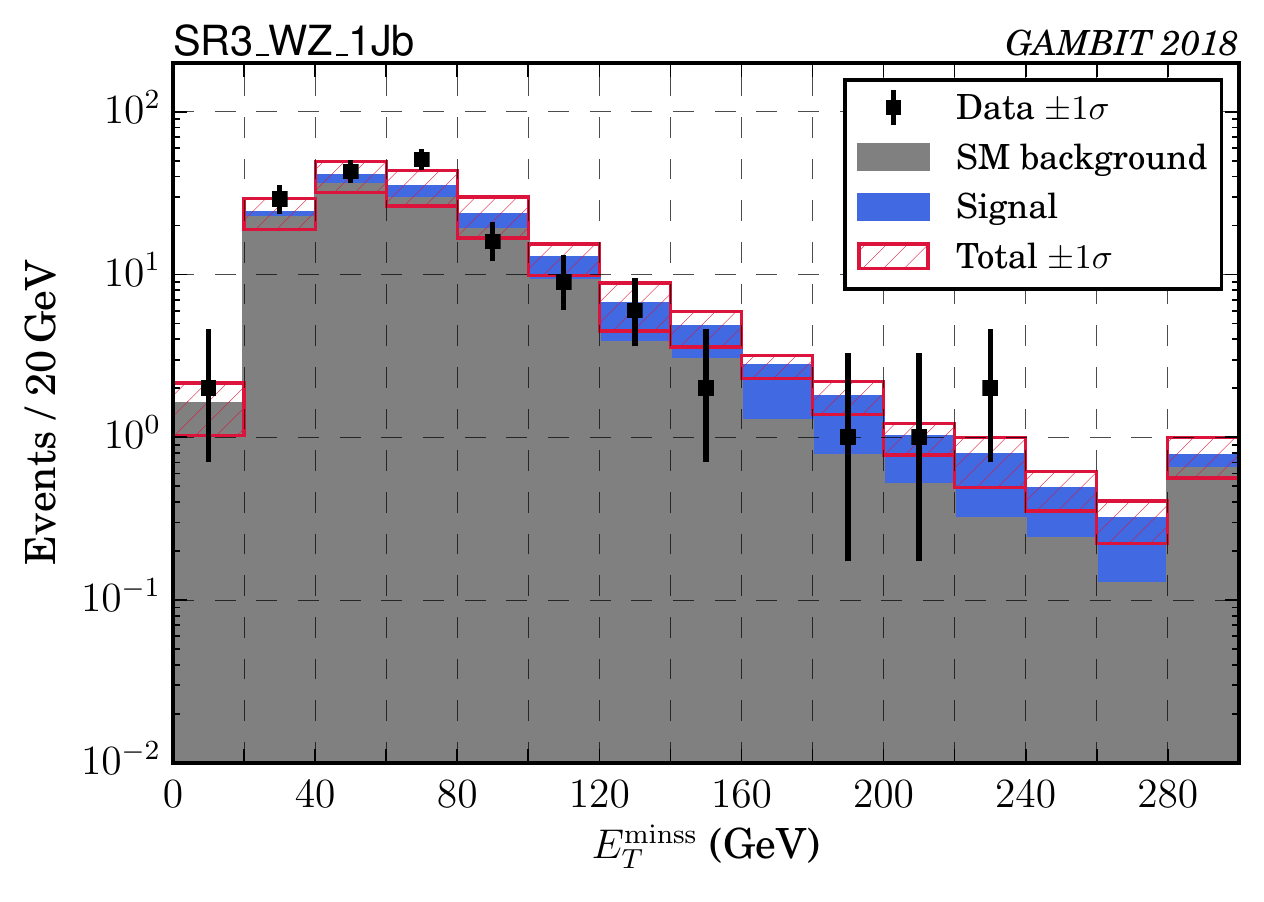}\hspace{0.1\columnwidth}
  \includegraphics[width=0.9\columnwidth]{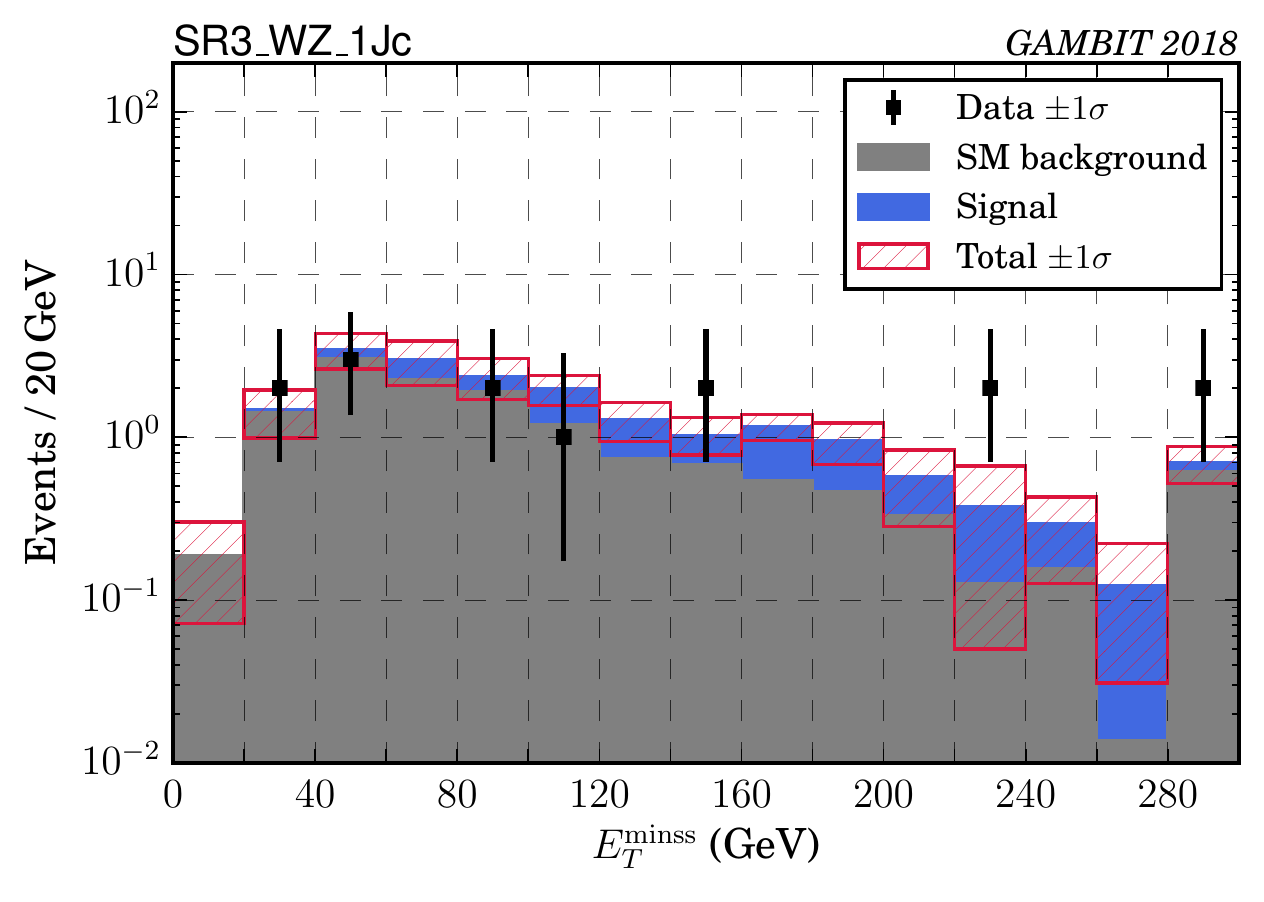}\\
  \caption{Distribution of missing transverse energy in the 3 lepton signal regions of the traditional ATLAS multilepton analysis, after applying all selection requirements. The grey bars show the total SM background (taken from Ref.~\cite{Aaboud:2018jiw}) and the stacked blue bars show the signal for our best-fit point. The hatched red bands show the $1\sigma$ uncertainty on the total number of expected events, found by summing in quadrature the background uncertainty and the signal statistical uncertainty for our best-fit point. The black points show the ATLAS data.}
  \label{fig:atlasComparison2}
\end{figure*}

\begin{figure*}[t]
  \centering
  \includegraphics[width=0.9\columnwidth]{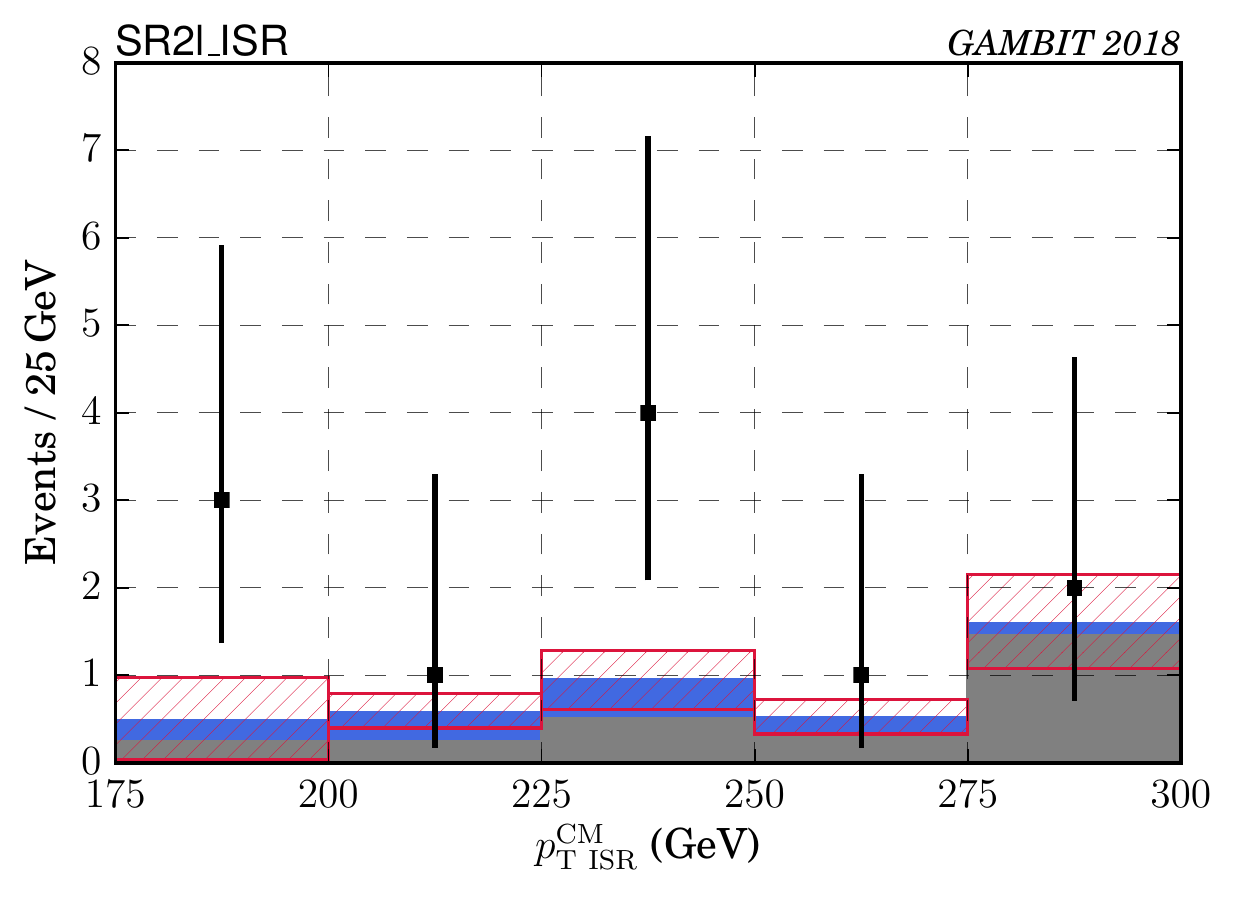}\hspace{0.1\columnwidth}
  \includegraphics[width=0.9\columnwidth]{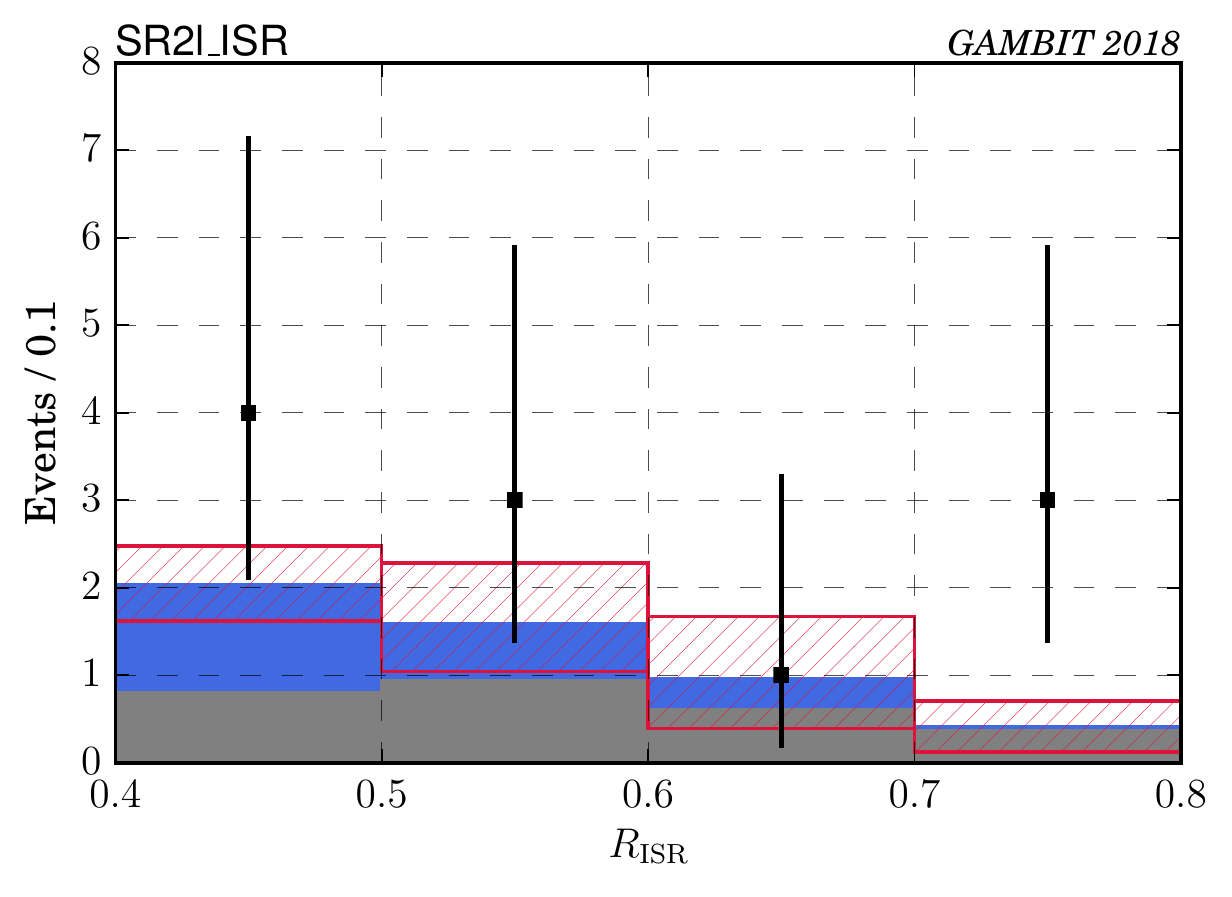}\\
  \includegraphics[width=0.9\columnwidth]{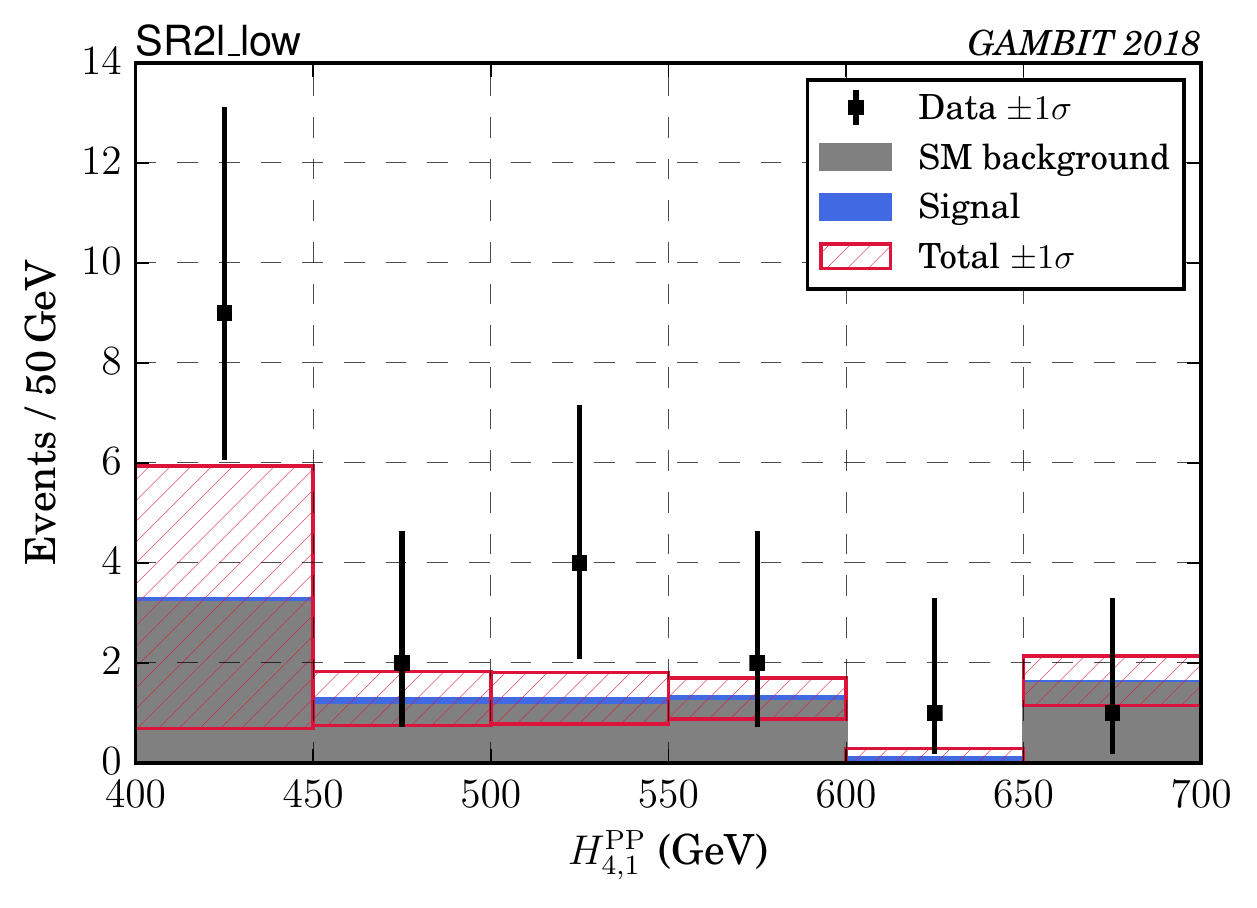}\hspace{0.1\columnwidth}
  \includegraphics[width=0.9\columnwidth]{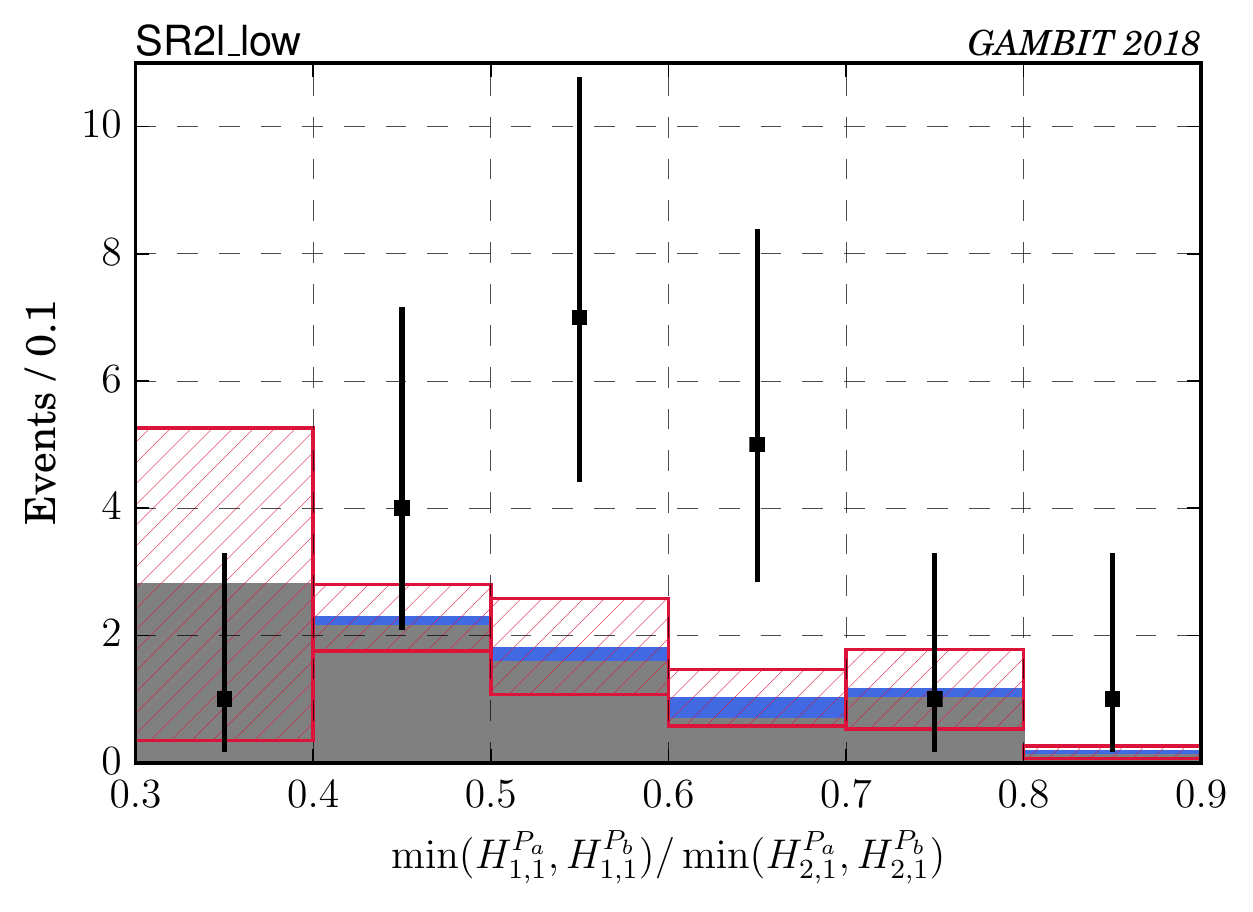}
  \caption{Distribution of kinematic variables in the 2 lepton signal regions for the ATLAS RJ analysis, after applying all selection requirements. The grey bars show the total SM background (taken from Ref.~\cite{Aaboud:2018sua}) and the stacked blue bars show the signal for our best-fit point. The hatched red bands show the $1\sigma$ uncertainty on the total number of expected events, found by summing in quadrature the background uncertainty and the signal statistical uncertainty for our best-fit point. The black points show the ATLAS data.}
  \label{fig:atlasComparison3}
\end{figure*}

\begin{figure*}[t]
  \centering
  \includegraphics[width=0.9\columnwidth]{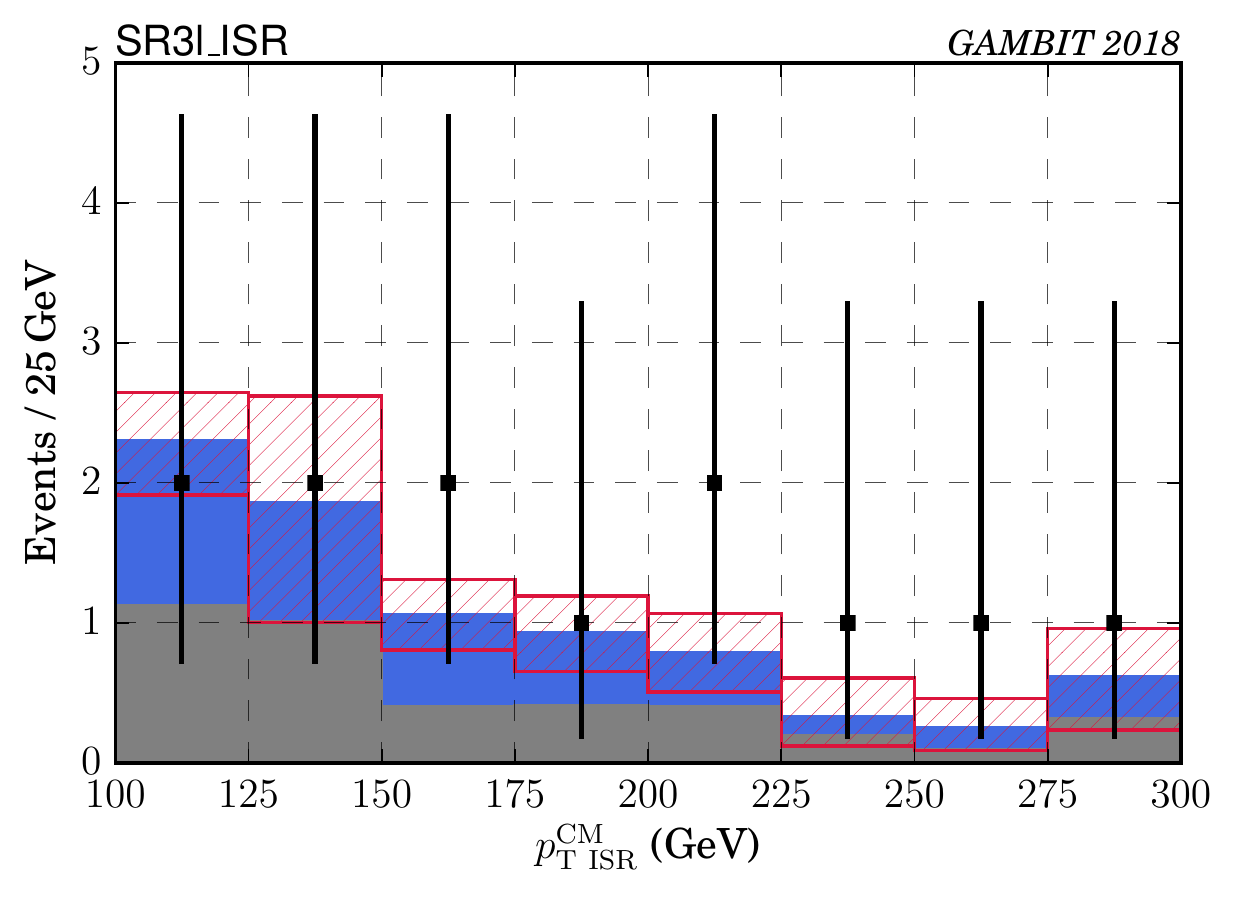}\hspace{0.1\columnwidth}
  \includegraphics[width=0.9\columnwidth]{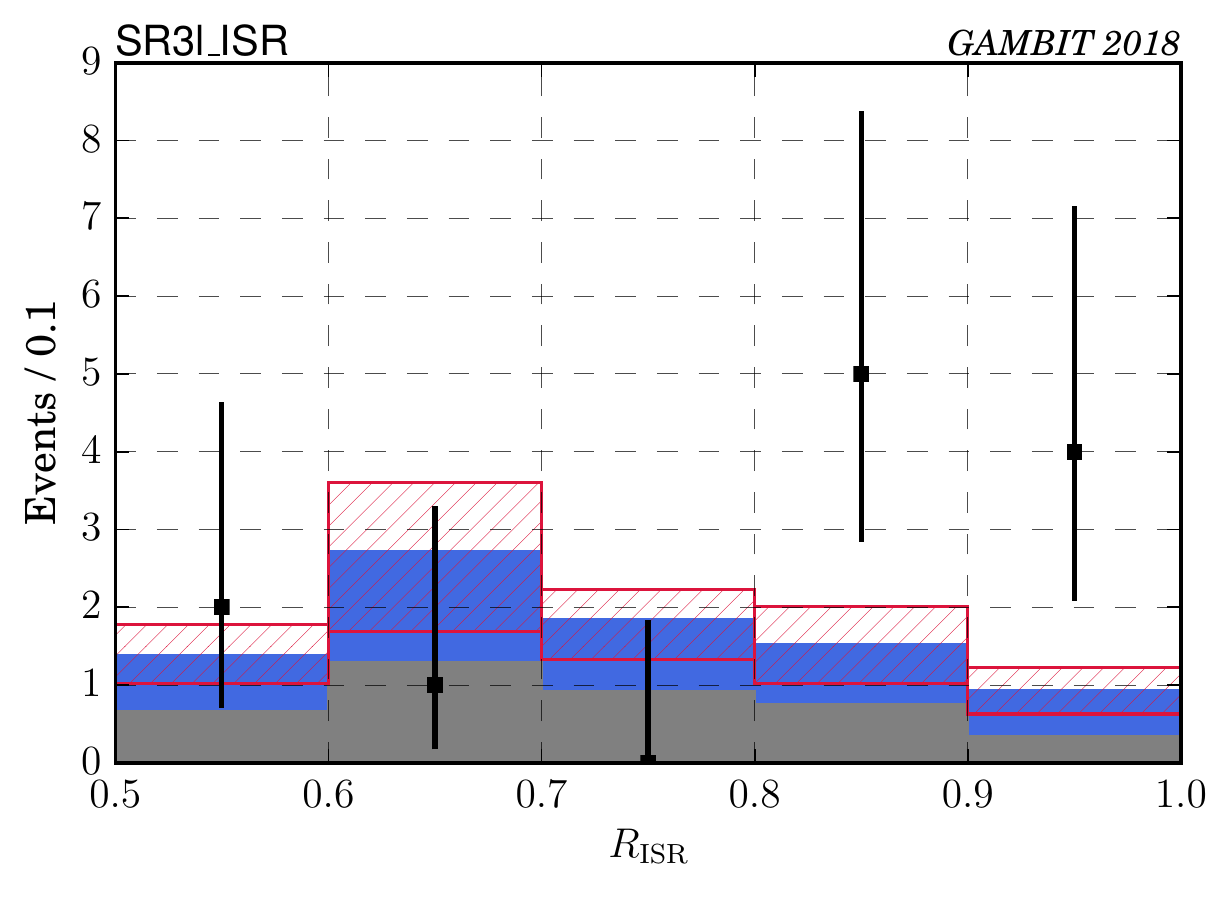}\\
  \includegraphics[width=0.9\columnwidth]{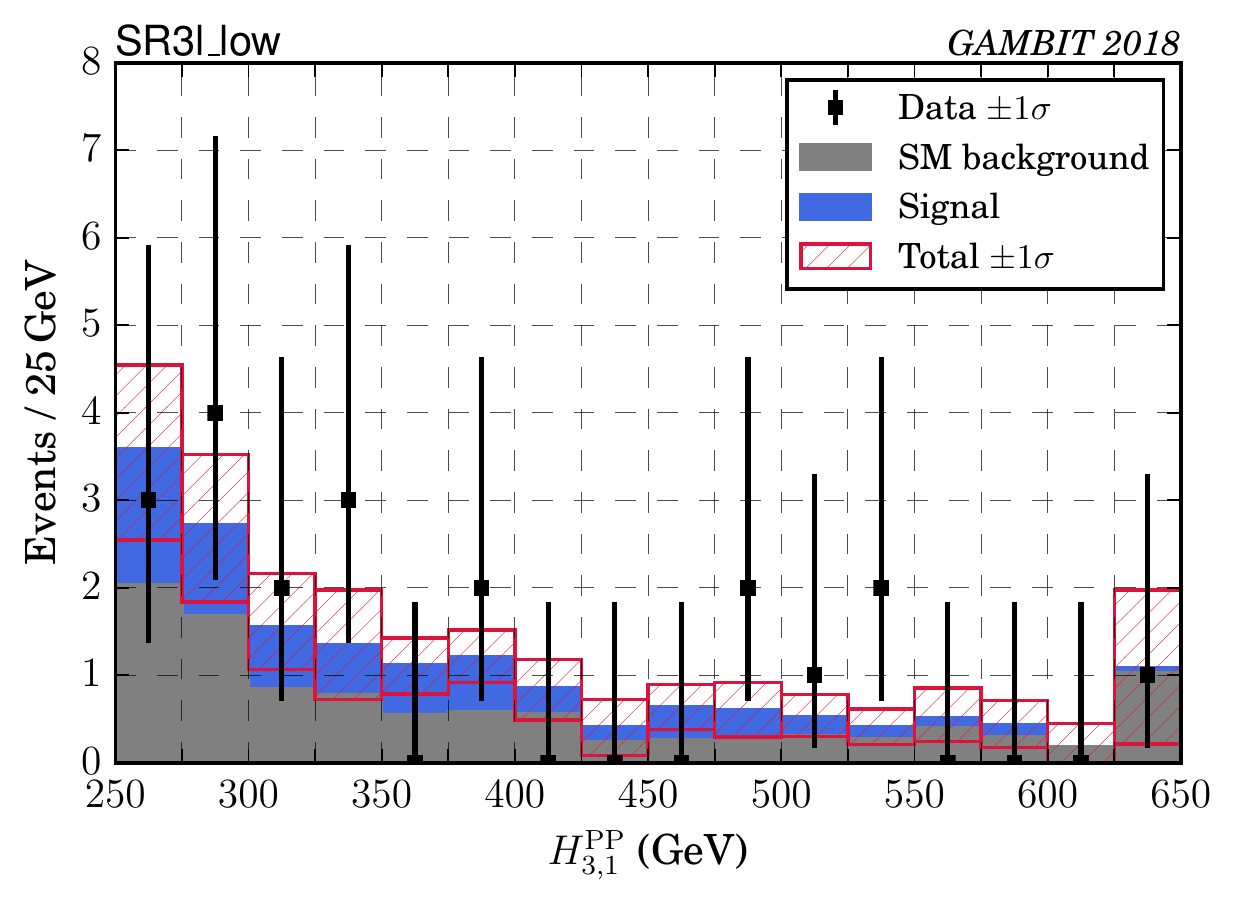}\hspace{0.1\columnwidth}
  \includegraphics[width=0.9\columnwidth]{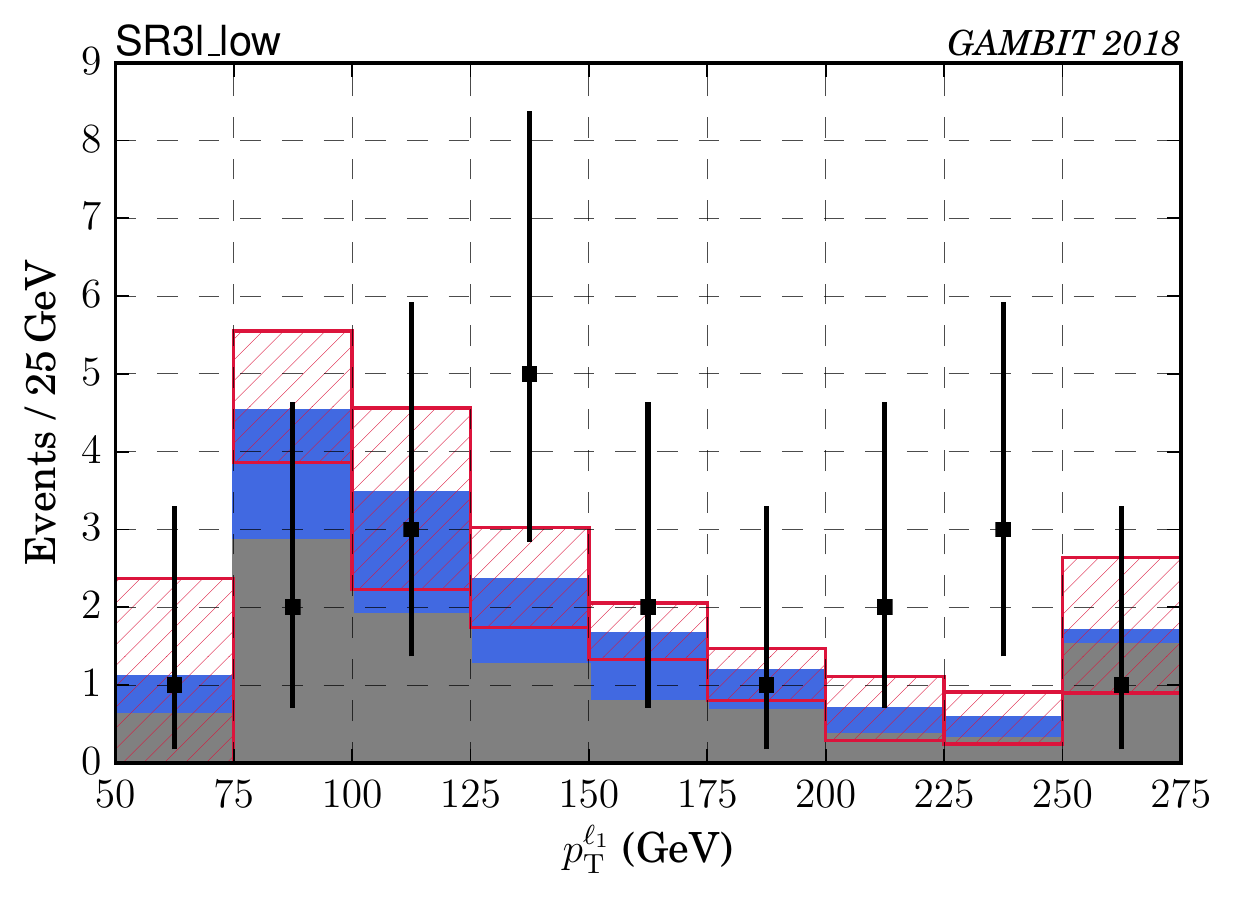}\\
  \includegraphics[width=0.9\columnwidth]{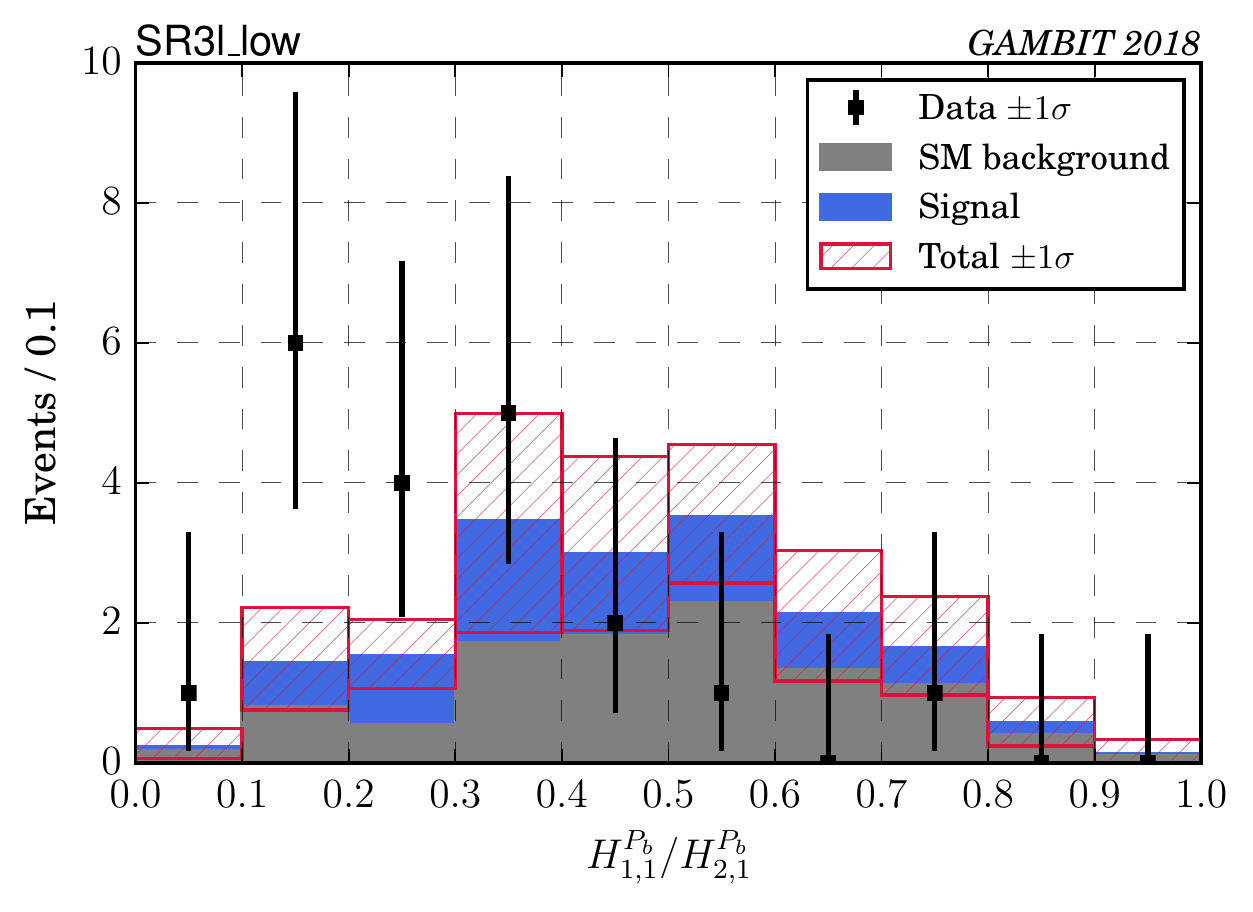}\\
  \caption{Distribution of kinematic variables in the 3 lepton signal regions for the ATLAS RJ analysis, after applying all selection requirements. The grey bars show the total SM background (taken from Ref.~\cite{Aaboud:2018sua}) and the stacked blue bars show the signal for our best-fit point. The hatched red bands show the $1\sigma$ uncertainty on the total number of expected events, found by summing in quadrature the background uncertainty and the signal statistical uncertainty for our best-fit point. The black points show the ATLAS data.}
  \label{fig:atlasComparison4}
\end{figure*}

As made clear in Sec.~\ref{sec:LHCanalyses}, we use LO cross
sections for our event generation. The increase in cross-sections
going to NLO (and beyond) would result in a shift in the best-fit
masses that give the same number of events.
Calculating cross-sections at LO and NLO for
our best-fit point using \prospino~\textsf{2.1}~\cite{Beenakker:1996ed,Beenakker:1999xh}, and
ignoring changes in efficiency -- which would be reasonably small when all the electroweakino
masses are changed by the same amount, giving very similar decay kinematics -- this corresponds to shifting all the neutralino and chargino masses upward by 7.0\,GeV. It is interesting to observe that this brings the mass up to slightly below half the Higgs boson mass.
Whatever
the continuing status of the small excesses in various signal regions,
it is interesting to note that such a light spectrum with an LSP of less than 60\,GeV is not
particularly constrained by current 13\,TeV LHC searches.

To finish this section, we compare the shapes of the signal
variable distributions in the ATLAS multilepton analyses for our best-fit point with those
published by the ATLAS experiment. The ATLAS simulation uses NLO cross-sections for normalization, and a full matching procedure including up to two extra hard jets in the matrix element.
We have checked the missing energy distributions for the two lepton plus jets and three lepton signal regions of the traditional multilepton analysis (Figures~\ref{fig:atlasComparison1} and \ref{fig:atlasComparison2}, respectively), as well as the distributions of several variables for the two lepton and three lepton signal regions of the recursive jigsaw analysis (Figures~\ref{fig:atlasComparison3} and \ref{fig:atlasComparison4}, respectively).
We see that the shape of the total expected contribution to data (SM background plus SUSY signal)
is well compatible with the observed data in all cases. This was, however, almost
inevitable given the limited numbers of events in these signal
regions. Clearly, the shapes of these distributions will offer a
powerful test of our best-fit hypothesis as more LHC data are collected.

\subsection{Extraction of underlying parameters}
\label{sec:underlyingparameters}
\begin{figure*}[t]
  \centering
  \includegraphics[height=0.8\columnwidth]{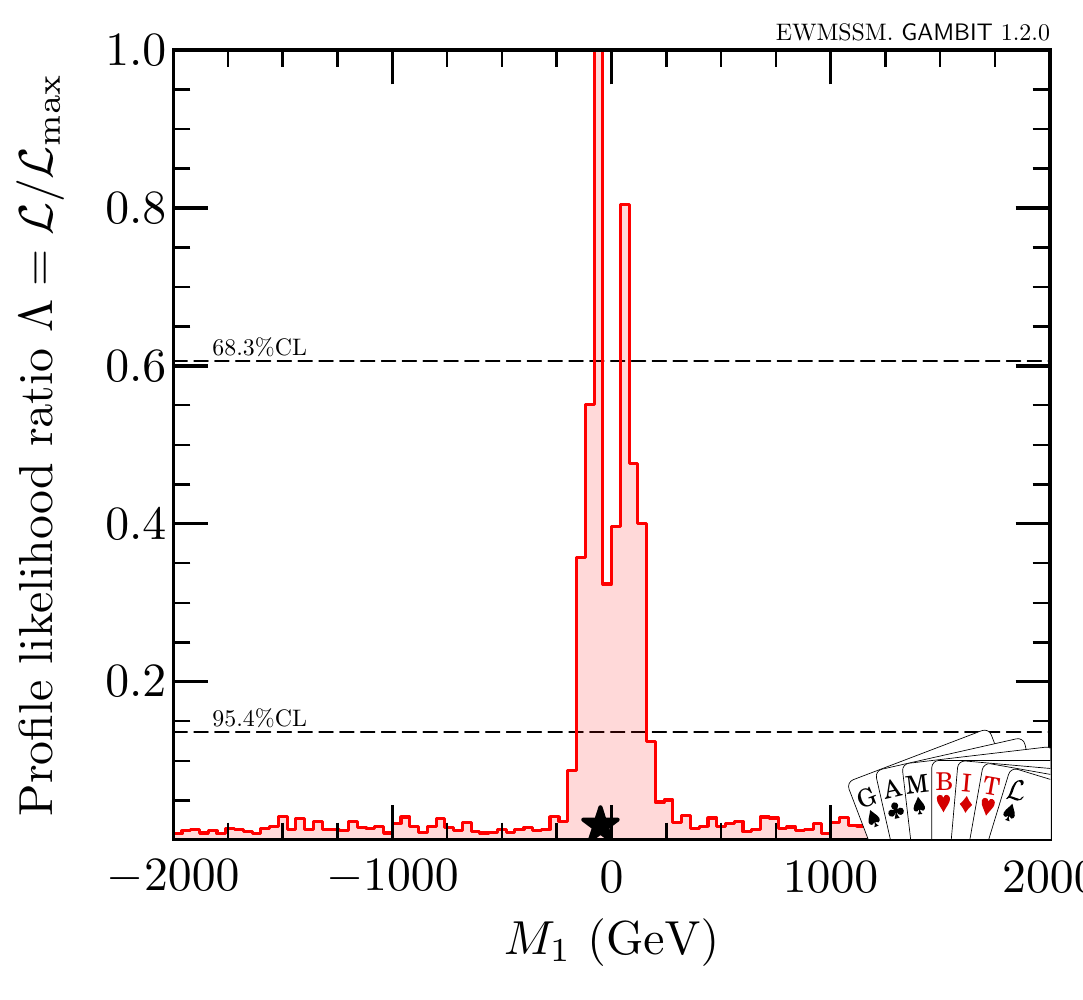}
  \includegraphics[height=0.8\columnwidth]{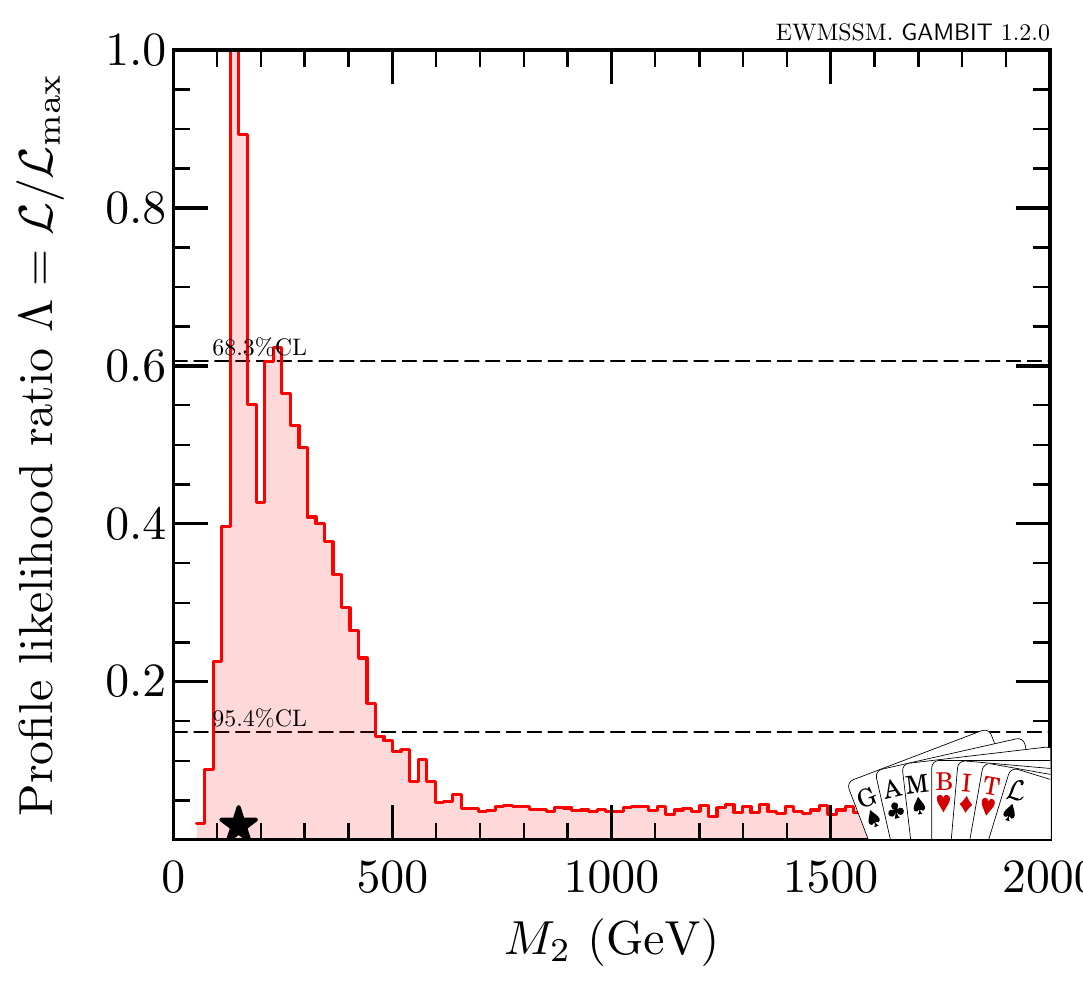} \\
\includegraphics[height=0.8\columnwidth]{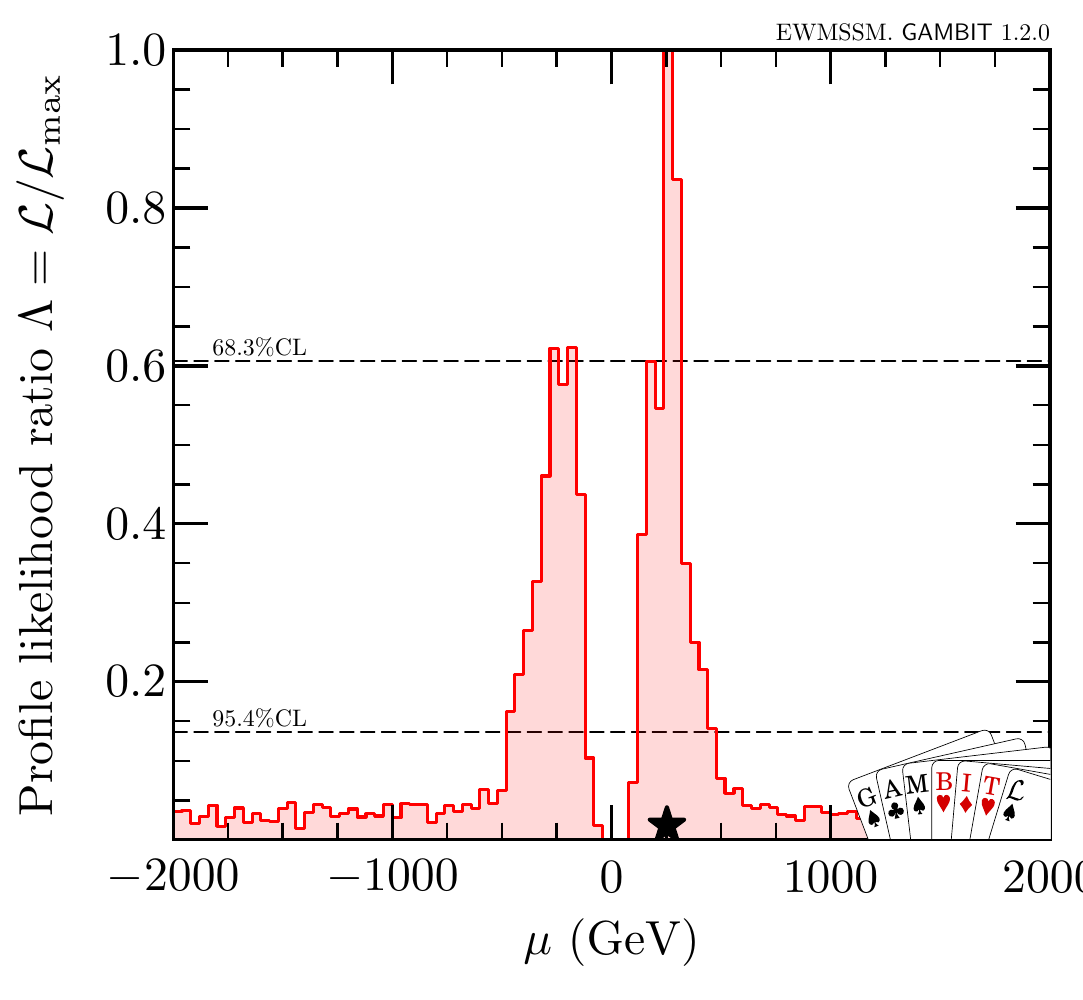}
  \includegraphics[height=0.8\columnwidth]{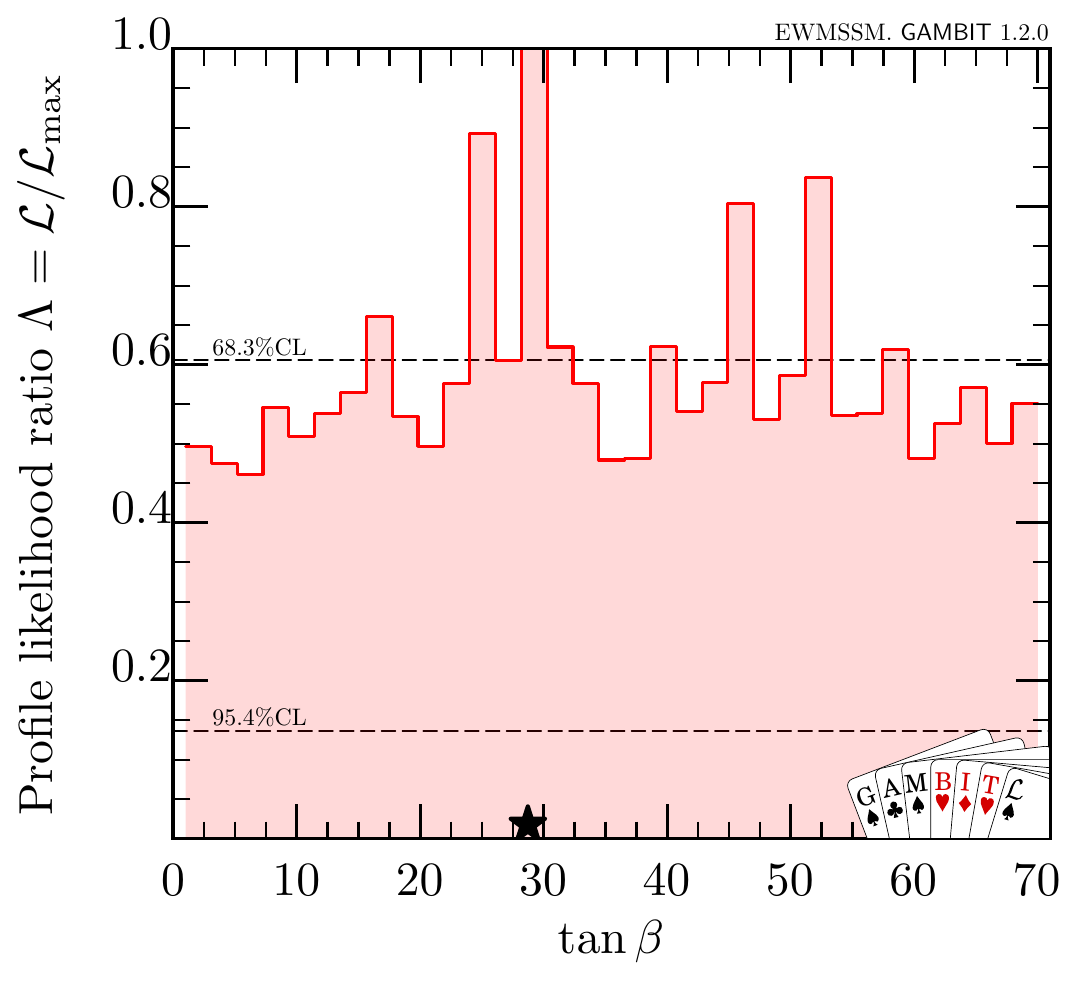}
\caption{One-dimensional profile likelihood for the electroweakino sector parameters: $M_1$ (top left), $M_2$ (top right), $\mu$ (bottom left) and $\tan\beta$ (bottom right). The dashed black lines show the $1\sigma$ and $2\sigma$ confidence limit and the black star marks the best-fit point.
  }
  \label{fig:mass_parameters_1D}
\end{figure*}

\begin{figure*}[t]
  \centering
  \includegraphics[height=0.8\columnwidth]{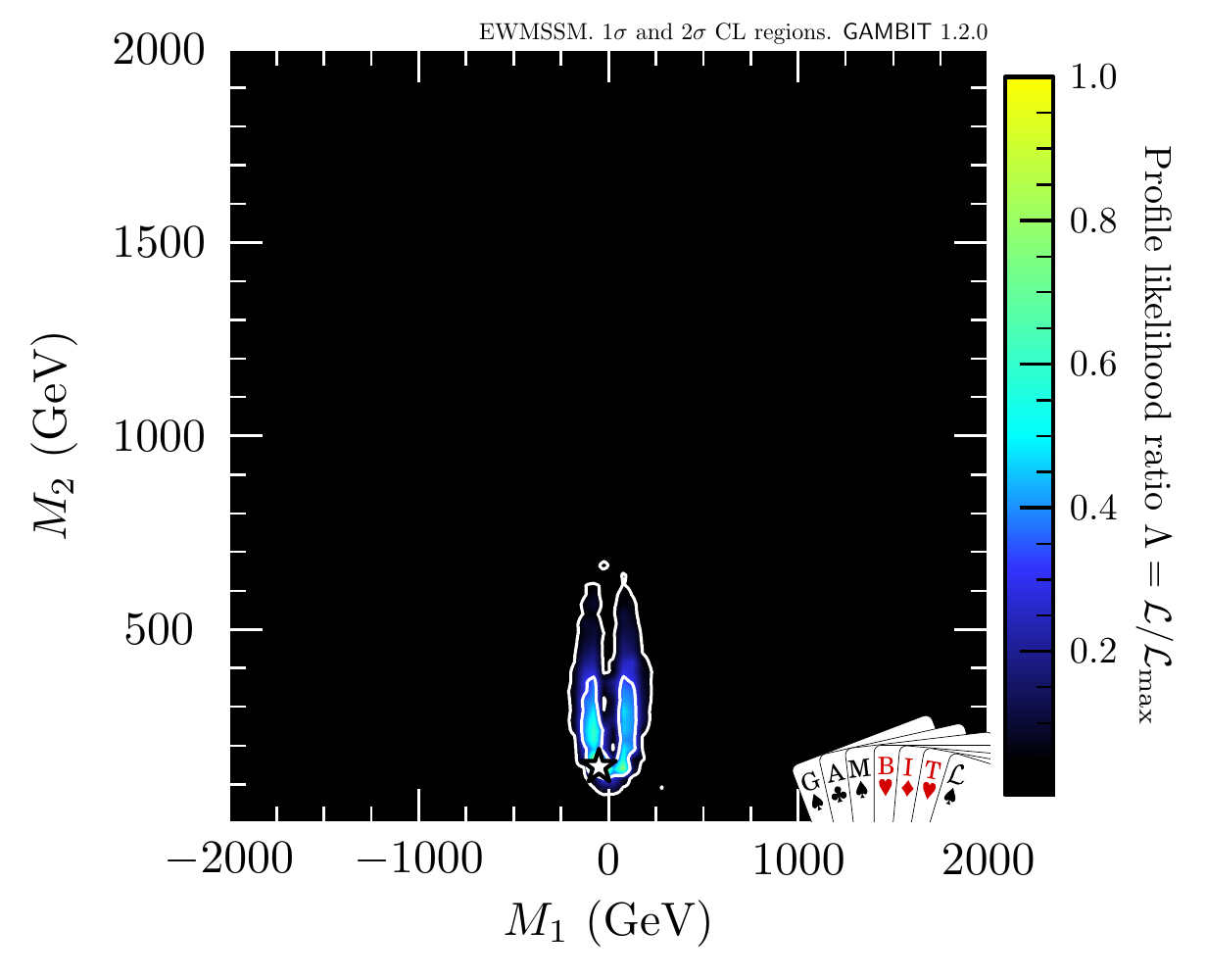}
  \includegraphics[height=0.8\columnwidth]{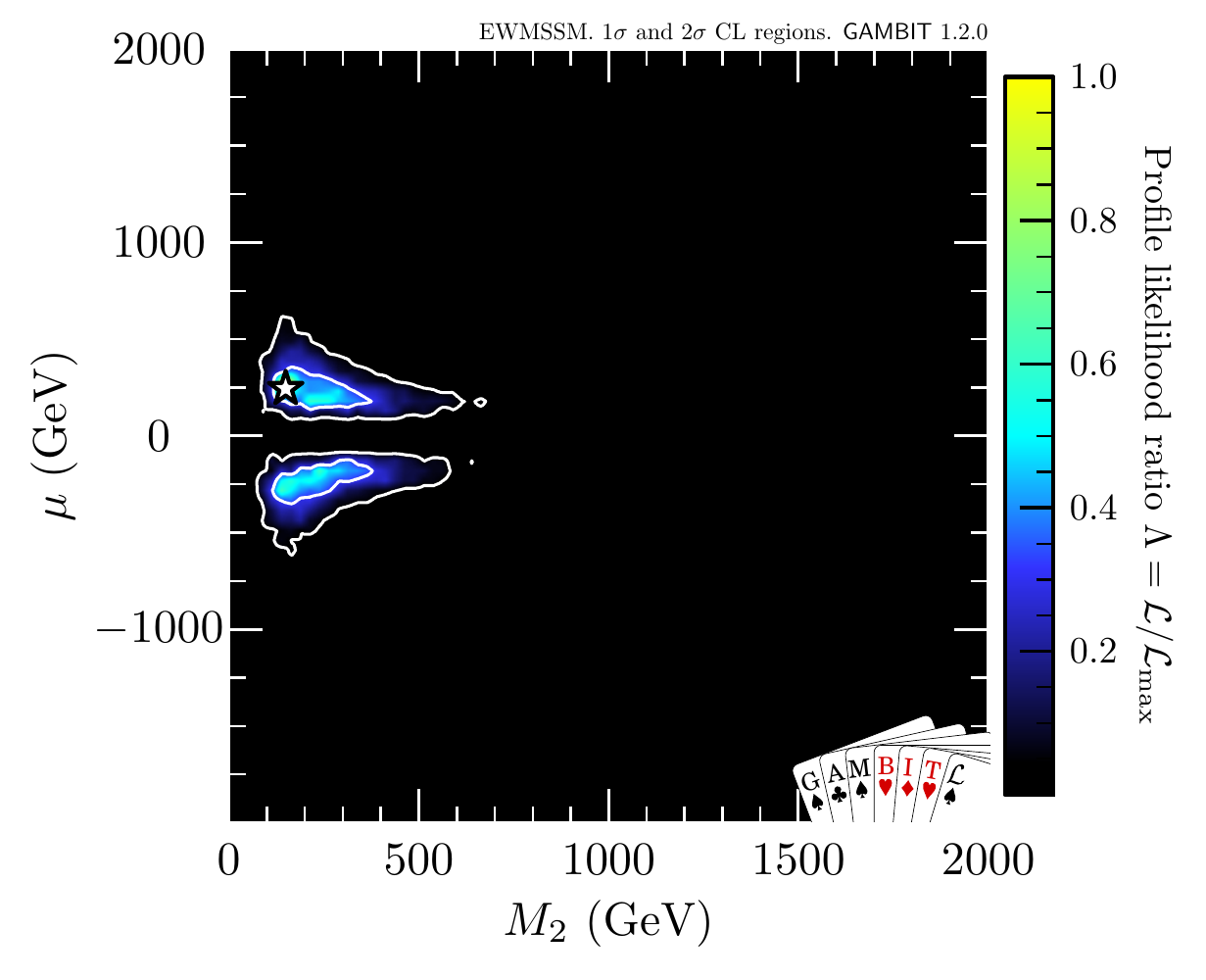}
  \caption{Profile likelihood in the $(M_1,M_2)$ plane (left) and the $(\mu,M_2)$ plane (right). The contour lines show the $1\sigma$ and $2\sigma$ confidence regions. The best-fit point is marked by the white star.
  }
  \label{fig:mass_parameters_2D}
\end{figure*}

The neutralino and chargino masses are fixed by a set of four parameters: $\{M_1, M_2, \mu, \tan\beta \}$.  If the excess is a real signal of new physics it will be very important to extract the underlying parameters from the data.  In Figure~\ref{fig:mass_parameters_1D} we show the profile likelihood of each parameter individually.  For the three dimensionful parameters, $M_1$ (top left panel), $M_2$ (top right panel) and $\mu$ (bottom left panel) a preference for a low mass scale can be seen, as one would expect from the fact that we have already seen a preference for all electroweakinos being light, with the current level of resolution similar to that with which we are able to determine the electroweakino pole masses. We do not observe any lower bound on $M_1$, allowing an extremely light bino, while $M_2$ and $\mu$ must be heavier than about $100$\,GeV.   We see no strong preference for any particular choice of $\tan\beta$ (bottom right panel) in the data, with the entire range from $1$ to $70$ permitted at $2\sigma$.  In Figure~\ref{fig:mass_parameters_2D}, we also show the profile likelihood in planes of the underlying dimensionful parameters, $(M_1,M_2)$ and $(M_2,\mu)$. We see that within the $2\sigma$ contours, all three parameters are light. This implies that two types of neutralino (wino, bino or Higgsino) are light. This was already suggested by Fig.\ \ref{fig:n1content}, which shows the mixing of the four neutralinos. Given collider constraints on the gluino mass, it appears that the excess is not compatible with high-scale unification of the gaugino masses, as assumed in e.g.\ constrained MSSM/mSUGRA scenarios.

\section{Implications for dark matter}
\label{sec:darkmatter}

With an indication that relatively light LSPs may
have been produced at the LHC, it becomes very interesting to consider
the possibility that they may constitute DM.  The most important
observables to check in this context are the thermal relic density of
DM, constraints on the interaction of the LSP with nuclei from
direct detection experiments and neutrino telescope observations of the Sun, and limits from
indirect searches for DM annihilation.

In this paper we have analysed electroweakino searches specifically in
an effective framework where the sfermions are decoupled, in order
to fully understand the implications of electroweak LHC searches for the
electroweakino sector.  Whilst this framework fully captures all
phenomenology relevant for those particular searches, it does not
cover all possible implications of the same electroweakinos for DM.
Light sfermions and/or non-SM Higgs bosons can provide
(co-)annihilation channels able to deplete the relic density of the
LSP and boost late-time annihilation signals, as well as impact
nuclear scattering rates. Modifications to the expansion history during freeze-out
could also significantly dilute the final relic density, as could decay of the lightest neutralino
to gravitino DM.
A full exploration of possible DM scenarios
involving the neutralinos and charginos involved in the putative LHC
signal is therefore beyond the scope of the current paper.  However,
we can at least consider a standard cosmological history along with the possible annihilation and scattering
processes that involve only the electroweakinos and SM particles, in
order to see if they might be able to explain DM alone.

The three relevant DM annihilation processes in the early Universe in
this context are efficient annihilation of Higgsino DM (potentially
also involving co-annihilation with similar-mass Higgsino charginos
and next-to-lightest neutralinos) or wino DM (potentially involving
co-annihilation with similar-mass wino lightest charginos), or
resonant annihilation of binos via the SM Higgs or $Z$ boson.  Whilst
all of these processes have been shown to be effective in the relevant
mass range in recent
studies \cite{2017JHEP...09..064C,2018arXiv180701476P,MSSM}, the
detailed mixture of the LSP plays a significant role in determining
whether the resulting relic density of DM is equal to the full
cosmological abundance, or some fraction of it.  At a mass of a few
tens or hundreds of GeV, pure winos and Higgsinos annihilate too
efficiently to produce the full relic density.  On the other hand,
annihilation of pure binos is too inefficient to bring the relic
density down to the observed value, unless assisted by a resonance.
Solutions that produce the full relic density of DM must therefore
either be predominantly bino with an LSP mass of $m_h/2$ or $m_Z/2$,
in order to trigger the resonance mechanism, or a mixture featuring a
significant bino component plus some Higgsino and/or wino
contribution(s).

In order to examine the potential of the models preferred by LHC
electroweakino searches to explain DM, we postprocessed all points
found in our scans to apply a series of additional DM likelihoods.
These likelihoods are based on the relic density measured
by \textit{Planck} \cite{Planck15cosmo} (applied as an upper limit),
constraints on the DM-nucleon scattering rate from LUX \cite{Akerib:2016vxi}, PandaX \cite{Tan:2016zwf, Cui:2017nnn},
XENON1T \cite{Aprile:2018dbl}, CDMSlite \cite{Agnese:2015nto}, CRESST-II \cite{Angloher:2015ewa}, PICO-60 \cite{PICO60},  DarkSide-50 \cite{Agnes:2018fwg} and
IceCube \cite{IC79,IC79_SUSY}, as well as gamma-ray limits from observations
of 15 Milky Way dwarf spheroidal galaxies by the \textit{Fermi} Large
Area Telescope (LAT; \cite{LATdwarfP8}).  More details of these
observable calculations and likelihood functions can be found in
Refs.\ \cite{DarkBit, HiggsPortal}.

\begin{figure*}[h!]
  \centering
  \includegraphics[height=0.8\columnwidth]{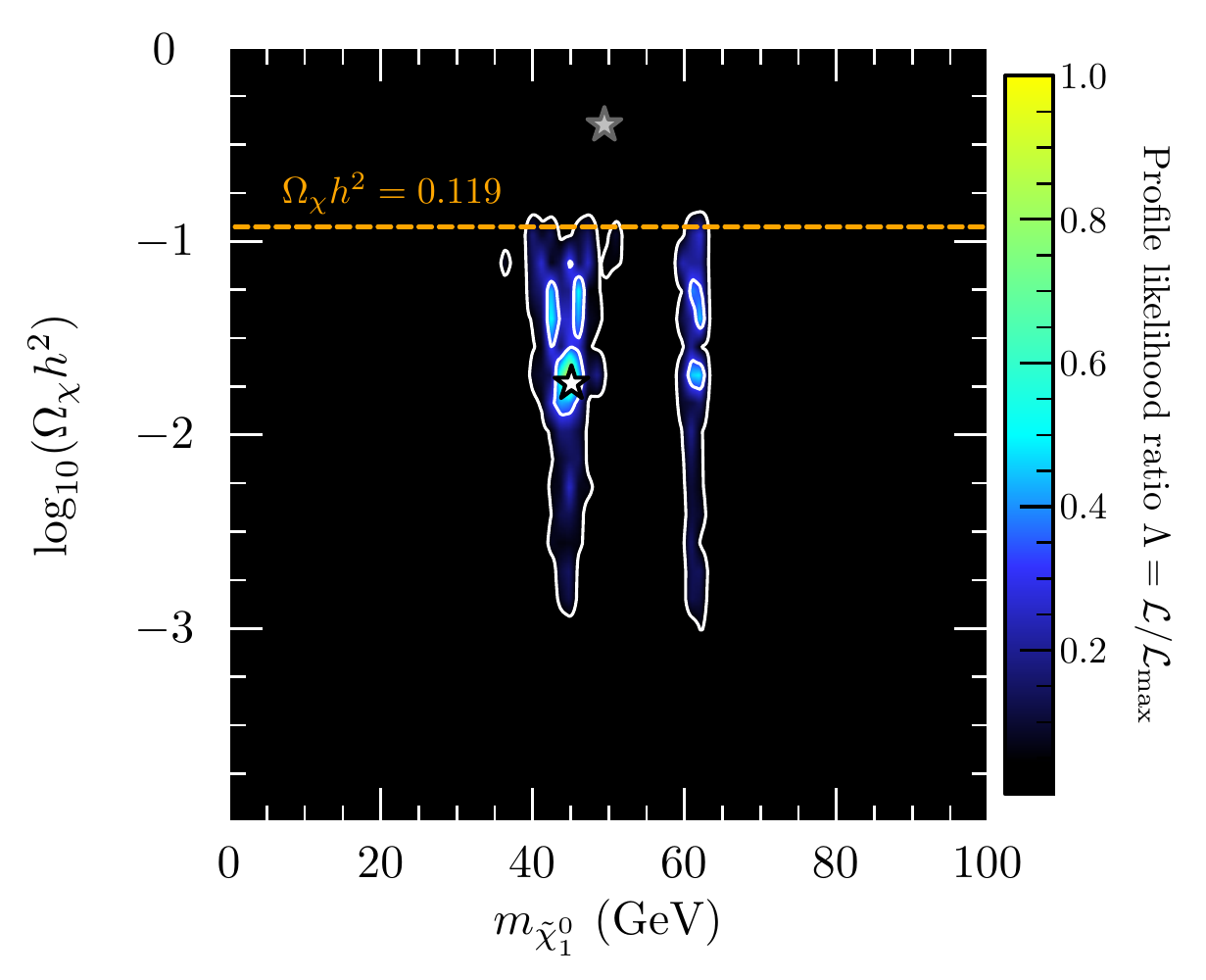}
  \includegraphics[height=0.8\columnwidth]{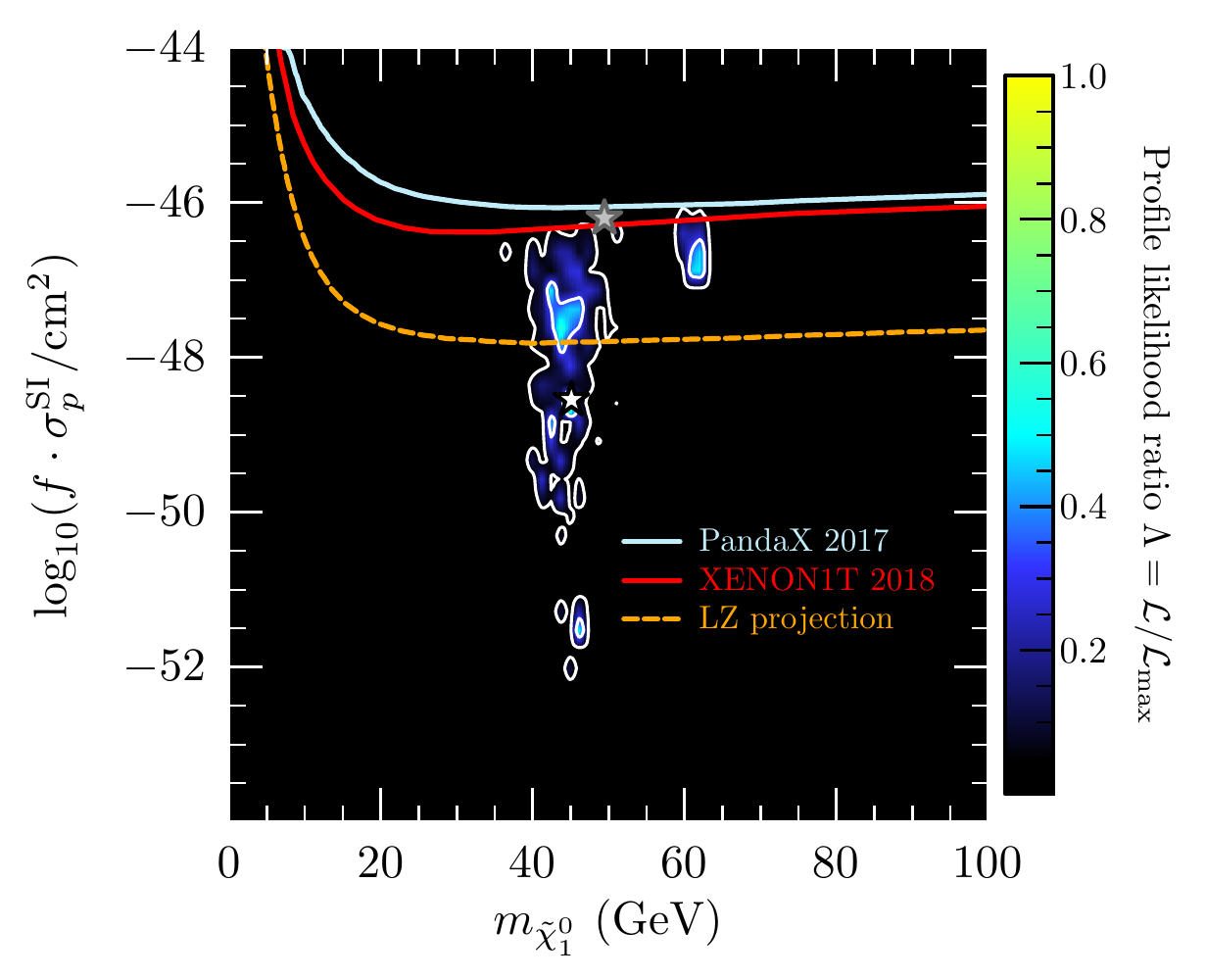}
  \caption{Combined collider and DM profile likelihood of the relic density of DM (left) and spin-independent direct detection cross-section (right), both plotted against the DM candidate mass. The contours show the $1\sigma$ and $2\sigma$ regions. The white star marks the point with the highest combined collider-DM likelihood, whereas the grey star marks our collider-only best-fit point.  For comparison, we show the latest limits from PandaX \cite{Cui:2017nnn} and XENON1T \cite{Aprile:2018dbl}, along with the projected sensitivity of the LZ experiment \cite{LZ}.}
  \label{fig:relic_density_direct_detection}
\end{figure*}

In the left panel of Fig.\ \ref{fig:relic_density_direct_detection}, we show the relic density of the models found in our scans, coloured according to their combined collider-DM profile likelihood. The two disconnected regions are those where the relic density is brought down to (or below) the observed value by resonant annihilation via either the $Z$ (left region) or Higgs boson (right region).  The $Z$ funnel in particular is mapped out quite clearly in our results, with both sides of the resonance clearly visible around $m_{\tilde{\chi}_1^0} = m_Z/2$. Whilst the total likelihood is highest when the lightest neutralino makes up only $\sim$10\% of the DM, it is interesting to see the possibility that the LSP makes up all of the DM is well within the 2$\sigma$ region. The small preference for the lower relic density is driven by the direct detection likelihoods, in particular those of PandaX, Xenon1T, and PICO-60.

The right panel of Fig.\ \ref{fig:relic_density_direct_detection} shows the spin-independent nuclear scattering cross-sections of the models found in our scans, compared to the latest limits from two leading experiments included in our likelihood (PandaX \cite{Cui:2017nnn} and XENON1T \cite{Aprile:2018dbl}), as well as the expected sensitivity of the LZ experiment \cite{LZ}.  We account for the fraction of the observed DM in neutralinos at each point in the scan by rescaling the cross-sections by $f=\Omega_{\tilde{\chi}_1^0} / \Omega_\mathrm{DM}$, so as to compare fairly with the experimental limits (which assume $f=1$).  We see that the $h$-funnel region already sits at the edge of the current experimental sensitivity, and will be probed in its entirety in the next generation of experiments.  A substantial part of the $Z$-funnel region will also be tested by LZ and similar experiments, but this does not include the best-fit point.

We do not show plots relevant for indirect searches for DM, as the preferred annihilation cross-sections in the EWMSSM (after the application of the collider and DM likelihoods) all lie at $f^2\langle\sigma v\rangle_0 < 10^{-28}\,\text{cm}^3\,\text{s}^{-1}$.  This is significantly below the sensitivity of any planned future indirect detection experiment. Although the preferred masses (around 45 or 62\,GeV) and dominant DM annihilation final states (mostly $\bar{b}b$) of our best-fit models are strikingly similar to those preferred in DM fits to the Galactic Centre gamma-ray excess (e.g.\ \cite{Calore:2014nla}), the annihilation cross-sections are too low to explain the excess without the presence of e.g.\ an additional light CP-odd Higgs boson to mediate additional late-time annihilation to $\bar{b}b$ \cite{Cheung:2014lqa,Cao:2014efa}.

\begin{figure*}[h!]
  \centering
  \includegraphics[height=0.8\columnwidth]{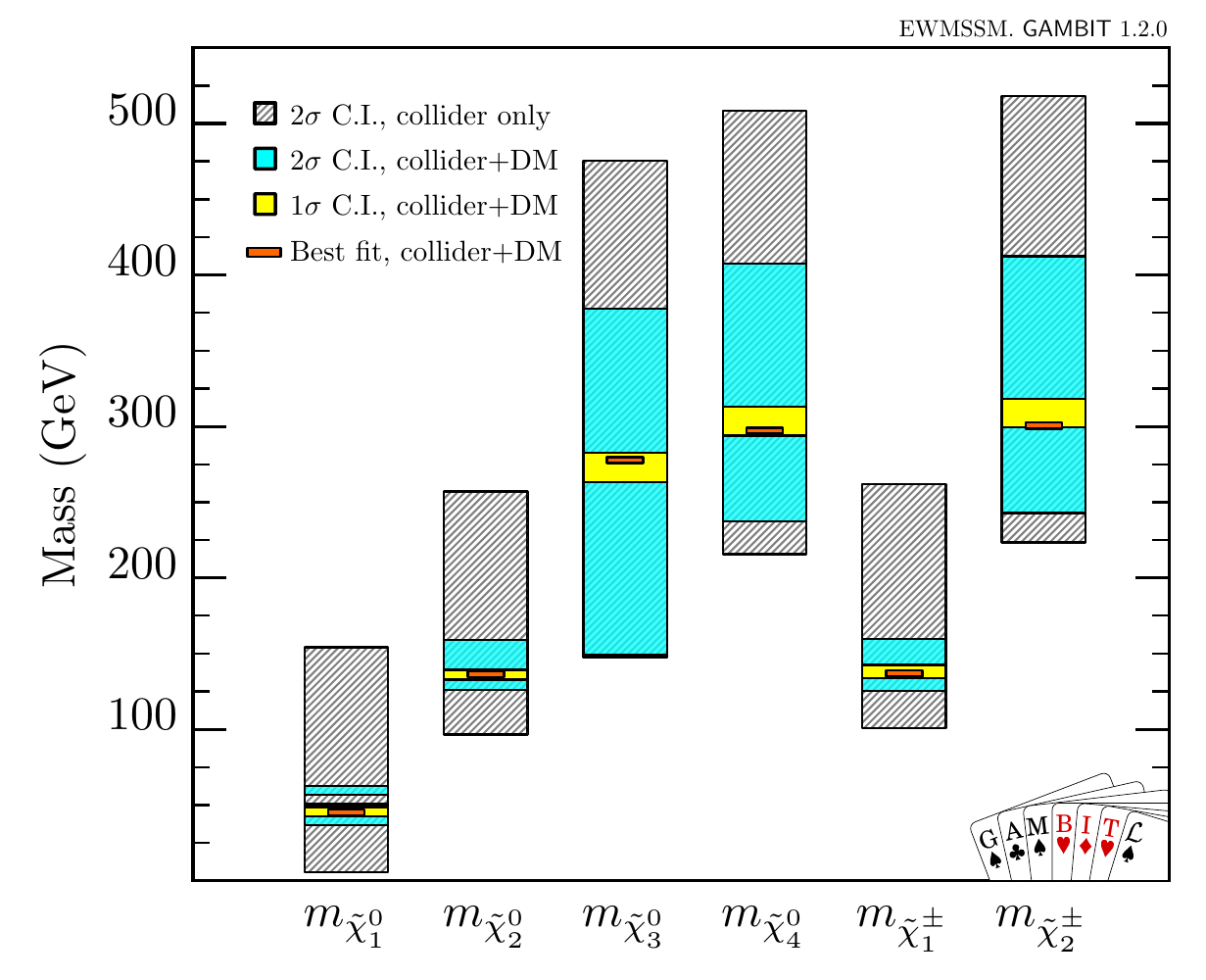}
  \caption{Summary of the one-dimensional $1\sigma$ (yellow) and $2\sigma$ (cyan) confidence intervals for the neutralino and chargino masses, resulting from the joint collider and dark matter likelihood. For comparison, the $2\sigma$ confidence intervals from the collider likelihood alone are shown in grey (hatched). The orange lines mark the best-fit values.}
  \label{fig:mass_summary_plot_DM}
\end{figure*}

We summarise our result for the joint collider and DM likelihood in Figure~\ref{fig:mass_summary_plot_DM}, where we show the one-dimensional
$1\sigma$ and $2\sigma$ confidence intervals for the six electroweakino masses.
This is compared to the $2\sigma$ confidence intervals resulting from the collider
likelihood alone, i.e.\ the $2\sigma$ intervals from Figure~\ref{fig:mass_summary_plot}.
The restriction of the EWMSSM to the $Z$-funnel and $h$-funnel regions by the DM likelihood not only restricts the LSP mass to the relevant resonance, but also significantly contracts the range of allowed $\tilde{\chi}^\pm_1$ and $\tilde{\chi}^0_2$ masses.  This is to be expected from the strong correlation between all three of $m_{\tilde{\chi}^0_1}$, $m_{\tilde{\chi}^0_2}$ and $m_{\tilde{\chi}^\pm_1}$ in Fig.\ \ref{fig:pole_masses_2D}.

Finally, we emphasise that these results are based on a scan that searched for regions of parameter space that could provide the best fit to the LHC likelihood. As a result, the parts of parameter space preferred at the 2$\sigma$ level by the combined DM and LHC likelihood are rather sparsely sampled, with only 541 points. Therefore, these results should be taken as a rough first check of the DM properties of the EWMSSM models consistent with the excesses seen at the LHC, and it should be kept in mind that a scan that seeks to map the combined likelihood could ultimately reveal more viable regions of parameter space.

%%%%%%%%%%%%%%%%%%%%%%%%%%%%%%%%%%%%%%%%%%%%%%%%%%%%%%%
%%%%%%%%%%%%%%%%%%%%%%%%%%%%%%%%%%%%%%%%%%%%%%%%%%%%%%%
\section{Conclusions}
\label{sec:conc}
We are in a very interesting period in the hunt for electroweakinos at the Large Hadron Collider, as the large accumulated datasets at a centre of mass energy of 13\,TeV offer real potential for discovering weakly-produced sparticles. In this paper, we have performed a comprehensive global statistical fit of a 4D MSSM model in which $M_1$, $M_2$, $\mu$ and $\tan\beta$ are varied, whilst other MSSM parameters are held at fixed values in order to decouple all sparticles except the electroweakinos. In interpreting the results, we have considered both the case where one assumes that supersymmetry is not realised at a scale accessible by the LHC (in which case we are testing the exclusion power of current LHC searches), and the case where one allows the presence of a possible signal in the LHC data.

In the case where we assume that the data are consistent with SM backgrounds only, we find that current LHC searches offer little power to exclude any point on the $(m_{\tilde{\chi}_1^{\pm}},m_{\tilde{\chi}_1^0})$ plane.  This is due to the differences between the simplified SUSY models used for optimisation and interpretation at the LHC, and the more realistic model that we employ here.  This model allows for richer final states, plus a much wider variation in the assumed electroweakino contents. Our results interpreted in this fashion can be used to generate insights into how to better optimise the LHC's ability to exclude sparticles.

In the case of a possible signal, we find that a series of excesses in the LHC data point towards a model with neutralino masses of $(m_{\tilde{\chi}_1^0}, m_{\tilde{\chi}_2^0}, m_{\tilde{\chi}_3^0}, m_{\tilde{\chi}_4^0})$ =  (8--155, 103--260, 130--473, 219--502)\,GeV, and chargino masses of $(m_{\tilde{\chi}_1^{\pm}}, m_{\tilde{\chi}_2^{\pm}})$ = (104--259, 224--507)\,GeV at the 95.4\% confidence level. The LSP is predominantly bino in our best-fit region, and the models are otherwise split into those that have the winos lighter than the Higgsinos, and those that have the Higgsinos lighter than the winos. Intriguingly, having all of the electroweakino spectrum light not only helps our best-fit model evade some LHC searches, but it also highlights a series of excesses that all contribute positively to our best-fit log-likelihood in the same mass region. Even if one does not take the pattern of current excesses seriously, this suggests that, at the very least, optimising analyses on simple one-step decay chains resulting from NLSP pair-production is not a good way to probe light electroweakino spectra.

Our best-fit point has neutralino masses of $(m_{\tilde{\chi}_1^0}, m_{\tilde{\chi}_2^0}, m_{\tilde{\chi}_3^0}, m_{\tilde{\chi}_4^0})\approx  (49.4, 141.6, 270.3, 290.2)$\,GeV, and chargino masses of $(m_{\tilde{\chi}_1^{\pm}}, m_{\tilde{\chi}_2^{\pm}}) \approx (142.1, 293.9)$\,GeV. We find a local significance of 3.3$\sigma$ for this excess. If there is indeed a supersymmetric signal resembling these properties the ATLAS and CMS experiments should be sensitive to it using the full LHC Run 2 dataset. If one includes LHC searches for charginos and neutralinos conducted with proton--proton collision data collected at a centre of mass energy of 8\,TeV, the local significance reduces to 2.9$\sigma$, but the general details of our best fit region apparently remain intact.

Analysis of the DM implications of our points is complicated by the fact that the particular values of the MSSM parameters that are held fixed in our analysis might influence the ability of our models to generate the correct relic density. Nevertheless, we find that a subset of the area around our best-fit point is very much consistent with both the observed relic density and constraints from direct and indirect searches for DM -- even assuming that only electroweakinos and SM particles play a role in setting the relic density.  Excellent fits to both the DM and collider likelihoods are possible in the so-called $Z$- and $h$-funnel regions, where the lightest neutralino has a mass equal to approximately half the mass of either the $Z$ or Higgs boson.  Many of these models will be accessible to the next generation of direct detection experiments, raising the possibility of a simultaneous confirmation of the putative LHC signal in future datasets from both the LHC and dark matter experiments.

\begin{acknowledgements}

We thank Will Handley for interfacing \textsf{polychord} in \GB \textsf{1.2.0}, Peter Skands for discussions on ISR jets, and the rest of the GAMBIT Community for helpful discussions and comments.
We acknowledge PRACE for awarding us access to the Marconi supercluster at CINECA, Italy. This work was supported by
STFC (UK; ST/K00414X/1, ST/N000838/1, ST/P000762/1),
the Royal Society (UK; UF110191, UF160548),
the Research Council of Norway (Norway; FRIPRO 230546/F20), NOTUR (Norway; NN9284K),
Red Espa\~nola de Supercomputaci\'on (Spain; FI-2016-1-0021),
the Horizon 2020 Marie Sk\l{}odowska-Curie actions (EU; H2020-MSCA-IF-2016-752162),
the Australian Research Council (Australia; CE110001004, DP180100031, FT130100018, FT140100244, FT160100274),
the Labex ILP part of the Idex SUPER (France; ANR-10-LABX-63), the Agence Nationale de la Recherche as part of the programme Investissements d'avenir (France; ANR-11-IDEX-0004-02),
and the National Sciences and Engineering Research Council (NSERC) of Canada.

\end{acknowledgements}

\appendix

\section{Impacts of 8\,TeV searches}
\label{app}

Although 13\,TeV LHC searches for electroweakinos are in general the most sensitive, ATLAS and CMS analyses of Run I data collected at centre-of-mass energies of 8\,TeV can also be relevant, particularly given the rather low electroweakino masses favoured in our fits.  In this appendix, we explore the impacts of 8\,TeV results on the regions preferred by the 13\,TeV data.  We also provide additional information about the predicted yields in all signal regions at our best-fit parameter combinations.

\begin{figure}[t]
  \centering
  \includegraphics[height=0.8\columnwidth]{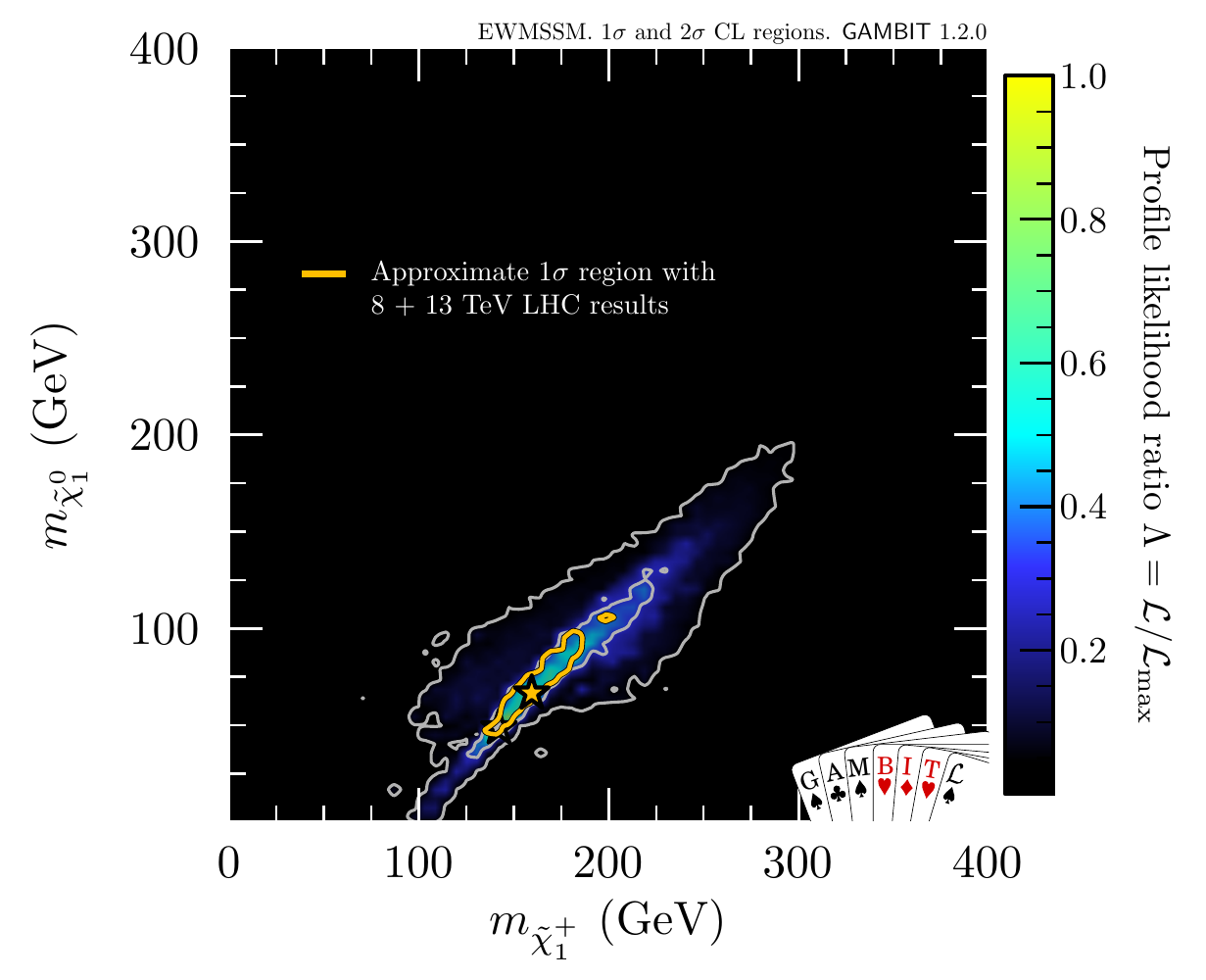}
  \caption{Profile likelihood in the $(m_{\tilde{\chi}_1^{\pm}},m_{\tilde{\chi}_1^0})$ plane of the main fit, with approximate $1\sigma$ contour after also applying 8\,TeV searches overlaid in orange. White contour lines show the $1\sigma$ and $2\sigma$ confidence regions of the main fit. The best-fit point based on 8 + 13\,TeV data is marked by the orange star, and the 13\,Tev-only point (partially obscured by the orange contour) is marked with a white star.}
  \label{fig:mass_plane_8TeV}
\end{figure}

We consider the following set of 8 TeV analyses, all based on 20\,fb$^{-1}$ of data: the ATLAS 1 lepton + 2 $b$-jet \cite{Aad:2015jqa}, 2 lepton \cite{ATLAS:2LEPEW_20invfb} and 3 lepton \cite{ATLAS:3LEPEW_20invfb} electroweak analyses, and the 20\,fb$^{-1}$ CMS 3 and 4 lepton electroweak analysis \cite{CMS:3LEPEW_20invfb}.  In order to determine the impacts of these searches, we have computed their additional contributions to the global likelihood for all parameter samples within the 1$\sigma$ preferred region of our main fit, generating 64 million Monte Carlo events per parameter point.  In the absence of correlation information for 8\,TeV searches, we computed the likelihoods based on the single signal region in each analysis with the best predicted sensitivity to each model.  The resulting approximate $1\sigma$ profile likelihood region can be seen in Fig.~\ref{fig:mass_plane_8TeV}.

\begin{table*}[t]
\begin{center}
\begin{tabular}{c c c c}
\hline
Parameter & \#5 Best fit & \#6 Heavy winos \\
\hline
$M_1(Q)$                   &$-$69.1\,GeV &      89.6\,GeV  \\
$M_2(Q)$                   & 162.8\,GeV  &     348.0\,GeV  \\
$\mu(Q)$                   & 281.7\,GeV  &  $-$173.2\,GeV  \\
$\tan\beta(m_Z)$           &  52.7       &     30.0        \\
\hline
$m_{\tilde{\chi}_1^{0}}$   &  67.3\,GeV  &      83.2\,GeV  \\
$m_{\tilde{\chi}_2^{0}}$   & 158.9\,GeV  &     174.7\,GeV  \\
$m_{\tilde{\chi}_3^{0}}$   & 299.0\,GeV  &     188.9\,GeV  \\
$m_{\tilde{\chi}_4^{0}}$   & 315.7\,GeV  &     392.4\,GeV  \\
$m_{\tilde{\chi}_1^{\pm}}$ & 159.4\,GeV  &     171.3\,GeV  \\
$m_{\tilde{\chi}_2^{\pm}}$ & 319.5\,GeV  &     392.8\,GeV  \\
\hline
Collider log-likelihood & 10.0 & 9.5 & \\
\end{tabular}
\caption{\label{tab:bestfit_8TeV} Parameter values and sparticle masses for new benchmark points obtained after applying 8\,TeV searches to the region preferred at $1\sigma$ by 13\,TeV searches. The first point (\#5) is the new best-fit model, for which the Higgsinos are heavier than the winos. The second point (\#6) is the new best-fitting solution to have winos heavier than the Higgsinos.}
\end{center}
\centering
\end{table*}

The combined impact of the 8\,TeV likelihoods on points within the region preferred at the $1\sigma$ level by 13\,TeV data ranges from $\ln \mathcal{L}_\mathrm{8\,TeV} = -2.9$ to $\ln \mathcal{L}_\mathrm{8\,TeV} = -0.2$.
As expected, the points that receive the strongest likelihood penalty from the 8\,TeV analyses are generally points with lower $\tilde{\chi}_1^0$ masses and winos lighter than the Higgsinos. On the other hand, points with $m_{\tilde{\chi}_1^0} \gtrsim 70$\,GeV and the Higgsinos lighter than the winos are largely unconstrained by the 8\,TeV results.\footnote{If we naively combine the likelihood contributions from all signal regions, neglecting the unknown correlations, we find that the combined 8\,TeV likelihood contribution ranges from $\ln \mathcal{L}_\mathrm{8\,TeV} = -6.3$ for the most strongly penalized point in this region, to a small positive contribution of $\ln \mathcal{L}_\mathrm{8\,TeV} = 0.5$ for a set of points with Higgsinos lighter than the winos and $m_{\tilde{\chi}_1^0} \gtrsim 90$\,GeV. The small positive log-likelihood contribution arises from some small excesses in the 8\,TeV CMS 3 lepton signal regions.}
As evidenced by Fig.~\ref{fig:mass_plane_8TeV}, the overall impact of 8\,TeV data on the best-fit region in the $(m_{\tilde{\chi}_1^{\pm}},m_{\tilde{\chi}_1^0})$ plane is relatively mild, disfavouring only the highest and lowest-mass ends of the region.  Note that the true $1\sigma$ region will be slightly larger than this, as the small suppression of the overall best fit ($\Delta \ln \mathcal{L} < 0.8$ between point \#1 and point \#5) means that were it computationally feasible to post-process \textit{all} samples from the original fit, some of the highest-likelihood points from the original $2\sigma$ region would move into the new $1\sigma$ region.

After applying the 8\,TeV analyses to our best-fit region, we identified two additional relevant benchmark points, given in Table \ref{tab:bestfit_8TeV}: the new overall best-fit point (\#5), and the new best point to have heavier winos than Higgsinos (\#6).  Compared to point \#1 from Table \ref{tab:bestfit}, the electroweakino masses of point \#5 are all shifted upwards by $\sim$$20$\,GeV. For point \#6 the lighter electroweakinos are $\sim$$10$\,GeV heavier than for point \#2 in Table \ref{tab:bestfit}, and the masses of the heavy winos are increased by $\sim$$100$\,GeV. The high-mass point (\#3) from Table \ref{tab:bestfit} remains the highest-mass point within the (nominal) $1\sigma$ region after the application of 8\,TeV data.

\begin{figure*}[t]
  \centering
  \includegraphics[width=0.9\columnwidth]{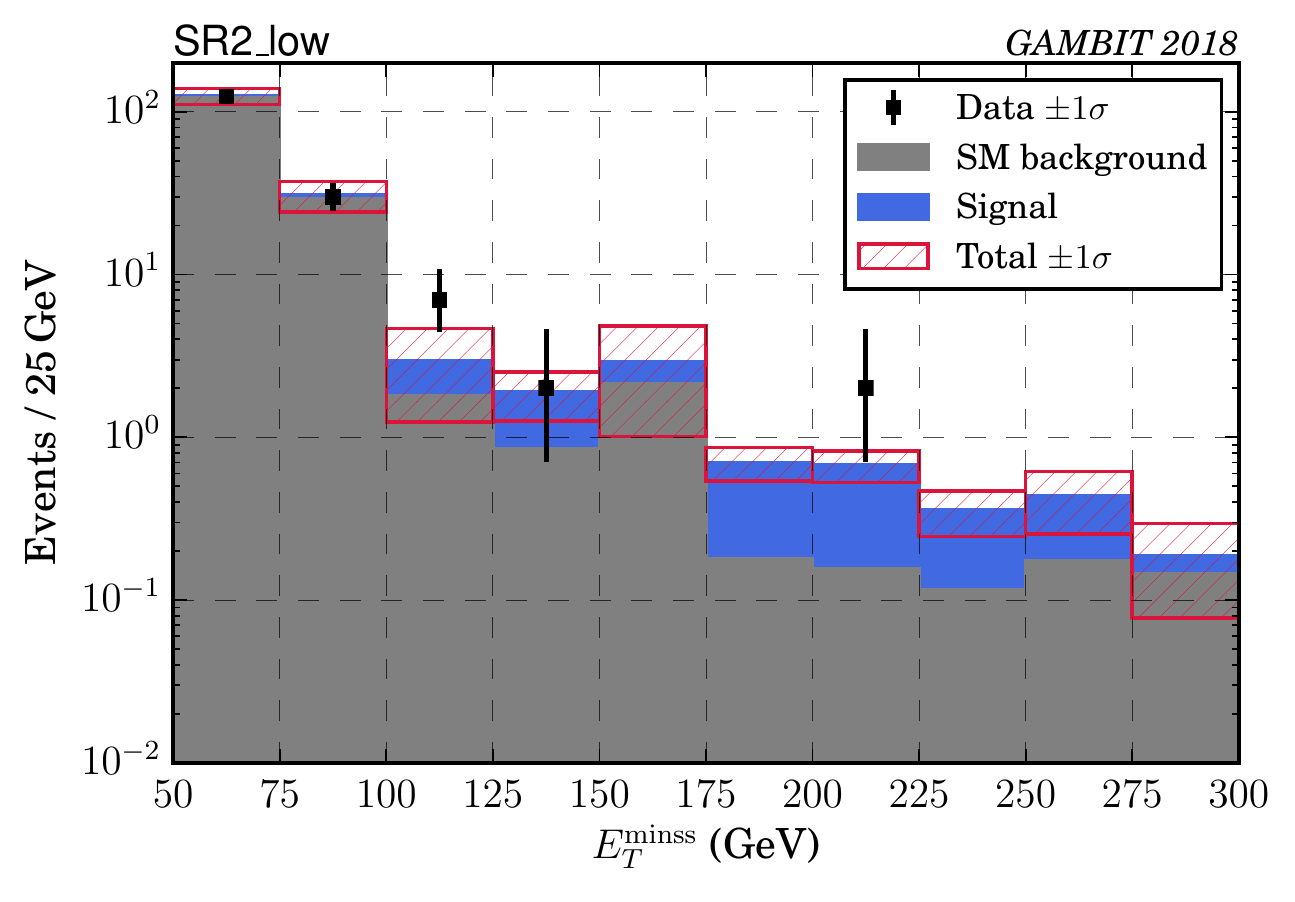}\hspace{0.1\columnwidth}
  \includegraphics[width=0.9\columnwidth]{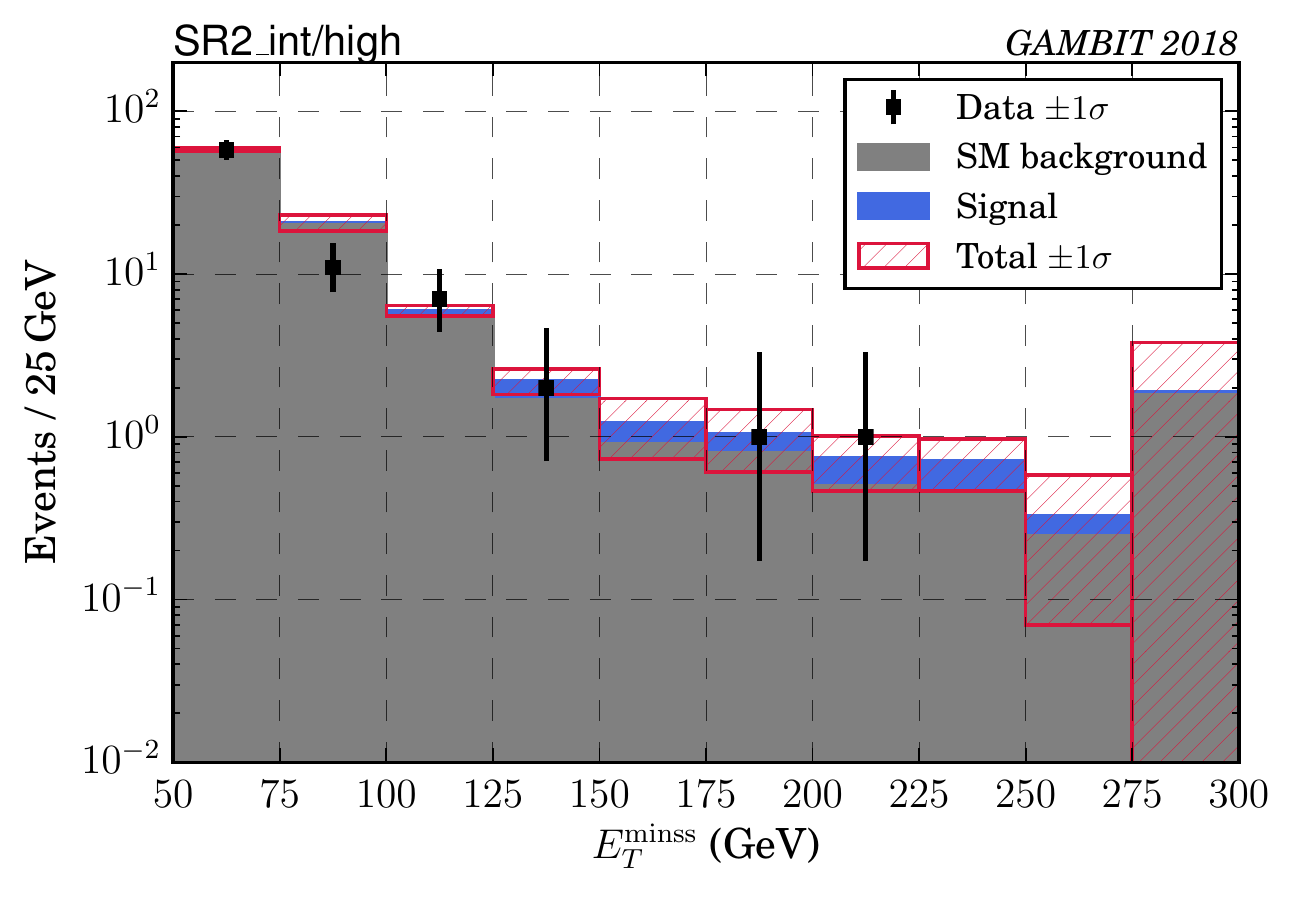}\\
  \caption{Distribution of missing transverse energy in the 2 lepton plus jets signal regions of the traditional traditional ATLAS multilepton analysis, after applying all selection requirements. The grey bars show the total SM background (taken from Ref.~\cite{Aaboud:2018jiw}) and the stacked blue bars show the signal for our best-fit point based on the combination of 8 and 13\,TeV data. The hatched red bands show the $1\sigma$ uncertainty on the total number of expected events, found by summing in quadrature the background uncertainty and the signal statistical uncertainty for our best-fit point. The black points show the ATLAS data.}
  \label{fig:atlasComparison1_8TeV}
\end{figure*}

\begin{figure*}[t]
  \centering
  \includegraphics[width=0.9\columnwidth]{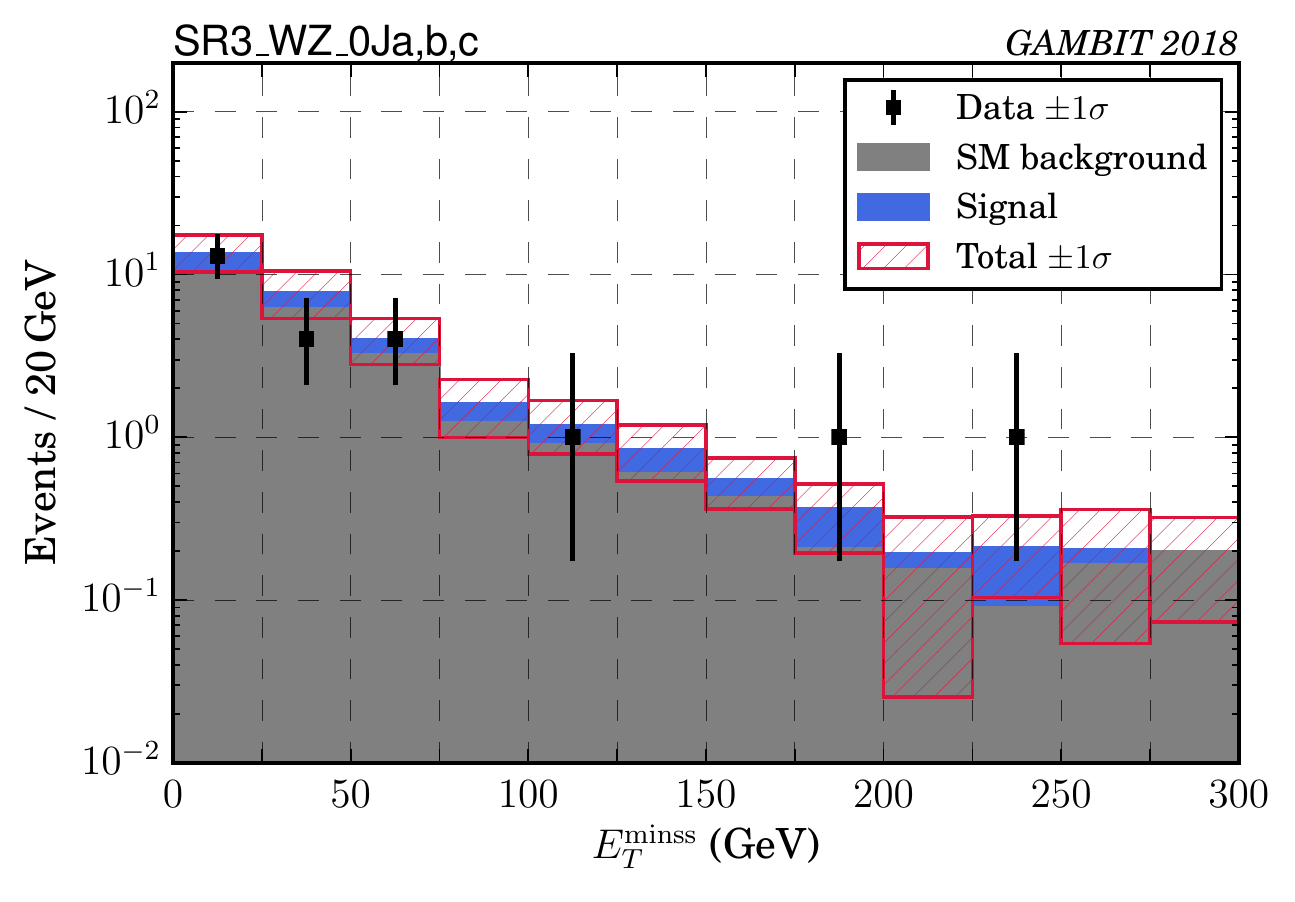}\hspace{0.1\columnwidth}
  \includegraphics[width=0.9\columnwidth]{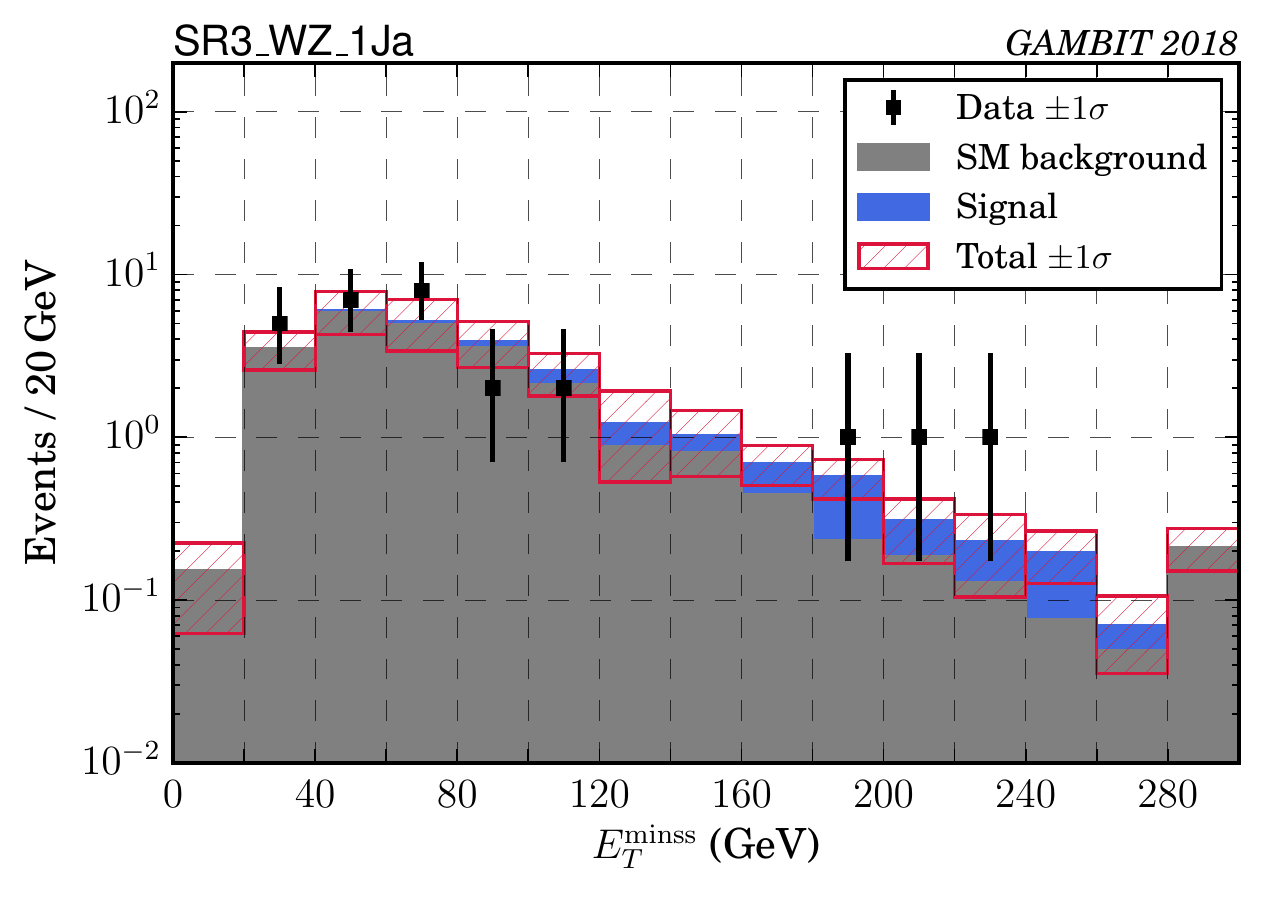}\\
  \includegraphics[width=0.9\columnwidth]{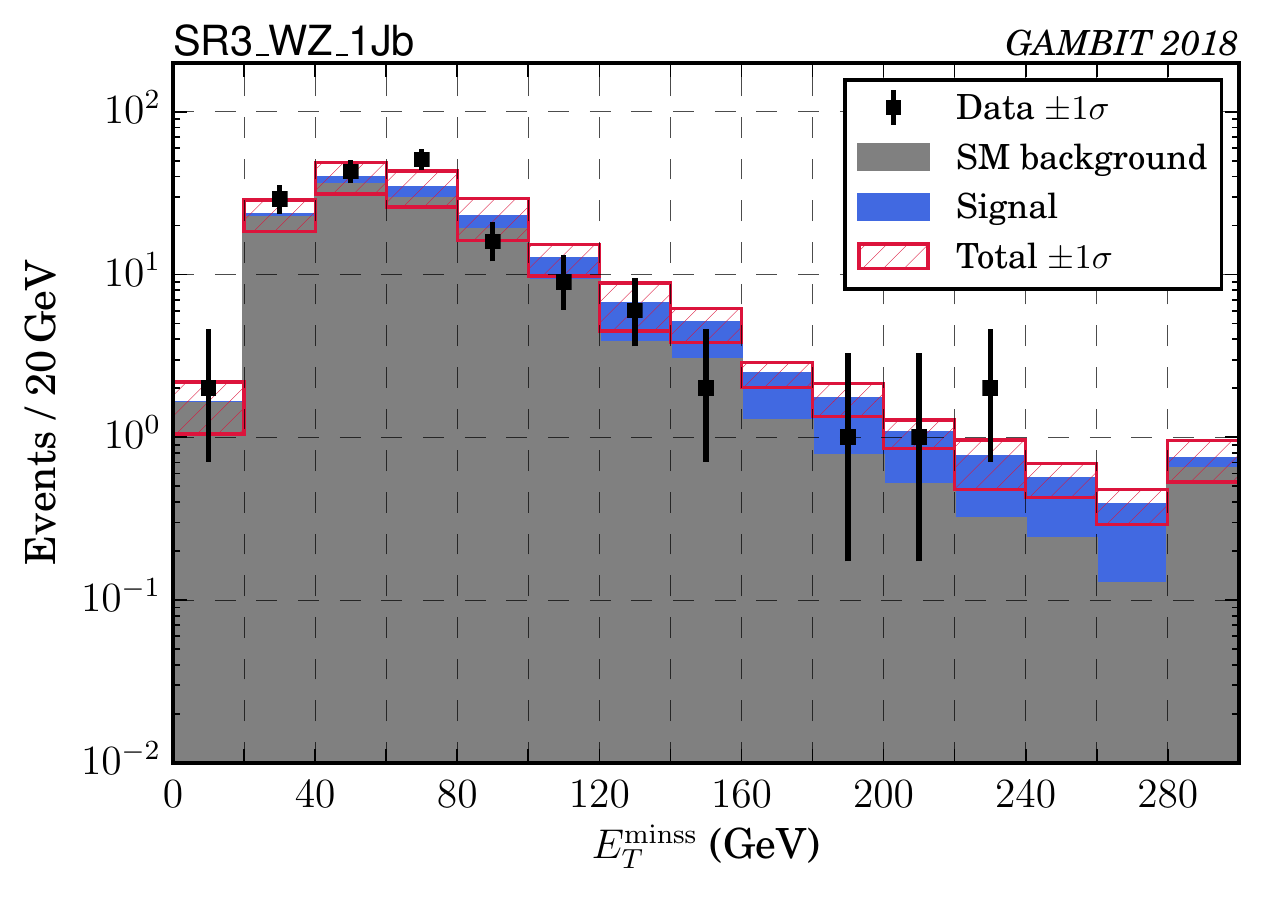}\hspace{0.1\columnwidth}
  \includegraphics[width=0.9\columnwidth]{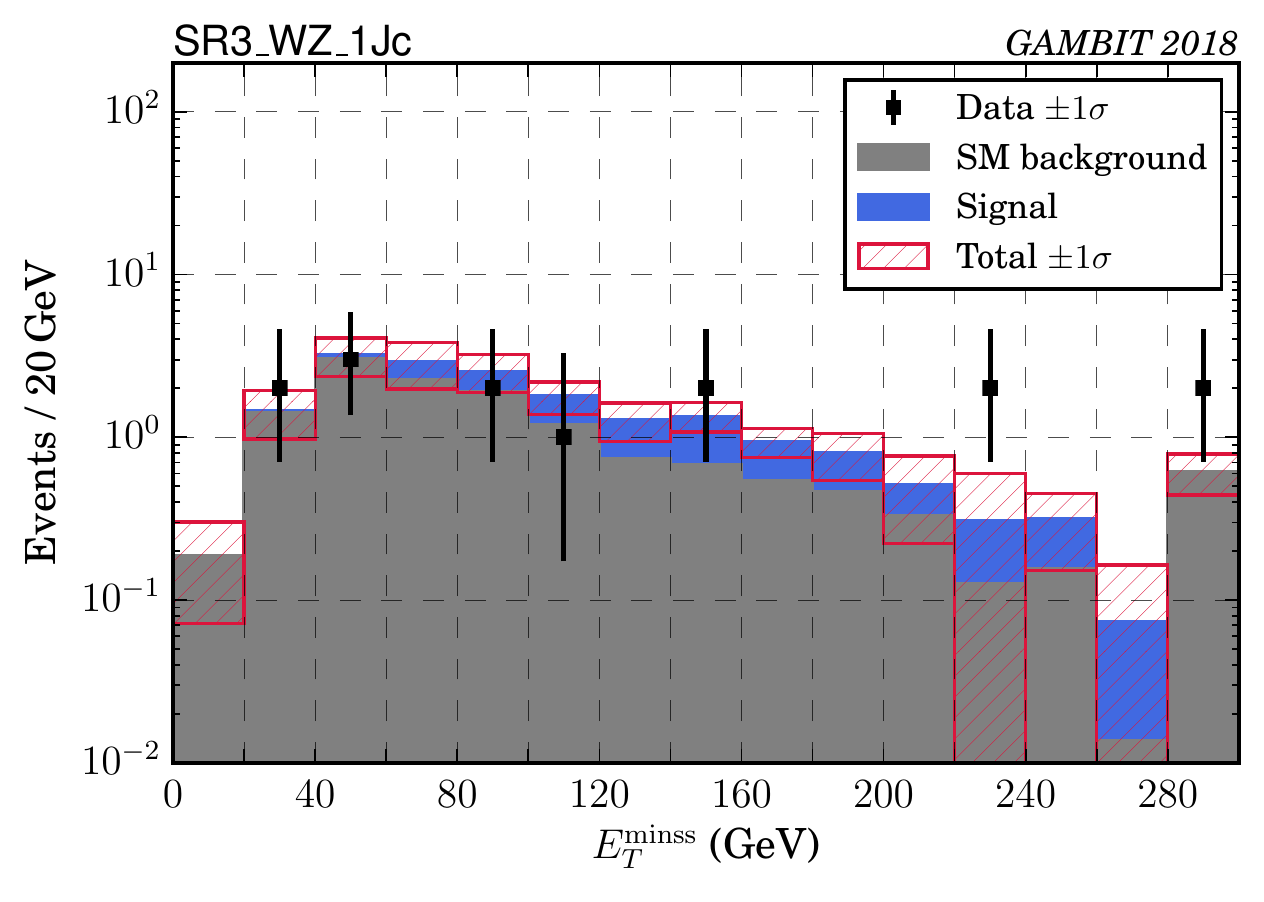}\\
  \caption{Distribution of missing transverse energy in the 3 lepton signal regions of the traditional ATLAS multilepton analysis, after applying all selection requirements. The grey bars show the total SM background (taken from Ref.~\cite{Aaboud:2018jiw}) and the stacked blue bars show the signal for our best-fit point based on the combination of 8 and 13\,TeV data. The hatched red bands show the $1\sigma$ uncertainty on the total number of expected events, found by summing in quadrature the background uncertainty and the signal statistical uncertainty for our best-fit point. The black points show the ATLAS data.}
  \label{fig:atlasComparison2_8TeV}
\end{figure*}

\begin{figure*}[t]
  \centering
  \includegraphics[width=0.9\columnwidth]{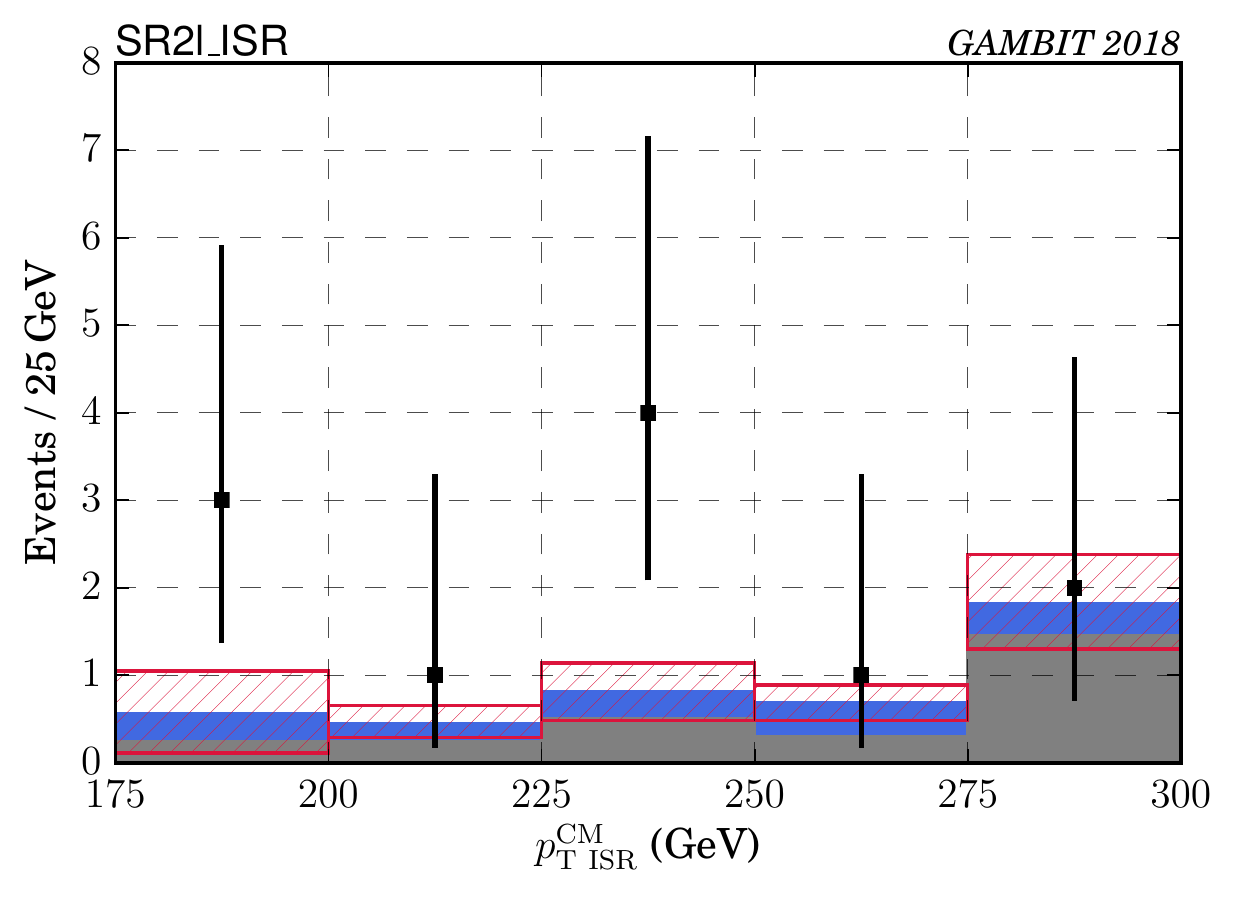}\hspace{0.1\columnwidth}
  \includegraphics[width=0.9\columnwidth]{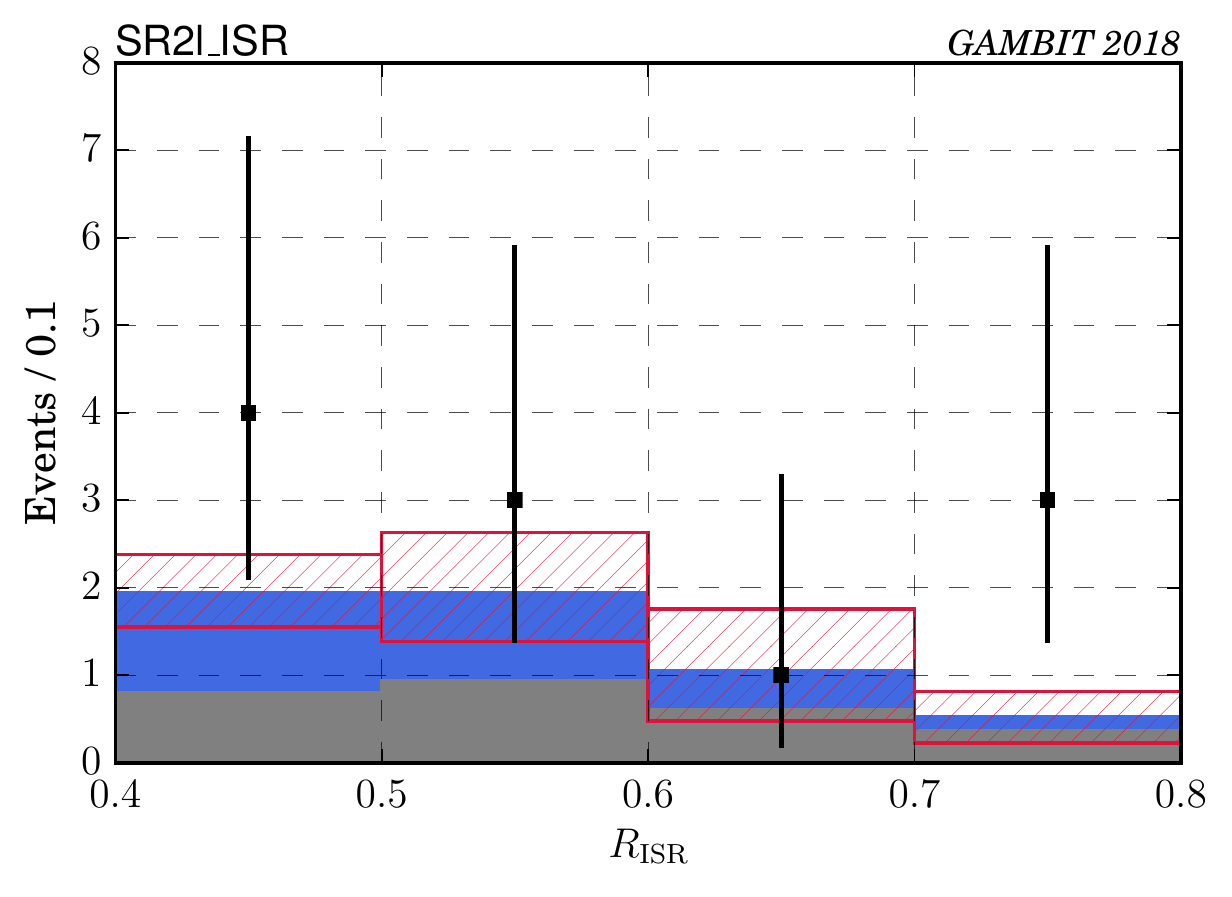}\\
  \includegraphics[width=0.9\columnwidth]{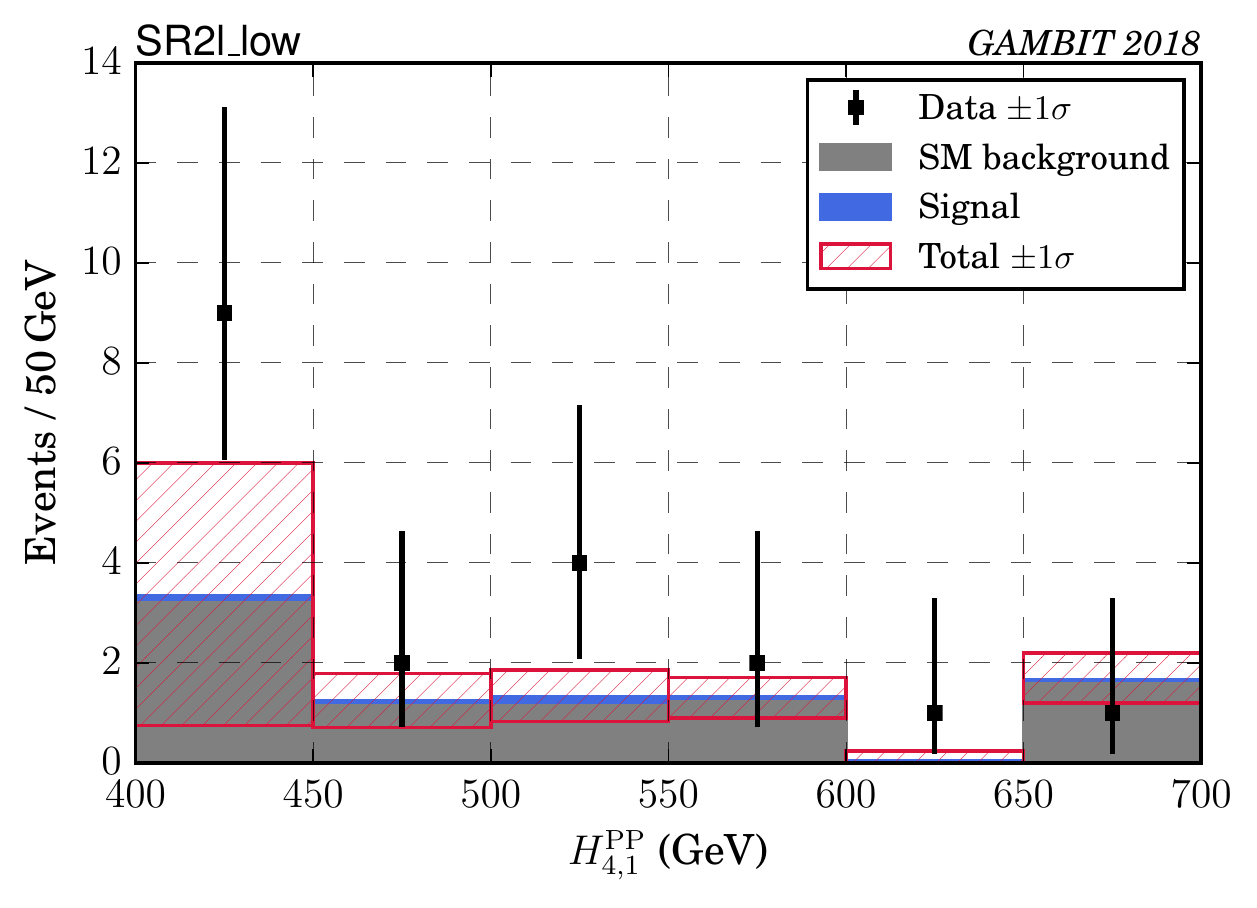}\hspace{0.1\columnwidth}
  \includegraphics[width=0.9\columnwidth]{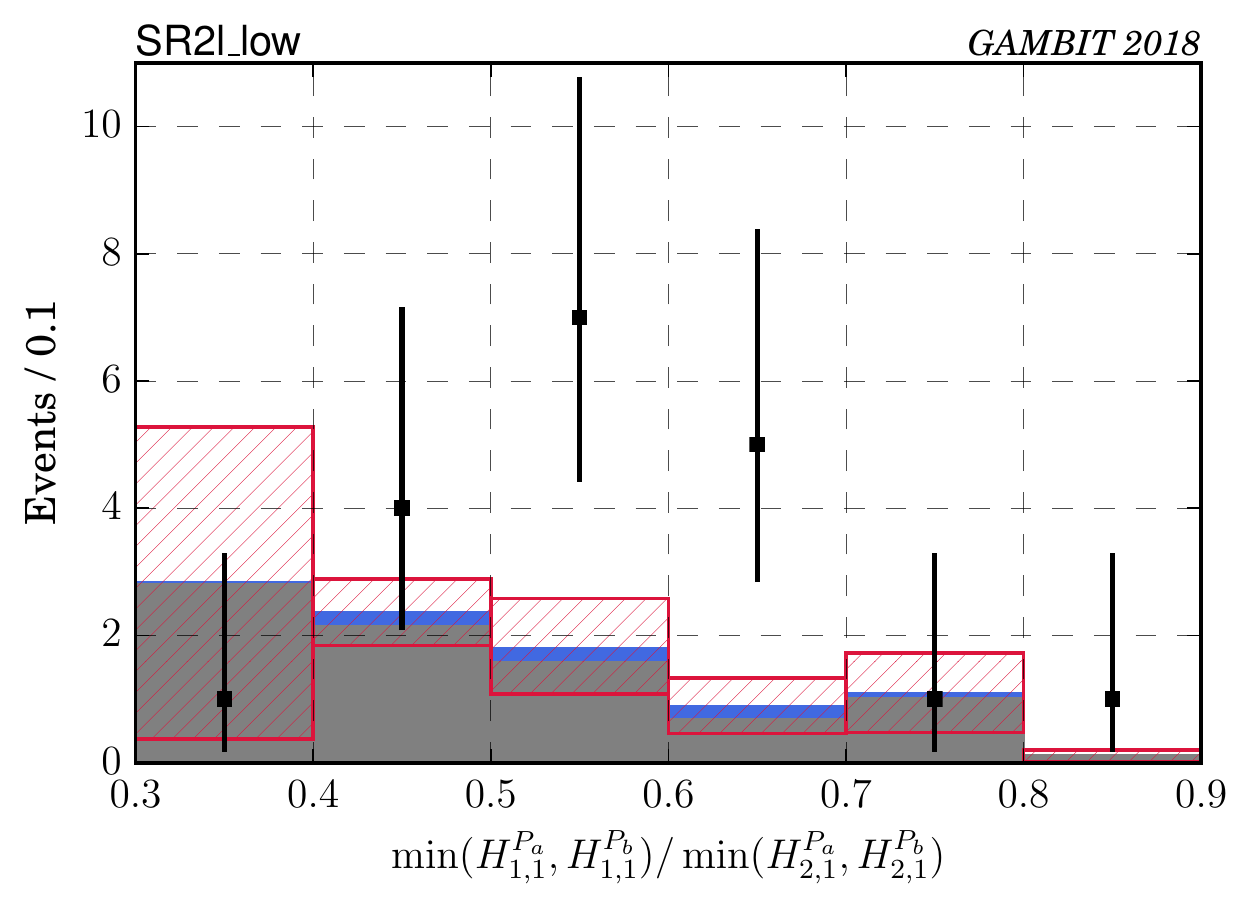}
  \caption{Distribution of kinematic variables in the 2 lepton signal regions for the ATLAS RJ analysis, after applying all selection requirements. The grey bars show the total SM background (taken from Ref.~\cite{Aaboud:2018sua}) and the stacked blue bars show the signal for our best-fit point based on the combination of 8 and 13\,TeV data. The hatched red bands show the $1\sigma$ uncertainty on the total number of expected events, found by summing in quadrature the background uncertainty and the signal statistical uncertainty for our best-fit point. The black points show the ATLAS data.}
  \label{fig:atlasComparison3_8TeV}
\end{figure*}

\begin{figure*}[t]
  \centering
  \includegraphics[width=0.9\columnwidth]{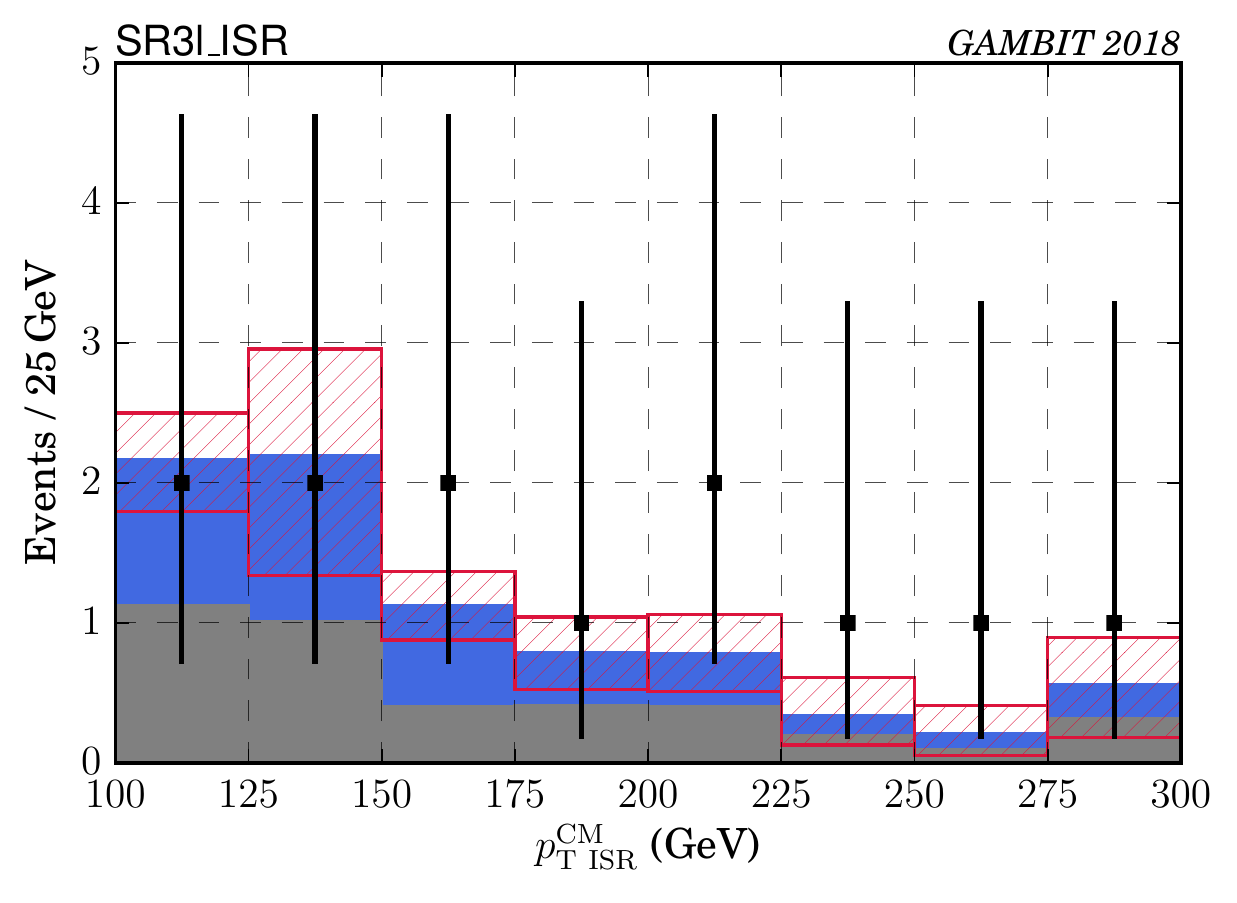}\hspace{0.1\columnwidth}
  \includegraphics[width=0.9\columnwidth]{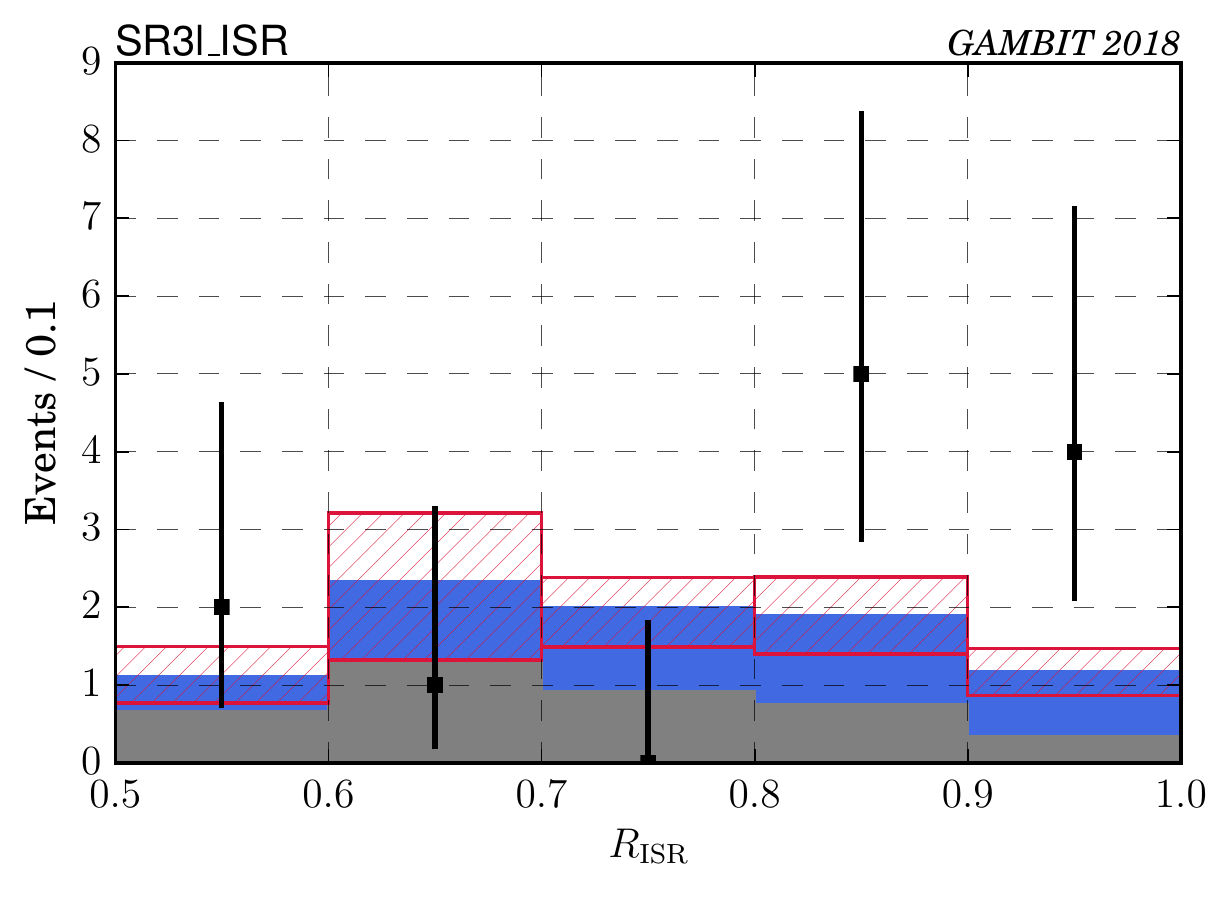}\\
  \includegraphics[width=0.9\columnwidth]{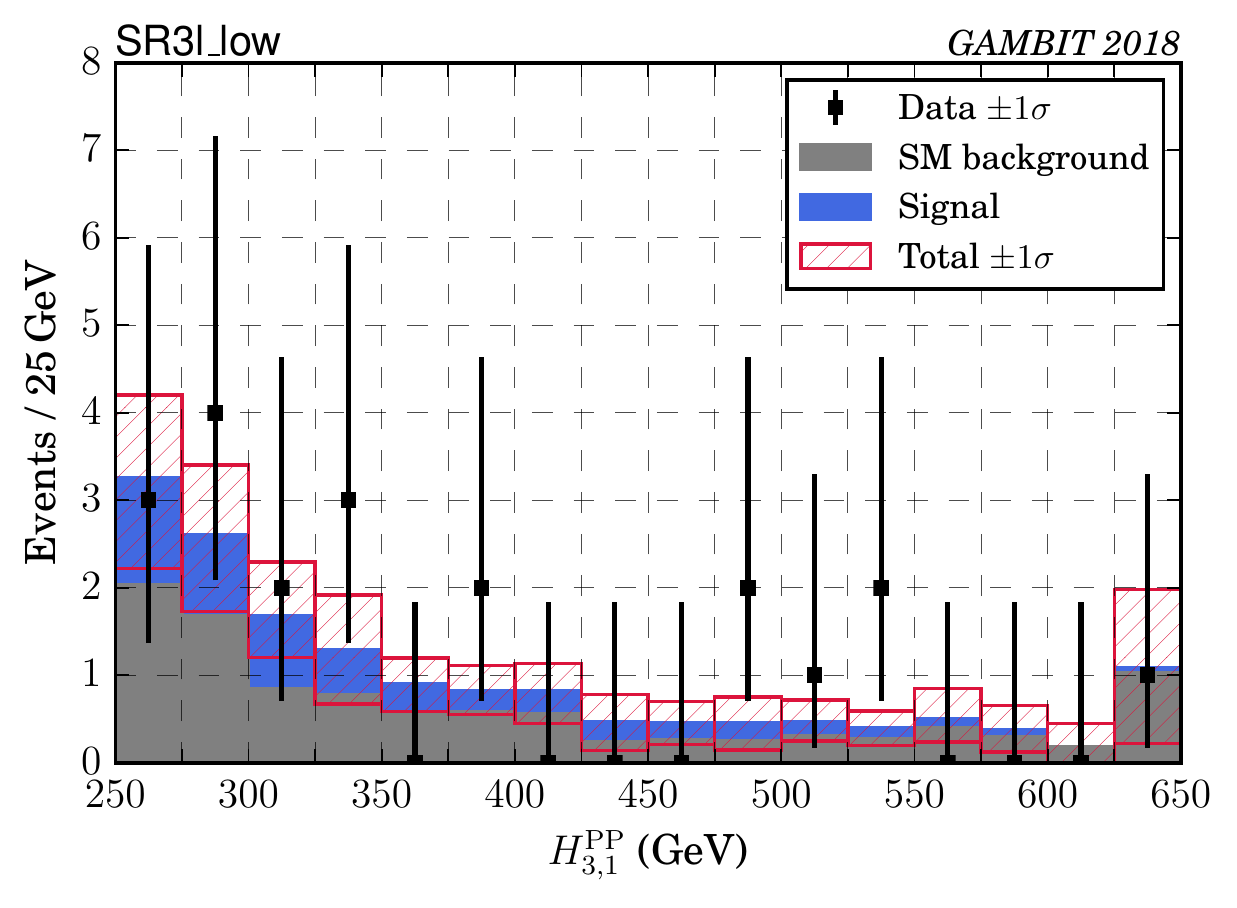}\hspace{0.1\columnwidth}
  \includegraphics[width=0.9\columnwidth]{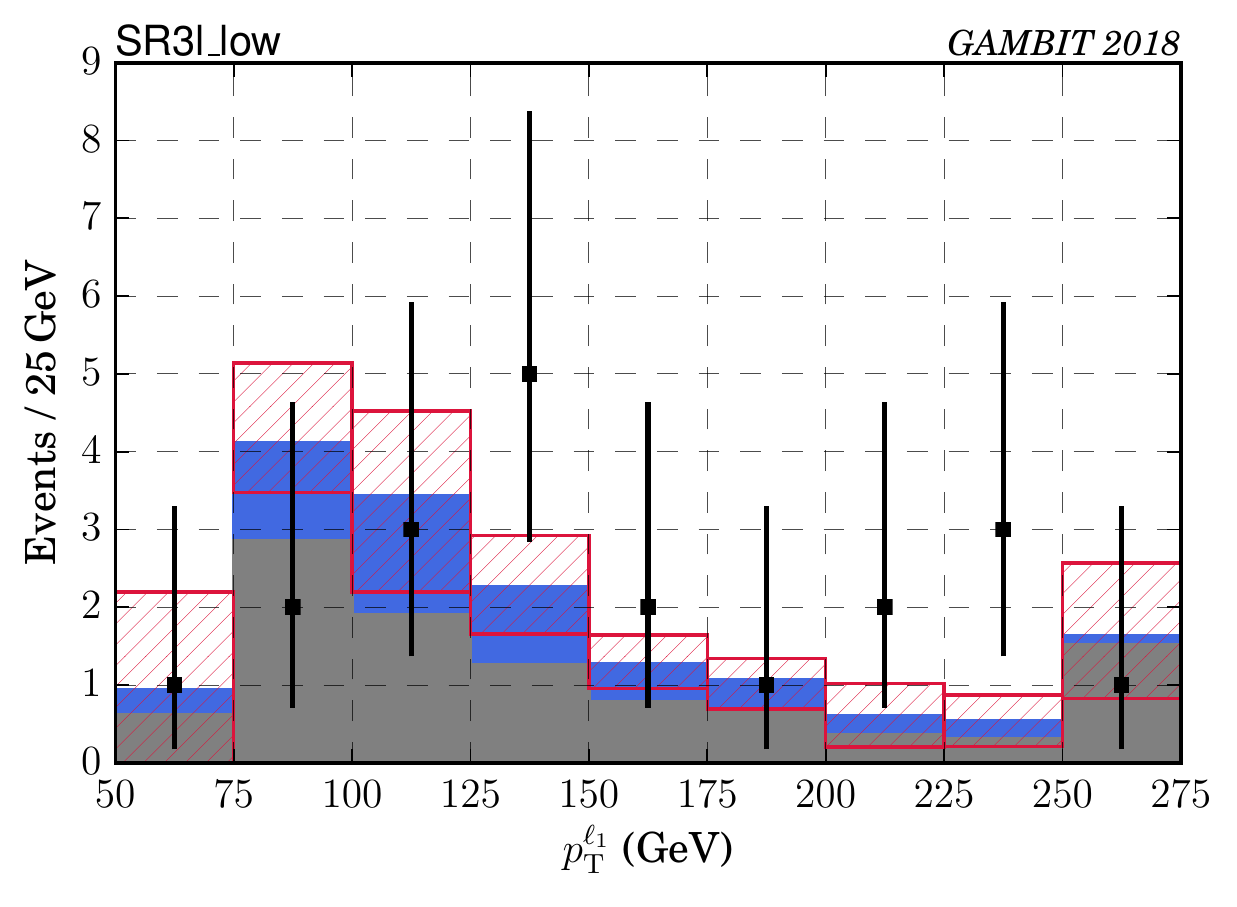}\\
  \includegraphics[width=0.9\columnwidth]{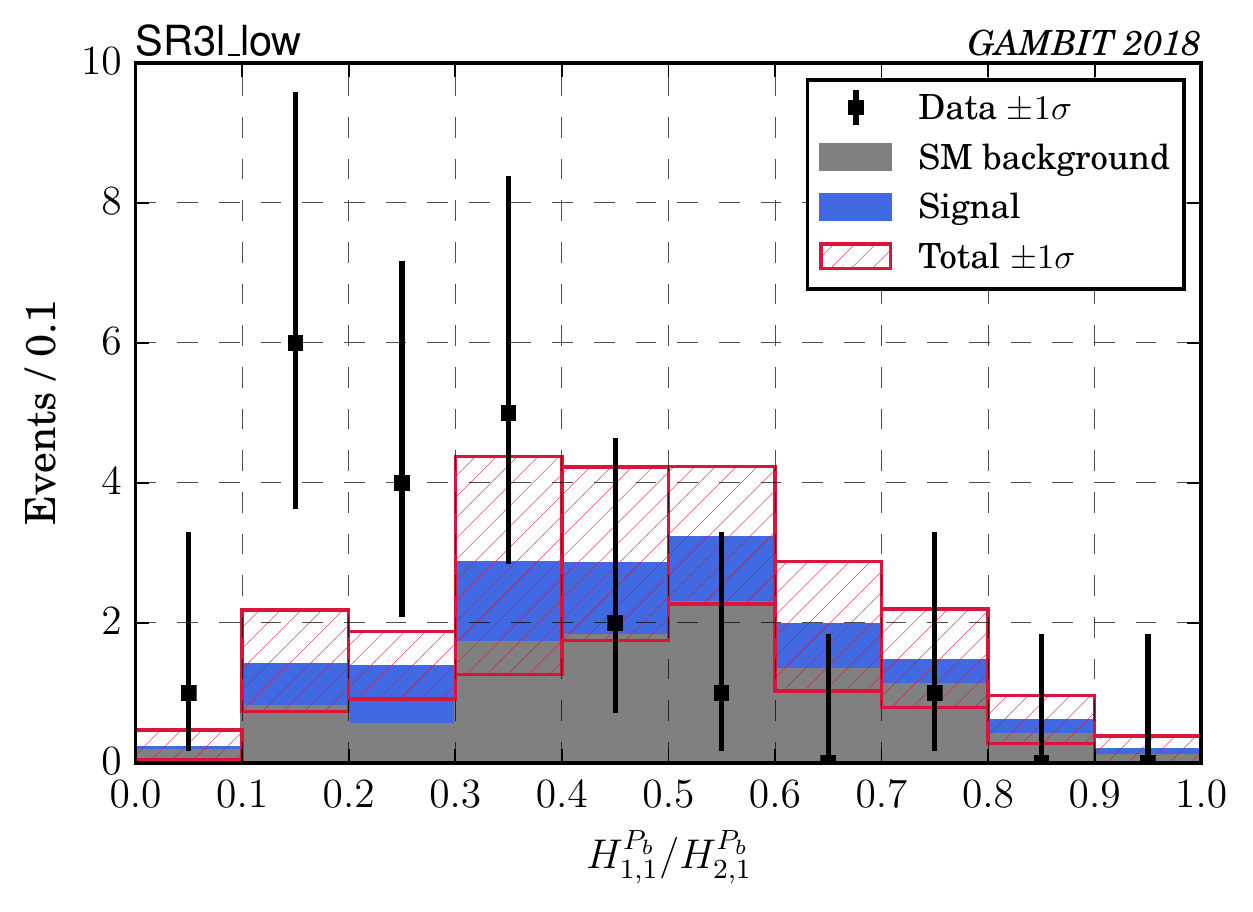}\\
  \caption{Distribution of kinematic variables in the 3 lepton signal regions for the ATLAS RJ analysis, after applying all selection requirements. The grey bars show the total SM background (taken from Ref.~\cite{Aaboud:2018sua}) and the stacked blue bars show the signal for our best-fit point based on the combination of 8 and 13\,TeV data. The hatched red bands show the $1\sigma$ uncertainty on the total number of expected events, found by summing in quadrature the background uncertainty and the signal statistical uncertainty for our best-fit point. The black points show the ATLAS data.}
  \label{fig:atlasComparison4_8TeV}
\end{figure*}

In analogy with Figs.\ \ref{fig:atlasComparison1}--\ref{fig:atlasComparison4}, we provide kinematic variable distributions relevant to the ATLAS multilepton searches for the new best-fit point in Figs.\ \ref{fig:atlasComparison1_8TeV}--\ref{fig:atlasComparison4_8TeV}.  The $\sim$$20$\,GeV heavier spectrum of point \#5 compared to point \#1 leads to slightly smaller integrated signals, but apart from this there is little difference with respect to the distributions in Figs.\ \ref{fig:atlasComparison1}--\ref{fig:atlasComparison4}.

Among the parameter samples in the approximate $1\sigma$ region with 8\,TeV results included, we find points both in the $Z$-funnel and $h$-funnel regions that are allowed by the dark matter likelihood. However, we do not attempt to map out the allowed parameter space in full, since neither the dark matter likelihood nor the 8\,TeV LHC likelihood was optimized in our original sampling of the EWMSSM parameter space

We have also repeated the \pvalue calculations of Sec.\ \ref{sec:pvalue} for the modified best-fit point, including all five 8\,TeV searches.  The corresponding significances can be found in Table\ \ref{tab:pvalues_8TeV}. The effect of including the 8 TeV analyses is to lower the combined local \pvalue to 2.9$\sigma$, and to lower our estimate of the significance with which the background-only hypothesis is excluded to 0.9$\sigma$. If we naively combine all SRs whilst neglecting correlations, we estimate the local \pvalue to be 3.6$\sigma$. The best-fit EWMSSM model remains a good fit to the data, with no significant tensions between analyses. The strongest, in the case of the ``best-expected SR'' analysis, result from the \textsf{ATLAS\_RJ\_ 3lep} analysis, at $1.1\sigma$, and the \textsf{ATLAS\_8TeV\_ 3lep} analysis, at $1.0\sigma$. We remind the reader of the caution with which these significance estimates should be treated.

\begin{table*}[t]
\centering
\begin{tabular}{lrrrrr|rrrr}
\toprule
 & & \multicolumn{4}{c}{Best expected SRs} & \multicolumn{4}{|c}{All SRs; neglect correlations} \\ 
\midrule
Analysis & \multicolumn{2}{c}{\makecell{Local \\ signif. ($\sigma$)}} & \makecell{SM \\ fit ($\sigma$)} & \makecell{MSSM4 \\ fit ($\sigma$)} & \#SRs & \makecell{Local \\ signif. ($\sigma$)} & \makecell{SM \\ fit ($\sigma$)} & \makecell{MSSM4 \\ fit ($\sigma$)} & \#SRs \\
\midrule
Higgs invisible width &                                                    & 0 & 0 & 0 &  1 & 0 & 0 & 0 &  1 \\
Z invisible width &                                                        & 0 & 1.3 & 1.3 &  1 & 0 & 1.3 & 1.3 &  1 \\
\textsf{ATLAS\_4b} &                                      & 0.7 & 0 & 0 &  1 & 1.5 & 0 & 0 &  46 \\
\textsf{ATLAS\_4lep} &                                    & 2.3 & 1.9 & 0 &  1 & 2.5 & 1.0 & 0 &  4 \\
\multicolumn{2}{l}{\textsf{ATLAS\_MultiLep\_2lep\_0jet}}    & 0.9 & 0.3 & 0.1 &  1 & 1.3 & 0 & 0 &  6 \\
\multicolumn{2}{l}{\textsf{ATLAS\_MultiLep\_2lep\_jet}} & 0 & 0 & 0.5 &  1 & 0.9 & 0.5 & 0.2 &  3 \\
\textsf{ATLAS\_MultiLep\_3lep} &                           & 1.8 & 1.5 & 0.7 &  1 & 1.1 & 0.4 & 0.3 &  11 \\
\textsf{ATLAS\_RJ\_2lep\_2jet} &                            & 0 & 0.3 & 0.5 &  1 & 1.5 & 1.8 & 1.5 &  4 \\
\textsf{ATLAS\_RJ\_3lep} &                                 & 2.7 & 2.4 & 1.1 &  1 & 3.4 & 2.6 & 0.7 &  4 \\
\textsf{CMS\_1lep\_2b} &                                    & 0.9 & 0.3 & 0.3 &  1 & 0 & 0 & 0 &  2 \\
\textsf{CMS\_2lep\_soft} &                                  & 0.1 & 0.2 & 0.2 &  12 & 0.1 & 0.2 & 0.2 &  12 \\
\textsf{CMS\_2OSlep} &                                    & 0.1 & 0.5 & 0.5 &  7 & 0.1 & 0.5 & 0.5 &  7 \\
\textsf{CMS\_MultiLep\_2SSlep} &                           & 0.2 & 0 & 0 &  1 & 0.2 & 0 & 0 &  2 \\
\textsf{CMS\_MultiLep\_3lep} &                             & 0 & 0 & 0.4 &  1 & 0 & 0 & 0 &  6 \\
\textsf{CMS\_8TeV\_3lep} &                              & 0 & 0 & 0 &  1 & 1.1 & 0 & 0 &  108 \\
\textsf{CMS\_8TeV\_4lep} &                              & 0.3 & 0 & 0 &  1 & 0.3 & 0 & 0 &  12 \\
\textsf{ATLAS\_8TeV\_1lep\_2b} &                                   & 0 & 0 & 0 &  1 & 0 & 0 & 0 &  2 \\
\textsf{ATLAS\_8TeV\_2lep} &                                   & 0 & 0 & 0 &  1 & 0 & 0 & 0 &  13 \\
\textsf{ATLAS\_8TeV\_3lep} &                                   & 0 & 0 & 1.0 &  1 & 0.4 & 0 & 0 &  23 \\
\midrule
    Combined &                                                                 & 2.9 & 0.9 & 0 &  36 & 3.6 & 0 & 0 &  267 \\
\bottomrule
\end{tabular}

\caption{
    As per Table \ref{tab:pvalues}, but for alternative best-fit point including 8\,TeV results. The combined local significance of excesses is reduced by approximately $0.4\sigma$ as compared to the previous best-fit point. The combined ``goodness of fit'' when correlations are ignored (but all signal regions are included) is approximately $0\sigma$ for both the EWMSSM and SM with the new 8 TeV analyses, however this is unfortunately not very meaningful due to the very high number of signal regions in the combined fit. The discrimination power of our ``goodness of fit'' test is very low for such a large combination. This is because the test uses a very general signal model in which signals may appear independently in any signal region with any strength, which introduces many more degrees of freedom than actually exist in the EWMSSM. In terms of the ``look-elsewhere'' effect, we are ``looking elsewhere'' to a far larger degree than is correct for the EWMSSM, diluting the significance of any excesses to zero. Unfortunately, a true look-elsewhere corrected p-value for a test of the SM against the EWMSSM is not yet technically feasible to calculate.
}
\label{tab:pvalues_8TeV}
\end{table*}

\section{Predicted signal counts}
\label{appb}

Finally, to save readers the trouble of extracting the information themselves from the public dataset provided with this paper \cite{the_gambit_collaboration_2018_1410335}, we provide full signal predictions for the 13\,TeV-only and the 8 + 13\,TeV benchmark points. Predicted signal counts for the 13\,TeV analyses are given in Tables \ref{tab:SR_predictions_13TeV_part1} and \ref{tab:SR_predictions_13TeV_part2}, while signal counts for the 8\,TeV analyses are listed in Tables \ref{tab:SR_predictions_8TeV_part1}--\ref{tab:SR_predictions_8TeV_part3}. We remind the reader that these signal predictions are based on LO+LL cross-sections.

% 13 TeV analyses, part 1
\begin{table*}[t]
\begin{center}
\setlength{\tabcolsep}{3pt} %% default is 6pt
\scriptsize
\begin{tabular}{l c X{1.8cm} X{1.8cm} X{1.8cm} X{1.8cm} X{1.8cm} X{1.8cm}}
\hline
\multirowcell{3}{} & \multirowcell{3}{} & \multirowcell{3}{\#1\\Best fit} & \multirowcell{3}{\#2\\Heavy winos} & \multirowcell{3}{\#3\\Highest mass} & \multirowcell{3}{\#4\\DM} & \multirowcell{3}{\#5\\Best fit\\incl.\ 8\,TeV} & \multirowcell{3}{\#6\\Heavy winos\\incl.\ 8\,TeV} \\
     & &              &                 &                  &            &   &  \\
     & &              &                 &                  &            &   &  \\
\hline
SR index, label & {\bf obs}, bkg  &   signal   &   signal   &   signal   &   signal   &   signal   &   signal  \\
\hline
\multicolumn{8}{l}{\textsf{ATLAS\_4b}~\cite{Aaboud:2018htj}}\\
% \textsf{ATLAS\_4b}~\cite{Aaboud:2018htj}\\
0, \textsf{meff160\_ETmiss0}   & $\mathbf{20} , 16.21 \pm 0.11$ &   $0.00 \pm 0.00$  &   $0.00 \pm 0.00$   &   $0.00 \pm 0.00$   &   $0.00 \pm 0.00$   &   $0.00 \pm 0.00$   &   $0.00 \pm 0.00$ \\
1, \textsf{meff160\_ETmiss20}   & $\mathbf{3} , 0.65 \pm 0.075$ &   $0.00 \pm 0.00$  &   $0.00 \pm 0.00$   &   $0.00 \pm 0.00$   &   $0.00 \pm 0.00$   &   $0.00 \pm 0.00$   &   $0.00 \pm 0.00$ \\
2, \textsf{meff200\_ETmiss0}   & $\mathbf{1503} , 1480 \pm 26$ &   $0.01 \pm 0.00$  &   $0.01 \pm 0.00$   &   $0.00 \pm 0.00$   &   $0.01 \pm 0.01$   &   $0.00 \pm 0.00$   &   $0.01 \pm 0.01$ \\
3, \textsf{meff200\_ETmiss20}   & $\mathbf{1137} , 1088 \pm 7$ &   $0.01 \pm 0.00$  &   $0.02 \pm 0.00$   &   $0.00 \pm 0.00$   &   $0.02 \pm 0.01$   &   $0.01 \pm 0.00$   &   $0.01 \pm 0.01$ \\
4, \textsf{meff200\_ETmiss45}   & $\mathbf{65} , 58.05 \pm 0.39$ &   $0.00 \pm 0.00$  &   $0.00 \pm 0.00$   &   $0.00 \pm 0.00$   &   $0.00 \pm 0.00$   &   $0.00 \pm 0.00$   &   $0.00 \pm 0.00$ \\
5, \textsf{meff200\_ETmiss70}   & $\mathbf{0} , 0.27 \pm 0.05$ &   $0.00 \pm 0.00$  &   $0.00 \pm 0.00$   &   $0.00 \pm 0.00$   &   $0.00 \pm 0.00$   &   $0.00 \pm 0.00$   &   $0.00 \pm 0.00$ \\
6, \textsf{meff260\_ETmiss0}   & $\mathbf{1329} , 1297 \pm 8$ &   $0.01 \pm 0.00$  &   $0.01 \pm 0.00$   &   $0.00 \pm 0.00$   &   $0.01 \pm 0.00$   &   $0.01 \pm 0.00$   &   $0.01 \pm 0.00$ \\
7, \textsf{meff260\_ETmiss20}   & $\mathbf{2877} , 2860 \pm 36$ &   $0.05 \pm 0.01$  &   $0.06 \pm 0.01$   &   $0.01 \pm 0.00$   &   $0.06 \pm 0.01$   &   $0.04 \pm 0.01$   &   $0.03 \pm 0.01$ \\
8, \textsf{meff260\_ETmiss45}   & $\mathbf{951} , 991 \pm 6.5$ &   $0.04 \pm 0.01$  &   $0.06 \pm 0.01$   &   $0.00 \pm 0.00$   &   $0.04 \pm 0.01$   &   $0.04 \pm 0.01$   &   $0.02 \pm 0.01$ \\
9, \textsf{meff260\_ETmiss70}   & $\mathbf{150} , 149.4 \pm 1.0$ &   $0.01 \pm 0.00$  &   $0.03 \pm 0.01$   &   $0.01 \pm 0.00$   &   $0.01 \pm 0.00$   &   $0.02 \pm 0.01$   &   $0.01 \pm 0.01$ \\
10, \textsf{meff260\_ETmiss100}   & $\mathbf{2} , 2.02 \pm 1.43$ &   $0.01 \pm 0.00$  &   $0.00 \pm 0.00$   &   $0.00 \pm 0.00$   &   $0.00 \pm 0.00$   &   $0.00 \pm 0.00$   &   $0.01 \pm 0.00$ \\
11, \textsf{meff340\_ETmiss0}   & $\mathbf{373} , 390.1 \pm 2.6$ &   $0.01 \pm 0.00$  &   $0.00 \pm 0.00$   &   $0.00 \pm 0.00$   &   $0.00 \pm 0.00$   &   $0.00 \pm 0.00$   &   $0.01 \pm 0.00$ \\
12, \textsf{meff340\_ETmiss20}   & $\mathbf{873} , 884.6 \pm 13.1$ &   $0.01 \pm 0.01$  &   $0.02 \pm 0.00$   &   $0.00 \pm 0.00$   &   $0.03 \pm 0.01$   &   $0.02 \pm 0.01$   &   $0.01 \pm 0.01$ \\
13, \textsf{meff340\_ETmiss45}   & $\mathbf{444} , 472.6 \pm 3.0$ &   $0.02 \pm 0.01$  &   $0.02 \pm 0.00$   &   $0.01 \pm 0.00$   &   $0.03 \pm 0.01$   &   $0.03 \pm 0.01$   &   $0.01 \pm 0.01$ \\
14, \textsf{meff340\_ETmiss70}   & $\mathbf{164} , 171.1 \pm 1.1$ &   $0.02 \pm 0.01$  &   $0.02 \pm 0.00$   &   $0.00 \pm 0.00$   &   $0.04 \pm 0.01$   &   $0.03 \pm 0.01$   &   $0.04 \pm 0.01$ \\
15, \textsf{meff340\_ETmiss100}   & $\mathbf{40} , 36.24 \pm 0.24$ &   $0.03 \pm 0.01$  &   $0.02 \pm 0.00$   &   $0.01 \pm 0.00$   &   $0.05 \pm 0.01$   &   $0.02 \pm 0.01$   &   $0.03 \pm 0.01$ \\
16, \textsf{meff340\_ETmiss150}   & $\mathbf{3} , 1.46 \pm 0.11$ &   $0.01 \pm 0.00$  &   $0.01 \pm 0.00$   &   $0.00 \pm 0.00$   &   $0.00 \pm 0.00$   &   $0.01 \pm 0.00$   &   $0.01 \pm 0.01$ \\
17, \textsf{meff340\_ETmiss200}   & $\mathbf{0} , 0.01 \pm 0.00$ &   $0.00 \pm 0.00$  &   $0.00 \pm 0.00$   &   $0.00 \pm 0.00$   &   $0.00 \pm 0.00$   &   $0.00 \pm 0.00$   &   $0.00 \pm 0.00$ \\
18, \textsf{meff440\_ETmiss0}   & $\mathbf{121} , 130.3 \pm 0.8$ &   $0.00 \pm 0.00$  &   $0.00 \pm 0.00$   &   $0.00 \pm 0.00$   &   $0.00 \pm 0.00$   &   $0.00 \pm 0.00$   &   $0.01 \pm 0.01$ \\
19, \textsf{meff440\_ETmiss20}   & $\mathbf{304} , 310.8 \pm 9.5$ &   $0.03 \pm 0.01$  &   $0.02 \pm 0.00$   &   $0.00 \pm 0.00$   &   $0.02 \pm 0.01$   &   $0.01 \pm 0.00$   &   $0.02 \pm 0.01$ \\
20, \textsf{meff440\_ETmiss45}   & $\mathbf{170} , 176.6 \pm 1.2$ &   $0.01 \pm 0.01$  &   $0.01 \pm 0.00$   &   $0.00 \pm 0.00$   &   $0.01 \pm 0.00$   &   $0.01 \pm 0.00$   &   $0.03 \pm 0.01$ \\
21, \textsf{meff440\_ETmiss70}   & $\mathbf{62} , 65.1 \pm 1.1$ &   $0.01 \pm 0.00$  &   $0.01 \pm 0.00$   &   $0.00 \pm 0.00$   &   $0.03 \pm 0.01$   &   $0.02 \pm 0.01$   &   $0.02 \pm 0.01$ \\
22, \textsf{meff440\_ETmiss100}   & $\mathbf{31} , 22.16 \pm 6.03$ &   $0.03 \pm 0.01$  &   $0.01 \pm 0.00$   &   $0.00 \pm 0.00$   &   $0.03 \pm 0.01$   &   $0.02 \pm 0.01$   &   $0.05 \pm 0.01$ \\
23, \textsf{meff440\_ETmiss150}   & $\mathbf{3} , 3.90 \pm 0.14$ &   $0.01 \pm 0.00$  &   $0.01 \pm 0.00$   &   $0.01 \pm 0.00$   &   $0.01 \pm 0.00$   &   $0.01 \pm 0.00$   &   $0.02 \pm 0.01$ \\
24, \textsf{meff440\_ETmiss200}   & $\mathbf{1} , 0.48 \pm 0.06$ &   $0.01 \pm 0.00$  &   $0.01 \pm 0.00$   &   $0.00 \pm 0.00$   &   $0.01 \pm 0.00$   &   $0.01 \pm 0.00$   &   $0.00 \pm 0.00$ \\
25, \textsf{meff560\_ETmiss0}   & $\mathbf{40} , 43.46 \pm 0.29$ &   $0.00 \pm 0.00$  &   $0.00 \pm 0.00$   &   $0.00 \pm 0.00$   &   $0.00 \pm 0.00$   &   $0.00 \pm 0.00$   &   $0.00 \pm 0.00$ \\
26, \textsf{meff560\_ETmiss20}   & $\mathbf{95} , 102.6 \pm 6.6$ &   $0.01 \pm 0.00$  &   $0.00 \pm 0.00$   &   $0.00 \pm 0.00$   &   $0.00 \pm 0.00$   &   $0.00 \pm 0.00$   &   $0.01 \pm 0.00$ \\
27, \textsf{meff560\_ETmiss45}   & $\mathbf{75} , 68.03 \pm 0.45$ &   $0.01 \pm 0.00$  &   $0.01 \pm 0.00$   &   $0.00 \pm 0.00$   &   $0.01 \pm 0.00$   &   $0.00 \pm 0.00$   &   $0.01 \pm 0.01$ \\
28, \textsf{meff560\_ETmiss70}   & $\mathbf{20} , 30.72 \pm 0.2$ &   $0.01 \pm 0.00$  &   $0.01 \pm 0.00$   &   $0.00 \pm 0.00$   &   $0.01 \pm 0.00$   &   $0.01 \pm 0.00$   &   $0.01 \pm 0.00$ \\
29, \textsf{meff560\_ETmiss100}   & $\mathbf{15} , 14.13 \pm 3.19$ &   $0.02 \pm 0.01$  &   $0.01 \pm 0.00$   &   $0.01 \pm 0.00$   &   $0.00 \pm 0.00$   &   $0.02 \pm 0.00$   &   $0.02 \pm 0.01$ \\
30, \textsf{meff560\_ETmiss150}   & $\mathbf{2} , 2.36 \pm 1.02$ &   $0.01 \pm 0.00$  &   $0.01 \pm 0.00$   &   $0.01 \pm 0.00$   &   $0.01 \pm 0.00$   &   $0.01 \pm 0.00$   &   $0.01 \pm 0.00$ \\
31, \textsf{meff560\_ETmiss200}   & $\mathbf{2} , 1.08 \pm 0.23$ &   $0.00 \pm 0.00$  &   $0.00 \pm 0.00$   &   $0.00 \pm 0.00$   &   $0.00 \pm 0.00$   &   $0.02 \pm 0.00$   &   $0.01 \pm 0.01$ \\
32, \textsf{meff700\_ETmiss0}   & $\mathbf{17} , 13.56 \pm 0.09$ &   $0.00 \pm 0.00$  &   $0.00 \pm 0.00$   &   $0.00 \pm 0.00$   &   $0.00 \pm 0.00$   &   $0.00 \pm 0.00$   &   $0.00 \pm 0.00$ \\
33, \textsf{meff700\_ETmiss20}   & $\mathbf{30} , 32.67 \pm 3.39$ &   $0.00 \pm 0.00$  &   $0.00 \pm 0.00$   &   $0.00 \pm 0.00$   &   $0.00 \pm 0.00$   &   $0.00 \pm 0.00$   &   $0.00 \pm 0.00$ \\
34, \textsf{meff700\_ETmiss45}   & $\mathbf{22} , 23.78 \pm 0.15$ &   $0.00 \pm 0.00$  &   $0.00 \pm 0.00$   &   $0.00 \pm 0.00$   &   $0.01 \pm 0.00$   &   $0.00 \pm 0.00$   &   $0.00 \pm 0.00$ \\
35, \textsf{meff700\_ETmiss70}   & $\mathbf{12} , 12.47 \pm 0.08$ &   $0.00 \pm 0.00$  &   $0.00 \pm 0.00$   &   $0.00 \pm 0.00$   &   $0.00 \pm 0.00$   &   $0.01 \pm 0.00$   &   $0.01 \pm 0.00$ \\
36, \textsf{meff700\_ETmiss100}   & $\mathbf{6} , 5.55 \pm 0.87$ &   $0.00 \pm 0.00$  &   $0.00 \pm 0.00$   &   $0.01 \pm 0.00$   &   $0.00 \pm 0.00$   &   $0.01 \pm 0.00$   &   $0.01 \pm 0.00$ \\
37, \textsf{meff700\_ETmiss150}   & $\mathbf{2} , 1.73 \pm 0.88$ &   $0.00 \pm 0.00$  &   $0.00 \pm 0.00$   &   $0.00 \pm 0.00$   &   $0.00 \pm 0.00$   &   $0.00 \pm 0.00$   &   $0.02 \pm 0.01$ \\
38, \textsf{meff700\_ETmiss200}   & $\mathbf{2} , 0.86 \pm 0.12$ &   $0.00 \pm 0.00$  &   $0.00 \pm 0.00$   &   $0.00 \pm 0.00$   &   $0.00 \pm 0.00$   &   $0.00 \pm 0.00$   &   $0.01 \pm 0.01$ \\
39, \textsf{meff860\_ETmiss0}   & $\mathbf{2} ,  2.82 \pm 0.25$ &   $0.00 \pm 0.00$  &   $0.00 \pm 0.00$   &   $0.00 \pm 0.00$   &   $0.00 \pm 0.00$   &   $0.00 \pm 0.00$   &   $0.00 \pm 0.00$ \\
40, \textsf{meff860\_ETmiss20}   & $\mathbf{7} , 7.77 \pm 2.11$ &   $0.00 \pm 0.00$  &   $0.00 \pm 0.00$   &   $0.00 \pm 0.00$   &   $0.00 \pm 0.00$   &   $0.00 \pm 0.00$   &   $0.00 \pm 0.00$ \\
41, \textsf{meff860\_ETmiss45}   & $\mathbf{10} , 8.97 \pm 2.33$ &   $0.00 \pm 0.00$  &   $0.00 \pm 0.00$   &   $0.00 \pm 0.00$   &   $0.00 \pm 0.00$   &   $0.00 \pm 0.00$   &   $0.01 \pm 0.00$ \\
42, \textsf{meff860\_ETmiss70}   & $\mathbf{5} , 4.30 \pm 0.34$ &   $0.00 \pm 0.00$  &   $0.00 \pm 0.00$   &   $0.00 \pm 0.00$   &   $0.00 \pm 0.00$   &   $0.00 \pm 0.00$   &   $0.01 \pm 0.00$ \\
43, \textsf{meff860\_ETmiss100}   & $\mathbf{2} , 2.79 \pm 0.29$ &   $0.00 \pm 0.00$  &   $0.01 \pm 0.00$   &   $0.00 \pm 0.00$   &   $0.01 \pm 0.00$   &   $0.00 \pm 0.00$   &   $0.00 \pm 0.00$ \\
44, \textsf{meff860\_ETmiss150}   & $\mathbf{4} , 0.93 \pm 0.23$ &   $0.01 \pm 0.00$  &   $0.00 \pm 0.00$   &   $0.00 \pm 0.00$   &   $0.00 \pm 0.00$   &   $0.00 \pm 0.00$   &   $0.00 \pm 0.00$ \\
45, \textsf{meff860\_ETmiss200}   & $\mathbf{1} , 0.43 \pm 0.07$ &   $0.00 \pm 0.00$  &   $0.01 \pm 0.00$   &   $0.00 \pm 0.00$   &   $0.01 \pm 0.00$   &   $0.00 \pm 0.00$   &   $0.02 \pm 0.01$ \\
\hline
\multicolumn{8}{l}{\textsf{ATLAS\_4lep}~\cite{Aaboud:2018zeb}}\\
% \textsf{ATLAS\_4lep}~\cite{Aaboud:2018zeb}\\
0, \textsf{SR0A}   & $\mathbf{13}, 10.2 \pm 2.1$ &   $0.85 \pm 0.05$  &   $0.70 \pm 0.03$   &   $0.25 \pm 0.02$   &   $0.79 \pm 0.05$   &   $0.64 \pm 0.04$   &   $0.82 \pm 0.06$ \\
1, \textsf{SR0B}   & $\mathbf{2}, 1.31 \pm 0.24$ &   $0.11 \pm 0.02$  &   $0.09 \pm 0.01$   &   $0.03 \pm 0.01$   &   $0.09 \pm 0.02$   &   $0.07 \pm 0.01$   &   $0.08 \pm 0.02$ \\
2, \textsf{SR0C}   & $\mathbf{47}, 37 \pm 9$     &   $11.20 \pm 0.19$  &   $16.89 \pm 0.15$   &   $7.49 \pm 0.13$   &   $10.28 \pm 0.19$   &   $8.35 \pm 0.13$   &   $12.12 \pm 0.22$ \\
3, \textsf{SR0D}   & $\mathbf{10}, 4.1 \pm 0.7$  &   $5.07 \pm 0.13$  &   $6.61 \pm 0.09$   &   $3.97 \pm 0.09$   &   $4.73 \pm 0.13$   &   $4.00 \pm 0.09$   &   $5.68 \pm 0.15$ \\
\hline
\multicolumn{8}{l}{\textsf{ATLAS\_MultiLep\_2lep\_0jet}~\cite{Aaboud:2018jiw}}\\
% \textsf{ATLAS\_MultiLep\_2lep\_0jet}~\cite{Aaboud:2018jiw}\\
0, \textsf{SR2\_SF\_loose}   & $\mathbf{153} , 133 \pm 22$ &   $2.25 \pm 0.08$  &   $2.54 \pm 0.06$   &   $1.82 \pm 0.06$   &   $2.13 \pm 0.09$   &   $1.81 \pm 0.06$   &   $0.67 \pm 0.05$ \\
1, \textsf{SR2\_SF\_tight}   & $\mathbf{9} , 9.8 \pm 2.9$ &   $0.03 \pm 0.01$  &   $0.06 \pm 0.01$   &   $0.02 \pm 0.01$   &   $0.04 \pm 0.01$   &   $0.03 \pm 0.01$   &   $0.04 \pm 0.01$ \\
2, \textsf{SR2\_DF\_100}   & $\mathbf{78} , 68 \pm 7$ &   $2.01 \pm 0.08$  &   $2.40 \pm 0.06$   &   $1.60 \pm 0.06$   &   $1.77 \pm 0.08$   &   $1.67 \pm 0.06$   &   $0.70 \pm 0.05$ \\
3, \textsf{SR2\_DF\_150}   & $\mathbf{11} , 11.5 \pm 3.1$ &   $0.34 \pm 0.03$  &   $0.36 \pm 0.02$   &   $0.14 \pm 0.02$   &   $0.36 \pm 0.04$   &   $0.32 \pm 0.03$   &   $0.17 \pm 0.03$ \\
4, \textsf{SR2\_DF\_200}   & $\mathbf{6} , 2.1 \pm 1.9$ &   $0.07 \pm 0.01$  &   $0.06 \pm 0.01$   &   $0.01 \pm 0.01$   &   $0.11 \pm 0.02$   &   $0.08 \pm 0.01$   &   $0.03 \pm 0.01$ \\
5, \textsf{SR2\_DF\_300}   & $\mathbf{2} , 0.6 \pm 0.6$ &   $0.02 \pm 0.01$  &   $0.01 \pm 0.00$   &   $0.00 \pm 0.00$   &   $0.01 \pm 0.01$   &   $0.01 \pm 0.01$   &   $0.00 \pm 0.00$ \\
\hline
\end{tabular}
\caption{\label{tab:SR_predictions_13TeV_part1} Predicted signal counts for the 13\,TeV analyses \textsf{ATLAS\_4b}, \textsf{ATLAS\_4lep} and \textsf{ATLAS\_MultiLep\_2lep\_0jet}, for all six benchmark points. For comparison, the second column lists the observed event count and background prediction for each signal region, taken from the corresponding ATLAS or CMS analysis. The signal region indices and labels correspond to those in the public dataset provided with this paper \cite{the_gambit_collaboration_2018_1410335}.}
\end{center}
\centering
\end{table*}

% 13 TeV analyses, part 2
\begin{table*}[t]
\begin{center}
\setlength{\tabcolsep}{2.9pt} %% default is 6pt
\scriptsize
\begin{tabular}{l c X{1.8cm} X{1.8cm} X{1.8cm} X{1.8cm} X{1.8cm} X{1.8cm}}
\hline
\multirowcell{3}{} & \multirowcell{3}{} & \multirowcell{3}{\#1\\Best fit} & \multirowcell{3}{\#2\\Heavy winos} & \multirowcell{3}{\#3\\Highest mass} & \multirowcell{3}{\#4\\DM} & \multirowcell{3}{\#5\\Best fit\\incl.\ 8\,TeV} & \multirowcell{3}{\#6\\Heavy winos\\incl.\ 8\,TeV} \\
     & &              &                 &                  &            &   &  \\
     & &              &                 &                  &            &   &  \\
\hline
SR index, label & {\bf obs}, bkg  &   signal   &   signal   &   signal   &   signal   &   signal   &   signal  \\
\hline
\multicolumn{8}{l}{\textsf{ATLAS\_MultiLep\_2lep\_jet}~\cite{Aaboud:2018jiw}}\\
% \textsf{ATLAS\_MultiLep\_2lep\_jet}~\cite{Aaboud:2018jiw}\\
0, \textsf{SR2\_int}   & $\mathbf{2} , 4.1 \pm 2.6$ &   $1.02 \pm 0.06$  &   $1.16 \pm 0.04$   &   $1.16 \pm 0.05$   &   $1.07 \pm 0.06$   &   $0.98 \pm 0.05$   &   $0.89 \pm 0.06$ \\
1, \textsf{SR2\_high}   & $\mathbf{0} , 1.6 \pm 1.6$ &   $0.29 \pm 0.03$  &   $0.29 \pm 0.02$   &   $0.33 \pm 0.03$   &   $0.26 \pm 0.03$   &   $0.23 \pm 0.02$   &   $0.22 \pm 0.03$ \\
2, \textsf{SR2\_low}   & $\mathbf{11} , 4.2 \pm 3.4$ &   $2.86 \pm 0.09$  &   $2.46 \pm 0.06$   &   $2.12 \pm 0.07$   &   $2.38 \pm 0.09$   &   $3.07 \pm 0.08$   &   $2.59 \pm 0.10$ \\
\hline
\multicolumn{8}{l}{\textsf{ATLAS\_MultiLep\_3lep}~\cite{Aaboud:2018jiw}}\\
% \textsf{ATLAS\_MultiLep\_3lep}~\cite{Aaboud:2018jiw}\\
0, \textsf{SR3\_slep\_a}   & $\mathbf{4} , 2.2 \pm 0.8$ &   $1.28 \pm 0.06$  &   $1.47 \pm 0.04$   &   $0.94 \pm 0.05$   &   $1.16 \pm 0.06$   &   $1.09 \pm 0.05$   &   $0.71 \pm 0.05$ \\
1, \textsf{SR3\_slep\_b}   & $\mathbf{3} , 2.8 \pm 0.4$ &   $1.74 \pm 0.07$  &   $2.02 \pm 0.05$   &   $1.21 \pm 0.05$   &   $1.44 \pm 0.07$   &   $1.34 \pm 0.05$   &   $0.75 \pm 0.06$ \\
2, \textsf{SR3\_slep\_c}   & $\mathbf{9} , 5.4 \pm 0.9$ &   $1.58 \pm 0.07$  &   $1.75 \pm 0.05$   &   $1.11 \pm 0.05$   &   $1.51 \pm 0.07$   &   $1.38 \pm 0.05$   &   $1.16 \pm 0.07$ \\
3, \textsf{SR3\_slep\_d}   & $\mathbf{0} , 1.4 \pm 0.4$ &   $0.56 \pm 0.04$  &   $0.60 \pm 0.03$   &   $0.33 \pm 0.03$   &   $0.65 \pm 0.05$   &   $0.50 \pm 0.03$   &   $0.36 \pm 0.04$ \\
4, \textsf{SR3\_slep\_e}   & $\mathbf{0} , 1.1 \pm 0.2$ &   $0.21 \pm 0.03$  &   $0.19 \pm 0.02$   &   $0.08 \pm 0.01$   &   $0.20 \pm 0.03$   &   $0.22 \pm 0.02$   &   $0.15 \pm 0.02$ \\
5, \textsf{SR3\_WZ\_0Ja}   & $\mathbf{21} , 21.7 \pm 2.9$ &   $4.02 \pm 0.11$  &   $4.65 \pm 0.08$   &   $4.97 \pm 0.10$   &   $2.98 \pm 0.10$   &   $3.62 \pm 0.09$   &   $2.41 \pm 0.10$ \\
6, \textsf{SR3\_WZ\_0Jb}   & $\mathbf{1} , 2.7 \pm 0.5$ &   $0.73 \pm 0.05$  &   $1.21 \pm 0.04$   &   $1.41 \pm 0.06$   &   $0.74 \pm 0.05$   &   $0.63 \pm 0.04$   &   $0.33 \pm 0.04$ \\
7, \textsf{SR3\_WZ\_0Jc}   & $\mathbf{2} , 1.6 \pm 0.3$ &   $0.47 \pm 0.04$  &   $0.71 \pm 0.03$   &   $0.79 \pm 0.04$   &   $0.51 \pm 0.04$   &   $0.43 \pm 0.03$   &   $0.21 \pm 0.03$ \\
8, \textsf{SR3\_WZ\_1Ja}   & $\mathbf{1} , 2.2 \pm 0.5$ &   $0.60 \pm 0.04$  &   $0.94 \pm 0.04$   &   $0.55 \pm 0.03$   &   $0.56 \pm 0.04$   &   $0.62 \pm 0.04$   &   $0.92 \pm 0.06$ \\
9, \textsf{SR3\_WZ\_1Jb}   & $\mathbf{3} , 1.8 \pm 0.3$ &   $1.53 \pm 0.07$  &   $1.72 \pm 0.05$   &   $1.64 \pm 0.06$   &   $1.57 \pm 0.08$   &   $1.56 \pm 0.06$   &   $1.47 \pm 0.08$ \\
10, \textsf{SR3\_WZ\_1Jc}   & $\mathbf{4} , 1.3 \pm 0.3$ &   $0.88 \pm 0.05$  &   $0.96 \pm 0.04$   &   $0.85 \pm 0.04$   &   $0.83 \pm 0.05$   &   $0.83 \pm 0.04$   &   $0.87 \pm 0.06$ \\
\hline
\textsf{ATLAS\_RJ\_2lep\_2jet}~\cite{Aaboud:2018sua}\\
0, \textsf{2L2JHIGH}   & $\mathbf{0} , 1.9 \pm 0.8$  &   $0.13 \pm 0.02$  &   $0.13 \pm 0.01$   &   $0.13 \pm 0.02$   &   $0.17 \pm 0.02$   &   $0.15 \pm 0.02$   &   $0.12 \pm 0.02$ \\
1, \textsf{2L2JINT}   & $\mathbf{1} , 2.4 \pm 0.9$  &   $0.31 \pm 0.03$  &   $0.26 \pm 0.02$   &   $0.24 \pm 0.02$   &   $0.33 \pm 0.03$   &   $0.27 \pm 0.02$   &   $0.20 \pm 0.03$ \\
2, \textsf{2L2JLOW}   & $\mathbf{19} , 8.4 \pm 5.8$  &   $0.57 \pm 0.04$  &   $0.81 \pm 0.03$   &   $1.00 \pm 0.05$   &   $0.41 \pm 0.04$   &   $0.47 \pm 0.03$   &   $0.39 \pm 0.04$ \\
3, \textsf{2L2JCOMP}   & $\mathbf{11} , 2.7 \pm 2.7$  &   $1.72 \pm 0.07$  &   $1.16 \pm 0.04$   &   $0.68 \pm 0.04$   &   $1.66 \pm 0.08$   &   $1.90 \pm 0.06$   &   $1.55 \pm 0.08$ \\
\hline
\textsf{ATLAS\_RJ\_3lep}~\cite{Aaboud:2018sua}\\
0, \textsf{3LHIGH}   & $\mathbf{2} , 1.1 \pm 0.5$  &   $0.32 \pm 0.03$  &   $0.27 \pm 0.02$   &   $0.21 \pm 0.02$   &   $0.27 \pm 0.03$   &   $0.23 \pm 0.02$   &   $0.23 \pm 0.03$ \\
1, \textsf{3LINT}   & $\mathbf{1} , 2.3 \pm 0.5$  &   $0.71 \pm 0.05$  &   $0.71 \pm 0.03$   &   $0.54 \pm 0.03$   &   $0.75 \pm 0.05$   &   $0.54 \pm 0.03$   &   $0.38 \pm 0.04$ \\
2, \textsf{3LLOW}   & $\mathbf{20} , 10 \pm 2.0$  &   $5.02 \pm 0.13$  &   $3.12 \pm 0.06$   &   $2.54 \pm 0.07$   &   $4.71 \pm 0.13$   &   $3.62 \pm 0.09$   &   $1.65 \pm 0.08$ \\
3, \textsf{3LCOMP}   & $\mathbf{12} , 3.9 \pm 1.0$  &   $3.48 \pm 0.10$  &   $2.75 \pm 0.06$   &   $1.85 \pm 0.06$   &   $3.15 \pm 0.11$   &   $3.27 \pm 0.08$   &   $2.02 \pm 0.09$ \\
\hline
\textsf{CMS\_1lep\_2b}~\cite{CMS:2017fth}\\
0, \textsf{SRA}   & $\mathbf{11}, 7.5 \pm 2.5$  &   $0.05 \pm 0.01$  &   $0.05 \pm 0.01$   &   $0.05 \pm 0.01$   &   $0.04 \pm 0.01$   &   $0.05 \pm 0.01$   &   $0.01 \pm 0.01$ \\
1, \textsf{SRB}   & $\mathbf{7}, 8.7 \pm 2.2$  &   $0.15 \pm 0.02$  &   $0.11 \pm 0.01$   &   $0.10 \pm 0.01$   &   $0.08 \pm 0.02$   &   $0.13 \pm 0.02$   &   $0.10 \pm 0.02$ \\
\hline
\textsf{CMS\_2lep\_soft}~\cite{Sirunyan:2018iwl}\\
0, \textsf{SR1}   & $\mathbf{2} , 3.5 \pm 1$  &   $0.03 \pm 0.01$  &   $0.02 \pm 0.00$   &   $0.01 \pm 0.00$   &   $0.02 \pm 0.01$   &   $0.02 \pm 0.01$   &   $0.01 \pm 0.01$ \\
1, \textsf{SR2}   & $\mathbf{15} , 12 \pm 2.3$  &   $0.11 \pm 0.02$  &   $0.09 \pm 0.01$   &   $0.04 \pm 0.01$   &   $0.10 \pm 0.02$   &   $0.08 \pm 0.01$   &   $0.05 \pm 0.01$ \\
2, \textsf{SR3}   & $\mathbf{19} , 17 \pm 2.4$  &   $0.12 \pm 0.02$  &   $0.10 \pm 0.01$   &   $0.05 \pm 0.01$   &   $0.11 \pm 0.02$   &   $0.09 \pm 0.01$   &   $0.05 \pm 0.01$ \\
3, \textsf{SR4}   & $\mathbf{18} , 11 \pm 2$  &   $0.11 \pm 0.02$  &   $0.08 \pm 0.01$   &   $0.05 \pm 0.01$   &   $0.11 \pm 0.02$   &   $0.08 \pm 0.01$   &   $0.05 \pm 0.01$ \\
4, \textsf{SR5}   & $\mathbf{1} , 1.6 \pm 0.7$  &   $0.02 \pm 0.01$  &   $0.01 \pm 0.00$   &   $0.00 \pm 0.00$   &   $0.01 \pm 0.01$   &   $0.01 \pm 0.00$   &   $0.02 \pm 0.01$ \\
5, \textsf{SR6}   & $\mathbf{0} , 3.5 \pm 0.9$  &   $0.04 \pm 0.01$  &   $0.05 \pm 0.01$   &   $0.03 \pm 0.01$   &   $0.04 \pm 0.01$   &   $0.07 \pm 0.01$   &   $0.06 \pm 0.02$ \\
6, \textsf{SR7}   & $\mathbf{3} , 2 \pm 0.7$  &   $0.05 \pm 0.01$  &   $0.04 \pm 0.01$   &   $0.03 \pm 0.01$   &   $0.04 \pm 0.01$   &   $0.03 \pm 0.01$   &   $0.04 \pm 0.01$ \\
7, \textsf{SR8}   & $\mathbf{1} , 0.51 \pm 0.52$  &   $0.05 \pm 0.01$  &   $0.02 \pm 0.01$   &   $0.02 \pm 0.01$   &   $0.02 \pm 0.01$   &   $0.03 \pm 0.01$   &   $0.02 \pm 0.01$ \\
8, \textsf{SR9}   & $\mathbf{2} , 1.4 \pm 0.7$  &   $0.02 \pm 0.01$  &   $0.01 \pm 0.00$   &   $0.01 \pm 0.01$   &   $0.02 \pm 0.01$   &   $0.01 \pm 0.01$   &   $0.00 \pm 0.00$ \\
9, \textsf{SR10}   & $\mathbf{1} , 1.5 \pm 0.6$  &   $0.05 \pm 0.01$  &   $0.05 \pm 0.01$   &   $0.04 \pm 0.01$   &   $0.05 \pm 0.01$   &   $0.05 \pm 0.01$   &   $0.05 \pm 0.01$ \\
10, \textsf{SR11}   & $\mathbf{2} , 1.5 \pm 0.8$  &   $0.04 \pm 0.01$  &   $0.04 \pm 0.01$   &   $0.01 \pm 0.01$   &   $0.03 \pm 0.01$   &   $0.03 \pm 0.01$   &   $0.01 \pm 0.01$ \\
11, \textsf{SR12}   & $\mathbf{0} , 1.2 \pm 0.6$  &   $0.04 \pm 0.01$  &   $0.01 \pm 0.00$   &   $0.01 \pm 0.00$   &   $0.04 \pm 0.01$   &   $0.02 \pm 0.01$   &   $0.02 \pm 0.01$ \\
\hline
\textsf{CMS\_2OSlep}~\cite{Sirunyan:2017qaj}\\
0, \textsf{SR1}   & $\mathbf{57} , 54.9 \pm 7$  &   $5.00 \pm 0.12$  &   $5.31 \pm 0.08$   &   $5.09 \pm 0.11$   &   $4.80 \pm 0.13$   &   $4.15 \pm 0.09$   &   $3.69 \pm 0.12$ \\
1, \textsf{SR2}   & $\mathbf{29} , 21.6 \pm 5.6$  &   $4.08 \pm 0.11$  &   $4.33 \pm 0.08$   &   $4.37 \pm 0.10$   &   $4.06 \pm 0.12$   &   $3.57 \pm 0.09$   &   $3.37 \pm 0.12$ \\
2, \textsf{SR3}   & $\mathbf{2} , 6 \pm 1.9$  &   $0.79 \pm 0.05$  &   $0.96 \pm 0.04$   &   $0.74 \pm 0.04$   &   $0.83 \pm 0.05$   &   $0.83 \pm 0.04$   &   $0.75 \pm 0.06$ \\
3, \textsf{SR4}   & $\mathbf{0} , 2.5 \pm 0.9$  &   $0.29 \pm 0.03$  &   $0.25 \pm 0.02$   &   $0.26 \pm 0.02$   &   $0.36 \pm 0.04$   &   $0.22 \pm 0.02$   &   $0.32 \pm 0.04$ \\
4, \textsf{SR5}   & $\mathbf{9} , 7.6 \pm 2.8$  &   $0.00 \pm 0.00$  &   $0.00 \pm 0.00$   &   $0.00 \pm 0.00$   &   $0.00 \pm 0.00$   &   $0.00 \pm 0.00$   &   $0.00 \pm 0.00$ \\
5, \textsf{SR6}   & $\mathbf{5} , 5.6 \pm 1.6$  &   $0.00 \pm 0.00$  &   $0.00 \pm 0.00$   &   $0.00 \pm 0.00$   &   $0.00 \pm 0.00$   &   $0.00 \pm 0.00$   &   $0.00 \pm 0.00$ \\
6, \textsf{SR7}   & $\mathbf{1} , 1.3 \pm 0.4$  &   $0.00 \pm 0.00$  &   $0.00 \pm 0.00$   &   $0.00 \pm 0.00$   &   $0.01 \pm 0.01$   &   $0.00 \pm 0.00$   &   $0.00 \pm 0.00$ \\
\hline
\textsf{CMS\_MultiLep\_2SSlep}~\cite{CMS-PAS-SUS-16-039}\\
0, \textsf{SR1}   & $\mathbf{13} , 12 \pm 3$  &   $1.85 \pm 0.08$  &   $2.33 \pm 0.06$   &   $1.40 \pm 0.06$   &   $1.80 \pm 0.08$   &   $1.47 \pm 0.06$   &   $0.66 \pm 0.05$ \\
1, \textsf{SR2}   & $\mathbf{18} , 18 \pm 4$  &   $2.18 \pm 0.08$  &   $1.82 \pm 0.05$   &   $0.96 \pm 0.05$   &   $2.01 \pm 0.09$   &   $1.64 \pm 0.06$   &   $0.71 \pm 0.05$ \\
\hline
\textsf{CMS\_MultiLep\_3lep}~\cite{CMS-PAS-SUS-16-039}\\
0, \textsf{SR3}   & $\mathbf{19} , 19 \pm 4$  &   $6.20 \pm 0.14$  &   $6.62 \pm 0.09$   &   $5.13 \pm 0.11$   &   $6.23 \pm 0.15$   &   $5.71 \pm 0.11$   &   $5.39 \pm 0.15$ \\
1, \textsf{SR4}   & $\mathbf{128} , 142 \pm 34$  &   $11.91 \pm 0.19$  &   $8.67 \pm 0.11$   &   $6.08 \pm 0.12$   &   $12.21 \pm 0.21$   &   $9.60 \pm 0.14$   &   $7.53 \pm 0.18$ \\
2, \textsf{SR5}   & $\mathbf{18} , 22 \pm 5$  &   $2.42 \pm 0.09$  &   $2.05 \pm 0.05$   &   $1.41 \pm 0.06$   &   $2.33 \pm 0.09$   &   $1.85 \pm 0.06$   &   $1.69 \pm 0.08$ \\
3, \textsf{SR6}   & $\mathbf{2} , 1 \pm 0.6$  &   $0.61 \pm 0.04$  &   $0.61 \pm 0.03$   &   $0.38 \pm 0.03$   &   $0.60 \pm 0.05$   &   $0.55 \pm 0.03$   &   $0.48 \pm 0.04$ \\
4, \textsf{SR7}   & $\mathbf{82} , 109 \pm 28$  &   $10.99 \pm 0.19$  &   $9.33 \pm 0.11$   &   $4.38 \pm 0.10$   &   $11.02 \pm 0.20$   &   $8.14 \pm 0.13$   &   $5.60 \pm 0.15$ \\
5, \textsf{SR8}   & $\mathbf{166} , 197 \pm 42$  &   $2.05 \pm 0.08$  &   $2.31 \pm 0.06$   &   $1.21 \pm 0.05$   &   $2.12 \pm 0.09$   &   $1.68 \pm 0.06$   &   $2.31 \pm 0.10$ \\
\hline
\end{tabular}
\caption{\label{tab:SR_predictions_13TeV_part2} Predicted signal counts for the 13\,TeV analyses \textsf{ATLAS\_MultiLep\_2lep\_jet}, \textsf{ATLAS\_MultiLep\_3lep}, \textsf{ATLAS\_RJ\_2lep\_2jet}, \textsf{ATLAS\_RJ\_3lep}, \textsf{CMS\_1lep\_2b}, \textsf{CMS\_2lep\_soft}, \textsf{CMS\_2OSlep}, \textsf{CMS\_MultiLep\_2SSlep} and \textsf{CMS\_MultiLep\_3lep}, for all six benchmark points. For comparison, the second column lists the observed event count and background prediction for each signal region, taken from the corresponding ATLAS or CMS analysis. The signal region indices and labels correspond to those in the public dataset provided with this paper \cite{the_gambit_collaboration_2018_1410335}.}
\end{center}
\centering
\end{table*}

% 8 TeV analyses, part 1
\begin{table*}[t]
\begin{center}
\setlength{\tabcolsep}{2.7pt} %% default is 6pt
\scriptsize
\begin{tabular}{l c X{1.8cm} X{1.8cm} X{1.8cm} X{1.8cm} X{1.8cm} X{1.8cm}}
\hline
\multirowcell{3}{} & \multirowcell{3}{} & \multirowcell{3}{\#1\\Best fit} & \multirowcell{3}{\#2\\Heavy winos} & \multirowcell{3}{\#3\\Highest mass} & \multirowcell{3}{\#4\\DM} & \multirowcell{3}{\#5\\Best fit\\incl.\ 8\,TeV} & \multirowcell{3}{\#6\\Heavy winos\\incl.\ 8\,TeV} \\
     & &              &                 &                  &            &   &  \\
     & &              &                 &                  &            &   &  \\
\hline
SR index, label & {\bf obs}, bkg  &   signal   &   signal   &   signal   &   signal   &   signal   &   signal  \\
\hline
\textsf{ATLAS\_8TeV\_1lep\_2b}~\cite{Aad:2015jqa}\\
0, \textsf{SRA}   & $\mathbf{4}, 5.69 \pm 1.10$  &   $0.01 \pm 0.00$  &   $0.04 \pm 0.00$   &   $0.01 \pm 0.00$   &   $0.02 \pm 0.00$   &   $0.01 \pm 0.00$   &   $0.01 \pm 0.00$ \\
1, \textsf{SRB}   & $\mathbf{3}, 2.67 \pm 0.69$  &   $0.03 \pm 0.01$  &   $0.03 \pm 0.00$   &   $0.03 \pm 0.00$   &   $0.03 \pm 0.01$   &   $0.03 \pm 0.00$   &   $0.02 \pm 0.00$ \\
\hline
\textsf{ATLAS\_8TeV\_2lep}~\cite{ATLAS:2LEPEW_20invfb}\\
0, \textsf{MT2\_90\_SF}   & $\mathbf{33} , 38.2 \pm 5.1$  &   $2.05 \pm 0.04$  &   $1.38 \pm 0.02$   &   $0.93 \pm 0.01$   &   $2.03 \pm 0.04$   &   $1.51 \pm 0.03$   &   $0.60 \pm 0.01$ \\
1, \textsf{MT2\_90\_DF}   & $\mathbf{21} , 23.3 \pm 3.7$  &   $1.89 \pm 0.04$  &   $1.24 \pm 0.02$   &   $0.95 \pm 0.01$   &   $1.74 \pm 0.04$   &   $1.47 \pm 0.03$   &   $0.26 \pm 0.01$ \\
2, \textsf{MT2\_120\_SF}   & $\mathbf{5} , 8.9 \pm 2.1$  &   $0.35 \pm 0.02$  &   $0.31 \pm 0.01$   &   $0.17 \pm 0.00$   &   $0.37 \pm 0.02$   &   $0.23 \pm 0.01$   &   $0.18 \pm 0.01$ \\
3, \textsf{MT2\_120\_DF}   & $\mathbf{5} , 3.6 \pm 1.2$  &   $0.17 \pm 0.01$  &   $0.19 \pm 0.01$   &   $0.12 \pm 0.00$   &   $0.15 \pm 0.01$   &   $0.12 \pm 0.01$   &   $0.02 \pm 0.00$ \\
4, \textsf{MT2\_150\_SF}   & $\mathbf{3} , 3.2 \pm 0.7$  &   $0.16 \pm 0.01$  &   $0.13 \pm 0.01$   &   $0.06 \pm 0.00$   &   $0.15 \pm 0.01$   &   $0.09 \pm 0.01$   &   $0.08 \pm 0.00$ \\
5, \textsf{MT2\_150\_DF}   & $\mathbf{2} , 1.0 \pm 0.5$  &   $0.07 \pm 0.01$  &   $0.07 \pm 0.00$   &   $0.04 \pm 0.00$   &   $0.06 \pm 0.01$   &   $0.04 \pm 0.00$   &   $0.01 \pm 0.00$ \\
6, \textsf{WWa\_SF}   & $\mathbf{73} , 86.5 \pm 7.4$  &   $5.79 \pm 0.07$  &   $2.50 \pm 0.03$   &   $1.38 \pm 0.01$   &   $6.42 \pm 0.08$   &   $3.95 \pm 0.05$   &   $1.41 \pm 0.02$ \\
7, \textsf{WWa\_DF}   & $\mathbf{70} , 73.6 \pm 7.9$  &   $6.70 \pm 0.07$  &   $2.63 \pm 0.03$   &   $1.56 \pm 0.01$   &   $7.26 \pm 0.08$   &   $4.71 \pm 0.05$   &   $1.14 \pm 0.02$ \\
8, \textsf{WWb\_SF}   & $\mathbf{26} , 30.2 \pm 3.5$  &   $1.92 \pm 0.04$  &   $1.27 \pm 0.02$   &   $0.84 \pm 0.01$   &   $1.90 \pm 0.04$   &   $1.42 \pm 0.03$   &   $0.58 \pm 0.01$ \\
9, \textsf{WWb\_DF}   & $\mathbf{17} , 18.1 \pm 2.6$  &   $1.73 \pm 0.04$  &   $1.11 \pm 0.02$   &   $0.85 \pm 0.01$   &   $1.61 \pm 0.04$   &   $1.35 \pm 0.03$   &   $0.24 \pm 0.01$ \\
10, \textsf{WWc\_SF}   & $\mathbf{10} , 20.3 \pm 3.5$  &   $0.79 \pm 0.03$  &   $0.72 \pm 0.02$   &   $0.53 \pm 0.01$   &   $0.76 \pm 0.03$   &   $0.59 \pm 0.02$   &   $0.34 \pm 0.01$ \\
11, \textsf{WWc\_DF}   & $\mathbf{11} , 9.0 \pm 2.2$  &   $0.53 \pm 0.02$  &   $0.59 \pm 0.01$   &   $0.49 \pm 0.01$   &   $0.44 \pm 0.02$   &   $0.44 \pm 0.02$   &   $0.06 \pm 0.00$ \\
12, \textsf{Zjets}   & $\mathbf{1} , 1.4 \pm 0.6$  &   $0.40 \pm 0.02$  &   $0.39 \pm 0.01$   &   $0.44 \pm 0.01$   &   $0.33 \pm 0.02$   &   $0.28 \pm 0.01$   &   $0.21 \pm 0.01$ \\
\hline
\textsf{ATLAS\_8TeV\_3lep}~\cite{ATLAS:3LEPEW_20invfb}\\
0, \textsf{SR0tau\_a\_bin\_1}   & $\mathbf{36} , 23 \pm 4$  &   $0.35 \pm 0.02$  &   $0.33 \pm 0.01$   &   $0.09 \pm 0.00$   &   $0.36 \pm 0.02$   &   $0.25 \pm 0.01$   &   $0.08 \pm 0.00$ \\
1, \textsf{SR0tau\_a\_bin\_2}   & $\mathbf{5} , 4.2 \pm 1.5$  &   $0.25 \pm 0.01$  &   $0.21 \pm 0.01$   &   $0.07 \pm 0.00$   &   $0.23 \pm 0.02$   &   $0.17 \pm 0.01$   &   $0.09 \pm 0.00$ \\
2, \textsf{SR0tau\_a\_bin\_3}   & $\mathbf{9} , 10.6 \pm 1.8$  &   $0.24 \pm 0.01$  &   $0.21 \pm 0.01$   &   $0.06 \pm 0.00$   &   $0.25 \pm 0.02$   &   $0.14 \pm 0.01$   &   $0.06 \pm 0.00$ \\
3, \textsf{SR0tau\_a\_bin\_4}   & $\mathbf{9} , 8.5 \pm 1.7$  &   $0.64 \pm 0.02$  &   $0.64 \pm 0.01$   &   $0.21 \pm 0.01$   &   $0.61 \pm 0.02$   &   $0.50 \pm 0.02$   &   $0.20 \pm 0.01$ \\
4, \textsf{SR0tau\_a\_bin\_5}   & $\mathbf{11} , 12.9 \pm 2.4$  &   $0.49 \pm 0.02$  &   $0.31 \pm 0.01$   &   $0.09 \pm 0.00$   &   $0.55 \pm 0.02$   &   $0.31 \pm 0.01$   &   $0.12 \pm 0.01$ \\
5, \textsf{SR0tau\_a\_bin\_6}   & $\mathbf{13} , 6.6 \pm 1.9$  &   $0.77 \pm 0.03$  &   $0.54 \pm 0.01$   &   $0.19 \pm 0.00$   &   $0.75 \pm 0.03$   &   $0.52 \pm 0.02$   &   $0.23 \pm 0.01$ \\
6, \textsf{SR0tau\_a\_bin\_7}   & $\mathbf{15} , 14.1 \pm 2.2$  &   $1.20 \pm 0.03$  &   $0.96 \pm 0.02$   &   $0.31 \pm 0.01$   &   $1.24 \pm 0.03$   &   $0.81 \pm 0.02$   &   $0.31 \pm 0.01$ \\
7, \textsf{SR0tau\_a\_bin\_8}   & $\mathbf{1} , 1.1 \pm 0.4$  &   $0.32 \pm 0.02$  &   $0.31 \pm 0.01$   &   $0.12 \pm 0.00$   &   $0.31 \pm 0.02$   &   $0.26 \pm 0.01$   &   $0.13 \pm 0.01$ \\
8, \textsf{SR0tau\_a\_bin\_9}   & $\mathbf{28} , 22.4 \pm 3.6$  &   $2.48 \pm 0.05$  &   $1.03 \pm 0.02$   &   $0.35 \pm 0.01$   &   $2.69 \pm 0.05$   &   $1.55 \pm 0.03$   &   $0.64 \pm 0.01$ \\
9, \textsf{SR0tau\_a\_bin\_10}   & $\mathbf{24} , 16.4 \pm 2.8$  &   $2.28 \pm 0.04$  &   $1.08 \pm 0.02$   &   $0.37 \pm 0.01$   &   $2.38 \pm 0.05$   &   $1.58 \pm 0.03$   &   $0.57 \pm 0.01$ \\
10, \textsf{SR0tau\_a\_bin\_11}   & $\mathbf{29} , 27 \pm 5$  &   $5.03 \pm 0.06$  &   $2.56 \pm 0.03$   &   $1.03 \pm 0.01$   &   $5.53 \pm 0.07$   &   $3.47 \pm 0.04$   &   $1.53 \pm 0.02$ \\
11, \textsf{SR0tau\_a\_bin\_12}   & $\mathbf{8} , 5.5 \pm 1.5$  &   $1.40 \pm 0.03$  &   $1.34 \pm 0.02$   &   $0.69 \pm 0.01$   &   $1.30 \pm 0.04$   &   $1.13 \pm 0.02$   &   $0.60 \pm 0.01$ \\
12, \textsf{SR0tau\_a\_bin\_13}   & $\mathbf{714} , 715 \pm 70$  &   $58.23 \pm 0.22$  &   $23.69 \pm 0.09$   &   $9.70 \pm 0.03$   &   $67.19 \pm 0.26$   &   $39.06 \pm 0.15$   &   $16.92 \pm 0.07$ \\
13, \textsf{SR0tau\_a\_bin\_14}   & $\mathbf{214} , 219 \pm 33$  &   $32.45 \pm 0.16$  &   $15.60 \pm 0.07$   &   $8.05 \pm 0.03$   &   $36.15 \pm 0.19$   &   $22.21 \pm 0.11$   &   $11.70 \pm 0.06$ \\
14, \textsf{SR0tau\_a\_bin\_15}   & $\mathbf{63} , 65 \pm 13$  &   $10.22 \pm 0.09$  &   $9.42 \pm 0.06$   &   $6.29 \pm 0.03$   &   $9.15 \pm 0.09$   &   $8.44 \pm 0.07$   &   $4.86 \pm 0.04$ \\
15, \textsf{SR0tau\_a\_bin\_16}   & $\mathbf{3} , 4.6 \pm 1.7$  &   $2.40 \pm 0.04$  &   $2.82 \pm 0.03$   &   $2.46 \pm 0.02$   &   $2.39 \pm 0.05$   &   $2.21 \pm 0.03$   &   $1.78 \pm 0.02$ \\
16, \textsf{SR0tau\_a\_bin\_17}   & $\mathbf{60} , 69 \pm 9$  &   $2.37 \pm 0.04$  &   $1.99 \pm 0.03$   &   $0.67 \pm 0.01$   &   $2.25 \pm 0.05$   &   $1.62 \pm 0.03$   &   $0.82 \pm 0.01$ \\
17, \textsf{SR0tau\_a\_bin\_18}   & $\mathbf{1} , 3.4 \pm 1.4$  &   $0.28 \pm 0.02$  &   $0.23 \pm 0.01$   &   $0.08 \pm 0.00$   &   $0.28 \pm 0.02$   &   $0.21 \pm 0.01$   &   $0.14 \pm 0.01$ \\
18, \textsf{SR0tau\_a\_bin\_19}   & $\mathbf{0} , 1.2 \pm 0.4$  &   $0.08 \pm 0.01$  &   $0.07 \pm 0.00$   &   $0.03 \pm 0.00$   &   $0.08 \pm 0.01$   &   $0.07 \pm 0.01$   &   $0.05 \pm 0.00$ \\
19, \textsf{SR0tau\_a\_bin\_20}   & $\mathbf{0} , 0.29 \pm 0.18$  &   $0.07 \pm 0.01$  &   $0.05 \pm 0.00$   &   $0.03 \pm 0.00$   &   $0.08 \pm 0.01$   &   $0.05 \pm 0.01$   &   $0.05 \pm 0.00$ \\
20, \textsf{SR1tau}   & $\mathbf{13} , 10.3 \pm 1.2$  &   $1.15 \pm 0.03$  &   $0.92 \pm 0.02$   &   $0.31 \pm 0.01$   &   $1.14 \pm 0.03$   &   $0.75 \pm 0.02$   &   $0.39 \pm 0.01$ \\
21, \textsf{SR2tau\_a}   & $\mathbf{6} , 6.9 \pm 0.8$  &   $0.86 \pm 0.03$  &   $0.69 \pm 0.02$   &   $0.32 \pm 0.01$   &   $0.91 \pm 0.03$   &   $0.62 \pm 0.02$   &   $0.40 \pm 0.01$ \\
22, \textsf{SR2tau\_b}   & $\mathbf{5} , 7.2 \pm 0.8$  &   $0.00 \pm 0.00$  &   $0.00 \pm 0.00$   &   $0.00 \pm 0.00$   &   $0.00 \pm 0.00$   &   $0.00 \pm 0.00$   &   $0.00 \pm 0.00$ \\
\hline
\textsf{CMS\_8TeV\_4lep}~\cite{CMS:3LEPEW_20invfb}\\
0, \textsf{1OSSF0tau\_ETmiss<30}   & $\mathbf{1} , 2.3 \pm 0.6$  &   $0.03 \pm 0.00$  &   $0.05 \pm 0.00$   &   $0.01 \pm 0.00$   &   $0.03 \pm 0.01$   &   $0.02 \pm 0.00$   &   $0.01 \pm 0.00$ \\
1, \textsf{1OSSF0tau\_ETmiss30-50}   & $\mathbf{3} , 1.2 \pm 0.3$  &   $0.05 \pm 0.01$  &   $0.09 \pm 0.01$   &   $0.02 \pm 0.00$   &   $0.05 \pm 0.01$   &   $0.03 \pm 0.00$   &   $0.02 \pm 0.00$ \\
2, \textsf{1OSSF0tau\_ETmiss50-100}   & $\mathbf{2} , 1.5 \pm 0.4$  &   $0.14 \pm 0.01$  &   $0.21 \pm 0.01$   &   $0.06 \pm 0.00$   &   $0.12 \pm 0.01$   &   $0.10 \pm 0.01$   &   $0.05 \pm 0.00$ \\
3, \textsf{1OSSF0tau\_ETmiss>100}   & $\mathbf{2} , 0.8 \pm 0.3$  &   $0.12 \pm 0.01$  &   $0.13 \pm 0.01$   &   $0.05 \pm 0.00$   &   $0.11 \pm 0.01$   &   $0.09 \pm 0.01$   &   $0.08 \pm 0.00$ \\
4, \textsf{1OSSF1tau\_ETmiss<30}   & $\mathbf{33} , 25 \pm 12$  &   $0.12 \pm 0.01$  &   $0.22 \pm 0.01$   &   $0.04 \pm 0.00$   &   $0.11 \pm 0.01$   &   $0.07 \pm 0.01$   &   $0.10 \pm 0.00$ \\
5, \textsf{1OSSF1tau\_ETmiss30-50}   & $\mathbf{11} , 11 \pm 3.1$  &   $0.18 \pm 0.01$  &   $0.30 \pm 0.01$   &   $0.07 \pm 0.00$   &   $0.17 \pm 0.01$   &   $0.12 \pm 0.01$   &   $0.11 \pm 0.01$ \\
6, \textsf{1OSSF1tau\_ETmiss50-100}   & $\mathbf{9} , 9.3 \pm 1.9$  &   $0.44 \pm 0.02$  &   $0.66 \pm 0.01$   &   $0.20 \pm 0.00$   &   $0.45 \pm 0.02$   &   $0.29 \pm 0.01$   &   $0.24 \pm 0.01$ \\
7, \textsf{1OSSF1tau\_ETmiss>100}   & $\mathbf{2} , 2.9 \pm 0.6$  &   $0.38 \pm 0.02$  &   $0.47 \pm 0.01$   &   $0.20 \pm 0.00$   &   $0.37 \pm 0.02$   &   $0.26 \pm 0.01$   &   $0.28 \pm 0.01$ \\
8, \textsf{2OSSF0tau\_ETmiss<30}   & $\mathbf{142} , 149 \pm 46$  &   $0.56 \pm 0.02$  &   $0.80 \pm 0.02$   &   $0.20 \pm 0.00$   &   $0.52 \pm 0.02$   &   $0.30 \pm 0.01$   &   $0.78 \pm 0.01$ \\
9, \textsf{2OSSF0tau\_ETmiss30-50}   & $\mathbf{25} , 28 \pm 11$  &   $0.84 \pm 0.03$  &   $1.21 \pm 0.02$   &   $0.32 \pm 0.01$   &   $0.74 \pm 0.03$   &   $0.46 \pm 0.02$   &   $0.97 \pm 0.02$ \\
10, \textsf{2OSSF0tau\_ETmiss50-100}   & $\mathbf{4} , 4.5 \pm 2.7$  &   $1.82 \pm 0.04$  &   $3.01 \pm 0.03$   &   $0.90 \pm 0.01$   &   $1.63 \pm 0.04$   &   $1.20 \pm 0.03$   &   $1.17 \pm 0.02$ \\
11, \textsf{2OSSF0tau\_ETmiss>100}   & $\mathbf{1} , 0.8 \pm 0.3$  &   $1.54 \pm 0.04$  &   $2.01 \pm 0.03$   &   $0.93 \pm 0.01$   &   $1.50 \pm 0.04$   &   $1.15 \pm 0.02$   &   $1.13 \pm 0.02$ \\
\hline
\end{tabular}
\caption{\label{tab:SR_predictions_8TeV_part1} Predicted signal counts for the 8\,TeV analyses \textsf{ATLAS\_8TeV\_1lep\_2b}, \textsf{ATLAS\_8TeV\_2lep}, \textsf{ATLAS\_8TeV\_3lep} and \textsf{CMS\_8TeV\_4lep}, for all six benchmark points. For comparison, the second column lists the observed event count and background prediction for each signal region, taken from the corresponding ATLAS or CMS analysis. The signal region indices and labels correspond to those in the public dataset provided with this paper \cite{the_gambit_collaboration_2018_1410335}.}
\end{center}
\centering
\end{table*}

% 8 TeV analyses, part 2
\begin{table*}[t]
\begin{center}
\setlength{\tabcolsep}{3pt} %% default is 6pt
\tiny
\begin{tabular}{l c X{1.6cm} X{1.6cm} X{1.6cm} X{1.6cm} X{1.6cm} X{1.6cm}}
\hline
\multirowcell{3}{} & \multirowcell{3}{} & \multirowcell{3}{\#1\\Best fit} & \multirowcell{3}{\#2\\Heavy winos} & \multirowcell{3}{\#3\\Highest mass} & \multirowcell{3}{\#4\\DM} & \multirowcell{3}{\#5\\Best fit\\incl.\ 8\,TeV} & \multirowcell{3}{\#6\\Heavy winos\\incl.\ 8\,TeV} \\
     & &              &                 &                  &            &   &  \\
     & &              &                 &                  &            &   &  \\
\hline
SR index, label & {\bf obs}, bkg     &  signal   &   signal   &   signal   &   signal   &   signal   &   signal  \\
\hline
\multicolumn{8}{l}{\textsf{CMS\_8TeV\_3lep}~\cite{CMS:3LEPEW_20invfb}}\\
% \textsf{CMS\_8TeV\_3lep}~\cite{CMS:3LEPEW_20invfb}\\
0, \textsf{OSSF\_mT<120\_ETmiss50-100\_mll<75}   & $\mathbf{138}, 132 \pm 19 $   &   $9.68 \pm 0.09$  &   $5.31 \pm 0.04$   &   $1.75 \pm 0.01$   &   $10.09 \pm 0.10$   &   $6.33 \pm 0.06$   &   $2.42 \pm 0.02$\\
1, \textsf{OSSF\_mT<120\_ETmiss50-100\_mll75-105}   & $\mathbf{821}, 776 \pm 125 $   &   $105.95 \pm 0.29$  &   $44.55 \pm 0.12$   &   $18.62 \pm 0.05$   &   $120.49 \pm 0.34$   &   $71.43 \pm 0.19$   &   $30.50 \pm 0.09$ \\
2, \textsf{OSSF\_mT<120\_ETmiss50-100\_mll>105}   & $\mathbf{49}, 45 \pm 7 $   &   $1.43 \pm 0.03$  &   $1.11 \pm 0.02$   &   $0.32 \pm 0.01$   &   $1.42 \pm 0.04$   &   $0.94 \pm 0.02$   &   $0.35 \pm 0.01$ \\
3, \textsf{OSSF\_mT<120\_ETmiss100-150\_mll<75}   & $\mathbf{16}, 20 \pm 4 $   &   $2.68 \pm 0.05$  &   $1.74 \pm 0.02$   &   $0.67 \pm 0.01$   &   $2.84 \pm 0.05$   &   $1.90 \pm 0.03$   &   $0.83 \pm 0.01$ \\
4, \textsf{OSSF\_mT<120\_ETmiss100-150\_mll75-105}   & $\mathbf{123}, 131 \pm 30 $   &   $27.65 \pm 0.15$  &   $13.92 \pm 0.07$   &   $7.15 \pm 0.03$   &   $30.58 \pm 0.17$   &   $19.21 \pm 0.10$   &   $9.64 \pm 0.05$ \\
5, \textsf{OSSF\_mT<120\_ETmiss100-150\_mll>105}   & $\mathbf{10}, 10.0 \pm 1.9 $   &   $0.57 \pm 0.02$  &   $0.47 \pm 0.01$   &   $0.15 \pm 0.00$   &   $0.59 \pm 0.02$   &   $0.41 \pm 0.01$   &   $0.20 \pm 0.01$ \\
6, \textsf{OSSF\_mT<120\_ETmiss150-200\_mll<75}   & $\mathbf{5}, 4.0 \pm 0.8 $   &   $0.83 \pm 0.03$  &   $0.52 \pm 0.01$   &   $0.22 \pm 0.01$   &   $0.84 \pm 0.03$   &   $0.56 \pm 0.02$   &   $0.29 \pm 0.01$ \\
7, \textsf{OSSF\_mT<120\_ETmiss150-200\_mll75-105}   & $\mathbf{34}, 34 \pm 8 $   &   $7.98 \pm 0.08$  &   $4.26 \pm 0.04$   &   $2.35 \pm 0.02$   &   $8.71 \pm 0.09$   &   $5.53 \pm 0.05$   &   $3.24 \pm 0.03$ \\
8, \textsf{OSSF\_mT<120\_ETmiss150-200\_mll>105}   & $\mathbf{4}, 2.5 \pm 0.5 $   &   $0.18 \pm 0.01$  &   $0.15 \pm 0.01$   &   $0.06 \pm 0.00$   &   $0.19 \pm 0.01$   &   $0.13 \pm 0.01$   &   $0.09 \pm 0.00$ \\
9, \textsf{OSSF\_mT<120\_ETmiss200-250\_mll<75}   & $\mathbf{2}, 1.9 \pm 0.4 $   &   $0.26 \pm 0.01$  &   $0.17 \pm 0.01$   &   $0.08 \pm 0.00$   &   $0.26 \pm 0.02$   &   $0.20 \pm 0.01$   &   $0.11 \pm 0.01$ \\
10, \textsf{OSSF\_mT<120\_ETmiss200-250\_mll75-105}   & $\mathbf{14}, 21 \pm 7 $   &   $2.57 \pm 0.05$  &   $1.45 \pm 0.02$   &   $0.84 \pm 0.01$   &   $2.84 \pm 0.05$   &   $1.90 \pm 0.03$   &   $1.16 \pm 0.02$ \\
11, \textsf{OSSF\_mT<120\_ETmiss200-250\_mll>105}   & $\mathbf{4}, 1.2 \pm 0.3 $   &   $0.09 \pm 0.01$  &   $0.06 \pm 0.00$   &   $0.02 \pm 0.00$   &   $0.07 \pm 0.01$   &   $0.05 \pm 0.01$   &   $0.04 \pm 0.00$ \\
12, \textsf{OSSF\_mT120-160\_ETmiss50-100\_mll<75}   & $\mathbf{8}, 9.6 \pm 1.7 $   &   $1.00 \pm 0.03$  &   $0.77 \pm 0.02$   &   $0.31 \pm 0.01$   &   $0.89 \pm 0.03$   &   $0.69 \pm 0.02$   &   $0.27 \pm 0.01$ \\
13, \textsf{OSSF\_mT120-160\_ETmiss50-100\_mll75-105}   & $\mathbf{29}, 23 \pm 5 $   &   $5.43 \pm 0.07$  &   $4.99 \pm 0.04$   &   $3.24 \pm 0.02$   &   $4.74 \pm 0.07$   &   $4.56 \pm 0.05$   &   $2.42 \pm 0.02$ \\
14, \textsf{OSSF\_mT120-160\_ETmiss50-100\_mll>105}   & $\mathbf{4}, 2.7 \pm 0.5 $   &   $0.22 \pm 0.01$  &   $0.19 \pm 0.01$   &   $0.05 \pm 0.00$   &   $0.23 \pm 0.01$   &   $0.14 \pm 0.01$   &   $0.06 \pm 0.00$ \\
15, \textsf{OSSF\_mT120-160\_ETmiss100-150\_mll<75}   & $\mathbf{2}, 3.3 \pm 0.8 $   &   $0.44 \pm 0.02$  &   $0.47 \pm 0.01$   &   $0.20 \pm 0.00$   &   $0.38 \pm 0.02$   &   $0.33 \pm 0.01$   &   $0.14 \pm 0.01$ \\
16, \textsf{OSSF\_mT120-160\_ETmiss100-150\_mll75-105}   & $\mathbf{4}, 3.4 \pm 0.7 $   &   $1.90 \pm 0.04$  &   $2.33 \pm 0.03$   &   $1.89 \pm 0.02$   &   $1.63 \pm 0.04$   &   $1.67 \pm 0.03$   &   $1.04 \pm 0.02$ \\
17, \textsf{OSSF\_mT120-160\_ETmiss100-150\_mll>105}   & $\mathbf{2}, 0.71 \pm 0.22 $   &   $0.13 \pm 0.01$  &   $0.13 \pm 0.01$   &   $0.05 \pm 0.00$   &   $0.14 \pm 0.01$   &   $0.11 \pm 0.01$   &   $0.05 \pm 0.00$ \\
18, \textsf{OSSF\_mT120-160\_ETmiss150-200\_mll<75}   & $\mathbf{0}, 0.26 \pm 0.10 $   &   $0.15 \pm 0.01$  &   $0.14 \pm 0.01$   &   $0.07 \pm 0.00$   &   $0.13 \pm 0.01$   &   $0.11 \pm 0.01$   &   $0.06 \pm 0.00$ \\
19, \textsf{OSSF\_mT120-160\_ETmiss150-200\_mll75-105}   & $\mathbf{1}, 0.72 \pm 0.19 $   &   $0.59 \pm 0.02$  &   $0.75 \pm 0.02$   &   $0.64 \pm 0.01$   &   $0.59 \pm 0.02$   &   $0.61 \pm 0.02$   &   $0.43 \pm 0.01$ \\
20, \textsf{OSSF\_mT120-160\_ETmiss150-200\_mll>105}   & $\mathbf{0}, 0.38 \pm 0.14 $   &   $0.06 \pm 0.01$  &   $0.04 \pm 0.00$   &   $0.02 \pm 0.00$   &   $0.06 \pm 0.01$   &   $0.04 \pm 0.00$   &   $0.02 \pm 0.00$ \\
21, \textsf{OSSF\_mT120-160\_ETmiss200-250\_mll<75}   & $\mathbf{0}, 0.29 \pm 0.11 $   &   $0.04 \pm 0.01$  &   $0.05 \pm 0.00$   &   $0.02 \pm 0.00$   &   $0.04 \pm 0.01$   &   $0.04 \pm 0.00$   &   $0.03 \pm 0.00$ \\
22, \textsf{OSSF\_mT120-160\_ETmiss200-250\_mll75-105}   & $\mathbf{1}, 0.36 \pm 0.12 $   &   $0.20 \pm 0.01$  &   $0.24 \pm 0.01$   &   $0.22 \pm 0.01$   &   $0.19 \pm 0.01$   &   $0.21 \pm 0.01$   &   $0.18 \pm 0.01$ \\
23, \textsf{OSSF\_mT120-160\_ETmiss200-250\_mll>105}   & $\mathbf{0}, 0.24 \pm 0.20 $   &   $0.01 \pm 0.00$  &   $0.01 \pm 0.00$   &   $0.01 \pm 0.00$   &   $0.02 \pm 0.00$   &   $0.01 \pm 0.00$   &   $0.01 \pm 0.00$ \\
24, \textsf{OSSF\_mT>160\_ETmiss50-100\_mll<75}   & $\mathbf{12}, 5.8 \pm 1.1 $   &   $0.48 \pm 0.02$  &   $0.40 \pm 0.01$   &   $0.13 \pm 0.00$   &   $0.43 \pm 0.02$   &   $0.31 \pm 0.01$   &   $0.13 \pm 0.01$ \\
25, \textsf{OSSF\_mT>160\_ETmiss50-100\_mll75-105}   & $\mathbf{13}, 7.5 \pm 1.4 $   &   $1.67 \pm 0.04$  &   $1.82 \pm 0.02$   &   $1.20 \pm 0.01$   &   $1.56 \pm 0.04$   &   $1.37 \pm 0.03$   &   $1.00 \pm 0.02$ \\
26, \textsf{OSSF\_mT>160\_ETmiss50-100\_mll>105}   & $\mathbf{1}, 2.6 \pm 1.2 $   &   $0.13 \pm 0.01$  &   $0.10 \pm 0.01$   &   $0.03 \pm 0.00$   &   $0.16 \pm 0.01$   &   $0.10 \pm 0.01$   &   $0.05 \pm 0.00$ \\
27, \textsf{OSSF\_mT>160\_ETmiss100-150\_mll<75}   & $\mathbf{3}, 4.5 \pm 1.1 $   &   $0.55 \pm 0.02$  &   $0.55 \pm 0.01$   &   $0.22 \pm 0.01$   &   $0.52 \pm 0.02$   &   $0.38 \pm 0.01$   &   $0.17 \pm 0.01$ \\
28, \textsf{OSSF\_mT>160\_ETmiss100-150\_mll75-105}   & $\mathbf{8}, 4.0 \pm 1.0 $   &   $1.45 \pm 0.03$  &   $1.89 \pm 0.02$   &   $1.44 \pm 0.01$   &   $1.32 \pm 0.04$   &   $1.21 \pm 0.03$   &   $0.94 \pm 0.02$ \\
29, \textsf{OSSF\_mT>160\_ETmiss100-150\_mll>105}   & $\mathbf{3}, 1.8 \pm 0.9 $   &   $0.20 \pm 0.01$  &   $0.15 \pm 0.01$   &   $0.06 \pm 0.00$   &   $0.19 \pm 0.01$   &   $0.14 \pm 0.01$   &   $0.08 \pm 0.00$ \\
30, \textsf{OSSF\_mT>160\_ETmiss150-200\_mll<75}   & $\mathbf{2}, 1.5 \pm 0.4 $   &   $0.28 \pm 0.01$  &   $0.27 \pm 0.01$   &   $0.12 \pm 0.00$   &   $0.27 \pm 0.02$   &   $0.23 \pm 0.01$   &   $0.12 \pm 0.01$ \\
31, \textsf{OSSF\_mT>160\_ETmiss150-200\_mll75-105}   & $\mathbf{3}, 1.5 \pm 0.5 $   &   $0.81 \pm 0.03$  &   $0.97 \pm 0.02$   &   $0.82 \pm 0.01$   &   $0.76 \pm 0.03$   &   $0.61 \pm 0.02$   &   $0.51 \pm 0.01$ \\
32, \textsf{OSSF\_mT>160\_ETmiss150-200\_mll>105}   & $\mathbf{0}, 0.7 \pm 0.4 $   &   $0.11 \pm 0.01$  &   $0.08 \pm 0.01$   &   $0.03 \pm 0.00$   &   $0.09 \pm 0.01$   &   $0.08 \pm 0.01$   &   $0.05 \pm 0.00$ \\
33, \textsf{OSSF\_mT>160\_ETmiss200-250\_mll<75}   & $\mathbf{0}, 0.81 \pm 0.21 $   &   $0.12 \pm 0.01$  &   $0.11 \pm 0.01$   &   $0.05 \pm 0.00$   &   $0.11 \pm 0.01$   &   $0.10 \pm 0.01$   &   $0.06 \pm 0.00$ \\
34, \textsf{OSSF\_mT>160\_ETmiss200-250\_mll75-105}   & $\mathbf{2}, 1.1 \pm 0.4 $   &   $0.34 \pm 0.02$  &   $0.37 \pm 0.01$   &   $0.33 \pm 0.01$   &   $0.33 \pm 0.02$   &   $0.29 \pm 0.01$   &   $0.23 \pm 0.01$ \\
35, \textsf{OSSF\_mT>160\_ETmiss200-250\_mll>105}   & $\mathbf{0}, 0.40 \pm 0.24 $   &   $0.04 \pm 0.01$  &   $0.03 \pm 0.00$   &   $0.01 \pm 0.00$   &   $0.04 \pm 0.01$   &   $0.04 \pm 0.00$   &   $0.02 \pm 0.00$ \\
36, \textsf{noOSSF\_mT<120\_ETmiss50-100\_mll<100}   & $\mathbf{29},  32 \pm 7 $   &   $1.47 \pm 0.03$  &   $1.35 \pm 0.02$   &   $0.38 \pm 0.01$   &   $1.40 \pm 0.04$   &   $0.93 \pm 0.02$   &   $0.28 \pm 0.01$ \\
37, \textsf{noOSSF\_mT<120\_ETmiss50-100\_mll>100}   & $\mathbf{1}, 1.7 \pm 0.4 $   &   $0.14 \pm 0.01$  &   $0.08 \pm 0.01$   &   $0.03 \pm 0.00$   &   $0.12 \pm 0.01$   &   $0.09 \pm 0.01$   &   $0.04 \pm 0.00$ \\
38, \textsf{noOSSF\_mT<120\_ETmiss100-150\_mll<100}   & $\mathbf{5}, 7.3 \pm 1.7 $   &   $0.51 \pm 0.02$  &   $0.47 \pm 0.01$   &   $0.14 \pm 0.00$   &   $0.46 \pm 0.02$   &   $0.33 \pm 0.01$   &   $0.13 \pm 0.01$ \\
39, \textsf{noOSSF\_mT<120\_ETmiss100-150\_mll>100}   & $\mathbf{0}, 0.30 \pm 0.11 $   &   $0.05 \pm 0.01$  &   $0.03 \pm 0.00$   &   $0.01 \pm 0.00$   &   $0.06 \pm 0.01$   &   $0.04 \pm 0.00$   &   $0.02 \pm 0.00$ \\
40, \textsf{noOSSF\_mT<120\_ETmiss150-200\_mll<100}   & $\mathbf{1}, 1.0 \pm 0.3 $   &   $0.14 \pm 0.01$  &   $0.14 \pm 0.01$   &   $0.04 \pm 0.00$   &   $0.15 \pm 0.01$   &   $0.10 \pm 0.01$   &   $0.06 \pm 0.00$ \\
41, \textsf{noOSSF\_mT<120\_ETmiss150-200\_mll>100}   & $\mathbf{0}, 0.14 \pm 0.09 $   &   $0.02 \pm 0.00$  &   $0.01 \pm 0.00$   &   $0.00 \pm 0.00$   &   $0.02 \pm 0.00$   &   $0.01 \pm 0.00$   &   $0.01 \pm 0.00$ \\
42, \textsf{noOSSF\_mT<120\_ETmiss200-250\_mll<100}   & $\mathbf{0}, 0.53 \pm 0.24 $   &   $0.06 \pm 0.01$  &   $0.05 \pm 0.00$   &   $0.01 \pm 0.00$   &   $0.05 \pm 0.01$   &   $0.04 \pm 0.00$   &   $0.02 \pm 0.00$ \\
43, \textsf{noOSSF\_mT<120\_ETmiss200-250\_mll>100}   & $\mathbf{0}, 0.03 \pm 0.03 $   &   $0.01 \pm 0.00$  &   $0.00 \pm 0.00$   &   $0.00 \pm 0.00$   &   $0.01 \pm 0.00$   &   $0.00 \pm 0.00$   &   $0.00 \pm 0.00$ \\
44, \textsf{noOSSF\_mT120-160\_ETmiss50-100\_mll<100}   & $\mathbf{3}, 5.5 \pm 1.2 $   &   $0.34 \pm 0.02$  &   $0.32 \pm 0.01$   &   $0.10 \pm 0.00$   &   $0.30 \pm 0.02$   &   $0.22 \pm 0.01$   &   $0.07 \pm 0.00$ \\
45, \textsf{noOSSF\_mT120-160\_ETmiss50-100\_mll>100}   & $\mathbf{1}, 0.25 \pm 0.07 $   &   $0.04 \pm 0.01$  &   $0.03 \pm 0.00$   &   $0.01 \pm 0.00$   &   $0.06 \pm 0.01$   &   $0.03 \pm 0.00$   &   $0.01 \pm 0.00$ \\
46, \textsf{noOSSF\_mT120-160\_ETmiss100-150\_mll<100}   & $\mathbf{1}, 1.9 \pm 0.5 $   &   $0.18 \pm 0.01$  &   $0.18 \pm 0.01$   &   $0.06 \pm 0.00$   &   $0.15 \pm 0.01$   &   $0.12 \pm 0.01$   &   $0.04 \pm 0.00$ \\
47, \textsf{noOSSF\_mT120-160\_ETmiss100-150\_mll>100}   & $\mathbf{0}, 0.19 \pm 0.10 $   &   $0.03 \pm 0.00$  &   $0.01 \pm 0.00$   &   $0.00 \pm 0.00$   &   $0.02 \pm 0.00$   &   $0.01 \pm 0.00$   &   $0.01 \pm 0.00$ \\
48, \textsf{noOSSF\_mT120-160\_ETmiss150-200\_mll<100}   & $\mathbf{1}, 0.46 \pm 0.18 $   &   $0.06 \pm 0.01$  &   $0.05 \pm 0.00$   &   $0.02 \pm 0.00$   &   $0.05 \pm 0.01$   &   $0.04 \pm 0.00$   &   $0.02 \pm 0.00$ \\
49, \textsf{noOSSF\_mT120-160\_ETmiss150-200\_mll>100}   & $\mathbf{0}, 0.03 \pm 0.03 $   &   $0.00 \pm 0.00$  &   $0.00 \pm 0.00$   &   $0.00 \pm 0.00$   &   $0.00 \pm 0.00$   &   $0.00 \pm 0.00$   &   $0.00 \pm 0.00$ \\
50, \textsf{noOSSF\_mT120-160\_ETmiss200-250\_mll<100}   & $\mathbf{0}, 0.10 \pm 0.05 $   &   $0.02 \pm 0.00$  &   $0.01 \pm 0.00$   &   $0.01 \pm 0.00$   &   $0.01 \pm 0.00$   &   $0.01 \pm 0.00$   &   $0.01 \pm 0.00$ \\
51, \textsf{noOSSF\_mT120-160\_ETmiss200-250\_mll>100}   & $\mathbf{0}, 0.008 \pm 0.010 $   &   $0.00 \pm 0.00$  &   $0.00 \pm 0.00$   &   $0.00 \pm 0.00$   &   $0.00 \pm 0.00$   &   $0.00 \pm 0.00$   &   $0.00 \pm 0.00$ \\
52, \textsf{noOSSF\_mT>160\_ETmiss50-100\_mll<100}   & $\mathbf{2}, 3.2 \pm 0.8 $   &   $0.24 \pm 0.01$  &   $0.19 \pm 0.01$   &   $0.05 \pm 0.00$   &   $0.19 \pm 0.01$   &   $0.16 \pm 0.01$   &   $0.05 \pm 0.00$ \\
53, \textsf{noOSSF\_mT>160\_ETmiss50-100\_mll>100}   & $\mathbf{0}, 0.44 \pm 0.33 $   &   $0.05 \pm 0.01$  &   $0.03 \pm 0.00$   &   $0.01 \pm 0.00$   &   $0.05 \pm 0.01$   &   $0.03 \pm 0.00$   &   $0.02 \pm 0.00$ \\
54, \textsf{noOSSF\_mT>160\_ETmiss100-150\_mll<100}   & $\mathbf{3}, 2.1 \pm 0.7 $   &   $0.33 \pm 0.02$  &   $0.28 \pm 0.01$   &   $0.10 \pm 0.00$   &   $0.34 \pm 0.02$   &   $0.21 \pm 0.01$   &   $0.08 \pm 0.00$ \\
55, \textsf{noOSSF\_mT>160\_ETmiss100-150\_mll>100}   & $\mathbf{0}, 0.42 \pm 0.19 $   &   $0.04 \pm 0.01$  &   $0.03 \pm 0.00$   &   $0.01 \pm 0.00$   &   $0.05 \pm 0.01$   &   $0.04 \pm 0.00$   &   $0.02 \pm 0.00$ \\
56, \textsf{noOSSF\_mT>160\_ETmiss150-200\_mll<100}   & $\mathbf{0}, 0.59 \pm 0.18 $   &   $0.15 \pm 0.01$  &   $0.15 \pm 0.01$   &   $0.05 \pm 0.00$   &   $0.14 \pm 0.01$   &   $0.12 \pm 0.01$   &   $0.05 \pm 0.00$ \\
57, \textsf{noOSSF\_mT>160\_ETmiss150-200\_mll>100}   & $\mathbf{0}, 0.10 \pm 0.06 $   &   $0.02 \pm 0.00$  &   $0.01 \pm 0.00$   &   $0.00 \pm 0.00$   &   $0.03 \pm 0.01$   &   $0.02 \pm 0.00$   &   $0.02 \pm 0.00$ \\
58, \textsf{noOSSF\_mT>160\_ETmiss200-250\_mll<100}   & $\mathbf{1}, 0.37 \pm 0.13 $   &   $0.08 \pm 0.01$  &   $0.05 \pm 0.00$   &   $0.02 \pm 0.00$   &   $0.07 \pm 0.01$   &   $0.05 \pm 0.00$   &   $0.03 \pm 0.00$ \\
59, \textsf{noOSSF\_mT>160\_ETmiss200-250\_mll>100}   & $\mathbf{0}, 0.16 \pm 0.14 $   &   $0.01 \pm 0.00$  &   $0.00 \pm 0.00$   &   $0.00 \pm 0.00$   &   $0.01 \pm 0.00$   &   $0.01 \pm 0.00$   &   $0.01 \pm 0.00$ \\
\hline
\end{tabular}
\caption{\label{tab:SR_predictions_8TeV_part2} Predicted signal counts for the 0-tau signal regions of the 8\,TeV analysis \textsf{CMS\_8TeV\_3lep}, for all six benchmark points. For comparison, the second column lists the observed event count and background prediction for each signal region, taken from the corresponding ATLAS or CMS analysis. The signal region indices and labels correspond to those in the public dataset provided with this paper \cite{the_gambit_collaboration_2018_1410335}.}
\end{center}
\centering
\end{table*}

% 8 TeV analyses, part 3
\begin{table*}[t]
\begin{center}
\setlength{\tabcolsep}{3pt} %% default is 6pt
\tiny
\begin{tabular}{l c X{1.6cm} X{1.6cm} X{1.6cm} X{1.6cm} X{1.6cm} X{1.6cm}}
\hline
\multirowcell{3}{} & \multirowcell{3}{} & \multirowcell{3}{\#1\\Best fit} & \multirowcell{3}{\#2\\Heavy winos} & \multirowcell{3}{\#3\\Highest mass} & \multirowcell{3}{\#4\\DM} & \multirowcell{3}{\#5\\Best fit\\incl.\ 8\,TeV} & \multirowcell{3}{\#6\\Heavy winos\\incl.\ 8\,TeV} \\
     & &              &                 &                  &            &   &  \\
     & &              &                 &                  &            &   &  \\
\hline
SR index, label & {\bf obs}, bkg     &  signal   &   signal   &   signal   &   signal   &   signal   &   signal  \\
\hline
\multicolumn{8}{l}{\textsf{CMS\_8TeV\_3lep}~\cite{CMS:3LEPEW_20invfb} \textit{(continued from Table \ref{tab:SR_predictions_8TeV_part2})} }\\
% \textsf{CMS\_8TeV\_3lep}~\cite{CMS:3LEPEW_20invfb}\\
60, \textsf{SS1tau\_mT<120\_ETmiss50-100\_mll<100}   & $\mathbf{46}, 51 \pm 8 $   &   $2.23 \pm 0.04$  &   $1.45 \pm 0.02$   &   $0.49 \pm 0.01$   &   $2.32 \pm 0.05$   &   $1.44 \pm 0.03$   &   $0.63 \pm 0.01$ \\
61, \textsf{SS1tau\_mT<120\_ETmiss50-100\_mll>100}   & $\mathbf{3}, 2.8 \pm 0.6 $   &   $0.07 \pm 0.01$  &   $0.07 \pm 0.00$   &   $0.01 \pm 0.00$   &   $0.09 \pm 0.01$   &   $0.06 \pm 0.01$   &   $0.03 \pm 0.00$ \\
62, \textsf{SS1tau\_mT<120\_ETmiss100-150\_mll<100}   & $\mathbf{1}, 6.0 \pm 1.3 $   &   $0.55 \pm 0.02$  &   $0.41 \pm 0.01$   &   $0.16 \pm 0.00$   &   $0.54 \pm 0.02$   &   $0.37 \pm 0.01$   &   $0.18 \pm 0.01$ \\
63, \textsf{SS1tau\_mT<120\_ETmiss100-150\_mll>100}   & $\mathbf{0}, 0.50 \pm 0.14 $   &   $0.02 \pm 0.00$  &   $0.03 \pm 0.00$   &   $0.01 \pm 0.00$   &   $0.04 \pm 0.01$   &   $0.03 \pm 0.00$   &   $0.02 \pm 0.00$ \\
64, \textsf{SS1tau\_mT<120\_ETmiss150-200\_mll<100}   & $\mathbf{0}, 2.0 \pm 0.4 $   &   $0.16 \pm 0.01$  &   $0.12 \pm 0.01$   &   $0.05 \pm 0.00$   &   $0.16 \pm 0.01$   &   $0.13 \pm 0.01$   &   $0.07 \pm 0.00$ \\
65, \textsf{SS1tau\_mT<120\_ETmiss150-200\_mll>100}   & $\mathbf{0}, 0.11 \pm 0.07 $   &   $0.01 \pm 0.00$  &   $0.01 \pm 0.00$   &   $0.00 \pm 0.00$   &   $0.01 \pm 0.00$   &   $0.01 \pm 0.00$   &   $0.01 \pm 0.00$ \\
66, \textsf{SS1tau\_mT<120\_ETmiss200-250\_mll<100}   & $\mathbf{0}, 0.90 \pm 0.24 $   &   $0.06 \pm 0.01$  &   $0.04 \pm 0.00$   &   $0.02 \pm 0.00$   &   $0.06 \pm 0.01$   &   $0.04 \pm 0.00$   &   $0.04 \pm 0.00$ \\
67, \textsf{SS1tau\_mT<120\_ETmiss200-250\_mll>100}   & $\mathbf{0}, 0.042 \pm 0.021 $   &   $0.01 \pm 0.00$  &   $0.00 \pm 0.00$   &   $0.00 \pm 0.00$   &   $0.01 \pm 0.00$   &   $0.00 \pm 0.00$   &   $0.00 \pm 0.00$ \\
68, \textsf{SS1tau\_mT120-160\_ETmiss50-100\_mll<100}   & $\mathbf{6}, 5.5 \pm 1.0 $   &   $0.42 \pm 0.02$  &   $0.31 \pm 0.01$   &   $0.11 \pm 0.00$   &   $0.45 \pm 0.02$   &   $0.28 \pm 0.01$   &   $0.12 \pm 0.01$ \\
69, \textsf{SS1tau\_mT120-160\_ETmiss50-100\_mll>100}   & $\mathbf{1}, 0.35 \pm 0.13 $   &   $0.03 \pm 0.00$  &   $0.02 \pm 0.00$   &   $0.00 \pm 0.00$   &   $0.03 \pm 0.00$   &   $0.02 \pm 0.00$   &   $0.01 \pm 0.00$ \\
70, \textsf{SS1tau\_mT120-160\_ETmiss100-150\_mll<100}   & $\mathbf{2}, 0.91 \pm 0.26 $   &   $0.13 \pm 0.01$  &   $0.14 \pm 0.01$   &   $0.05 \pm 0.00$   &   $0.12 \pm 0.01$   &   $0.09 \pm 0.01$   &   $0.04 \pm 0.00$ \\
71, \textsf{SS1tau\_mT120-160\_ETmiss100-150\_mll>100}   & $\mathbf{0}, 0.06 \pm 0.05 $   &   $0.01 \pm 0.00$  &   $0.01 \pm 0.00$   &   $0.00 \pm 0.00$   &   $0.01 \pm 0.00$   &   $0.01 \pm 0.00$   &   $0.01 \pm 0.00$ \\
72, \textsf{SS1tau\_mT120-160\_ETmiss150-200\_mll<100}   & $\mathbf{0}, 0.15 \pm 0.10 $   &   $0.04 \pm 0.01$  &   $0.03 \pm 0.00$   &   $0.02 \pm 0.00$   &   $0.03 \pm 0.01$   &   $0.03 \pm 0.00$   &   $0.02 \pm 0.00$ \\
73, \textsf{SS1tau\_mT120-160\_ETmiss150-200\_mll>100}   & $\mathbf{0}, 0 \pm 0.008 $   &   $0.00 \pm 0.00$  &   $0.00 \pm 0.00$   &   $0.00 \pm 0.00$   &   $0.00 \pm 0.00$   &   $0.00 \pm 0.00$   &   $0.00 \pm 0.00$ \\
74, \textsf{SS1tau\_mT120-160\_ETmiss200-250\_mll<100}   & $\mathbf{0}, 0.06 \pm 0.08 $   &   $0.01 \pm 0.00$  &   $0.01 \pm 0.00$   &   $0.01 \pm 0.00$   &   $0.02 \pm 0.00$   &   $0.01 \pm 0.00$   &   $0.01 \pm 0.00$ \\
75, \textsf{SS1tau\_mT120-160\_ETmiss200-250\_mll>100}   & $\mathbf{0}, 0.011 \pm 0.012 $   &   $0.00 \pm 0.00$  &   $0.00 \pm 0.00$   &   $0.00 \pm 0.00$   &   $0.00 \pm 0.00$   &   $0.00 \pm 0.00$   &   $0.00 \pm 0.00$ \\
76, \textsf{SS1tau\_mT>160\_ETmiss50-100\_mll<100}   & $\mathbf{2}, 3.1 \pm 0.6 $   &   $0.30 \pm 0.02$  &   $0.19 \pm 0.01$   &   $0.06 \pm 0.00$   &   $0.31 \pm 0.02$   &   $0.19 \pm 0.01$   &   $0.09 \pm 0.00$ \\
77, \textsf{SS1tau\_mT>160\_ETmiss50-100\_mll>100}   & $\mathbf{1}, 0.50 \pm 0.21 $   &   $0.02 \pm 0.00$  &   $0.02 \pm 0.00$   &   $0.00 \pm 0.00$   &   $0.02 \pm 0.00$   &   $0.01 \pm 0.00$   &   $0.01 \pm 0.00$ \\
78, \textsf{SS1tau\_mT>160\_ETmiss100-150\_mll<100}   & $\mathbf{1}, 2.3 \pm 0.5 $   &   $0.25 \pm 0.01$  &   $0.21 \pm 0.01$   &   $0.10 \pm 0.00$   &   $0.24 \pm 0.01$   &   $0.16 \pm 0.01$   &   $0.08 \pm 0.00$ \\
79, \textsf{SS1tau\_mT>160\_ETmiss100-150\_mll>100}   & $\mathbf{1}, 0.40 \pm 0.17 $   &   $0.02 \pm 0.00$  &   $0.02 \pm 0.00$   &   $0.00 \pm 0.00$   &   $0.03 \pm 0.01$   &   $0.02 \pm 0.00$   &   $0.02 \pm 0.00$ \\
80, \textsf{SS1tau\_mT>160\_ETmiss150-200\_mll<100}   & $\mathbf{0}, 0.52 \pm 0.16 $   &   $0.11 \pm 0.01$  &   $0.10 \pm 0.01$   &   $0.04 \pm 0.00$   &   $0.10 \pm 0.01$   &   $0.09 \pm 0.01$   &   $0.06 \pm 0.00$ \\
81, \textsf{SS1tau\_mT>160\_ETmiss150-200\_mll>100}   & $\mathbf{0}, 0.21 \pm 0.11 $   &   $0.01 \pm 0.00$  &   $0.01 \pm 0.00$   &   $0.00 \pm 0.00$   &   $0.01 \pm 0.00$   &   $0.01 \pm 0.00$   &   $0.01 \pm 0.00$ \\
82, \textsf{SS1tau\_mT>160\_ETmiss200-250\_mll<100}   & $\mathbf{2}, 0.41 \pm 0.12 $   &   $0.05 \pm 0.01$  &   $0.04 \pm 0.00$   &   $0.02 \pm 0.00$   &   $0.03 \pm 0.01$   &   $0.03 \pm 0.00$   &   $0.02 \pm 0.00$ \\
83, \textsf{SS1tau\_mT>160\_ETmiss200-250\_mll>100}   & $\mathbf{0}, 0.06 \pm 0.05 $   &   $0.00 \pm 0.00$  &   $0.00 \pm 0.00$   &   $0.00 \pm 0.00$   &   $0.00 \pm 0.00$   &   $0.00 \pm 0.00$   &   $0.01 \pm 0.00$ \\
84, \textsf{OS1tau\_mT<120\_ETmiss50-100\_mll<100}   & $\mathbf{290}, 259 \pm 93 $   &   $2.14 \pm 0.04$  &   $1.54 \pm 0.02$   &   $0.48 \pm 0.01$   &   $2.13 \pm 0.04$   &   $1.42 \pm 0.03$   &   $0.52 \pm 0.01$ \\
85, \textsf{OS1tau\_mT<120\_ETmiss50-100\_mll>100}   & $\mathbf{27}, 30 \pm 13 $   &   $0.24 \pm 0.01$  &   $0.16 \pm 0.01$   &   $0.04 \pm 0.00$   &   $0.24 \pm 0.02$   &   $0.17 \pm 0.01$   &   $0.07 \pm 0.00$ \\
86, \textsf{OS1tau\_mT<120\_ETmiss100-150\_mll<100}   & $\mathbf{62}, 60 \pm 25 $   &   $0.51 \pm 0.02$  &   $0.45 \pm 0.01$   &   $0.15 \pm 0.00$   &   $0.51 \pm 0.02$   &   $0.34 \pm 0.01$   &   $0.17 \pm 0.01$ \\
87, \textsf{OS1tau\_mT<120\_ETmiss100-150\_mll>100}   & $\mathbf{8}, 5.9 \pm 2.6 $   &   $0.10 \pm 0.01$  &   $0.06 \pm 0.00$   &   $0.02 \pm 0.00$   &   $0.10 \pm 0.01$   &   $0.06 \pm 0.01$   &   $0.03 \pm 0.00$ \\
88, \textsf{OS1tau\_mT<120\_ETmiss150-200\_mll<100}   & $\mathbf{10}, 11 \pm 5 $   &   $0.13 \pm 0.01$  &   $0.12 \pm 0.01$   &   $0.04 \pm 0.00$   &   $0.17 \pm 0.01$   &   $0.11 \pm 0.01$   &   $0.07 \pm 0.00$ \\
89, \textsf{OS1tau\_mT<120\_ETmiss150-200\_mll>100}   & $\mathbf{0}, 2.3 \pm 1.4 $   &   $0.02 \pm 0.00$  &   $0.02 \pm 0.00$   &   $0.01 \pm 0.00$   &   $0.03 \pm 0.01$   &   $0.03 \pm 0.00$   &   $0.01 \pm 0.00$ \\
90, \textsf{OS1tau\_mT<120\_ETmiss200-250\_mll<100}   & $\mathbf{2}, 2.9 \pm 1.4 $   &   $0.06 \pm 0.01$  &   $0.05 \pm 0.00$   &   $0.02 \pm 0.00$   &   $0.06 \pm 0.01$   &   $0.04 \pm 0.00$   &   $0.03 \pm 0.00$ \\
91, \textsf{OS1tau\_mT<120\_ETmiss200-250\_mll>100}   & $\mathbf{0}, 1.1 \pm 0.6 $   &   $0.01 \pm 0.00$  &   $0.01 \pm 0.00$   &   $0.00 \pm 0.00$   &   $0.01 \pm 0.00$   &   $0.01 \pm 0.00$   &   $0.00 \pm 0.00$ \\
92, \textsf{OS1tau\_mT120-160\_ETmiss50-100\_mll<100}   & $\mathbf{41}, 42 \pm 16 $   &   $0.41 \pm 0.02$  &   $0.34 \pm 0.01$   &   $0.11 \pm 0.00$   &   $0.39 \pm 0.02$   &   $0.28 \pm 0.01$   &   $0.11 \pm 0.01$ \\
93, \textsf{OS1tau\_mT120-160\_ETmiss50-100\_mll>100}   & $\mathbf{7}, 8.3 \pm 2.9 $   &   $0.06 \pm 0.01$  &   $0.05 \pm 0.00$   &   $0.01 \pm 0.00$   &   $0.07 \pm 0.01$   &   $0.04 \pm 0.00$   &   $0.02 \pm 0.00$ \\
94, \textsf{OS1tau\_mT120-160\_ETmiss100-150\_mll<100}   & $\mathbf{18}, 17 \pm 9 $   &   $0.15 \pm 0.01$  &   $0.15 \pm 0.01$   &   $0.06 \pm 0.00$   &   $0.14 \pm 0.01$   &   $0.12 \pm 0.01$   &   $0.05 \pm 0.00$ \\
95, \textsf{OS1tau\_mT120-160\_ETmiss100-150\_mll>100}   & $\mathbf{4}, 2.3 \pm 1.3 $   &   $0.02 \pm 0.00$  &   $0.01 \pm 0.00$   &   $0.01 \pm 0.00$   &   $0.02 \pm 0.00$   &   $0.01 \pm 0.00$   &   $0.01 \pm 0.00$ \\
96, \textsf{OS1tau\_mT120-160\_ETmiss150-200\_mll<100}   & $\mathbf{2}, 2.0 \pm 1.2 $   &   $0.05 \pm 0.01$  &   $0.04 \pm 0.00$   &   $0.02 \pm 0.00$   &   $0.05 \pm 0.01$   &   $0.04 \pm 0.00$   &   $0.02 \pm 0.00$ \\
97, \textsf{OS1tau\_mT120-160\_ETmiss150-200\_mll>100}   & $\mathbf{0}, 0.27 \pm 0.32 $   &   $0.01 \pm 0.00$  &   $0.00 \pm 0.00$   &   $0.00 \pm 0.00$   &   $0.01 \pm 0.00$   &   $0.00 \pm 0.00$   &   $0.00 \pm 0.00$ \\
98, \textsf{OS1tau\_mT120-160\_ETmiss200-250\_mll<100}   & $\mathbf{1}, 0.8 \pm 0.5 $   &   $0.01 \pm 0.00$  &   $0.01 \pm 0.00$   &   $0.01 \pm 0.00$   &   $0.02 \pm 0.00$   &   $0.01 \pm 0.00$   &   $0.01 \pm 0.00$ \\
99, \textsf{OS1tau\_mT120-160\_ETmiss200-250\_mll>100}   & $\mathbf{0}, 0.5 \pm 0.4 $   &   $0.00 \pm 0.00$  &   $0.00 \pm 0.00$   &   $0.00 \pm 0.00$   &   $0.00 \pm 0.00$   &   $0.00 \pm 0.00$   &   $0.00 \pm 0.00$ \\
100, \textsf{OS1tau\_mT>160\_ETmiss50-100\_mll<100}   & $\mathbf{19}, 15 \pm 8 $   &   $0.24 \pm 0.01$  &   $0.17 \pm 0.01$   &   $0.06 \pm 0.00$   &   $0.23 \pm 0.01$   &   $0.16 \pm 0.01$   &   $0.07 \pm 0.00$ \\
101, \textsf{OS1tau\_mT>160\_ETmiss50-100\_mll>100}   & $\mathbf{2}, 5.7 \pm 2.3 $   &   $0.05 \pm 0.01$  &   $0.04 \pm 0.00$   &   $0.01 \pm 0.00$   &   $0.06 \pm 0.01$   &   $0.04 \pm 0.00$   &   $0.03 \pm 0.00$ \\
102, \textsf{OS1tau\_mT>160\_ETmiss100-150\_mll<100}   & $\mathbf{14}, 14 \pm 9 $   &   $0.26 \pm 0.01$  &   $0.24 \pm 0.01$   &   $0.09 \pm 0.00$   &   $0.28 \pm 0.02$   &   $0.20 \pm 0.01$   &   $0.08 \pm 0.00$ \\
103, \textsf{OS1tau\_mT>160\_ETmiss100-150\_mll>100}   & $\mathbf{3}, 4.0 \pm 2.2 $   &   $0.07 \pm 0.01$  &   $0.05 \pm 0.00$   &   $0.01 \pm 0.00$   &   $0.05 \pm 0.01$   &   $0.04 \pm 0.00$   &   $0.03 \pm 0.00$ \\
104, \textsf{OS1tau\_mT>160\_ETmiss150-200\_mll<100}   & $\mathbf{1}, 3.7 \pm 2.1 $   &   $0.13 \pm 0.01$  &   $0.11 \pm 0.01$   &   $0.04 \pm 0.00$   &   $0.11 \pm 0.01$   &   $0.09 \pm 0.01$   &   $0.06 \pm 0.00$ \\
105, \textsf{OS1tau\_mT>160\_ETmiss150-200\_mll>100}   & $\mathbf{3}, 1.3 \pm 1.0 $   &   $0.02 \pm 0.00$  &   $0.02 \pm 0.00$   &   $0.01 \pm 0.00$   &   $0.03 \pm 0.01$   &   $0.03 \pm 0.00$   &   $0.02 \pm 0.00$ \\
106, \textsf{OS1tau\_mT>160\_ETmiss200-250\_mll<100}   & $\mathbf{2}, 1.5 \pm 1.0 $   &   $0.04 \pm 0.01$  &   $0.04 \pm 0.00$   &   $0.02 \pm 0.00$   &   $0.04 \pm 0.01$   &   $0.04 \pm 0.00$   &   $0.03 \pm 0.00$ \\
107, \textsf{OS1tau\_mT>160\_ETmiss200-250\_mll>100}   & $\mathbf{1}, 0.7 \pm 0.4 $   &   $0.02 \pm 0.00$  &   $0.01 \pm 0.00$   &   $0.00 \pm 0.00$   &   $0.01 \pm 0.00$   &   $0.01 \pm 0.00$   &   $0.01 \pm 0.00$ \\
\hline
\end{tabular}
\caption{\label{tab:SR_predictions_8TeV_part3} Predicted signal counts for the 1-tau signal regions of the 8\,TeV analysis \textsf{CMS\_8TeV\_3lep}, for all six benchmark points. For comparison, the second column lists the observed event count and background prediction for each signal region, taken from the corresponding ATLAS or CMS analysis. The signal region indices and labels correspond to those in the public dataset provided with this paper \cite{the_gambit_collaboration_2018_1410335}.}
\end{center}
\centering
\end{table*}

\bibliography{R1.5}

\end{document}